%% file: PPNP_EHM_F.tex
\documentclass[12pt, aps, floatfix, nofootinbib,  superscriptaddress]{revtex4-1}
\usepackage{latexsym}
\usepackage{amsmath}
\usepackage{amssymb}
\usepackage{amsfonts}
\usepackage{bm}

\usepackage[vcentermath]{youngtab}
\usepackage{mathtools}
\usepackage[squaren]{SIunits}

\usepackage[margin=2cm]{geometry}
\usepackage{makecell}

\usepackage{color}

\usepackage{longtable}
\usepackage{multirow}
\usepackage{supertabular}
\usepackage{graphicx}

\usepackage[mathscr,scaled=1.15]{urwchancal}
\DeclareFontFamily{OT1}{pzc}{}
\DeclareFontShape{OT1}{pzc}{m}{it}%
{<-> s * [1.15] pzcmi7t}{}
\DeclareMathAlphabet{\mathpzc}{OT1}{pzc}{m}{it}

\definecolor{purple}{rgb}{0.5,0,0.5}
\definecolor{blue}{rgb}{0.0,0,0.9}
\definecolor{prdblue}{rgb}{0.133,0.118,0.498}
\usepackage[colorlinks=true, pdfstartview=FitV, linkcolor=prdblue, citecolor= prdblue, urlcolor=prdblue]{hyperref}
\newlength{\Wfockuud}

\begin{document}

% Use the \preprint command to place your local institutional report
% number in the upper righthand corner of the title page in preprint mode.
% Multiple \preprint commands are allowed.
% Use the 'preprintnumbers' class option to override journal defaults
% to display numbers if necessary
%\preprint{}

% Title of paper
\title{$\,$\\[-7ex]\hspace*{\fill}{\normalsize{\sf\emph{Preprint no}. NJU-INP 034/21}}\\[1ex]
Insights into the Emergence of Mass \\
from Studies of Pion and Kaon Structure}
%

%%%%%%%%%%%%%%%%%%%%%%%%%%%%%%%%%%%%%%%%%%%%%%%%%%%%%%%%%%%%%%%%%%%%%%%%%%%%%%%%
% Authors

% Authors

\author{Craig D. Roberts}
\email{cdroberts@nju.edu.cn}
\affiliation{School of Physics, Nanjing University, Nanjing, Jiangsu 210093, China}
\affiliation{Institute for Nonperturbative Physics, Nanjing University, Nanjing, Jiangsu 210093, China}

\author{David G. Richards}
\email{dgr@jlab.org}
\affiliation{Thomas Jefferson National Accelerator Facility, Newport News, Virginia 23606, USA}

\author{Tanja Horn}
\email{hornt@cua.edu}
\affiliation{Catholic University of America, Washington, D.C. 20064, USA}
\affiliation{Thomas Jefferson National Accelerator Facility, Newport News, Virginia 23606, USA}

\author{Lei Chang}
\email[]{leichang@nankai.edu.cn}
\affiliation{School of Physics, Nankai University, Tianjin 300071, China}

%%%%%%%%%%%%%%%%%%%%%%%%%%%%%%%%%%%%%%%%%%%%%%%%%%%%%%%%%%%%%%%%%%%%%%%%%%%%%%%%
% Date
\date{2021 April 08}
%\date{2021 March 22}
%\date{2021 February 02}
%\date{2021 January 14}
%\date{2020 November 22}
%\date{2020 August 31}

% Abstract
\begin{abstract}
\leftline{\bf Abstract:}
\vspace*{-1.6em}

{\footnotesize There are two mass generating mechanisms in the standard model of particle physics (SM).  One is related to the Higgs boson and fairly well understood.  The other is embedded in quantum chromodynamics (QCD), the SM's strong interaction piece; and although responsible for emergence of the roughly 1 GeV mass scale that characterises the proton and hence all observable matter, the source and impacts of this emergent hadronic mass (EHM) remain puzzling.  As bound states seeded by a valence-quark and -antiquark, pseudoscalar mesons present a simpler problem in quantum field theory than that associated with the nucleon.  Consequently, there is a large array of robust predictions for pion and kaon properties whose empirical validation will provide a clear window onto many effects of both mass generating mechanisms and the constructive interference between them.  This has now become significant because new-era experimental facilities, in operation, construction, or planning, are capable of conducting such tests and thereby contributing greatly to resolving the puzzles of EHM.
These aspects of experiment, phenomenology, and theory, along with contemporary successes and challenges, are reviewed herein.  In addition to providing an overview of the experimental status, we focus on recent progress made using continuum Schwinger function methods and lattice-regularized QCD.  Advances made using other theoretical tools are also sketched.
Our primary goal is to highlight the potential gains that can accrue from a coherent effort aimed at finally reaching an understanding of the character and structure of Nature's Nambu-Goldstone modes.}

\bigskip

\leftline{\bf Keywords:}
\vspace*{-1.65em}

$\;\;${\footnotesize continuum Schwinger function methods;
electromagnetic form factors -- elastic and transition;
emergence of mass;
Higgs boson;
lattice regularised QCD;
Nambu-Goldstone modes -- pions and kaons;
nonperturbative quantum field theory;
parton distributions;
strong interactions in the standard model of particle physics}

\end{abstract}

\maketitle

\section*{Contents}
\unskip
\begin{itemize}
\item Sec.\,\ref{sec:Introduction} -- Emergence of Mass \dotfill\ \pageref{sec:Introduction}
\item Sec.\,\ref{sec:CTP} -- Masses, Coupling, and the Emergence of Nambu-Goldstone Modes \dotfill\ \pageref{sec:CTP}\\[-5ex]
\begin{itemize}
\item Sec.\,\ref{GluonMass} -- Gluon Mass Scale \dotfill\ \pageref{GluonMass}
\item Sec.\,\ref{SecPICharge} -- Process-Independent Effective Charge \dotfill\ \pageref{SecPICharge}
\item Sec.\,\ref{SecDCSB} -- Dynamical Chiral Symmetry Breaking \dotfill \pageref{SecDCSB}
\item Sec.\,\ref{SecNGBosons} -- Nambu-Goldstone Bosons \dotfill \pageref{SecNGBosons}
\end{itemize}
\item Sec.\,\ref{sec:PiKWF} -- Pion and Kaon Distribution Amplitudes \dotfill\ \pageref{sec:PiKWF}\\[-5ex]
\begin{itemize}
\item Sec.\,\ref{sec:ELFWFs} -- Essentials of Light-Front Wave Functions \dotfill\ \pageref{sec:ELFWFs}
\item Sec.\,\ref{SecDApion} -- Pion Distribution Amplitude \dotfill\ \pageref{SecDApion}
\item Sec.\,\ref{SecDAkaon} -- Kaon Distribution Amplitude \dotfill\ \pageref{SecDAkaon}
\end{itemize}
\item Sec.\,\ref{sec:EADAs} -- Empirical Access to Pseudoscalar Meson DAs \dotfill\ \pageref{sec:EADAs}\\[-5ex]
\begin{itemize}
\item Sec.\,\ref{sec:ETFFs} -- Electromagnetic Transition Form Factors \dotfill\ \pageref{sec:ETFFs}
\item Sec.\,\ref{sec:EEFFs} -- Elastic Electromagnetic Form Factors \dotfill\ \pageref{sec:EEFFs}
\item Sec.\,\ref{SecDiffractive} -- Diffractive Dissociation \dotfill \pageref{SecDiffractive}
\end{itemize}
\item Sec.\,\ref{sec:PiKDFs} -- Pion Distribution Functions \dotfill\ \pageref{sec:PiKDFs}\\[-5ex]
\begin{itemize}
\item Sec.\,\ref{SecFCSA} -- Forward Compton Scattering Amplitude \dotfill\ \pageref{SecFCSA}
\item Sec.\,\ref{SecHSPDF} -- Hadron Scale Pion Distribution Function \dotfill\ \pageref{SecHSPDF}
\item Sec.\,\ref{SecpiDF2} -- Pion Distribution Function at \mbox{$\zeta_2=\,$}2\,GeV \dotfill\ \pageref{SecpiDF2}
\item Sec.\,\ref{SecpiDF2} -- Pion Distribution Function at \mbox{$\zeta_5=\,$}5.2\,GeV \dotfill\ \pageref{SecpiDF5}
\end{itemize}
\item Sec.\,\ref{sec:kaonDFs} -- Kaon Distribution Functions \dotfill\ \pageref{sec:kaonDFs}
\item Sec.\,\ref{3DNG} -- Three Dimensional Structure of Nambu-Goldstone Modes \dotfill \pageref{3DNG} \\[-5ex]
\begin{itemize}
\item Sec.\,\ref{SecGTMD} -- Generalised Transverse-Momentum Dependent Distribution Functions \dotfill\ \pageref{SecGTMD}
\item Sec.\,\ref{SecTTGPD} -- Twist-Two Generalised Parton Distribution Functions \dotfill\ \pageref{SecTTGPD}
\item Sec.\,\ref{SecMFFs} -- Meson Fragmentation Functions \dotfill\ \pageref{SecMFFs}
\end{itemize}
\item Sec.\,\ref{SeclQCD} -- Developments in Lattice QCD \dotfill\ \pageref{SeclQCD}\\[-5ex]
\begin{itemize}
\item Sec.\,\ref{SecFormalism} -- Formulation \dotfill\ \pageref{SecFormalism}
\item Sec.\,\ref{lQCDpionFF} -- Pion Form Factor \dotfill\ \pageref{lQCDpionFF}
\item Sec.\,\ref{SeclQCDDF} -- Parton Distribution Functions \dotfill\ \pageref{SeclQCDDF}
\item Sec.\,\ref{SeclQCDDA} -- Quark Distribution Amplitude \dotfill\ \pageref{SeclQCDDA}
\item Sec.\,\ref{Sec3DlQCD} -- Three-Dimensional Imaging of Mesons \dotfill\ \pageref{Sec3DlQCD}
\item Sec.\,\ref{SecGluelQCD} -- Gluon and Flavour-Singlet Computations \dotfill\ \pageref{SecGluelQCD}
\item Sec.\,\ref{SecExa} -- Promise of the Exascale Era \dotfill\ \pageref{SecExa}
\end{itemize}
\item Sec.\,\ref{sec:Experiments} -- Experiments Completed or In Train \dotfill\ \pageref{sec:Experiments}\\[-5ex]
\begin{itemize}
\item Sec.\,\ref{sec:sullivan} -- Sullivan Process \dotfill\ \pageref{sec:sullivan}
\item Sec.\,\ref{sec:meson-form-factors-exp} -- Pion and Kaon Form Factors \dotfill\ \pageref{sec:meson-form-factors-exp}
\item Sec.\,\ref{sec:meson-distribution-functions-exp} -- Empirical Information on Parton Distribution Functions \dotfill\ \pageref{sec:meson-distribution-functions-exp}
\item Sec.\,\ref{SecGPDExp} -- GPDs: Experimental Meson Data \dotfill\ \pageref{SecGPDExp}
\end{itemize}
\item Sec.\,\ref{Sec:Prospects} -- Measurements on the Horizon \dotfill\ \pageref{Sec:Prospects}
\item Sec.\,\ref{Sec:Epilogue} -- Epilogue \dotfill\ \pageref{Sec:Epilogue}
\item Abbreviations \dotfill\ \pageref{abbreviations}
\item References \dotfill\ \pageref{references}
\end{itemize}

\pagenumbering{arabic}
\setcounter{page}{1}

\input{S1_Introduction}
\input{S2_Continuum}
\input{S3_Continuum}

\input{S4_Continuum}
\input{S5_Continuum}

\input{S6_Continuum}
\input{S7_Continuum}
\input{S8_Lattice}  %%% <<<--- problem file
\input{S910_Experiment}
\input{S11_Epilogue}

%%%%%%%%%%%%%%%%%%%%%%%%%%%%%%%%%%%%%%%%%%%%%%%%%%%%%%%%%%%%%%%%%%%%%%%%%%%%%%%%
% If you have acknowledgments, this puts in the proper section head.

\vspace*{0.10cm}
%\begin{acknowledgments}
\section*{Acknowledgments}
We are grateful for assistance and insightful comments from
D.~Binosi,
S.\,J.~Brodsky,
K.-L.~Cai,
Z.-F.~Cui,
O.~Denisov,
R.~Ent,
T.~Frederico,
J.~Friedrich,
J.~Karpie,
N.~Karthik,
C.~Mezrag,
W.-D.~Nowak,
S.~Platchkov,
J.~Qiu,
C.~Quintans,
J.~Rodr{\'{\i}}guez-Quintero,
G.~Salm\`e,
J.~Segovia,
R.~Sufian,
S.-S.~Xu,
J.-L.~Zhang.
%
%Work supported by:
% C.D. Roberts
CDR is suported in part by the Jiangsu Province \emph{Hundred Talents Plan for Professionals};
% D.G. Richards
TH and DGR by the U.S. Department of Energy, Office of Science, Office of Nuclear Physics, under contract DE-AC05-06OR23177;
% T. Horn
TH by the US National Science Foundation under grants PHY-1714133 and PHY-2012430;
% Lei Chang
LC by the Chinese Government's \emph{Thousand Talents Plan for Young Professionals}.
%\end{acknowledgments}

%%%%%%%%%%%%%%%%%%%%%%%%%%%%%%%%%%%%%%%%%%%%%%%%%%%%%%%%%%%%%%%%%%%%%%%%%%%%%%%%

\section*{Abbreviations}
\unskip
\label{abbreviations}
The following abbreviations are used in this manuscript:\\[-1ex]

\begin{longtable}{ll}
2PI & two particle irreducible \\
%BFM & background field method \\
BS (BSE) & Bethe-Salpeter (equation)\\
CEA & Cambridge (Massachusetts) electron accelerator \\
CERN & European Laboratory for Particle Physics \\
CLAS & detector in Hall-B at JLab\\
DA & distribution amplitude \\
DCSB & dynamical chiral symmetry breaking \\
DESY & Deutsches Elektronen-Synchrotron (accelerator in Hamburg) \\
DF & distribution function \\
DSE & Dyson-Schwinger equation\\
DVCS & deeply virtual Compton scattering \\
DVMP & deeply virtual meson production \\
DY & Drell-Yan (process) \\
EHM & emergent hadronic mass\\
EIC & electron ion collider in the USA\\
EicC & electron ion collider in China \\
Fermilab (FNAL) & Fermi National Accelerator Facility \\
FF & fragmentation function \\
FNC & forward neutron calorimeter \\
GFF & generalised form factor (related to a given GPD)\\
GLCS & good lattice cross-section(s) \\
GPD & generalised parton distribution \\
H1 & detector at HERA \\
HB & Higgs boson \\
HERA & particle accelerator in Hamburg (\emph{Hadron-Elektron-Ringanlage} \\
$\,$         & \rule{\parindent}{0ex} or Hadron-Electron Ring Accelerator) \\
HERMES & detector and associated collaboration at HERA \\
HMS & high momentum spectrometer (detector at JLab)\\
ISR &  Intersecting Storage Rings (accelerator at CERN) \\
JLab & Thomas Jefferson National Accelerator Facility\\
JLab\,12 & Thomas Jefferson National Accelerator Facility with 12\,GeV $e^-$ beams\\
LaMET & large-momentum effective theory \\
LFWF & light-front wave function \\
LHC & large hadron collider \\
LHCb & LHC beauty experiment \\
LO & leading order (in a controlled expansion or truncation) \\
lQCD (LQCD)  & lattice-regularised quantum chromodynamics\\
NG (boson/mode) & Nambu-Goldstone (boson/mode) \\
NLL & next-to-leading logarithms \\
NLO & next-to-leading order (in a controlled expansion or truncation) \\
PD & process dependent (effective charge or running coupling)\\
PDF & parton distribution function \\
PI & process independent (effective charge or running coupling)\\
pQCD & perturbative QCD \\
PTIR & perturbation theory integral representation \\
QED & quantum electrodynamics \\
QCD & quantum chromodynamics \\
RGI & renormalisation group invariant \\
RL (truncation) & rainbow-ladder (truncation) \\
SCHC & $s$-channel helicity conservation \\
SHMS & superconducting super high momentum spectrometer (detector at JLab) \\
SM & Standard Model of Particle Physics \\
SPS & super proton synchrotron \\
STI & Slavnov-Taylor identity\\
TMD &  transverse momentum dependent parton distribution \\
ZEUS & detector at HERA
%%%
\end{longtable}

\medskip

% Create the reference section using BibTeX:
%\addcontentsline{toc}{toc}{\mbox{\hspace*{-\parindent}\textbf{References}} \dotfill\ }
\section*{References}
\label{references}

%\bibliographystyle{h-physrev4}
%%\bibliographystyle{../../../zProc/z10/z10KITPC/h-physrev4}
%%\bibliography{CollectedBib_EHM}

\end{document}

%% file: S1_Introduction.tex
\section{Emergence of Mass}
\label{sec:Introduction}
%
%% Outline:
%% Intro and motivation (Craig)
%Then there was Ref.\,\cite{Horn:2016rip}
%
When considering the origin of mass in the standard model of particle physics (SM), thoughts typically turn to the Higgs boson because couplings to the Higgs are responsible for every mass-scale that appears in the SM Lagrangian.  The notion behind this Higgs mechanism for mass generation was introduced more than fifty years ago \cite{Higgs:1964ia, Englert:1964et, Higgs:1964pj} and it became an essential piece of the SM.  In the ensuing years, all the particles in the SM Lagrangian were found; although the Higgs boson proved elusive, escaping detection until 2012 \cite{Aad:2012tfa, Chatrchyan:2012xdj}.  With discovery of something possessing all the anticipated properties of the Higgs boson, the SM became complete and the Nobel Prize in physics was awarded to Englert and Higgs \cite{Englert:2014zpa, Higgs:2014aqa} ``\ldots for the theoretical discovery of a mechanism that contributes to our understanding of the origin of mass of subatomic particles \ldots''

Higgs boson physics is characterised by an explicit mass scale $v_H = 1/\sqrt{G_F \surd 2} = 246\,$GeV, where $G_F$ is the Fermi coupling for muon decay, which also controls the rate of neutron $\beta$ decay.  Governed by this scale, the SM's weak bosons disappeared from the equilibrium mix when the age of the Universe was roughly $10^{-12}\,$seconds.  The large value of $v_H$ is known through comparisons between SM predictions and contemporary experiment; it is not a SM prediction.  The size of $v_H$ is critical to the character of the Universe as it is seen today because it fixes the weak boson mass-scale to be commensurate in magnitude; thereby, \emph{e.g}.\ protecting Universe evolution from the destabilising influence of electrically charged gauge-bosons that propagate over great distances.  Evidently, the Higgs mechanism, or something practically identical at all length scales which have thus far been probed, is a crucial piece in the puzzle that explains our existence.

It is therefore peculiar that Higgs couplings to those fermions which are most important to everyday existence, \emph{i.e}.\ the electron $(e^-)$ and the up $(u)$ and down $(d)$ quarks, produce such small mass values \cite{Zyla:2020zbs}: $m_{e}=0.511\,$MeV, $m_u \approx 4m_e \approx 2.2 \,$MeV, $m_d \approx 2 m_u$.  These particles combine to form the hydrogen atom, the most abundant element in the Universe, whose mass is 939\,MeV.  Somehow one electron, two $u$ quarks and one $d$ quark, with a total Higgs-generated mass of $\sim 13 m_e \approx 6.6\,$MeV, combine to form an object whose mass is 140-times greater.

The energy levels of the hydrogen atom are measured in units of $m_e \alpha_{e}^2$, where $\alpha_{e}\approx 1/137$ is the fine structure constant in quantum electrodynamics (QED).  Both quantities appear explicitly in the SM Lagrangian; and all of atomic and molecular physics are unified when these two fundamental parameters have their small values.   The solution to the mystery of the missing mass of the Hydrogen atom must therefore lie somewhere else; and there is only one other place to look, \emph{viz}.\ the proton at the heart of the atom, whose mass is $m_p=0.938\,$GeV.

Protons and other hadrons began to appear following another key step in the evolution of the Universe.  Namely, after the cross-over from the quark-gluon plasma phase into the domain of hadron matter, which occurred when the Universe was approximately $10^{-6}\,$seconds old \cite{Busza:2018rrf, Bazavov:2019lgz, Bzdak:2019pkr}.  This marks the beginning of the emergence of hadron mass (EHM).  

Within the SM, the proton is supposed to be explained by quantum chromodynamics (QCD), a Poincar\'e-invariant local quantum gauge field theory with interactions based upon the non-Abelian group SU$(3)$.  The QCD Lagrangian is simple to write:\footnote{We use a Euclidean metric throughout; so, \emph{e.g}.\ $\{\gamma_\mu,\gamma_\nu\} = 2\delta_{\mu\nu}$; $\gamma_\mu^\dagger = \gamma_\mu$; $\gamma_5= \gamma_4\gamma_1\gamma_2\gamma_3$, tr$[\gamma_5\gamma_\mu\gamma_\nu\gamma_\rho\gamma_\sigma]=-4 \epsilon_{\mu\nu\rho\sigma}$; $a \cdot b = \sum_{i=1}^4 a_i b_i$; and $Q_\mu$ timelike $\Rightarrow$ $Q^2<0$.}
\begin{subequations}
\label{QCDdefine}
\begin{align}
{\mathpzc L}_{\rm QCD} & = \sum_{{\mathpzc f}=u,d,s,\ldots}
\bar{q}_{\mathpzc f} [\gamma\cdot\partial
    + i g \tfrac{1}{2} \lambda^a\gamma\cdot A^a+ m_{\mathpzc f}] q_{\mathpzc f}
    + \tfrac{1}{4} G^a_{\mu\nu} G^a_{\mu\nu},\\
%
%D_{\mu} & = \partial_\mu + i g \tfrac{1}{2} \lambda^a A^a_\mu\,, \\
%
\label{gluonSI}
G^a_{\mu\nu} & = \partial_\mu A^a_\nu + \partial_\nu A^a_\mu -
\underline{{g f^{abc}A^b_\mu A^c_\nu}},
\end{align}
\end{subequations}
where $\{q_{\mathpzc f}\,|\,{\mathpzc f}=u,d,\ldots\}$ are the quark fields, of which six flavours are currently known, and $\{m_{\mathpzc f}\}$ are their Higgs-generated current-quark masses; $\{A_\mu^a\,|\,a=1,\ldots,8\}$ are the  gluon fields, whose matrix structure is encoded in $\{\lambda^a\}$, the generators of SU$(3)$ in the fundamental representation; and $g$ is the \emph{unique} QCD coupling.  Similar to QED, one conventionally defines $\alpha = g^2/[4\pi^2]$.

It is worth noting here that when $\{m_{\mathpzc f}\equiv 0\}$, Eq.\,\eqref{QCDdefine} divides into two separate pieces, one describing massless left-handed fermions, $q_L = \tfrac{1}{2}({\mathbf I}-\gamma_5) q$, and the other describing right-handed fermions, $q_R = \tfrac{1}{2}({\mathbf I}+\gamma_5) q$.  No interactions in the Lagrangian can distinguish between $q_L$ and $q_R$; hence, the Lagrangian possesses a chiral symmetry.

Eq.\,\eqref{QCDdefine} is \emph{almost} identical to the QED Lagrangian.  The principal difference is the underlined term in Eq.\,\eqref{gluonSI}, which generates self interactions amongst the gluons.  
Yet, whereas QED is probably only an effective field theory, currently being ill-defined owing to the presence of a Landau pole at some (huge) spacelike momentum (see, \emph{e.g}.\ Ref.\,\cite[Ch.\,13]{IZ80} and Refs.\,\cite{Rakow:1990jv, Gockeler:1994ci, Reenders:1999bg, Kizilersu:2014ela}), QCD appears empirically to be well-defined at all momenta.
Theoretically, asymptotic freedom \cite{Politzer:2005kc, Wilczek:2005az, Gross:2005kv} ensures that QCD's ultraviolet behaviour is under control.  At the other extreme, \emph{i.e}.\ the infrared domain, all circumstantial evidence, including our existence, indicates that there are no issues either.  Gluon self interactions are certainly the origin of asymptotic freedom; and, logically, if QCD is also infrared complete, then the underlined term in Eq.\,\eqref{gluonSI} must be the effecting agent.

So, where is the proton's mass in Eq.\,\eqref{QCDdefine}?  As noted, already $m_p$ is not contained in the sum of the light-quark current masses that appear explicitly.  It is therefore worth exploring the character of the theory defined without them, \emph{i.e}.\ QCD with all quark couplings to the Higgs boson turned off.   The resulting Lagrangian is scale invariant; and it is readily established that in such a theory, compact bound states are impossible.  For suppose the field equations admit a nontrivial solution for a bound state with size ``$r$'', then simple dilation transformations, under which the theory is invariant, can be used to inflate $r\to \infty$, \emph{i.e}.\ to eliminate the bound state.  Consequently, scale invariant theories do not support dynamics, only kinematics.  Plainly, therefore, if Eq.\,\eqref{QCDdefine} is really the basis for, \emph{inter alia}, an explanation of the proton's mass and size, then something remarkable must happen in completing the definition of QCD.

In a Poincar\'e invariant quantum field theory, observables are independent of spacetime translations.  This places a constraint on the theory's energy-momentum tensor, $T_{\mu\nu}$:
\begin{equation}
\label{ConsEP}
\partial_\mu T_{\mu\nu} = 0 \,.
\end{equation}
Focus now on QCD and consider a global scale transformation in the classical action defined by Eq.\,\eqref{QCDdefine}:
\begin{equation}
\label{scaleT}
x \to x^\prime = {\rm e}^{-\sigma}x\,, \;
A_\mu^a(x)  \to A_\mu^{a\prime}(x^\prime) = {\rm e}^{-\sigma} A_\mu^a({\rm e}^{-\sigma}x ) \,,\;
q(x)  \to q^\prime(x^\prime) = {\rm e}^{- (3/2) \sigma} q({\rm e}^{-\sigma}x )\,.
\end{equation}
The Noether current connected with these transformations is
\begin{equation}
{\mathpzc D}_\mu = T_{\mu\nu} x_\nu\,,
\end{equation}
the dilation current.  In the absence of Higgs couplings into QCD, the classical action is invariant under dilations; hence, using Eq.\,\eqref{ConsEP},
\begin{equation}
\partial_\mu {\mathpzc D}_\mu = 0
= [\partial_\mu T_{\mu\nu} ] x_\nu + T_{\mu\nu} \delta_{\mu\nu}  = T_{\mu\mu}\,.
\label{SIcQCD}
\end{equation}
This proves that the trace of the energy-momentum tensor is zero in any theory that is truly scale invariant.  So much for the classical theory: in the absence of an explicit mass-scale, none can emerge.

Quantisation preserves Poincar\'e invariance.  It also entails the appearance of loop diagrams, which typically possess ultraviolet divergences.  A workable mathematical definition of such loop integrals requires introduction of a regularisation procedure and associated mass-scale, $\bar\nu$.  Following regularisation, a systematic renormalisation scheme must be introduced in order to eliminate from all computed quantities any dependence on the arbitrary scale $\bar\nu$, which can differ between loop integrals, and replace it by a dependence of Lagrangian parameters on a single scale $\zeta$, \emph{i.e}.\ the renormalisation scale \cite{tarrach}.  This outcome is known as ``dimensional transmutation'': everything in the QCD action comes to depend on $\zeta$, even those quantities which are dimensionless.

Dimensional transmutation has consequences.  Under the scale transformation in Eq.\,\eqref{scaleT}, the renormalisation mass-scale $\zeta \to {\rm e}^\sigma \zeta$.  For infinitesimal transformations of this type:
\begin{equation}
\label{TA1}
\alpha \to \sigma\,  \alpha  \beta(\alpha)\,,\;
{\mathpzc L}_{\rm QCD} \to \sigma \, \alpha \beta(\alpha) \, \frac{\delta {\mathpzc L}_{\rm QCD}}{\delta \alpha}
\Rightarrow \partial_\mu {\cal D}_\mu^{\rm QCD} =  \frac{\delta {\mathpzc L}_{\rm QCD}}{\delta \sigma} = \alpha \beta(\alpha) \, \frac{\delta{\mathpzc L}_{\rm QCD}}{\delta \alpha}\,,
\end{equation}
where $\beta(\alpha)$ is the QCD $\beta$-function, which measures the response rate of the coupling to changes in $\zeta$.  To compute the final product in Eq.\,\eqref{TA1}, one can first absorb the gauge coupling into the gluon field, \emph{i.e}.\ express the action in terms of $\tilde A_\mu^{a} = g A_\mu^a$, whereafter the running coupling appears only in the pure-gauge term as an inverse multiplicative factor:
\begin{equation}
\label{TA2}
 {\mathpzc L}_{\rm QCD}(\alpha)  = - \frac{1}{4\pi\alpha}\frac{1}{4} \tilde G_{\mu\nu}^a \tilde G_{\mu\nu}^a + \alpha\mbox{-independent~terms,}
\end{equation}
where $\tilde G_{\mu\nu}^a$ is the field-strength tensor expressed using $\tilde A_\mu^{a}$.  Eqs.\,\eqref{TA1}, \eqref{TA2} then yield
 \begin{equation}
 \label{EqAnomaly}
T_{\mu\mu}^{\rm QCD} = \partial_\mu {\cal D}_\mu^{\rm QCD} = \alpha \beta(\alpha) \frac{\delta {\mathpzc L}_{\rm QCD}}{\delta \alpha} =  \alpha \beta(\alpha)\frac{1}{4\pi\alpha^2} \frac{1}{4} \tilde G_{\mu\nu}^a \tilde G_{\mu\nu}^a = \beta(\alpha)  \tfrac{1}{4} G^{a}_{\mu\nu}G^{a}_{\mu\nu}\,.
\end{equation}
Evidently, following regularisation and renormalisation, Eq.\,\eqref{SIcQCD} is broken because the trace of the QCD stress-energy tensor is nonzero; so these necessary steps have introduced the chiral-limit trace anomaly into the dilation current:
\begin{equation}
\label{SIQCD}
\Theta_0 = \beta(\alpha)  \tfrac{1}{4} G^{a}_{\mu\nu}G^{a}_{\mu\nu}\,.
\end{equation}

Switching on the Higgs couplings into QCD, Eq.\,\eqref{SIQCD} becomes
\begin{equation}
\label{SIQCDm}
\Theta:= T_{\mu\mu}^{\rm QCD} = \tfrac{1}{4} \beta(\alpha(\zeta)) G^{a}_{\mu\nu}G^{a}_{\mu\nu}
+ [1+\gamma(\alpha(\zeta))]\sum_{{\mathpzc f}=u,d,\ldots} m_{\mathpzc f}^\zeta \, \bar q_{\mathpzc f} q_{\mathpzc f}\,,
\end{equation}
where $\gamma(\alpha)$ is the anomalous dimension of the now scale-dependent current-quark mass, $m_{\mathpzc f}^\zeta$.
Notably, the trace anomaly in Eq.\,\eqref{SIQCDm} is not homogeneous in the running coupling, $\alpha(\zeta)$; so, renormalisation-group-invariance does not imply form invariance of the right-hand-side \cite{Tarrach:1981bi}.  This is a material point because many discussions implicitly assume that all operators and associated identities are expressed with reference to a partonic basis, \emph{i.e}.\ using elementary field operators that can be renormalised perturbatively, wherewith the state-vector for any hadron must be an extremely complicated wave function.  A different perspective is required at renormalisation scales $\zeta \lesssim m_p$, whereupon a metamorphosis from parton-basis to quasiparticle-basis occurs.  Under such reductions in $\zeta$, light partons evolve into heavy dressed objects, corresponding to complex and highly nonlinear superpositions of partonic operators; and using the associated quasiparticle degrees-of-freedom, the wave functions can be expressed in a relatively simple form.  These statements are illustrated, \emph{e.g}.\ in Refs.\,\cite{Finger:1981gm, Adler:1984ri, Szczepaniak:2001rg}, and will be discussed further in Sec.\,\ref{sec:CTP}.

\begin{figure}[!t]
\hspace*{-1ex}\begin{tabular}{l}
{\sf A} \\[-1.1ex]
\includegraphics[clip, width=0.44\textwidth]{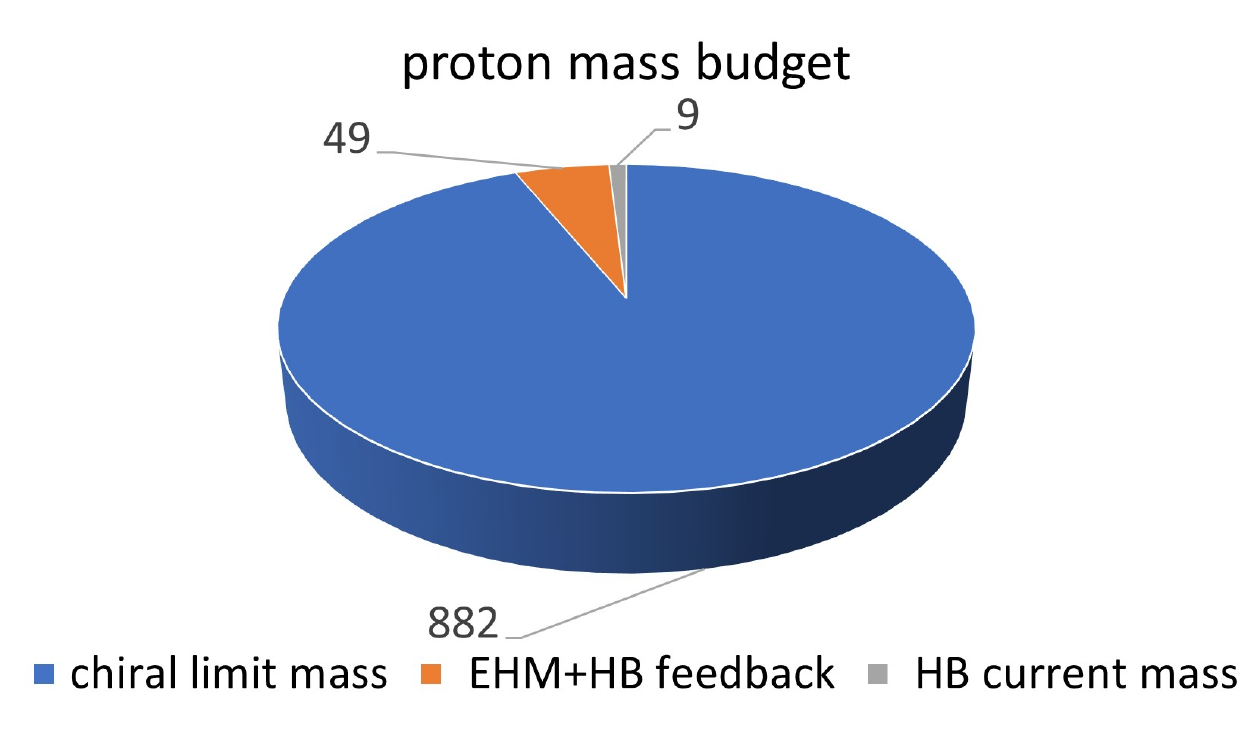}
\end{tabular}
\hspace*{-1ex}\begin{tabular}{ll}
{\sf B} & {\sf C} \\[-1.1ex]
\includegraphics[clip, width=0.44\textwidth]{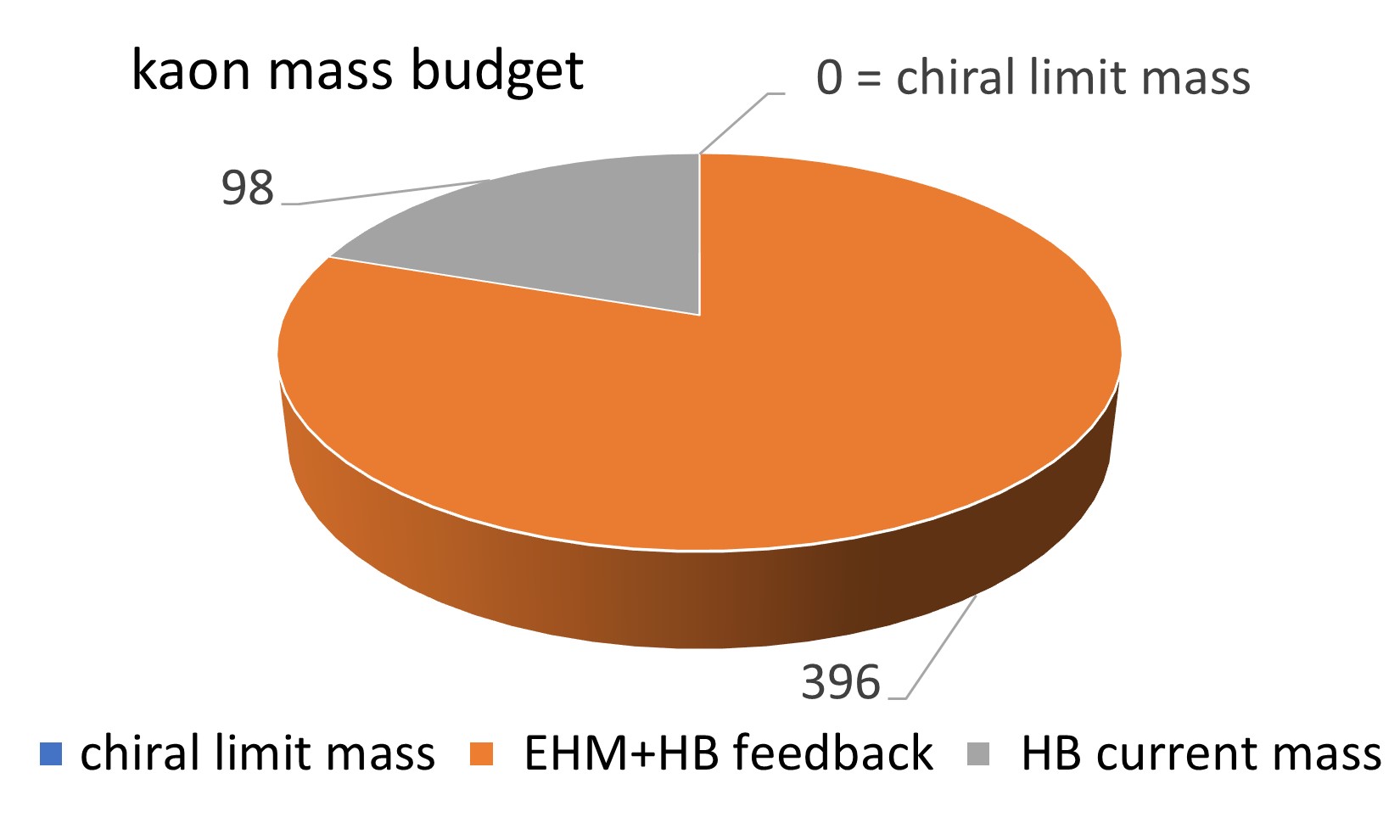} &
\includegraphics[clip, width=0.44\textwidth]{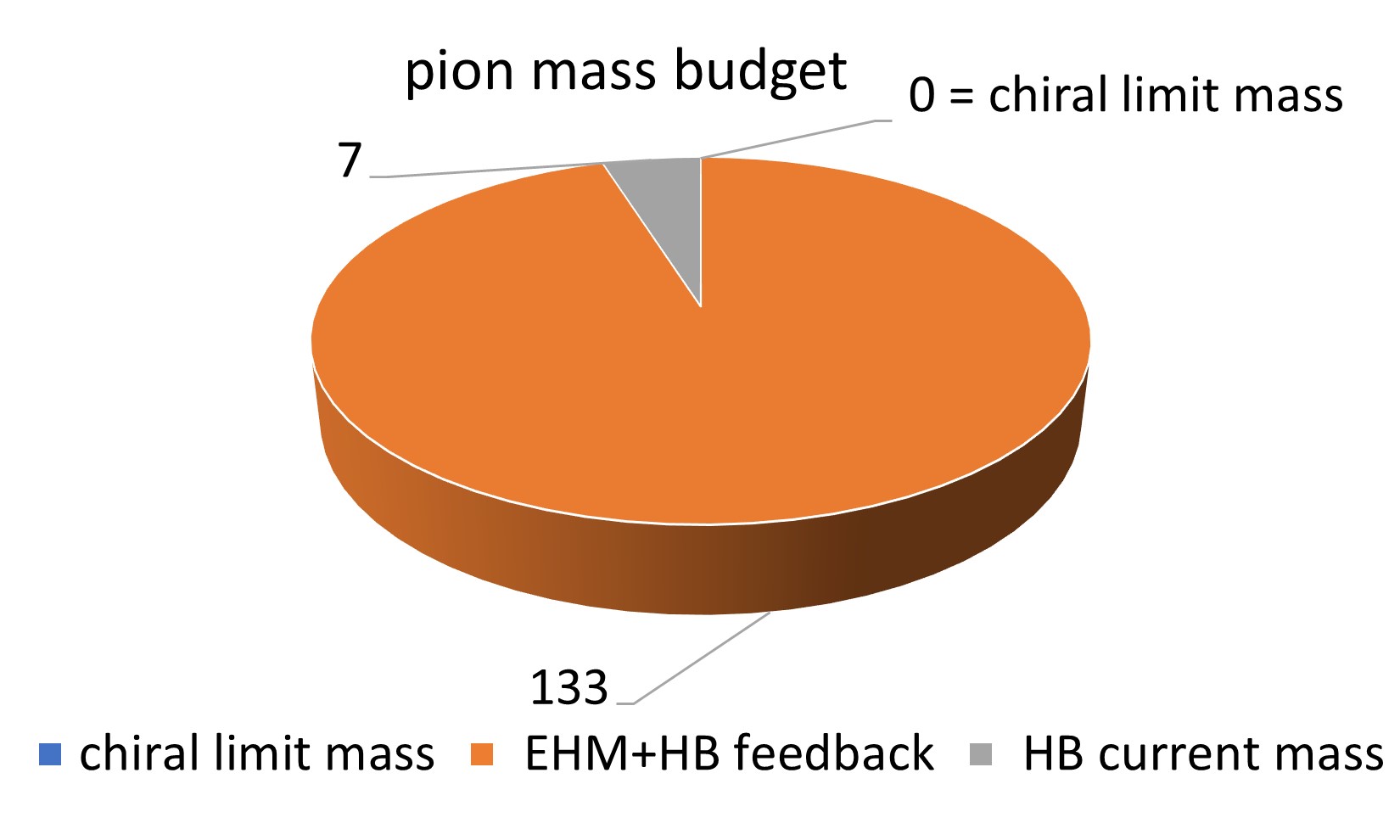}
\end{tabular}
%
%%\hspace*{-1ex}\begin{tabular}{lll}
%%{\sf A} & {\sf B} & {\sf C} \\[-1.1ex]
%%\includegraphics[clip, width=0.34\textwidth]{F1ACDR.jpg} &
%%\includegraphics[clip, width=0.32\textwidth]{F1BCDR.jpg} &
%%\includegraphics[clip, width=0.33\textwidth]{F1CCDR.jpg}
%%\end{tabular}
%
\caption{\label{F1CDR}
Mass budgets for {\sf A}\,--\,proton, {\sf B}\,--\,kaon and {\sf C}\,--\,pion, drawn using a Poincar\'e invariant decomposition.  There are crucial differences.
The proton's mass is large in the chiral limit, \emph{i.e}.\ even in the absence of Higgs boson (HB) couplings into QCD.  This nonzero chiral-limit component is an expression of emergent hadronic mass (EHM) in the SM.
Conversely and yet still owing to EHM via its dynamical chiral symmetry breaking (DCSB) corollary, the kaon and pion are massless in the chiral limit -- they are the SM's Nambu-Goldstone modes \cite{Nambu:1960tm, Goldstone:1961eq, GellMann:1968rz, Horn:2016rip}.  (See Eq.\,\eqref{MGMOR} below.)
(Units MeV, separation at $\zeta=2\,$GeV, produced using information from Refs.\,\cite{Flambaum:2005kc, RuizdeElvira:2017stg, Aoki:2019cca, Zyla:2020zbs}.)
}
\end{figure}

The existence of the trace anomaly means that QCD can support a mass-scale and potentially explain the origin of $m_p$ even in the absence of Higgs couplings and even though that scale is not evident in Eq.\,\eqref{QCDdefine}.  It is therefore natural to seek information on the size of the trace anomaly's contribution to $m_p$.  So consider the forward limit of the expectation value of QCD's energy momentum tensor in the proton state (hereafter, the superscript ``QCD'' is omitted):
\begin{equation}
\label{EPTproton}
\langle p(P) | T_{\mu\nu} | p(P) \rangle = - P_\mu P_\nu\,,
\end{equation}
where the equations-of-motion for an asymptotic one-particle proton state have been used to obtain the right-hand-side.  Now it is clear that, in the chiral limit,
\begin{equation}
\label{anomalyproton}
\langle p_0(P) | T_{\mu\mu} | p_0(P) \rangle  = - P^2  = m_{p_0}^2
 = \langle p_0(P) |  \Theta_0 | p_0(P) \rangle\,.
\end{equation}
From this perspective, the size of the trace anomaly is measured by the magnitude of the proton's mass in the chiral limit, $m_{p_0}$.  Many analyses have sought to determine this value using a variety of theoretical techniques \cite{Flambaum:2005kc, RuizdeElvira:2017stg, Aoki:2019cca}, with the result $m_{p_0} \approx 0.89\,$GeV$\approx 0.94 \,m_p$, illustrated by the blue domain in Fig.\,\ref{F1CDR}A.  Evidently, a very large fraction of the measured proton mass emerges as a consequence of the trace anomaly; and viewed from a perspective built on partonic degrees of freedom, this fraction appears to be generated entirely by gluon partons and the interactions between them because these things define $\Theta_0$ in Eq.\,\eqref{SIQCD}.

The proton is a basic building block of nuclei; but one cannot bind neutrons and protons into any nucleus without the pion, which is responsible for, \emph{inter alia}, long-range attraction and tensor forces within all nuclei.
%% Intermediate range tensor = 2 pi ~ rho
Hence, in nuclear physics terms, the pion, the proton and neutron are all equally important.  Drawing a connection with ${\mathpzc L}_{\rm QCD}$ in Eq.\,\eqref{QCDdefine}, the pion is seemingly the simplest of these bound states, being constituted from a single valence quark partnered by a valence antiquark, \emph{e.g}.\ $\pi^+ = u \bar d$.  It is therefore natural to consider the pion analogue of Eq.\,\eqref{anomalyproton}:
\begin{equation}
\label{anomalypion}
\langle \pi_0(q) | T_{\mu\mu} | \pi_0(q) \rangle  =\langle \pi_0(P) |  \Theta_0 | \pi_0(P) \rangle
= - q^2  = m_{\pi_0}^2 = 0
 \,,
\end{equation}
where the last identity follows because the chiral-limit pion is the Nambu-Goldstone (NG) mode associated with dynamical chiral symmetry breaking (DCSB) \cite{Nambu:1960tm, Goldstone:1961eq, GellMann:1968rz, Horn:2016rip}.  Comparing Eqs.\,\eqref{anomalyproton}, \eqref{anomalypion}, one is presented with a dilemma: how can it be that the trace anomaly evaluates to a $\sim 1\,$GeV mass-scale in the proton; yet, despite the anomaly, scale invariance is seemingly preserved in the pion?  It has been argued that Eq.\,\eqref{anomalypion} means, \emph{e.g}.\ that the gluon energy and quark energy in the pion separately vanish \cite{Yang:2014xsa}.  However, such an explanation would actually compound the puzzle because the pion's valence-quark and -antiquark are supposed to be bound together by gluon-mediated attraction: how could a bound state survive in the absence of any gluon energy?

Restoring Higgs boson couplings to light quarks, then using Eq.\,\eqref{SIQCDm}, Eq.\,\eqref{anomalyproton} becomes
\begin{equation}
\langle p(P) | \Theta | p(P) \rangle  \stackrel{\zeta \gg m_p}{=}
\langle p(P) |\left[ \tfrac{1}{4} \beta(\alpha(\zeta)) G^{a}_{\mu\nu}G^{a}_{\mu\nu}
+ [1+\gamma(\alpha(\zeta))]\sum_{{\mathpzc f}=u,u,d,} m_{\mathpzc f}^\zeta \, \bar q_{\mathpzc f} q_{\mathpzc f} \right] | p(P) \rangle \,.
%
%%%& \stackrel{\zeta \lesssim m_p}{=}
%%%\langle \tilde p(P) |\left[
%%%{\mathpzc D}_3 + {\mathpzc I}_3 \right] | \tilde p (P) \rangle \\
%
%%%{\mathpzc D}_3 & = \sum_{i=u,u,d} M_f(\zeta) \bar{\mathpzc Q}_{\,f}(\zeta) {\mathpzc Q}_{\,f}(\zeta)\,,
%%%\quad
%%%{\mathpzc I}_3 = \tfrac{1}{4} \left[\beta(\alpha(\zeta)) {\mathpzc G}^{a}_{\mu\nu}(\zeta){\mathpzc G}^{a}_{\mu\nu}(\zeta)\right]_{2PI}
\end{equation}
Consequently, two other slices appear in the pies drawn in Fig.\,\ref{F1CDR}.  The grey wedge in Fig.\,\ref{F1CDR}A shows the sum of the proton's valence-quark current-masses, which appear in perturbative analyses of QCD phenomena: the sum amounts to just $0.01 \times m_p$.  The remaining slice (orange) expresses the fraction of $m_p$ generated by constructive interference between EHM and Higgs-boson (HB) effects.  It is largely determined by the in-proton expectation value of the chiral condensate operator \cite{Brodsky:2010xf, Chang:2011mu, Brodsky:2012ku}: $\langle p(P) |\bar q_{\mathpzc f} q_{\mathpzc f}| p(P) \rangle $, and responsible for 5\% of $m_p$.
Unsurprisingly given Eq.\,\eqref{anomalypion}, the picture for the pion is completely different: in Fig.\,\ref{F1CDR}C, EHM+HB interference is seen to be responsible for 95\% of the pion's mass.
The kaon lies somewhere between these two poles.  It is a would-be NG mode; hence, there is no blue-domain in Fig.\,\ref{F1CDR}B.  On the other hand, the sum of valence-quark and -antiquark current-masses in the kaon accounts for 20\% of its measured mass, which is four times more than in the pion; and EHM+HB interference produces 80\%.

The mass budgets drawn in Fig.\,\ref{F1CDR}, and Eqs.\,\eqref{anomalyproton}, \eqref{anomalypion}
demand interpretation.  They highlight that any answer to the question ``How does the mass of the proton arise?'' will only explain one part of a greater puzzle.  It will be incomplete unless it simultaneously clarifies Eq.\,\eqref{anomalypion}.  Moreover, whilst not manifest in Eq.\,\eqref{QCDdefine}, Eqs.\,\eqref{anomalyproton}, \eqref{anomalypion} are coupled with confinement, \emph{i.e}.\ the fact that no gluon- or quark-like object has been seen to propagate over a length scale which exceeds the proton radius.  These observations stress the ubiquitous influence of emergent mass.  Consequently, the SM will remain incomplete until verified explanations are provided for the emergence of nuclear-size masses, its many attendant corollaries, and the modulations of these effects by Higgs boson interactions.  All these things are basic to forming an understanding of the evolution of our Universe.

In approaching these questions, many observables can be used to draw insights \cite{Burkert:2017djo, Brodsky:2020vco, Carman:2020qmb, Barabanov:2020jvn}.  However, unique opportunities are provided by studies of the properties of the SM's (pseudo-)\,Nambu-Goldstone modes, \emph{viz}.\ pions and kaons.  A diverse range of phenomenological and theoretical frameworks are now being employed in order to develop a cogent description of these bound states.  Progress in this direction is profiting from the formation of tight links between dynamics in QCD's gauge sector and properties of the light-front wave functions (LFWFs) which enable a probabilistic interpretation of pion and kaon structure.  In this connection, pion and kaon elastic form factors and distribution amplitudes and functions all play a prominent role \cite{Aguilar:2019teb, Horn:2020ces, Roberts:2020udq, Roberts:2020hiw, Chen:2020ijn}.  These efforts have a special resonance today because an array of upgraded and anticipated experimental facilities promise to provide new, high precision data on kinematic domains that have never before been reached or have not been plumbed for more than thirty years \cite{E12-06-101, E12-07-105, E12-09-011, Petrov:2011pg, JlabTDIS1, JlabTDIS2, Denisov:2018unjF, Aguilar:2019teb, Chen:2020ijn}.  There is much to be learnt: the internal structure of pions and kaons is far more complex than often imagined; and their properties provide the clearest windows onto EHM and its modulation by Higgs-boson interactions.

At this point it is worth remarking that \emph{measurements} of distribution amplitudes and functions, form factors, spectra, charge radii, polarisabilities, \emph{etc}., are all on the same footing.  Theory delivers predictions for such quantities. Good experiments measure precise cross-sections; and cross-sections are expressed, using truncations that sometimes have the quality of approximations, in terms of a given desired quantity.  At issue is the reliability of the truncation/approximation used in relating the measured cross-section to this quantity.  The phenomenology challenge is to ensure that all contributions known to have a potentially material effect are included in building the bridge.  The quality of the phenomenology cannot alter either that of the experiment or the theory; but inadequate phenomenology can deliver results that mislead interpretation.   The reverse is also true; so, progress demands the building of a constructive synergy between all subbranches of the programme.

As it was known five years ago, the status of experiment and theory relevant to pion and kaon elastic electromagnetic form factors is reviewed in Ref.\,\cite{Horn:2016rip}.  One must look ten years back to find a comparable overview of pion and kaon distribution functions \cite{Holt:2010vj}.
In this review, therefore, we focus on more recent associated developments in experiment and also in theory, paying particular attention to continuum Schwinger function methods and lattice-regularised QCD, but also noting advances made using other theory tools, because much has changed in the past decade.
%Herein, therefore, we focus on more recent associated developments because much has changed in the past decade.
%In Sec.\,\ref{sec:CTP}
In fact, during this period, many threads of experiment and theory have been drawn together; so that they are now recognised as facets and expressions of EHM, the understanding of which defines one of the last SM frontiers.  The approaching few decades should see that border crossed as high-profile initiatives in experiment deliver data whose sound interpretation will deliver full comprehension of EHM and its modulation by Higgs-boson couplings into QCD; to wit, complete understanding of the SM's two mass generating mechanisms, the interplay between them, and all the observable consequences thereof.
The potential for an array of pion and kaon structure studies to play a critical role in achieving these goals is the focus of the material which follows. 

%% file: S2_Continuum.tex
%\section{Continuum Theory and Phenomenology}
%\section{QCD's One-Body Sector and Nambu-Goldstone Modes}
\section{Masses, coupling, and the Emergence of Nambu-Goldstone Modes}
\label{sec:CTP}
\subsection{Gluon Mass Scale}
\label{GluonMass}
The QCD trace anomaly exerts a material influence on every one of QCD's Schwinger functions; but for those unfamiliar with analyses of QCD's gauge sector, the most striking impact, perhaps, is that expressed in the gluon two-point function.  Interaction induced dressing of a gauge boson is expressed through the appearance of a nonzero polarisation tensor, $\Pi_{\mu\nu}(k)$.  The generalisation of gauge symmetry to the quantised theory is expressed in Slavnov-Taylor identities (STIs) \cite{Taylor:1971ff, Slavnov:1972fg}.  Regarding gluons, a crucial STI requires $k_\mu \Pi_{\mu\nu}(k) = 0 = \Pi_{\mu\nu}(k) k_\nu$, which entails
\begin{equation}
\label{EqPiGB}
\Pi_{\mu\nu}(k) = [\delta_{\mu\nu} - k_\mu k_\nu / k^2] \Pi(k^2).
\end{equation}
Namely, in a quantised gauge theory, no interaction may introduce a longitudinal component to the polarisation tensor.  In these terms, the fully dressed gluon propagator takes the form
\begin{equation}
D_{\mu\nu}(k) = [\delta_{\mu\nu} - k_\mu k_\nu / k^2] \frac{1}{k^2[1 + \Pi(k^2)]},
\end{equation}
where any gauge parameter dependence is trivial; hence, omitted here.

In setting the QCD stage, it is useful to recall that the gauge boson propagator in two-dimensional quantum electrodynamics (QED), defined with massless fermions, was analysed in Ref.\,\cite{Schwinger:1962tp}.  Owing to the peculiar kinematic character of two dimensions, this theory is confining, \emph{i.e}.\ effectively strongly coupled.  In computing $\Pi_{\mu\nu}(k)$, one must sum a countable infinity of loop diagrams, each of which involves massless fermion+antifermion pairs.  Such pairs provide screening; and because the screening fields are massless and there are infinitely many loops, the screening is a long-range effect.  Hence, the complete vacuum polarisation acquires a mass-scale: $\lim_{k^2\to 0} k^2 \Pi(k^2) =m_\gamma^2$; and the gauge boson acquires a mass with no cost to gauge invariance.  This effect is now called the Schwinger mechanism of gauge-boson mass generation.  A qualitatively similar outcome is found in three-dimensional QED \cite{Bashir:2008fk, Bashir:2009fv, Braun:2014wja}.  However, there is a difference between both these cases and QCD, \emph{viz}.\ the Lagrangian coupling possesses a mass dimension in lower dimensional theories; hence, scale invariance is broken even at the classical level and the size of the gauge-sector mass is fixed by that of a parameter in the Lagrangian.

Against this backdrop, it was first suggested forty years ago that a Schwinger-like mechanism is active in QCD \cite{Cornwall:1981zr}.   Using QCD's Dyson-Schwinger equations (DSEs), it was argued that gauge sector dynamics transforms the massless gluon partons in Eq.\,\eqref{QCDdefine} into complex quasiparticles, characterised by a momentum-dependent mass-function whose value is large at infrared momenta: $m_g(0)= 0.5 \pm 0.2\,$GeV.  The intervening years have seen this first sketch refined into a detailed picture, with an important step along the way being the unification of bottom-up (matter sector based) and top-down (gauge sector focused) approaches to understanding QCD's interactions \cite{Binosi:2014aea}.  Comprehensive perspectives are provided elsewhere \cite{Aguilar:2015bud, Huber:2018ned}.  Nevertheless, it is worth remarking here that the Schwinger-like $1/k^2$ pole in $\Pi(k^2)$ can only emerge in QCD because a long-range (massless) longitudinally-coupled coloured correlation is dynamically generated in the three-gluon vertex.  Since the correlations are longitudinally coupled, they do not contribute to any directly measurable amplitude.

A combination of tools, capitalising on the various strengths of continuum and lattice formulations of QCD, have today arrived at a precise determination of the $\zeta$-independent gluon mass scale \cite{Cui:2019dwv}:
\begin{equation}
\label{gluonmass}
m_0 = 0.43(1)\,{\rm GeV}.
\end{equation}
This value was obtained using lattice configurations generated with three domain-wall fermions at a physical pion mass.  The lattice scale was set by computing the mass of the $\rho$- and $\omega$-mesons \cite{Blum:2014tka, Boyle:2015exm, Boyle:2017jwu}.  Ref.\,\cite{Zafeiropoulos:2019flq} tested the scheme by verifying that it simultaneously produces a value of the QCD running coupling at the $Z$-boson mass that agrees with the world average \cite{Zyla:2020zbs}.

\begin{figure}[t]
\centerline{%
\includegraphics[clip, width=0.5\textwidth]{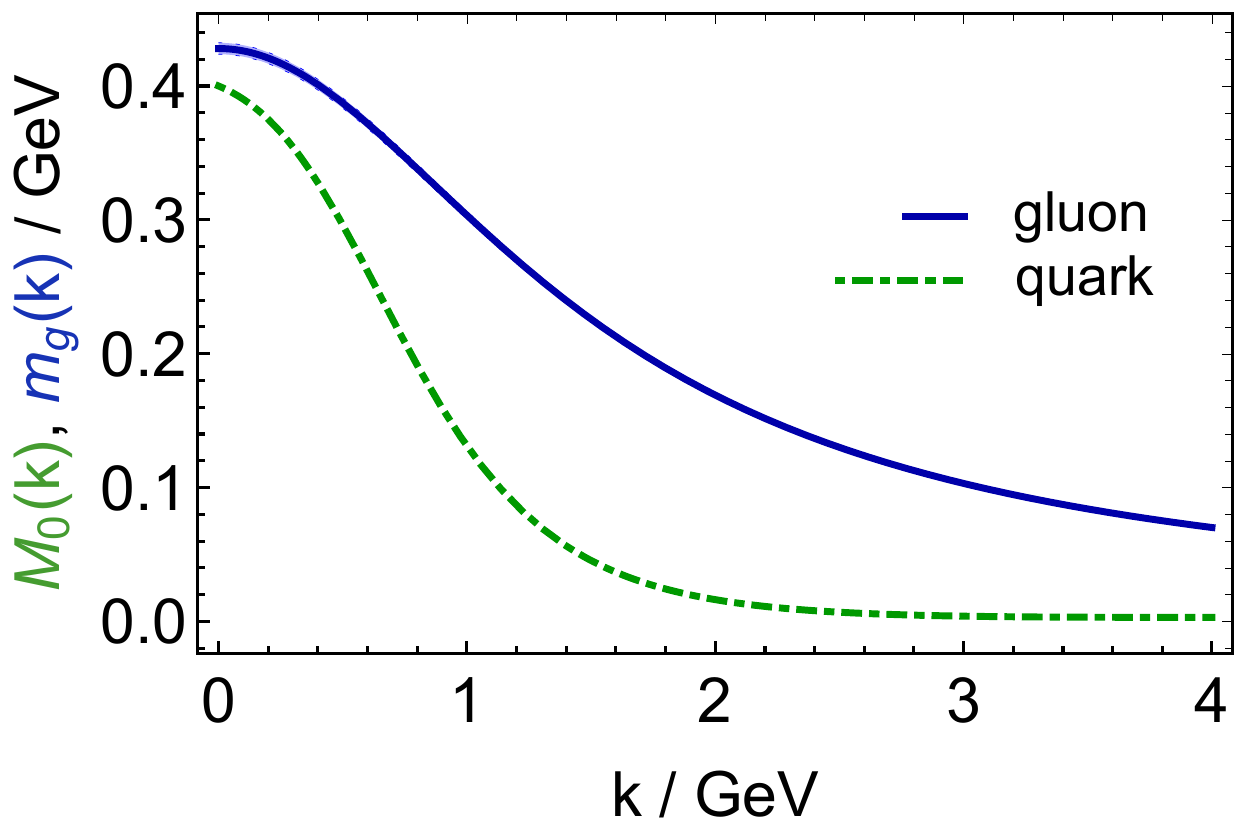}}
\caption{\label{Figmk}
$m_g(k)$ -- solid blue curve: renormalisation-group-invariant (RGI) gluon running-mass obtained, following the method described in Ref.\,\cite{Aguilar:2019uob}, from the gluon 2-point Schwinger function computed using the lattice-QCD configurations in Refs.\,\cite{Blum:2014tka, Boyle:2015exm, Boyle:2017jwu}.  (The barely visible bracketing band expresses extraction uncertainty from all sources. Curve provided by J.~Rodr{\'{\i}}guez-Quintero.)
 %generated with three domain-wall fermions at a physical pion mass and a lattice scale set by the mass of the $\rho$-meson \cite{Blum:2014tka, Boyle:2015exm, Boyle:2017jwu}.
 %
$M_0(k)$ -- dot-dashed green curve: for comparison, RGI chiral-limit dressed-quark running-mass, discussed in Sec.\,\ref{SecDCSB}.
}
\end{figure}

The calculated renormalisation group invariant (RGI, $\zeta$-independent) momentum-dependent gluon mass is depicted in Fig.\,\ref{Figmk}.  This curve is arguably the cleanest expression of EHM in the SM.  It  shows that the massless gluon parton in Eq.\,\eqref{QCDdefine} evolves into a mass-carrying dressed object, whose structure derives from complex and highly nonlinear superpositions of partonic operators.  No finite sum of diagrams in perturbation theory can recover the result in Fig.\,\ref{Figmk}.  The existence of $m_g(k)\neq 0$ is enabled by Eq.\,\eqref{EqAnomaly}, but is not a guaranteed outcome; and its infrared magnitude, Eq.\,\eqref{gluonmass}, is precisely that required to produce the measured mass of the $\rho$-meson from QCD.  Moreover, with the $\rho$ being like the proton, in the sense that it fits neatly into the standard hadron spectroscopic pattern, then the value of $m_0$ must also be a material part of any solution to the puzzle of the origin of the proton mass.

The existence and magnitude of $m_g(k)$ have been firmly demonstrated by forty years of theory.  New opportunities and challenges are now located in the need to elucidate a diverse array of observable consequences so that this basic manifestation of EHM can be confirmed empirically.
%Direct access is difficult because gluons are electrically neutral

%
\subsection{Process-Independent Effective Charge}
\label{SecPICharge}
Owing to dimensional transmutation, the QCD coupling depends on the scale at which it is measured.  In perturbation theory, within a given renormalisation scheme, this running coupling is unique.  A familiar example is found within QED.  At first glance, the renormalisation group flow of the QED coupling would appear to be governed by three renormalisation constants.  However, the Ward identity \cite{Ward:1950xp} ensures equality between the renormalisation constants for the fermion-photon vertex and fermion field operator.  Hence, $\alpha_e(\zeta)$ is completely determined by the flow of the photon field operator; equivalently, by the single renormalisation constant that survives in the expression for the photon polarisation tensor.  As apparent from Eq.\,\eqref{EqPiGB}, $\Pi(k^2)$ is a function of one momentum variable; so, QED possesses a unique running coupling whose momentum dependence is prescribed by that of the renormalised photon vacuum polarisation.  This is the Gell-Mann--Low effective charge \cite{GellMann:1954fq}, commonly known as the QED running coupling.

As a non-Abelian theory, QCD is more complicated: there are four individual interaction vertices; three associated STIs; no method in the usual treatments of diagram resummations by which any of the vertex couplings can be related to the gluon vacuum polarisation; and, of course, a renormalisation scheme must be chosen in addition.  From this position, one has four choices for the vertex that can be used to define a running coupling.  The simplest is the ghost-gluon vertex, but even that depends on two independent momenta; so one momentum pairing must be selected from an uncountable infinity of choices in order to define a single momentum with which the coupling can flow.  All such schemes produce the same running coupling on any domain within which perturbation theory is valid; but, unsurprisingly, there are great differences between the behaviours at infrared momenta.  A compendium of these results is presented elsewhere \cite[Ch.\,4]{Deur:2016tte}.

Such ambiguities are removed if one approaches the problem of diagram resummation by combining the pinch technique \cite{Cornwall:1981zr, Cornwall:1989gv, Pilaftsis:1996fh, Binosi:2009qm} and background field method \cite{Abbott:1980hw}.  This framework enables one to systematically rearrange both classes and sums of diagrams and thereby obtain modified Schwinger functions that satisfy linear STIs, \emph{i.e}.\ to make QCD appear Abelian in some important ways.  In the gauge sector, the approach leads to a modified gluon polarisation tensor whose renormalisation is identical to that of the coupling; hence, analogous to QED, one arrives at a unique RGI running coupling, $\hat\alpha(k^2)$, with momentum dependence prescribed by that of the renormalised gluon vacuum polarisation \cite{Binosi:2016nme}.  Additionally, thus defined, the coupling is process independent (PI); to wit, irrespective of the scattering process considered, gluon+gluon$\,\to\,$gluon+gluon, quark+quark$\,\to\,$quark+quark, \emph{etc}., precisely the same result is obtained.

\begin{figure}[t]
\centerline{%
\includegraphics[clip, width=0.52\textwidth]{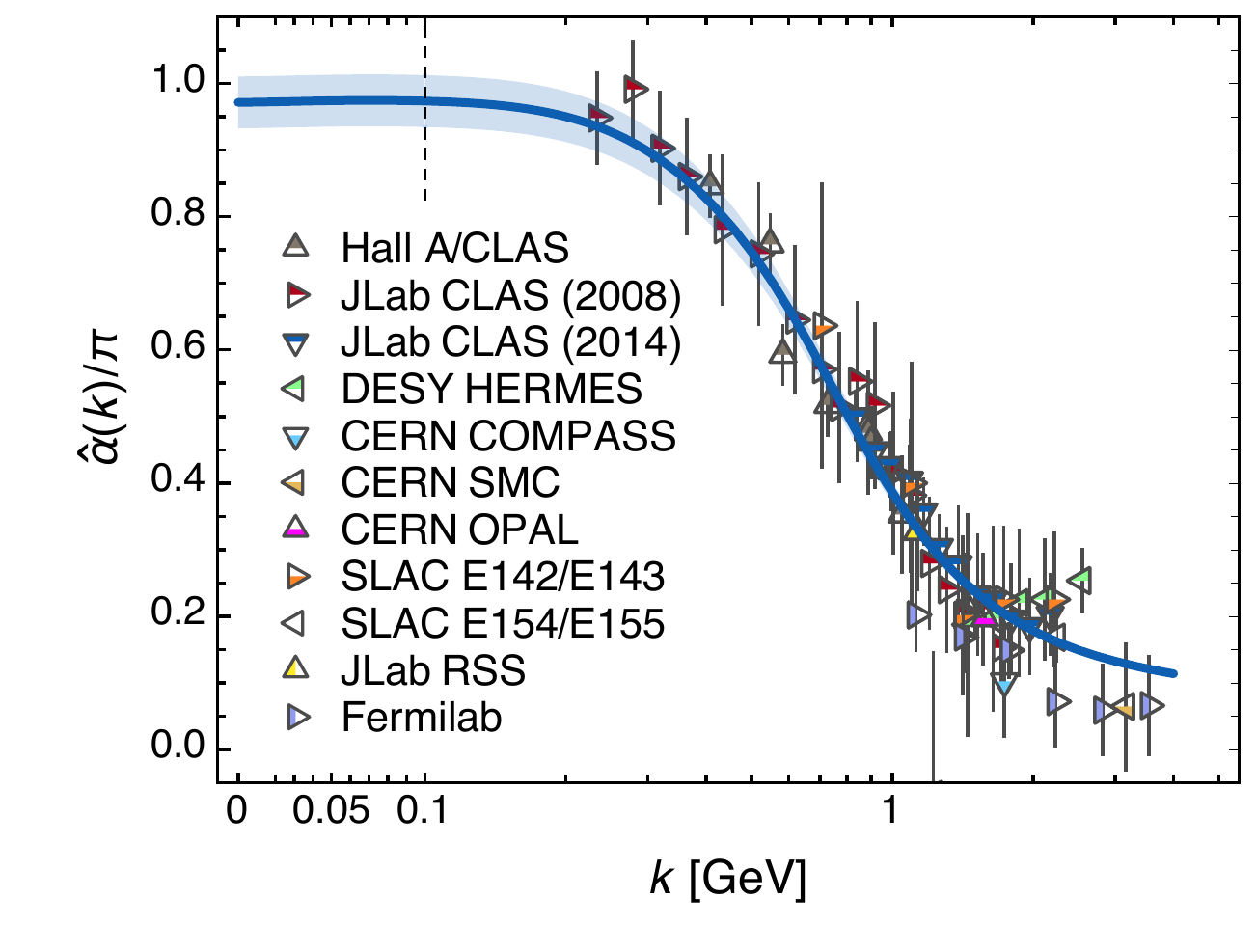}}
\caption{\label{FigalphaPI}
Process-independent running-coupling, $\hat{\alpha}(k^2)/\pi$, obtained by combining the best available results from continuum and lattice analyses \cite{Cui:2019dwv}.
%
%This running coupling saturates at infrared momenta: $\hat\alpha(0)/\pi=0.97(4)$ owing to the dynamical breakdown of scale invariance, expressed through emergence of a gluon mass-scale, with calculated value $m_0/{\rm GeV} = 0.43(1)$.
%
World data on the process-dependent charge $\alpha_{g_1}$ \cite{Deur:2016tte}, defined via the Bjorken sum rule, is presented for comparison
\cite{%
%%CLAS
Deur:2005cf, Deur:2008rf, Deur:2014vea,
%% Hermes
Ackerstaff:1997ws, Ackerstaff:1998ja, Airapetian:1998wi, Airapetian:2002rw, Airapetian:2006vy,
%% Fermilab
Kim:1998kia,
%% CERN:
Alexakhin:2006oza, Alekseev:2010hc, Adolph:2015saz,
%% SLAC:
Anthony:1993uf, Abe:1994cp, Abe:1995mt, Abe:1995dc, Abe:1995rn, Anthony:1996mw, Abe:1997cx, Abe:1997qk, Abe:1997dp, Abe:1998wq, Anthony:1999py, Anthony:1999rm, Anthony:2000fn, Anthony:2002hy}.
(Image courtesy of D.\,Binosi.)
}
\end{figure}

The key to a reliable determination of $\hat\alpha(k^2)$ is an accurate result for the dressed-gluon two-point function.  Such is available from Refs.\,\cite{Blum:2014tka, Boyle:2015exm, Boyle:2017jwu}, employed to compute $m_g(k)$ in Fig.\,\ref{Figmk}.  Using this input, Ref.\,\cite{Cui:2019dwv} delivered the parameter-free prediction depicted in Fig.\,\ref{FigalphaPI}, an interpolation of which is provided by
\begin{align}
\label{Eqhatalpha}
\hat{\alpha}(k^2) & = \frac{\gamma_m \pi}{\ln\left[\frac{{\mathpzc K}^2(k^2)}{\Lambda_{\rm QCD}^2}\right]}
%\frac{a_0^2 + a_1 k^2 + k^4}{[b_0 + k^2]\Lambda_{\rm QCD}^2}\right]}\,,
\,,\; {\mathpzc K}^2(y=k^2) = \frac{a_0^2 + a_1 y + y^2}{b_0 + y}\,,
\end{align}
%%
%\begin{equation}
%\label{Eqhatalpha}
%\hat{\alpha}(k^2) = \frac{\gamma_m \pi}{\ln\left[\Delta^{-1}(k^2)/\Lambda_{\rm QCD}^2\right]
%
%\frac{a_0^2 + a_1 k^2 + k^4}{[b_0 + k^2]\Lambda_{\rm QCD}^2}\right]}\,,
%\end{equation}
%%
$\gamma_m=4/[11 - (2/3)n_f]$, with (in GeV$^2$): $a_0=0.104(1)$, $a_1=0.0975$, $b_0=0.121(1)$.  The curve was obtained using a momentum-subtraction renormalisation scheme: $\Lambda_{\rm QCD} = 0.52\,$GeV when $n_f=4$.  The following features of $\hat{\alpha}$ deserve to be highlighted.

\begin{description}
\label{alphaPIlist}
\item[No Landau pole]
The PI charge is a smooth function, which saturates in the infrared: $\hat\alpha(s=0)/\pi = 0.97(4)$.  Hence, whereas the perturbative running coupling exhibits a Landau pole at $k^2=\Lambda_{\rm QCD}^2$, the PI charge is finite.  The value of ${\mathpzc K}(k^2=\Lambda^2_{\rm QCD}) $ defines a screening mass $\zeta_H \approx 1.4 \Lambda_{\rm QCD}$ because $\hat{\alpha}(k^2)$ is approximately $k^2$-independent on $k^2\lesssim \zeta_H^2$; consequently, the theory is effectively conformal on this domain.  These outcomes owe to EHM as expressed in Eq.\,\eqref{gluonmass}: the existence of $m_0 \approx m_p/2$ ensures that long wavelength gluons are screened, playing practically no dynamical role.
From this standpoint, $\zeta_H$ marks a border between nonperturbative/soft and perturbative/hard physics.  Hence, it is a natural choice for the ``hadronic scale'', \emph{i.e}.\ the renormalisation scale whereat one formulates and solves the bound state problem in terms of quasiparticle degrees-of-freedom \cite{Cui:2019dwv, Cui:2020dlm, Cui:2020tdf}.
%%% hadronic scale
%$\zeta_H$ marks a border between soft (nonperturbative) and hard (perturbative) physics.  Hence, it is a natural choice for the ``hadronic scale'', \emph{viz}.\ the renormalisation scale at which one formulates and solves the continuum bound state problem in terms of quasiparticle degrees-of-freedom \cite{Ding:2019qlr, Ding:2019lwe, Cui:2019dwv, Cui:2020dlm, Cui:2020tdf}.

\item[Match with Bjorken charge] Along with $\hat{\alpha}(k)$, Fig.\,\ref{FigalphaPI} also depicts data relating to $\alpha_{g_1}(k)$, a process-\emph{dependent} effective charge defined via the Bjorken sum rule \cite{Bjorken:1966jh, Bjorken:1969mm}, which expresses a central constraint on measurements of nucleon spin structure in deep inelastic scattering.  The concept of a process dependent charge was introduced in Ref.\,\cite{Grunberg:1982fw}: ``\ldots to each physical quantity depending on a single scale variable is associated an effective charge, whose corresponding St\"uckelberg -- Peterman -- Gell-Mann--Low function is identified as the proper object on which perturbation theory applies."  Such charges have subsequently been widely discussed and employed \cite{Dokshitzer:1998nz, Prosperi:2006hx, Deur:2016tte}.   So far as extant data can show, the predicted form of $\hat\alpha$ is practically identical to $\alpha_{g_1}$.  This feature may be attributed to the fact that the Bjorken sum rule is an isospin non-singlet relation, which eliminates many physical contributions that might distinguish it from $\hat\alpha$.  It is highlighted by the following result \cite{Binosi:2016nme}: on any domain within which perturbation theory is valid, $\alpha_{g_1}-\hat\alpha \lesssim \, 0.05 \alpha_{\overline{\rm MS}}^2$, where $\alpha_{\overline{\rm MS}}$ is the textbook one-loop coupling computed in the $\overline{\rm MS}$ renormalisation scheme.
%%Thus, the link between $\hat\alpha(s)$ and $\alpha_{g_1}(s)$ points to a potentially important role for $\hat\alpha(s)$ in connecting data with calculations of hadron light-front distribution amplitudes and functions \cite{Ding:2019qlr, Ding:2019lwe, Cui:2019dwv, Cui:2020dlm, Cui:2020tdf}.
%
(The gluon mass in Eq.\,\eqref{gluonmass} is commensurate with the scale $\kappa=m_p/2$ obtained in a light-front holographic approach to connecting the infrared and ultraviolet domains of $\alpha_{g_1}(s)$ \cite{Brodsky:2020ajy}.  This may point to a deeper connection.)

\item[Infrared completion] As a process independent charge, $\hat\alpha(s)$ fulfills a wide range of purposes and unifies numerous observables; hence, it is a strong candidate for that function which describes QCD's interaction strength at any accessible momentum scale \cite{Dokshitzer:1998nz}.  Furthermore, its features support a conclusion that QCD is a well-defined four dimensional quantum field theory.  As such, QCD emerges as a candidate for use in extending the SM by attributing compositeness to particles that may today seem elementary.  For instance, it was suggested long ago that all spin-$J=0$ bosons may be \cite{Schwinger:1962tp} ``\ldots secondary dynamical manifestations of strongly coupled primary fermion fields and vector gauge fields \ldots'.  Adopting this standpoint, the SM's Higgs boson might also be composite.

\end{description}

\subsection{Dynamical Chiral Symmetry Breaking}
\label{SecDCSB}
Nuclear and particle physics began roughly 100 years ago, following discovery of the proton \cite{RutherfordI, RutherfordII, RutherfordIII, RutherfordIV}.  The neutron followed thirteen years later \cite{Chadwick:1932ma}; then the pion and kaon, fifteen years after that \cite{Lattes:1947mw, Rochester:1947mi}.  Subsequently, the expanding use of particle accelerators revealed many more particles; so many, in fact, that Enrico Fermi is widely believed to have said ``\ldots if I could remember the names of these particles, I would have been a botanist.''  At this point, order was restored through development of the constituent quark model (CQM) \cite{GellMann:1964nj, Zweig:1981pd}, which made apparent that many gross features of the hadron spectrum could be explained by supposing
the existence of constituent-quarks with proton-scale masses \cite{Giannini:2015zia, Plessas:2015mpa, Eichmann:2016yit}: $M_U \approx M_D \approx 0.4\,$GeV, $M_S \approx 0.5\,$GeV, \emph{etc}.  Given the remarkable array of CQM successes, it is necessary to ask whether the idea has a connection with QCD.  An affirmative answer has emerged in the past vicennium and it can be made via the two-point quark Schwinger function.

\begin{figure}[t]
\centerline{%
\includegraphics[clip, width=0.42\textwidth]{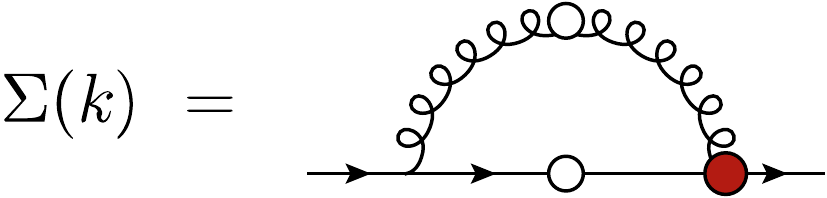}}
\caption{\label{FigQuarkDyson}
Gap (Dyson) equation for the dressed-quark self energy, Eq.\,\eqref{genSigma}:
solid line with open circle, dressed-quark propagator; open-circle ``spring'' , dressed-gluon propagator, $D_{\mu\nu}(k-q)$; and (red) shaded circle, dressed gluon-quark vertex, $\Gamma_\mu^a(k,q)$.  Here, $q$ is the loop momentum.
}
\end{figure}

The dressed-quark two-point function (propagator) can be written
\begin{subequations}
\label{EqQuarkDyson}
\begin{align}
S(k;\zeta) & = Z_2(\zeta,\Lambda) \,(i\gamma\cdot k + m_{\rm bm}) + \Sigma(k;\zeta)\,,  \\
\label{genSigma}
\Sigma(k;\zeta) &= Z_1(\zeta,\Lambda) \int^\Lambda_{dq}\!\! g^2 D_{\mu\nu}(k-q) \Gamma_\mu^a(k,q)
S(q)\frac{\lambda^a}{2}\gamma_\nu \,,
\end{align}
\end{subequations}
where $\int^\Lambda_{dq}$ represents a Poincar\'e invariant regularisation of the four-dimensional integral, with $\Lambda$ the regularization mass-scale;
$m_{\rm bm}$ is the Lagrangian current-quark (parton) mass;
$\Gamma_\mu^a(k,q)$ is the dressed gluon-quark vertex; 
and $Z_{1,2}(\zeta,\Lambda)$, are, respectively, the gluon-quark vertex and quark wave function renormalisation constants.  This fully dressed propagator is mathematically connected to the current-quark in Eq.\,\eqref{QCDdefine} via summation of the Dyson series of quark self-energy diagrams depicted in Fig.\,\ref{FigQuarkDyson}.  The solution has the following Poincar\'e covariant form:
\begin{equation}
S(k;\zeta) = \frac{1}{i \gamma\cdot k A(k^2;\zeta) + B(k^2;\zeta)}
= \frac{Z(k^2;\zeta)}{i\gamma\cdot k + M(p^2)}\,.
\end{equation}

Attempts to compute $S(k;\zeta)$ for light-quarks began with the birth of QCD \cite{Lane:1974he, Politzer:1976tv}.  They became increasingly sophisticated as proficiency grew with formulating and solving Eq.\,\eqref{EqQuarkDyson} \cite{Huber:2018ned, Fischer:2018sdj, Roberts:2020udq, Roberts:2020hiw, Qin:2020rad}; and the first computations using lattice QCD were completed roughly twenty years ago \cite{Skullerud:2000un}.  Today, continuum and lattice QCD agree that even in the absence of Higgs couplings into QCD, the then massless partonic quarks in Eq.\,\eqref{QCDdefine} acquire a momentum dependent mass function which is large at infrared momenta, see \emph{e.g}.\, Refs.\,\cite{Zhang:2009jf, Aguilar:2018epe, Oliveira:2018lln, Gao:2020qsj, Yang:2020crz}.  This is dynamical chiral symmetry breaking (DCSB): perturbatively massless quarks acquire a large infrared mass through interactions with their own gluon field.  The potential for such an outcome to be realised has been known for sixty years \cite{Nambu:1961tp, Nambu:2011zz}, but it is no less important for that because this is the first time the phenomena has been demonstrated in a fully-interacting four-dimensional quantum field theory that is possibly well-defined.

\begin{figure}[!t]
\begin{center}
\includegraphics[width=0.55\textwidth]{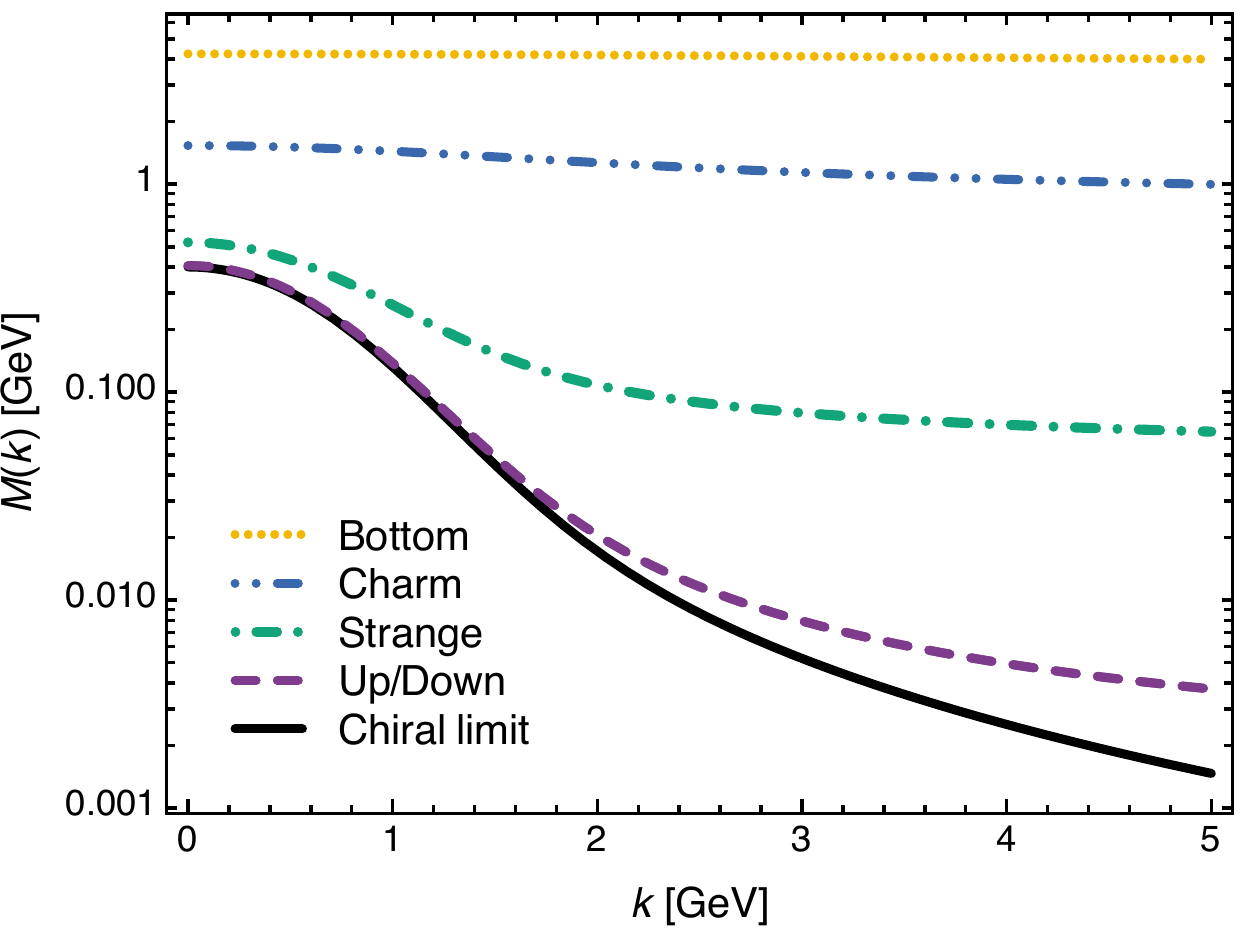}
\end{center}
\caption{\label{FigMp2}
Dressed-quark mass function, $M(k)$, obtained as the nonperturbative solution of the QCD gap equation using a modern kernel \cite{Binosi:2016wcx}: with $\zeta_2 = 2\,$GeV,
$M_0(0) =0.40\,$GeV;
$M_{u/d}(0) =0.406\,$GeV, $M_{u/d}(\zeta_2) =M_0(\zeta_2)+0.0034\,$GeV;
$M_{s}(0) =0.526\,$GeV, $M_{s}(\zeta_2) =M_0(\zeta_2)+0.095\,$GeV;
$M_{c}(\zeta_2) =1.27\,$GeV;
$M_{b}(\zeta_2) =4.18\,$GeV.
(Curves and image courtesy of D.\,Binosi.)
}
\end{figure}
%The vertex I used is in the G4 class as defined in Eq. (11) of https://arxiv.org/pdf/1609.02568.pdf. Everything is renormalised at 2 GeV using the values you have provided: 0.0034, 0.095, 1.27, 4.18 GeV.
%%M_0(0)=0.400 [GeV]
%%M_u(0)=0.406 [GeV]
%%M_s(0)=0.526 [GeV]
%%M_c(0)=1.53 [GeV]
%%M_b(0)=4.24 [GeV]

The chiral limit mass function obtained using a modern kernel for the quark gap equation \cite{Binosi:2016wcx} is drawn in Fig.\,\ref{FigMp2}.  This function is essentially nonperturbative: no sum of a finite number of perturbative diagrams can produce $M_(0)(k^2) \not\equiv 0$ \cite[Sec.\,2.3]{Roberts:2015dea}.  Kindred families of curves have been obtained in many analyses, \emph{e.g}.\ Refs.\,\cite{Jain:1993qh, Ivanov:1998ms, Williams:2006vva, Serna:2018dwk}.  In all such studies, $M_0(0) \approx 0.4\,$GeV, which is a typical scale for the constituent quark mass used in phenomenologically successful quark models \cite{Giannini:2015zia, Plessas:2015mpa, Eichmann:2016yit}.   When Higgs couplings are reintroduced, the mass function becomes flavour dependent and its $k^2=0$ value is roughly the sum of $M_0(0)$ and the appropriate current-quark mass, as illustrated in Fig.\,\ref{FigMp2}.

Interesting, too, is a comparison between the quark and gluon RGI running masses: $M_0(k)$ and $m_g(k)$, respectively, which is made in Fig.\,\ref{Figmk}.  Evidently, scale breaking in the one-body sectors, enabled by the trace anomaly and driven by gauge sector dynamics, is expressed in commensurate infrared values for these mass functions.  Naturally, since it is the gauge-boson mass-squared which has scaling power $2$, \emph{i.e}.\ $m_g^2(k) \sim 1/k^2$ at ultraviolet momenta, compared with the quark mass function itself, $M_0(k)$ runs more quickly to zero.

Indeed, for subsequent use, it is important to highlight here that the chiral-limit dressed-quark mass function has the following ultraviolet behaviour \cite{Politzer:1976tv}:
\begin{equation}
\label{M0OPE}
M_0(k^2) \stackrel{k^2 \gg \zeta_H^2}{=}
\frac{2\pi^2 \gamma_m}{3}
\frac{-\langle \bar q q\rangle_0}{k^2 \ln\left[\frac{k^2}{\Lambda_{\rm QCD}^2}\right]^{1-\gamma_m}}\,,
\end{equation}
where $\langle \bar q q\rangle_0$ is the RGI chiral-limit quark condensate \cite{Brodsky:2010xf, Chang:2011mu, Brodsky:2012ku}.  On this large-$k^2$ domain, $B_0(k^2)\approx M_0(k^2)$.  The behaviour in Eq.\,\eqref{M0OPE} is uniquely determined by the interaction in QCD: no other interaction can produce this behaviour.
% and the behaviour is different with any other interaction.  
For instance, if the interaction is momentum independent, then $M_0(k^2) =\,$constant \cite{GutierrezGuerrero:2010md}; and if the exchanged boson propagates as $D^n(k^2)=[1/k^2]^{n>1}$ on $k^2 \gg \hat m_0^2$, then $M_0(k^2) \sim D^n(k^2)$ on this same domain.

It is now possible to explain the general spectroscopic success of the constituent-quark picture.  The mass of a hadron is a global, volume-integrated property.  Hence, calculated using bound-state methods in quantum field theory, its value is largely determined by the infrared size of the mass function of the hadron's defining valence quarks \cite{Qin:2019hgk}: integrating over volume focuses resolution on infrared properties of the quasiparticle constituents.  This feature is underscored by the fact that even a judiciously formulated momentum-independent interaction produces a fair description of hadron spectra \cite{Yin:2019bxe, Gutierrez-Guerrero:2019uwa}.  The necessary infrared scales are provided by the mass functions in Fig.\,\ref{Figmk}; and those scales are generated by the effective charge in Fig.\,\ref{FigalphaPI} augmented by the Higgs-generated current-quark masses.

\subsection{Nambu-Goldstone Bosons}
\label{SecNGBosons}
Amongst the ground-state pseudoscalar mesons, the $\pi$ and $K$ mesons are NG modes.  The $\eta$ and $\eta^\prime$ would also be NG modes if it were not for the non-Abelian anomaly \cite{Christos:1984tu}, whose magnitude is set by the scale of EHM \cite[Eq.\,(20)]{Bhagwat:2007ha}.  Hence, the magnitude of the $\eta^\prime-\eta$ mass splitting is a direct measure of emergent mass: $m_{\eta^\prime} - m_\eta \approx 0.41\,{\rm GeV} \sim m_0$.

It has been known for more than fifty years that the SM's NG bosons do not fit naturally into a mass pattern typical of CQMs.  For instance, whereas pseudoscalar meson masses in quark models increase linearly with growth of the explicit chiral symmetry breaking term in the CQM Hamiltonian, just like the mass of every other system, in the neighbourhood of QCD's chiral limit, it is the mass-squared of NG modes that rises linearly with the current-quark mass in Eq.\,\eqref{QCDdefine} \cite{GellMann:1968rz}.  In modern terms, for a NG mode defined by $f$, $\bar g$ valence-quark degrees-of-freedom \cite{Maris:1997tm, Qin:2014vya}:
\begin{equation}
\label{MGMOR}
f_{NG} \, m_{NG}^2 = (m_f^\zeta + m_g^\zeta) \rho_{NG}^\zeta\,,
\end{equation}
where $f_{NG}$ is the meson's leptonic decay constant, \emph{i.e}.\ the pseudovector projection of the meson's wave function onto the origin in configuration space, and $\rho_{NG}^\zeta$ is the pseudoscalar analogue.  For ground-state pseudoscalar mesons, both $f_{NG}$, $\rho_{NG}^\zeta$ are order parameters for chiral symmetry breaking.  The Poincar\'e-covariant wave function of a pseudoscalar meson can be written
\begin{subequations}
\label{NGBSWF}
\begin{align}
\chi_{PS}^{f\bar g}(k,P) & = S_f(p_1) \Gamma_{PS}^{f\bar g}(k,P) S_{g}(p_2) \,, \\
\Gamma_{PS}^{f\bar g}(k,P) & = i \gamma_5 \left[ E_{PS}^{f\bar g}(k,P) +
\gamma\cdot P F_{PS}^{f\bar g}(k,P) + \gamma\cdot k G_{PS}^{f\bar g}(k,P) + \sigma_{\mu\nu}k_\mu P_\nu H_{PS}^{f\bar g}(k,P) \right]\,,
\end{align}
\end{subequations}
where
$P=p_1-p_2$ is the bound-state total-momentum and $k=(1-\eta) p_1 + \eta p_2$, $\eta\in[0,1]$, is the relative-momentum.  (Hereafter, for notational simplicity, the dependence on $\zeta$ is not indicated explicitly unless for a specific purpose.)

\begin{figure}[t]
\centerline{%
\includegraphics[clip, width=0.67\textwidth]{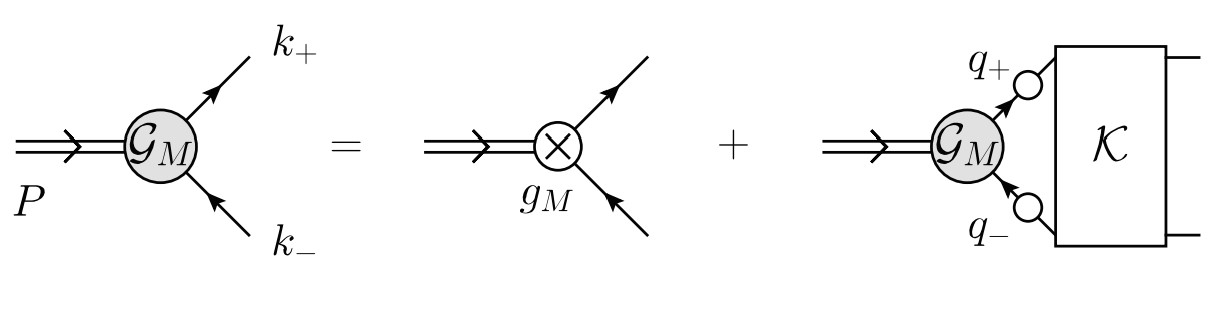}}
\caption{\label{FigMesonBSE}
Inhomogeneous Bethe-Salpeter equation (BSE) for a colour-singlet vertex, $\mathpzc{G}_M(k;P)$.  The channel is defined by the inhomogeneity, \emph{e.g}.\ with $\mathpzc{g}_M = \tfrac{1}{2} \tau^i \gamma_5\gamma_\mu$, where $\tau^i$ is a Pauli matrix associated with SU$(2)$ flavour, one gains access to all states that communicate with an isovector axial-vector probe, such as the pion and $a_1$ meson.  The interaction between the dressed valence-constituents is completely described by the two-particle-irreducible (2PI) scattering kernel, $\mathpzc{K}$.  The appearance of the dressed-quark propagator (solid line with open circle) highlights that the BSE must be solved in tandem with the quark gap equation, Fig.\,\ref{FigQuarkDyson}.
%, whose solution in a given channel yields the masses and Bethe-Salpeter amplitudes for every channel-coupled bound-state:
}
\end{figure}

Eq.\,\eqref{MGMOR} has many corollaries, \emph{e.g}.\ Refs.\,\cite{Holl:2004fr, Holl:2005vu, Qin:2014vya, Ballon-Bayona:2014oma}, not least of which is the fact that NG boson masses must vanish in the absence of Higgs couplings into QCD.  As highlighted by Fig.\,\ref{F1CDR}, this last feature means that the mass-scale which characterises all visible matter is hidden in $\pi$ and $K$ mesons and its manifestation in the physical $\pi$ and $K$ mesons is very different from that in all other hadrons.

These two quite particular consequences of EHM can be understood by studying the colour-singlet axial-vector vertex, which may be obtained by solving an inhomogeneous Bethe-Salpeter equation (BSE) \cite{Salpeter:1951sz, Nakanishi:1969ph} of the type depicted in Fig.\,\ref{FigMesonBSE}.  Both owe to the Ward-Green-Takahashi identity satisfied by the axial-vector vertex, which is a basic expression of chiral symmetry and the pattern by which it is broken in QCD:
\begin{equation}
P_\mu \Gamma_{5\mu}^{fg}(k,P)
= S_f^{-1}(p_1) i \gamma_5 +  i \gamma_5 S_g^{-1}(p_2)
 - \, i\,[m_f^\zeta+m_g^\zeta] \,\Gamma_5^{fg}(k,P)
\,,
\label{avwtimN}
\end{equation}
where:
%$P=p_1-p_2$ is the total-momentum entering the vertex and $k=(1-\eta) p_1 + \eta p_2$, $\eta\in[0,1]$, is the relative-momentum; and
$\Gamma_5^{fg}(k,P)$ is the associated pseudoscalar vertex (four-point Schwinger function).
Eq.\,\eqref{MGMOR} states that in the presence of Higgs-quark couplings, the actual mass of any pseudoscalar meson results from constructive interference between Higgs-boson effects and EHM.

Considered in the chiral limit, Eq.\,\eqref{avwtimN} can be used to show that a necessary and sufficient condition for the existence of NG modes is \cite{Maris:1997tm, Qin:2014vya}
\begin{equation}
\label{GTRE}
f_{PS}^0 E_{PS}^0(k;0) = B_0(k^2)\,.
\end{equation}
A rudimentary form of this identity can be found in Ref.\,\cite{Nambu:1961tp} and the first sketch of a proof appropriate to QCD was given in Ref.\,\cite{Delbourgo:1979me}.  This identity is remarkable and revealing.
First, it is a mathematical statement of equivalence between the pseudoscalar two-body and matter-sector one-body problems in chiral-limit QCD.  These problems are normally considered to be completely independent.
Second, it shows that the most direct expressions of EHM in the SM are located in the properties of the massless NG modes.
It is worth reiterating here that $\pi$- and $K$-mesons are indistinguishable in the absence of Higgs couplings.  At realistic Higgs couplings, $\pi$ and $K$ observables are windows onto EHM and its modulation by the Higgs boson.  Phrased differently, there are two mass generating mechanisms in the SM and $\pi$ and $K$ properties provide clear and direct access to both.

At this point, it is worth returning to Eq.\,\eqref{anomalypion}.  If one insists on working with a partonic basis, then a straightforward understanding of this identity and its reconciliation with Eq.\,\eqref{anomalyproton} seems impossible; at least, no approach from that direction has yet achieved the goal.

A different track is described in Ref.\,\cite{Roberts:2016vyn}.  Namely, $m_\pi^2$ can be calculated by solving a Bethe-Salpeter equation (BSE) of the type illustrated in Fig.\,\ref{FigMesonBSE}.  This is a scattering problem.  In the chiral limit and considering partonic degrees of freedom, two massless fermions interact via massless-gluon exchange, \emph{viz}.\ the initial system is massless; and it stays massless at every order in perturbation theory.
However, any complete analysis of the scattering process involves the summation of a countable infinity of one-body dressings, using Eq.\,\eqref{EqQuarkDyson}, and two$\,\to\,$two scatterings, via the BSE in Fig.\,\ref{FigMesonBSE}.  At $\zeta=\zeta_H$, the kernels are naturally built using a dressed-parton basis, \emph{i.e}.\ from valence-quark quasiparticles interacting via the exchange of quasiparticle gluons, each of which has a dynamically generated running mass.
Now using Eq.\,\eqref{GTRE}, one can prove algebraically \cite{Munczek:1994zz, Bender:1996bb} that in the chiral limit, at any order in a symmetry-preserving construction of the kernels for the gap- and BS-equations, there is an exact cancellation between the mass-generating effect of dressing the valence-quark and -antiquark, which produces the chiral limit mass function in Fig.\,\ref{FigMp2} for both fermions, and the attraction produced by the scattering events.  This mathematical identity guarantees that the simple, originally massless system becomes a complex bound system, with a nontrivial wave function attached to a pole in the scattering matrix which remains at $P^2=0$.  This entails
\begin{equation}
\Gamma_{5\mu}^0(k,P) \stackrel{P^2\simeq 0}{=} 2 f_{PS}^0 \frac{P_\mu}{P^2} \Gamma_{PS}^0(k,P)\,;
\end{equation}
hence, the bound-state is also massless.

%%%\begin{subequations}
%%%\begin{align}
%%%\langle \pi(q) | \Theta | \pi(q) \rangle & \stackrel{\zeta \gg m_0}{=}
%%%\langle \pi(q) |\left[ \tfrac{1}{4} \beta(\alpha(\zeta)) G^{a}_{\mu\nu}G^{a}_{\mu\nu}
%%%+ [1+\gamma(\alpha(\zeta))]\sum_{i=u,u,d,} m_i^\zeta \, \bar q_i q_i \right] | \pi(q) \rangle \\
%
%%%& \stackrel{\zeta \lesssim m_p}{=}
%%%\langle \tilde p(P) |\left[
%%%{\mathpzc D}_3 + {\mathpzc I}_3 \right] | \tilde p (P) \rangle \\
%
%%%{\mathpzc D}_3 & = \sum_{i=u,u,d} M_f(\zeta) \bar{\mathpzc Q}_{\,f}(\zeta) {\mathpzc Q}_{\,f}(\zeta)\,,
%%%\quad
%%%{\mathpzc I}_3 = \tfrac{1}{4} \left[\beta(\alpha(\zeta)) {\mathpzc G}^{a}_{\mu\nu}(\zeta){\mathpzc G}^{a}_{\mu\nu}(\zeta)\right]_{2PI}
%%%\end{align}
%%%\end{subequations}

These statements can be written as follows:
\begin{subequations}
\begin{align}
\langle \pi_0(P) | \Theta_0 | \pi_0(P) \rangle & \stackrel{\zeta \gg \zeta_H}{=}
\langle \pi_0(P) | \tfrac{1}{4} \beta(\alpha(\zeta)) G^{a}_{\mu\nu}G^{a}_{\mu\nu}
 | \pi_0(P) \rangle \to
 \stackrel{\zeta \simeq \zeta_H }{=}
\langle \pi_0(P) |\left[
{\mathpzc D}_2 + {\mathpzc I}_2 \right] |  \pi_0 (P) \rangle \\
{\mathpzc D}_2 & = \sum_{{\mathpzc f}=u,\bar d} M_{\mathpzc f}(\zeta) \bar{\mathpzc Q}_{\,{\mathpzc f}}(\zeta) {\mathpzc Q}_{\,{\mathpzc f}}(\zeta)\,,
\quad
{\mathpzc I}_2 = \tfrac{1}{4} \left[\beta(\alpha(\zeta)) {\mathpzc G}^{a}_{\mu\nu}(\zeta){\mathpzc G}^{a}_{\mu\nu}(\zeta)\right]_{2PI}\,,
\end{align}
\end{subequations}
which describes the transformation of the parton-basis chiral-limit expression into a new structure, written in terms of nonperturbatively-dressed quasiparticles, with dressed-quarks denoted by ${\cal Q}$ and the dressed-gluon field strength tensor by ${\cal G}$.
Here, the first term is positive: it realises the one-body-dressing content of the trace anomaly, whose reality is demonstrated by the chiral-limit mass function in Fig.\,\ref{FigMp2}.  The second term is negative because the net effect of interactions between the quark quasiparticles is attraction.  This term, too, has acquired a mass scale from the gluon- and quark-propagators; and owing to Eq.\,\eqref{GTRE}, it precisely cancels $\langle \pi_0(P)|{\mathpzc D}_2|  \pi_0 (P) \rangle$.

Away from the chiral limit for NG modes, the cancellation is incomplete and one arrives at Eq.\,\eqref{MGMOR}.  Similar destructive interference takes place in other systems, like the $\rho$-meson and proton; but in these cases, no symmetry ensures complete cancellation.  Consequently, as revealed mathematically when solving bound-state integral equations, all other hadron masses have values that are commensurate in magnitude with the strength of the scale anomaly in the solution of the gluon and quark one-body problems, \emph{i.e}.\ accounting for the number of valence quasiparticles, on the GeV scale. The combination of outcomes described here resolves the dichotomy expressed by the union of Eqs.\,\eqref{anomalyproton} and \eqref{anomalypion} and its analogues.

%% file: S3_Continuum.tex
\section{Pion and Kaon Distribution Amplitudes}
\label{sec:PiKWF}
\subsection{Essentials of Light-Front Wave Functions}
\label{sec:ELFWFs}
\begin{figure}[t]
\centerline{%
\includegraphics[clip, width=0.4\textwidth]{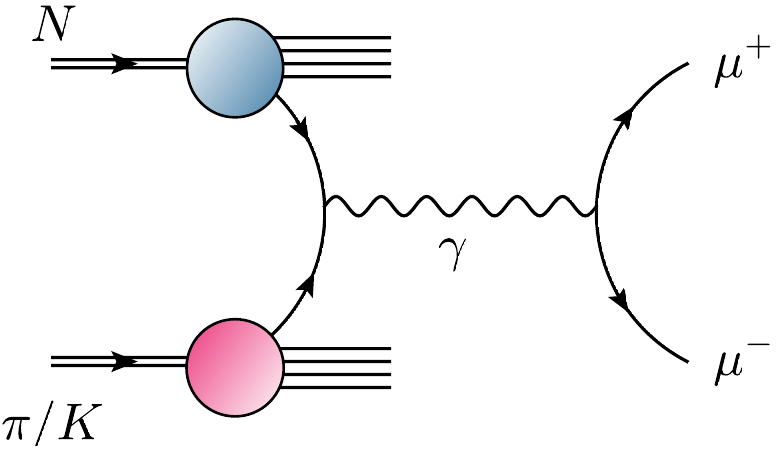}}
\caption{\label{FigDY}
Drell-Yan process: with appropriately chosen kinematics \cite{Drell:1970wh}, meson-nucleon collisions that produce lepton pairs with large invariant mass provide access to momentum distribution functions within the initial-state hadrons. (Image courtesy of D.\,Binosi.)}
\end{figure}

If one seeks to describe a given hadron's measurable properties in terms of the probabilities typical of quantum mechanics, then the hadron's LFWF, $\psi_H(x,\vec{k}_\perp;P)$, takes a leading role.  Here \cite{Coester:1992cg, Brodsky:1997de}: $P$ is the total four-momentum of the system, $x$ is the light-front longitudinal fraction of this momentum, and $\vec{k}_\perp$ is the light-front perpendicular component of $P$.
In principle, this LFWF is an eigenfunction of a QCD Hamiltonian defined at fixed light-front time and may be obtained by diagonalisation thereof \cite{Brodsky:1989pv}.  It is also invariant under Lorentz boosts \cite{Coester:1992cg, Brodsky:1997de}.  This means that when solving bound-state scattering problems using a light-front formulation, one never encounters compressed or contracted objects \cite{PhysicsTodayWeisskopf}.  As an example, the cross-section for the meson+proton Drell-Yan (DY) process illustrated in Fig.\,\ref{FigDY} is the same whether the proton is at rest or moving.

A primary obstacle on the path to a direct computation of a hadron's LFWF is the need to construct a sound approximation to QCD's light-front Hamiltonian.  This is made complicated by, \emph{inter alia}, the necessity of solving complex constraint equations along the way \cite{Heinzl:2000ht}.  The challenge is amplified if one elects to tackle the problem of expressing $\psi$ using a partonic basis, maintaining a connection to perturbative QCD, in which case a Fock-space decomposition of the LFWF is typically introduced.  The coefficient function attached to a given $n$-particle basis vector in that expansion  represents the probability amplitude for finding these $n$ partons in the hadron with momenta $\{ (x_i,k_{\perp i}) \, |\, i=1,\ldots,n\}$, constrained by requiring conservation of total momentum.  As noted above, such methods have not yet succeeded in describing EHM in QCD's gauge and matter sectors.  A contemporary perspective on the direct approach is presented in Ref.\,\cite{Hiller:2016itl}.

An alternative is to use the covariant DSE framework, compute the hadron's Poincar\'e-covariant Bethe-Salpeter wave function, $\chi$, and then project this object onto the light front.  Such an approach was used elsewhere \cite{tHooft:1974pnl} in analysing a local U$(N_c)$ gauge theory in two dimensions, with $N_c$ very large.  This scheme was shown to be practicable for QCD in Ref.\,\cite{Chang:2013pq}.  It delivers a LFWF expressed in the quasiparticle basis defined by the choice of renormalisation scale, $\zeta$.

One of the strengths of an approach that draws connections with hadron LFWFs is made manifest by observing that the distribution amplitudes (DAs) which feature in formulae describing hard exclusive processes and the distribution functions (DFs) that characterise hard inclusive reactions can both be written directly in terms of the LFWF \cite{Brodsky:1989pv}, respectively:
\begin{equation}
\label{PDFsPDAsLFWF}
\varphi_H(x;\zeta)  \propto \int^\zeta d^2k_\perp\, \psi_H(x,\vec{k}_\perp;P)\,,\quad
{\mathpzc q}^H(x;\zeta)  \propto \int^\zeta d^2k_\perp\, |\psi_H(x,\vec{k}_\perp;P)|^2,
\end{equation}
where $\zeta$ is the scale at which the hadron is being resolved.  This $\zeta$-dependence highlights that the perceived attributes of a hadron depend upon the scale at which it is observed.  It does not affect measured cross-sections; instead, this energy scale decides the optimal choice for the degrees of freedom required to solve the problem and express an insightful interpretation.

\begin{figure}[!t]
\hspace*{-1ex}\begin{tabular}{ll}
{\sf A} & {\sf B} \\[-2ex]
\hspace*{-4em}\includegraphics[clip, width=0.58\textwidth]{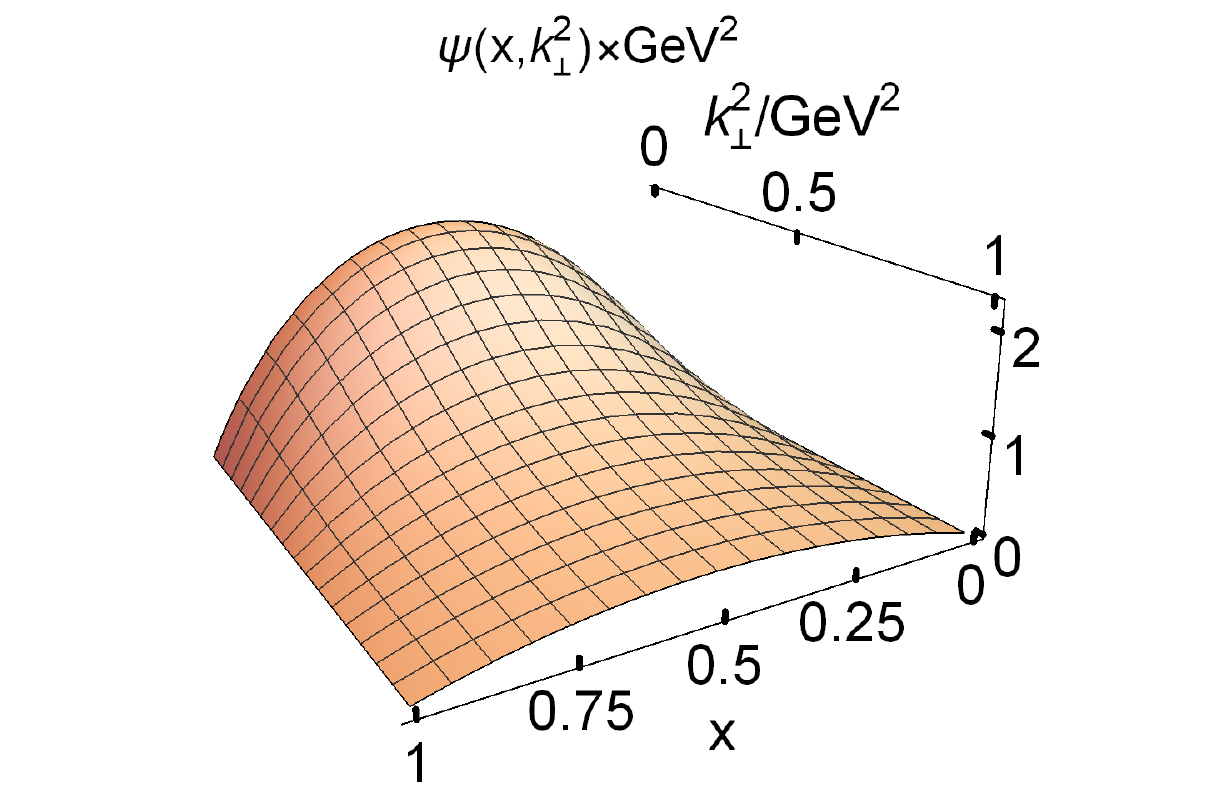} &
\includegraphics[clip, width=0.48\textwidth]{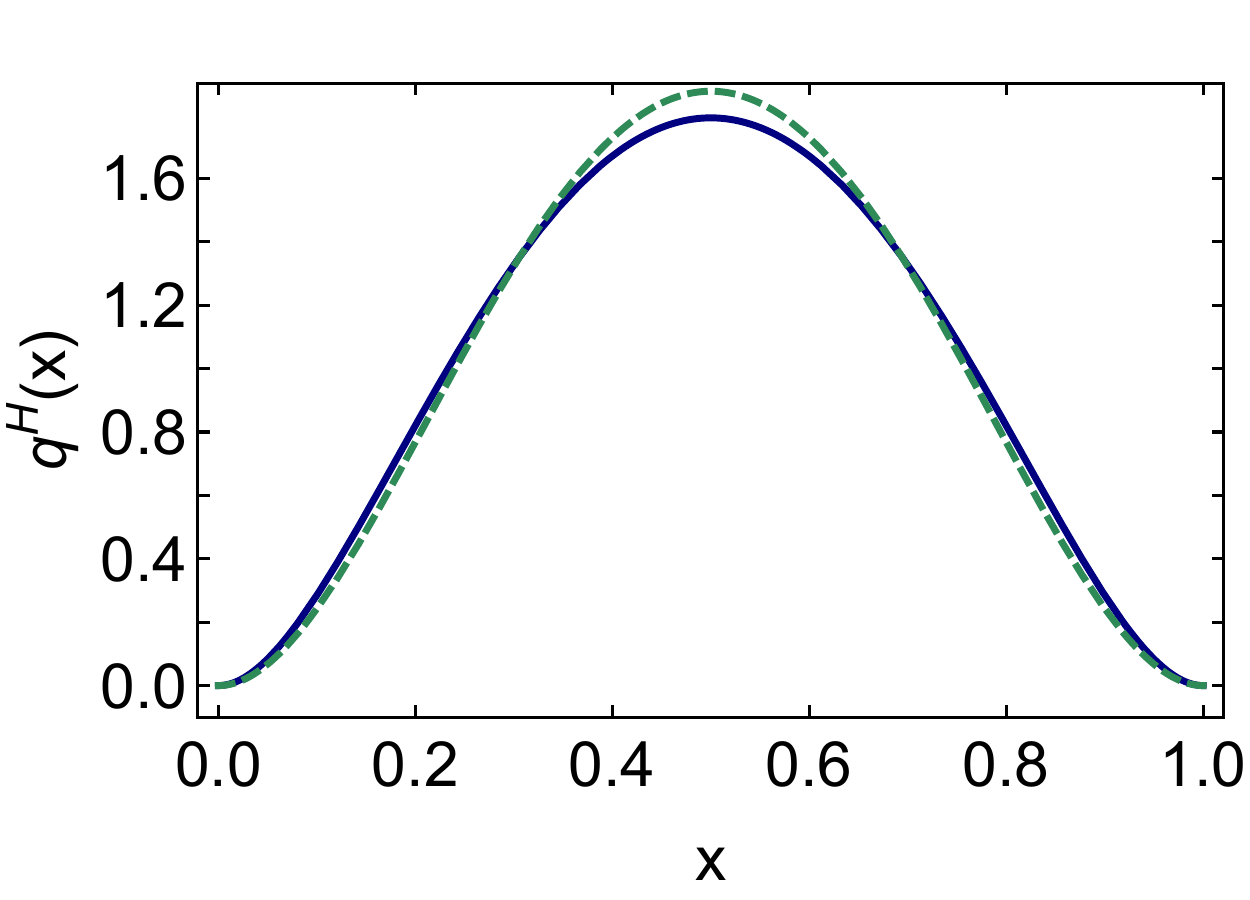}
\end{tabular}
\caption{\label{FigPsi}
\emph{Left panel}\,--\,{\sf A}.  Light-front wave function in Eq.\,\eqref{HLFWF}.
\emph{Right panel}\,--\,{\sf B}.  Associated DF, Eq.\,\eqref{qHDF} -- solid blue curve; and square of the DA, Eq.\,\eqref{qHDA} -- dashed green curve.
Curves in both panels computed with $M=0.4\,$GeV, $\delta \simeq 0$.
}
\end{figure}

The discussion herein focuses chiefly on the properties of $\pi$ and $K$ mesons; consequently, two-body quasiparticle LFWFs are of primary importance.  Thus, for future use, consider the following model:
\begin{equation}
\label{HLFWF}
\psi_H(x,k_\perp;\zeta_H) = \frac{n_\psi\, x(1-x)}{[M^2 x(1-x) + M^2 + k_\perp^2]^{1+\delta}}\,,
\quad
\int \frac{dx d^2 k_\perp}{16\pi^3} | \psi_H(x,k_\perp;\zeta_H) |^2 = 1\,,
\end{equation}
where $M$ is a mass whose size is assumed to be set by EHM and $n_\psi$ is the normalisation constant.  For $\delta=0$, this LFWF exhibits the large-$k_\perp^2$ scaling behaviour of a leading-twist two-body wave function in QCD \cite[Eq.\,(2.15)]{Lepage:1980fj}.  The result obtained with $M=0.4\,$GeV and $\delta\simeq 0$ is drawn in Fig.\,\ref{FigPsi}\,A.

Working with Eq.\,\eqref{HLFWF}, the hadron's DF is
\begin{equation}
\label{qHDF}
{\mathpzc q}^H(x;\zeta_H) = \int \frac{d^2 k_\perp}{16\pi^3} | \psi_H(x,k_\perp;\zeta_H) |^2\, ;
\end{equation}
and the result obtained with $\delta\simeq 0$ is drawn as the solid blue curve in Fig.\,\ref{FigPsi}\,B.  For comparison, Fig.\,\ref{FigPsi}\,B also depicts $\tilde \varphi_H^2(x;\zeta_H)$ as the dashed green curve, where $\tilde \varphi_H(x;\zeta_H)$ is the associated DA after normalisation adjustment:
\begin{equation}
\label{qHDA}
\tilde \varphi_H(x;\zeta_H) = n_{\varphi^2} \int\frac{d^2 k_\perp}{16\pi^3} \psi_H(x,k_\perp;\zeta_H) \,, \quad \int_0^1 dx\, \tilde \varphi_H^2(x;\zeta_H) = 1\,.
\end{equation}
The two curves in Fig.\,\ref{FigPsi}\,B possess the same functional $x$-dependence at the endpoints and, using a ${\mathpzc L}_1$ measure \cite{Rudin:1987}, they differ by just 4.1\%.  Using $\delta=1, 2$, the differences are, respectively, 4.4\% and 4.7\%.

The purpose of these comparisons is to illustrate an important fact.  Namely, a factorised approximation to $\psi_{H}(x,k_\perp;\zeta_H)$ is reliable for integrated quantities when the wave function has fairly uniform support \cite{Xu:2018eii}.  It is worth recalling here that $\zeta_H$ is that scale at which the dressed quasiparticles obtained mathematically from the valence quark-parton and antiquark-parton degrees of freedom embody all properties of a given hadron; in particular, they carry all its light-front momentum.  (This understanding of $\zeta_H$ has long been a characteristic of well-founded models, \emph{e.g}.\ Refs.\,\cite{Bentz:1999gx, Dorokhov:2000gu, Davidson:2001cc, Nam:2012vm, Lan:2019rba}.) Consequently, at the hadronic scale, one can reliably exploit the approximation
\begin{equation}
\psi_H(x,k_\perp^2;\zeta_H) \approx \tilde\varphi_{H}(x;\zeta_H) \tilde\psi_{H}(k_\perp^2;\zeta_H)\,,
\end{equation}
where the optimal choice for $ \tilde\psi_{H}(k_\perp^2;\zeta_H)$ is influenced by the application, to arrive at the result
\begin{equation}
\label{PDFeqPDA2}
{\mathpzc q}^H(x;\zeta_H) \approx \tilde\varphi_{H}^2(x;\zeta_H)\,,
\end{equation}
and be certain that the level of accuracy exceeds the precision of foreseeable experiments.

Parton splitting effects entail that Eq.\,\eqref{PDFeqPDA2} is not valid on $\zeta>\zeta_H$.  Nonetheless, since the evolution equations for both DFs and DAs are known \cite{Dokshitzer:1977sg, Gribov:1972ri, Lipatov:1974qm, Altarelli:1977zs, Lepage:1979zb, Efremov:1979qk, Lepage:1980fj}, the changing connection is readily tracked.  It follows that DAs and DFs are complementary; and in being accessed via different processes, they open different windows onto similar fields of view.  Hence, the simultaneous analysis of both, yielding predictions for seemingly disparate observables, provides opportunities for independent checks on the framework employed and insights drawn.  For instance, any prediction of EHM-induced broadening in the leading-twist DA of a NG boson must be matched by kindred manifestations in its DF.

\subsection{Pion Distribution Amplitude}
\label{SecDApion}
After its introduction \cite{Lepage:1979zb, Efremov:1979qk, Lepage:1980fj}, interest in the pion's leading-twist DA rapidly became intense.  Summaries of the story may be found in several topical reviews \cite{Chernyak:2014wra, Horn:2016rip} augmented by recent analyses \cite{Stefanis:2020rnd, Qian:2020utg}.  Today, following these forty years of effort, continuum phenomenology and theory agree that the pion's DA at hadronic scales is a broad, concave function, possessing greater support in the neighbourhood of its endpoints and therefore flatter than the asymptotic profile \cite{Lepage:1979zb, Efremov:1979qk, Lepage:1980fj}:
\begin{equation}
\label{phias}
\varphi_{\rm as}(x) = 6 x(1-x)\,.
\end{equation}
Quantitative differences in the pointwise expression of these features do remain; but those discrepancies are likely to disappear when all computational frameworks are required to provide a sound, unified description of an equally diverse array of phenomena.

In order to explicate these observations, suppose that one has obtained the solution of the homogeneous BSE derived from the equation drawn in Fig.\,\ref{FigMesonBSE}, \emph{i.e}.\ $\chi_\pi(k,P;\zeta_H)$ in Eq.\,\eqref{NGBSWF}, then the leading-twist DA for the $u$-quark in the $\pi^+$ may be obtained as follows \cite{Chang:2013pq}:
\begin{equation}
f_\pi \,\varphi^u_{\pi}(x;\zeta_H)= N_c {\rm tr}_{\rm D}Z_2(\zeta_H,\Lambda) \int_{dk}^\Lambda \delta_n^x(k_\eta)\gamma_5 \gamma\cdot n \chi_M(k_{\eta\bar\eta},P ;\zeta_H)\,. \label{varphiresult}
\end{equation}
Here
$N_c=3$; the trace is over spinor indices;
$\int_{dk}^\Lambda$ is a symmetry-preserving regularisation of the four-dimen\-sio\-nal integral, with $\Lambda$ the regularisation scale;
$\delta_n^x(k_\eta) = \delta(n\cdot k_\eta - x n\cdot P)$, $n$ is a light-like four-vector, $n^2=0$, with $n\cdot P = -m_\pi$ in the meson rest frame;
$k_{\eta\bar\eta}=[k_\eta + k_{\bar \eta}]/2$, $k_\eta = k+\eta P$, $k_{\bar\eta}=k-(1-\eta)P$;
and $f_\pi$ is the pion's leptonic decay constant, so
\begin{align}
\int_0^1 dx\, \varphi^u_{\pi}(x;\zeta_H) = 1\,.
\end{align}
The companion DA for the $d$-antiquark is
\begin{equation}
\varphi_{\pi}^{\bar d}(x;\zeta_H) =  \varphi_{\pi}^{u}(1-x;\zeta_H)\,.
\label{phisymmetry}
\end{equation}

\begin{figure}[!t]
\hspace*{-1ex}\begin{tabular}{lcl}
{\sf A} & \hspace*{1em} & {\sf B} \\[-2ex]
\includegraphics[clip, width=0.46\textwidth]{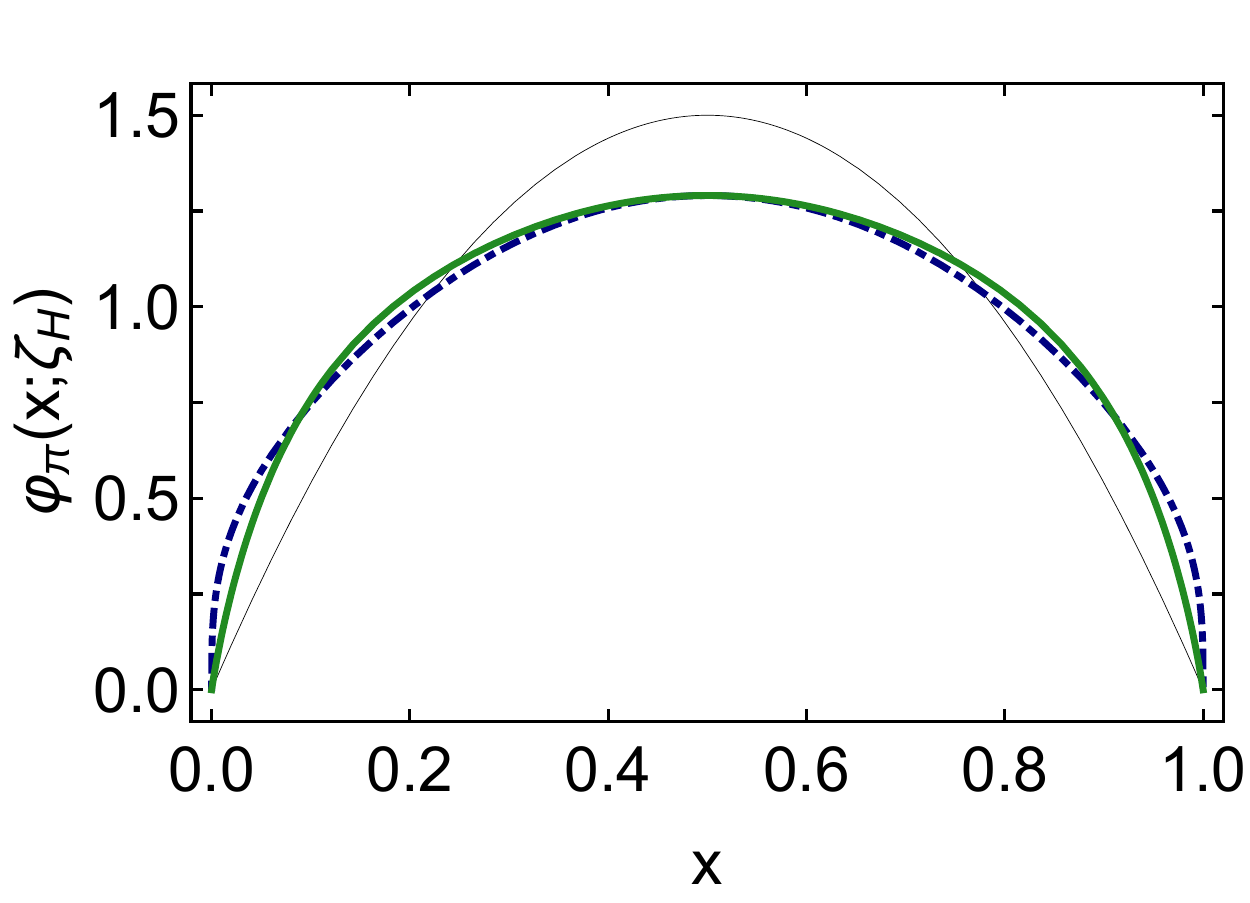} & &
\includegraphics[clip, width=0.46\textwidth]{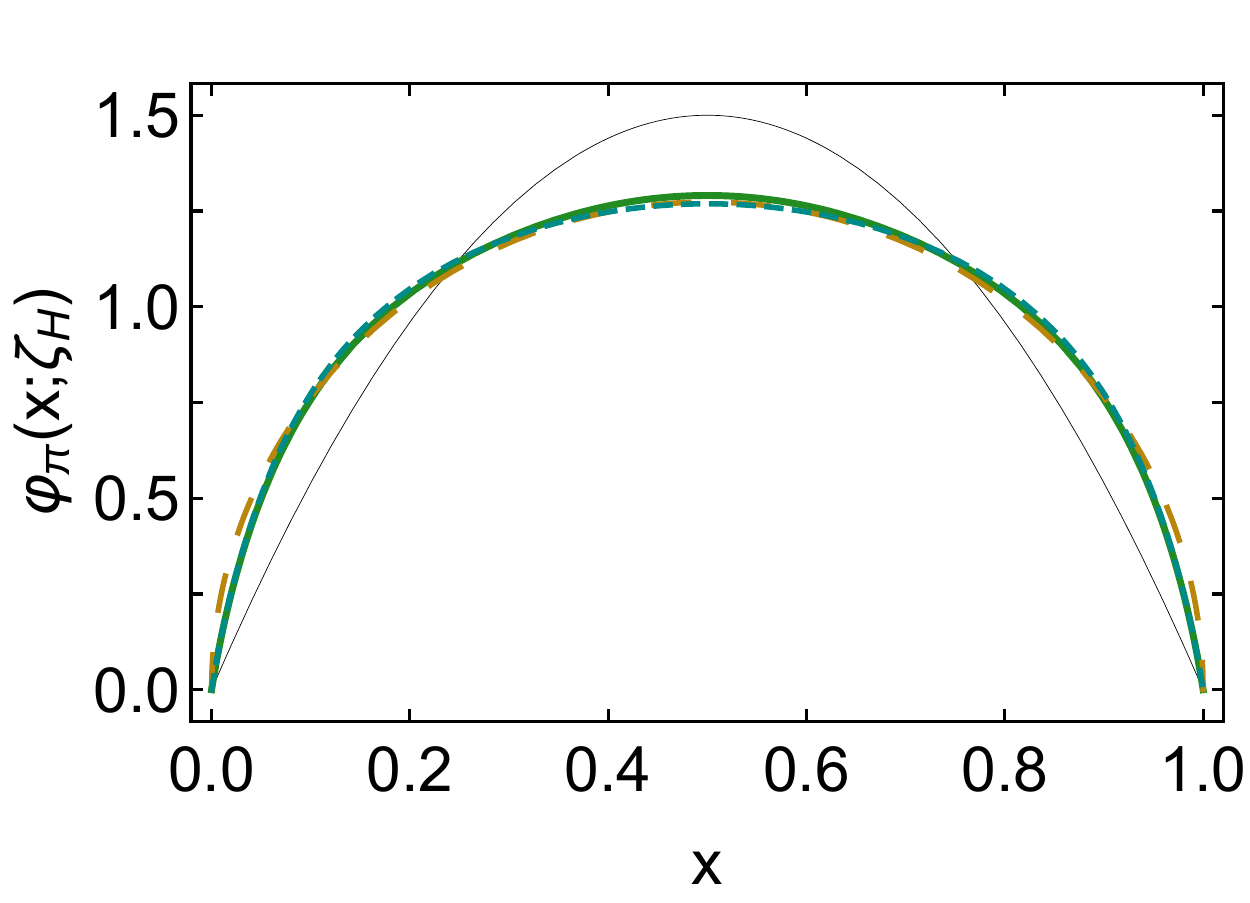}
\end{tabular}
\caption{\label{FigphiDB}
\emph{Left panel}\,--\,{\sf A}.  Dot-dashed blue curve -- pion DA in Eq.\,\eqref{varphipialpha}; and solid green curve -- DA in Eq.\,\eqref{varphipilinear}.
\emph{Right panel}\,--\,{\sf B}.
%Associated DF, Eq.\,\eqref{qHDF} -- solid blue curve; and square of the DA, Eq.\,\eqref{qHDA} -- dashed green curve.
Again, solid green curve is the DA in Eq.\,\eqref{varphipilinear}.  Comparisons are drawn with:
AdS/QCD model, $\varphi(x) = (8/\pi)\sqrt{x(1-x)}$ -- long-dashed gold curve;
and
QCD sum rules result, Eq.\,\eqref{varphipiNS} -- short-dashed cyan curve.
In both panels, the thin black curve is $\varphi_{\rm as}(x) = 6 x (1-x)$.
}
\end{figure}

Naturally, the form of $\chi_\pi(k,P;\zeta_H)$ is determined by the intimately connected kernels of the gap and Bethe-Salpeter equations.  Much has been learnt about their structure in QCD during the past twenty-five years, with key steps along the road being marked by Refs.\,\cite{Chang:2009zb, Fischer:2009jm, Chang:2011ei, Binosi:2014aea, Williams:2015cvx, Binosi:2016rxz, Binosi:2016wcx, Qin:2020jig}.  The kernel used to compute a spectrum of mesons in Ref.\,\cite{Chang:2011ei}, which expresses crucial consequences of DCSB, was employed in Ref.\,\cite{Chang:2013pq} to predict the pion's simplest two-body dressed-quark DA.  Projection onto the light-front was achieved by exploiting perturbation theory integral representations (PTIRs) \cite{Nakanishi:1969ph} for the dressed-quark propagators and meson Bethe-Salpeter amplitude.  The DA was subsequently reconstructed from (typically) fifty Mellin moments:
\begin{equation}
\langle f(x)^m \rangle_\varphi = \int_0^1dx\, f(x)^m \,\varphi(x)\,,
\end{equation}
$f(x)=x$, using a basis of Gegenbauer polynomials whose degree was included in the optimisation procedure so as to minimise the number of basis vectors with a material contribution.  This procedure yielded convergence using just two polynomials in the series:
\begin{equation}
\label{varphipialpha}
\varphi_\pi^{\alpha_\pi}(x;\zeta_H) = 1.81 [x(1-x)]^{\alpha_\pi} \left[[1+a_2^\pi C_2^{(\alpha_\pi+1/2)}(1-2x)\right]\,,
\end{equation}
with degree $\alpha_\pi = 0.31+1/2$ and coefficient $a_2^\pi=-0.12$, which is drawn as the dot-dashed blue curve in Fig.\,\ref{FigphiDB}\,A.  Using a ${\mathpzc L}_1$ measure, this curve differs from $\varphi_{\rm as}(x)$ by 15\%.

Having thus arrived at a pointwise-accurate approximation to $\varphi_\pi(x;\zeta_H)$, one can readily reexpress it using any other basis.  QCD predicts that this DAs endpoint behaviour should by linear, \emph{viz}.\ the same as that of $\varphi_{\rm as}(x)$.  Accounting for this, Refs.\,\cite{Cui:2020dlm, Cui:2020tdf} used a functional form suggested by fits to distribution functions in order to obtain an improved representation:
\begin{equation}
\label{varphipilinear}
\varphi_\pi(x;\zeta_H) = 18.2 \, x (1-x) \left[ 1 - 2.33 \sqrt{x(1-x)} + 1.79 x(1-x) \right]\,,
\end{equation}
%% 18.2043 (1-x) x \left(1.7889 (1-x) x-2.32734 \sqrt{1-x} \sqrt{x}+1\right)
which is drawn as the solid green curve in Fig.\,\ref{FigphiDB}\,A.  
Using a ${\mathpzc L}_1$ measure, this curve differs from that in Eq.\,\eqref{varphipialpha} by 3.3\%. Furthermore, their low-order Mellin moments compare as follows:
\begin{equation}
\label{piDAmoments}
\begin{array}{l|cc}
 & \langle (1-2x)^2\rangle & \langle (1-2x)^4\rangle  \\ \hline
\mbox{Eq.\,\eqref{varphipialpha}} &  0.251 &  0.128 \\
\mbox{Eq.\,\eqref{varphipilinear}} & 0.242  &  0.117
\end{array}\,.
\end{equation}
Consequently, so far as foreseeable experiments are concerned, the curves are practically identical; and the reconstruction in Eq.\,\eqref{varphipilinear} is to be preferred because its endpoint behaviour is consistent with QCD.

It is worth comparing the prediction in Eq.\,\eqref{varphipilinear} with several other determinations.  To that end, the result obtained in an AdS/QCD model \cite{Brodsky:2006uqa}, $\varphi(x) = (8/\pi)\sqrt{x(1-x)}$, is drawn as the long-dashed gold curve in Fig.\,\ref{FigphiDB}\,B.  This model omits the physics of perturbative QCD; hence, the endpoint behaviour does not match that of $\varphi_{\rm as}(x)$.  In this case, compared with Eq.\,\eqref{varphipilinear}, the ${\mathpzc L}_1$ difference is 2.3\%.

QCD sum rules have also been used to estimate the pion DA via analyses of the neutral-pion electromagnetic transition form factor; and working within a Gegenbauer polynomial basis of degree $3/2$, a broad band of results is possible \cite{Stefanis:2020rnd}.  Considering this, one may begin with a favoured representation (Table~I, line~1 in Ref.\,\cite{Stefanis:2020rnd}), which yields these values for the DA moments in Eq.\,\eqref{piDAmoments}: $(0.269,0.117)$;
%%{0.268571, 0.116883}
then implement the procedure that leads from Eq.\,\eqref{varphipialpha} to Eq.\,\eqref{varphipilinear}, thereby obtaining the following form:
\begin{equation}
\label{varphipiNS}
\varphi_\pi^{\rm SR}(x;\zeta_H) = 18.2 \, x (1-x) \left[ 1 - 2.24 \sqrt{x(1-x)} + 1.59 x(1-x) \right]\,,
%%%18.2025 (1-x) x \left(1.58713 (1-x) x-2.23592 \sqrt{1-x} \sqrt{x}+1\right)
\end{equation}
which is depicted as the short-dashed cyan curve in Fig.\,\ref{FigphiDB}\,B.  The moments of this function are $(0.245,0.119)$, well within the uncertainty of the original estimate;
%%{0.244863, 0.118596}
and the ${\mathpzc L}_1$ difference between $\varphi_\pi^{\rm SR}$ and Eq.\,\eqref{varphipilinear} is just 1.2\%.  Evidently, the two results, obtained using very different means, are practically indistinguishable.  Moreover, the DA in Eq.\,\eqref{varphipilinear} is also associated with a sound description of the neutral-pion electromagnetic transition form factor \cite{Raya:2015gva}, something canvassed further below.

Complementing such conclusions, drawn from forty years of continuum analyses, the past three years have seen lattice-regularised QCD (lQCD) deliver preliminary results for the pointwise behaviour of pion and kaon DAs \cite{Zhang:2017bzy, Zhang:2020gaj}: the DAs obtained also show the dilation evident in Fig.\,\ref{FigphiDB}.  Earlier and continuing studies of lQCD results for low-order Mellin moments of pion and kaon DAs, \emph{e.g}.\, Refs.\,\cite{Segovia:2013eca, Braun:2015axa, Bali:2019dqc}, yield results that are indicative of such dilation, too.  Additional examination of these ongoing developments is provided in Sec.\,\ref{SeclQCDDA}.

\begin{figure}[!t]
\hspace*{-1ex}\begin{tabular}{lcl}
{\sf A} & \hspace*{1em} & {\sf B} \\[-2ex]
\includegraphics[clip, width=0.46\textwidth]{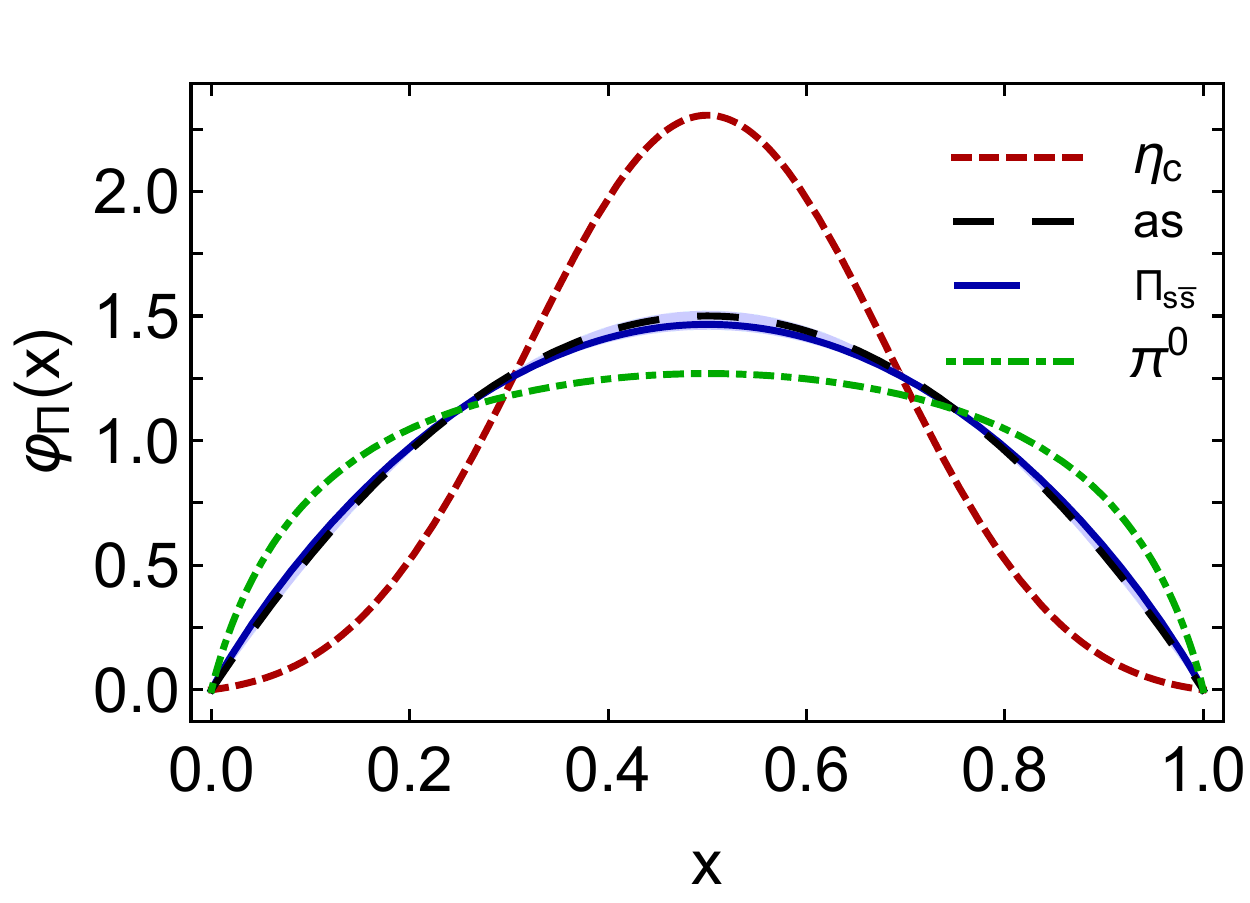} & &
\includegraphics[clip, width=0.46\textwidth]{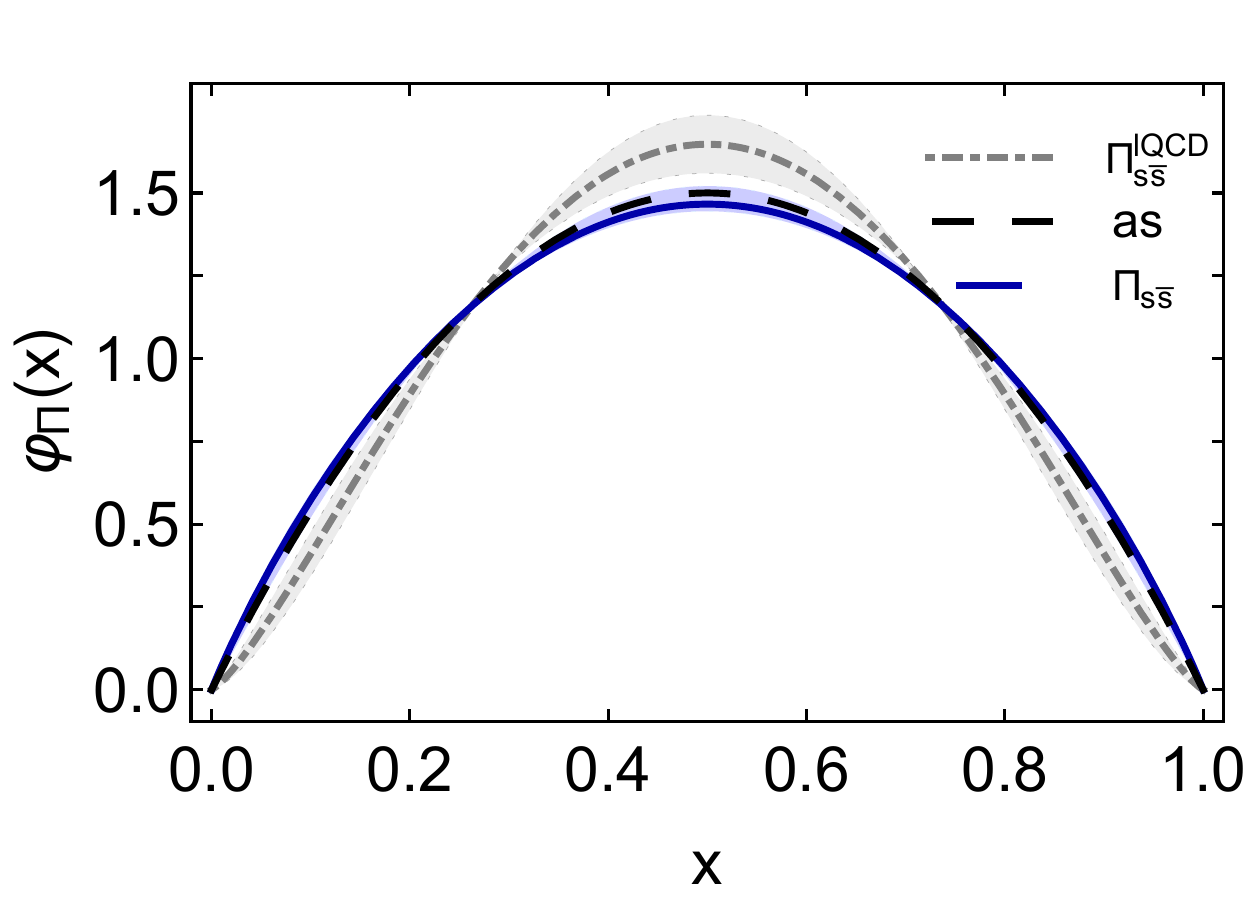}
\end{tabular}
\caption{\label{FigDACriticalMass}
\emph{Left panel}\,--\,{\sf A}.  Pseudoscalar meson DAs, $\varphi_\Pi(x;\zeta_H)$, computed for a range of current-quark masses.  The results are drawn from Refs.\,\cite{Ding:2015rkn, Chen:2018rwz, Cui:2020dlm, Cui:2020tdf}.  
Legend.
Dashed green ($\pi^0$) curve: Eq.\,\eqref{varphipilinear}.
Solid blue ($\Pi_{s\bar s}$) curve: DA of a fictitious pseudoscalar meson built from a valence-quark and -antiquark whose masses are chosen to match that of the $s$-quark, $\hat m_s$.  Namely, the mass of the analogous ground-state vector meson is approximately that of the $\phi$-meson.  (Reconstructed from Ref.\,\cite[Table~1]{Chen:2018rwz}: the associated light-blue band expresses the slightly asymmetric theory uncertainty in the leading Mellin moments.)
Dashed red curve: DA of the $\eta_c$ meson, built from a $c$-quark and its antimatter partner.
\emph{Right panel}\,--\,{\sf B}.  Comparison between continuum and lattice QCD results for the DA of the $\Pi_{s\bar s}$ system.
In both panels, the long-dashed black curve is $\varphi_{\rm as}(x) = 6 x (1-x)$.
}
\end{figure}

In order to learn something more about NG modes, it is useful to emphasise that the DAs in Fig.\,\ref{FigphiDB} are those for pseudoscalar mesons constituted from light valence degrees-of-freedom, whose properties are chiefly determined by the physics of EHM.  It is natural to study the response of these DAs to increasing the strength of Higgs-boson couplings into QCD.  This exercise was undertaken in Ref.\,\cite{Ding:2015rkn}, which considered the current-quark mass dependence of $S$-wave $q\bar q$ meson DAs and uncovered an important feature.  As highlighted by Fig.\,\ref{FigphiDB}, light-meson DAs are broad, concave functions.  At the other extreme, \emph{i.e}.\ mesons constituted from a valence-quark and -antiquark with degenerate current-quark masses that are far greater than $\Lambda_{\rm QCD}$, one has $\varphi_{Q\bar Q}(x;\zeta) \approx \delta(x-1/2)$.  Since meson DAs are smooth functions with unit normalisation, which must respond smoothly to increasing current-quark mass, it is reasonable to expect that there exists a current-quark mass, $m_{\rm cr}$, for which $\varphi_{q_{m_{\rm cr}}\bar q_{m_{\rm cr}}}(x;\zeta) \approx \varphi_{\rm as}(x)$.  Ref.\,\cite{Ding:2015rkn} verified this conjecture and found that $m_{\rm cr}$ lies in the neighbourhood of the $s$-quark current-mass, as may be seen in Fig.\,\ref{FigDACriticalMass}\,A.  The results were confirmed in studies of the elastic electromagnetic form factors of pseudoscalar mesons \cite{Chen:2018rwz} and $\gamma^\ast  \gamma \to \eta, \eta^\prime$ transition form factors \cite{Ding:2018xwy}.

More recently, lQCD calculations using large-momentum effective theory have delivered results for pseudoscalar meson DAs \cite{Zhang:2020gaj}.  Of particular interest in the present context is the DA obtained when the current-quark mass is set in the neighbourhood of the $s$-quark value, producing a bound-state mass $m_{s\bar s}=0.69\,$GeV.  The DA for this system is drawn as the grey curve within like colour bands in Fig.\,\ref{FigDACriticalMass}\,B.  Also depicted is the continuum prediction for the DA of a pseudoscalar meson bound-state with $m_{s\bar s}=0.69\,$GeV \cite{Chen:2018rwz}.  Plainly, continuum and lattice analyses in QCD agree upon the existence and value of $m_{\rm cr}$.  (Additional details are provided in connection with Fig.\,\ref{fig:x_dep_DA}.)

The curves in Fig.\,\ref{FigDACriticalMass}\,A answer a question, \emph{viz}.\ When does the Higgs mechanism begin to influence mass generation?  As already stated, the pointwise behaviour of the DAs for QCD's NG modes is largely formed by the mechanism of EHM.   On the other hand, the $\eta_c$ meson, built from a $c$-quark and its antimatter partner and with its DA being much narrower than $\varphi_{\rm as}$, feels the Higgs mechanism strongly.  Built from valence constituents with mass $\hat m_s$, the $\Pi_{s\bar s}$ system lies at the boundary: with a DA very similar to $\varphi_{\rm as}$, EHM and Higgs-boson couplings are playing a roughly equal role in forming the wave function.  It follows that comparisons between observables associated with truly light-quark bound-states and those involving $s$ quarks are ideally suited to exposing measurable signals of EHM in counterpoint
to Higgs-driven effects, \emph{i.e}.\ revealing Higgs-boson modulation of emergent mass.

\subsection{Kaon Distribution Amplitude}
\label{SecDAkaon}
With this appreciation of the importance of such comparisons, it is natural to turn toward the kaon.  Consider, therefore, the $K^+$, which is formed by one light valence $u$-quark and a heavier valence $\bar s$-quark.  The best available analyses indicate that $\hat m_s/\hat m_{\overline{ud}} \approx 27$, $m_{\overline{ud}} = (\hat m_u + \hat m_d)/2$ \cite{Zyla:2020zbs}.  On the other hand, regarding Fig.\,\ref{FigMp2}, $M_s(0)/M_{\overline{ud}}(0) \approx 1.3$.  Both these ratios differ from unity because of the Higgs mechanism for mass generation, but the ratio of current-quark masses is roughly 21-times larger than the ratio of constituent-like masses.   So whilst Higgs couplings into QCD have an enormous impact on partonic masses, they only appear to produce small modulations in the realm of EHM dominance, \emph{e.g}. $M_s(0)/M_{\overline{ud}}(0) \approx f_K/f_\pi$, where $f_{K,\pi}$ are the mesons' leptonic decay constants.  This being the case, how are Higgs couplings expressed in kaon DAs?

Attempts to constrain the kaon DAs, $\varphi_K^{u,\bar s}(x)$, have a long history, reaching back almost forty years \cite{Chernyak:1982it}.  Several qualitative features may be anticipated: (\emph{a}) whilst isospin symmetry in QCD means $\langle (1-2x) \rangle_\pi =0$, the large disparity between $u$- and $s$-quark current-masses entails $\langle (1-2x) \rangle_{K^+}^u >0$; and (\emph{b}) since the kaon is heavier than the pion, then $\langle (1-2x)^2 \rangle_{K^+}^u \leq \langle (1-2x)^2 \rangle_{\pi}$.  There has been measurable progress since the early analyses, and a survey of continuum and lattice results from the past decade \cite{Arthur:2010xf, Segovia:2013eca, Shi:2014uwa, Shi:2015esa, Horn:2016rip, Gao:2017mmp, Bali:2019dqc} supports the following conclusions $(\xi=1-2x)$:
\begin{equation}
\label{pionkaonPDA}
\langle [\xi,\xi^2]\rangle_\pi^{u_{\zeta_H}}  = [0,0.25]\,, \quad
\langle [\xi,\xi^2]\rangle_K^{u_{\zeta_H}}  = [0.035(5) , 0.24(1)]\,.
\end{equation}
Such skewing is also seen in the lQCD calculation of the pointwise behaviour of $\varphi_K(x)$ reported in Ref.\,\cite{Chen:2017gck}, but not within the precision of a more recent study \cite{Zhang:2020gaj}.  (Additional discussion presented in connection with Fig.\,\ref{fig:da_P_K}, which depicts $\varphi_K^{\bar s}(x)$.)

\begin{table}[t]
\caption{\label{parameterskaonDA}
Coefficients and powers that specify the kaon DA defined by Eq.\,\eqref{NewKPDAForm}.  Upper, middle, lower refer to the values of $\langle \xi^2\rangle_{K}^{u_{\zeta_H}}$ produced by the identified coefficients.  The ``upper'' parameter values produce the curve in Fig.\,\ref{FigNewKPDAForm} with the smallest magnitude at $x=0.5$, etc.
}
\begin{center}
\begin{tabular*}%{|c|c|c|c|c|c|c|}\hline
{\hsize}
{
l@{\extracolsep{0ptplus1fil}}|
c@{\extracolsep{0ptplus1fil}}
c@{\extracolsep{0ptplus1fil}}
c@{\extracolsep{0ptplus1fil}}
c@{\extracolsep{0ptplus1fil}}
c@{\extracolsep{0ptplus1fil}}}\hline\hline
& ${\mathpzc n}_{\varphi_K}\ $ & $\rho\ $ & $\gamma\ $ & ${\mathpzc a}\ $ & ${\mathpzc b}\ $\\\hline
{\rm upper}\ & $16.2\ $ & $4.92\ $ & $-6.00\ $ & $0.0946\ $ & $0.0731\ $\\
{\rm middle}$\ $&$18.2\ $ & $5.00\ $ & $-5.97\ $ & $0.0638\ $ & $0.0481\ $\\
{\rm lower}\ & $20.2\ $ & $5.00\ $ & $-5.90\ $ & $0.0425\ $ & $0.0308\ $\\
\hline\hline
\end{tabular*}
\end{center}
\end{table}

Following the procedures described in Refs.\,\cite{Segovia:2013eca, Shi:2014uwa, Cui:2020tdf}, the results in Eq.\,\eqref{pionkaonPDA} can be used to obtain the following pointwise form for the kaon's DA:
\begin{equation}
\varphi_K^u(x;\zeta_H) = {\mathpzc n}_{\varphi_K} \, x(1-x)  \left[1 + \rho x^{\frac{\mathpzc a}{2}} (1-x)^\frac{{\mathpzc b}}{2} + \gamma x^{\mathpzc a} (1-x)^{\mathpzc b}\right]\,, \label{NewKPDAForm}
\end{equation}
where ${\mathpzc n}_{\varphi_K}$ ensures unit normalisation.  The interpolation coefficients are listed in Table~\ref{parameterskaonDA}: ``upper'' indicates the curve that produces the largest value of $\langle\xi^2\rangle_K^{u_{\zeta_H}}$ and lower, the smallest.
$\varphi_K^{\bar s}(x;\zeta_H)$ is obtained using Eq.\,\eqref{phisymmetry}.

\begin{figure}[t]
\includegraphics[clip, width=0.5\textwidth]{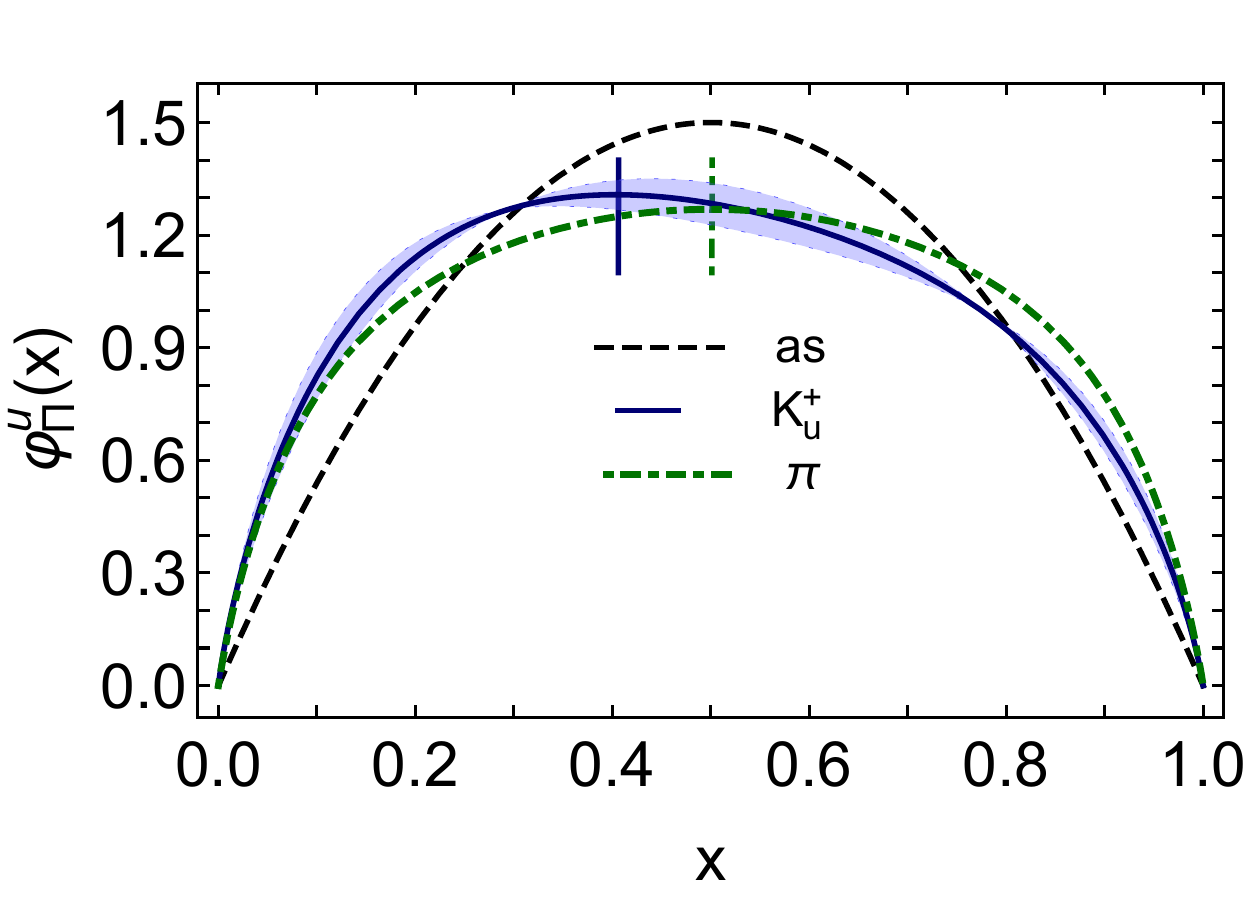}
\caption{\label{FigNewKPDAForm}
Kaon DA, $\varphi_K^u(x;\zeta_H)$, described by Eq.\,\eqref{NewKPDAForm} and the ``middle'' coefficients in Table~\ref{parameterskaonDA} -- solid blue curve.  The embedding band marks the domain bounded by the ``upper'' and ``lower''  coefficients in Table~\ref{parameterskaonDA}.
Pion DA in Eq.\,\eqref{varphipilinear} -- dot-dashed green curve; and asymptotic profile -- dashed black curve.
The vertical lines mark the peak position of the $K^+$ and $\pi$ DAs, \emph{viz}.\ $x=0.4, 0.5$, respectively.
}
\end{figure}

The family of DAs described by Eq.\,\eqref{NewKPDAForm} and the coefficients in Table~\ref{parameterskaonDA} is drawn in Fig.\,\ref{FigNewKPDAForm}: solid blue curve within blue shading.  It is slightly distorted when compared with the pion DA in Eq.\,\eqref{varphipilinear}, with a peak shifted to $x=0.4$, \emph{i.e}.\ 20\% to the left.  These features expose Higgs-boson modulation of EHM.  (Recall $f_K/f_\pi \approx 1.2\approx M_s(0)/M_{\overline{ud}}(0)$.)  In this connection, it is also worth remarking that the $K$ and $\pi$ DAs are unit-normalised; namely, in each case, an overall multiplicative factor of $f_{K,\pi}$, respectively, has been factorised.  What remains in the comparison between $K$ and $\pi$ DAs, therefore, is an essentially local expression of EHM and Higgs-related interference effects.

It may be anticipated from the curves in Fig.\,\ref{FigNewKPDAForm} that with increasing current mass of the heavier quark the distortion of this DA becomes more pronounced and its peak location, $\hat x_{\zeta_H}$, moves toward $x=0$ \cite{Binosi:2018rht, Tang:2019gvn, Serna:2020txe}.  However, consistent with notions of heavy-quark symmetry \cite{Neubert:1993mb}, there is a lower bound: $\lim_{m_P\to\infty} \hat x_{\zeta_H}= \hat x^0_{\zeta_H}$, $\hat x^0_{\zeta_H} \approx 0.05$.  This corresponds to a lower bound on the light-front momentum fraction stored with the lighter quark \cite{Binosi:2018rht}: $\langle x\rangle_0^{\zeta_H} \approx 0.12$.  Tracing from the pion, one has $\langle x \rangle_{\pi}^{\zeta_H} \approx 0.5$, $\langle x \rangle_{K^+}^{\zeta_H} \approx 0.48$, $\langle x \rangle_{D}^{\zeta_H} \approx 0.32$, $\langle x \rangle_{\bar B}^{\zeta_H} \approx 0.19$.  Evidently, the $D$-meson lies roughly at the halfway point; and in this case, drawn as in Fig.\,\ref{F1CDR}, the mass budget is EHM$+$HB $\approx 40$\%, half as much as in the kaon, and HB current mass $\approx 60$\%.

%% file: S4_Continuum.tex
\section{Empirical Access to Pseudoscalar Meson Distribution Amplitudes}
\label{sec:EADAs}
\subsection{Electromagnetic Transition Form Factors}
\label{sec:ETFFs}
There are few rigorous QCD predictions for processes that involve strong dynamics, like hadron elastic and transition form factors.  The cleanest are linked to $\gamma^\ast \gamma^{(\ast)} \to \Pi$ transition form factors, $G_\Pi(Q^2)$, where $\Pi$ is a charge-neutral pseudoscalar meson and $Q$ is the virtual photon momentum.  With the second photon being real and isolating a given $q\bar q$ component of $\Pi$, then there exists  $Q_0>\Lambda_{\rm QCD}$ such that \cite{Lepage:1980fj}
\begin{equation}
\label{EqHardScattering}
%Q^2 G_P^q(Q^2) \stackrel{Q^2 > Q_0^2}{\approx} 4 \pi^2 \, f_{P}^q\, N_c\, {\mathpzc e}_q^2\, \tilde{\mathpzc{w}}_{P}^q(Q^2),
Q^2 G_\Pi^q(Q^2) \stackrel{Q^2 > Q_0^2}{\approx} 4 \pi^2 \, f_{\Pi}^q\, {\mathpzc e}_q^2\, {\mathpzc{w}}_{\Pi}^q(Q^2),
\end{equation}
%% check pion case for normalisation
%% 4 \pi^2 f_\pi N_c (1/3) Integrate [phi/x] = factor of 3 too large because I have not used the unit-normalised <1/x>
%% Thus, remove Nc factor.
where:
$f_\Pi^q$ is the pseudovector projection of the $q\bar q$ piece of the meson's wave function onto the origin in configuration space, \emph{i.e}.\ a decay constant;
${\mathpzc e}_q$ is the quark's electric charge;
and
\begin{equation}
\label{wphi}
 {\mathpzc{w}}_{\Pi}^q(Q^2) = \int_0^1 dx\, \frac{1}{x} \,\varphi_{\Pi}^q(x;Q)\,,
\end{equation}
where $\varphi_{\Pi}^q(x;Q=\sqrt{Q^2})$ is the dressed-valence $q$-parton contribution to the meson's DA.  $f_\Pi^q$ is an order parameter for the strength of chiral symmetry breaking, which is driven by EHM in the light-quark sector.

Evidently, $G_\Pi(Q^2)$ presents an almost ideal case of power-law scaling in QCD.  Scaling violations are only expressed in the meson DA's evolution \cite{Lepage:1979zb, Efremov:1979qk, Lepage:1980fj}; but since it is the $\langle 1/x\rangle$ -moment which appears and evolution is logarithmic, then such effects can be observable on a large domain above $Q^2 \approx 10\,$GeV$^2$.  In this, one has indirect access to the DAs of neutral pseudoscalar mesons.  Notably, owing to isospin symmetry, the DAs of charged and neutral pions are the same; similarly for kaons.  It is worth remarking that when $\varphi_{\Pi}^q(x;Q) = \varphi_{\rm as}(x)$, ${\mathpzc{w}}_{\Pi}^q=3$.  On the other hand, the DAs drawn in Fig.\,\ref{FigphiDB} yield ${\mathpzc{w}}_{\pi}=3.7(3)$.  (In this case, the sum over electric charge states produces $1/3$.)

For the $\gamma^\ast \gamma^{(\ast)} \to \eta, \eta^\prime$ transitions, data are available on the domain $Q^2\in [0,112]\,$GeV$^2$ \cite{Gronberg:1997fj, Aubert:2006cy, BABAR:2011ad}.  More recently, data has become available for $\gamma^\ast(Q_1) \gamma^{\ast}(Q_2) \to \eta^\prime$ on $Q_1^2, Q_2^2\in [2,60]\,$GeV$^2$ \cite{BaBar:2018zpn}.  Contemporary analyses of these processes can be found in Refs.\,\cite{Agaev:2014wna, Ahmady:2018muv, Ding:2018xwy, Ji:2019som}.  They find that available data are consistent with the QCD prediction in Eq.\,\eqref{EqHardScattering} and the non-Abelian anomaly has a noticeable impact on $\eta^\prime$ physics, \emph{e.g}.\ \cite{Ding:2018xwy}: the topological charge content of the $\eta^\prime$ is more than twice as large as that of the $\eta$, which is itself also significant.

The $\gamma^\ast \gamma \to \pi^0$ case is somewhat less clear.  Data exists on the domain $Q^2/{\rm GeV}^2 \in [0.68,35]$ \cite{Behrend:1990sr, Gronberg:1997fj, Aubert:2009mc, Uehara:2012ag}.  All data agree on $Q^2\lesssim 10\,$GeV$^2$ and are compatible with Eq.\,\eqref{EqHardScattering}; but thereafter the two available sets \cite{Aubert:2009mc, Uehara:2012ag} exhibit conflicting trends in their evolution with photon virtuality \cite{Stefanis:2012yw}.  This issue has attracted much attention, as may be seen by following the trails identified in Refs.\,\cite{Raya:2015gva, Nedelko:2016vpj, Eichmann:2017wil, Choi:2020xsr, Stefanis:2020rnd}.

It is worth remarking here that the framework employed in Ref.\,\cite{Raya:2015gva} generates the broad, concave pion DA illustrated in Fig.\,\ref{FigphiDB}; expresses the QCD asymptotic limit, Eq.\,\eqref{EqHardScattering}; and has also been applied successfully to unifying $\gamma^\ast \gamma \to \eta, \eta^\prime, \eta_c, \eta_b$ transition form factors \cite{Raya:2016yuj, Ding:2018xwy}.  Hence, the following comparisons have some weight:
\begin{equation}
\begin{array}{l|c|c|c}
{\rm sources} & \mbox{Refs.\,\cite{Behrend:1990sr, Gronberg:1997fj, Aubert:2009mc}}
& \mbox{Refs.\,\cite{Behrend:1990sr, Gronberg:1997fj, Uehara:2012ag}}
& \mbox{Refs.\,\cite{Behrend:1990sr, Gronberg:1997fj, Aubert:2009mc, Uehara:2012ag}} \\\hline
\chi^2/{\rm datum} & 2.97 & 1.78 & 2.34
\end{array}\,.
\end{equation}
Evidently, the BaBar Collaboration data \cite{Aubert:2009mc} deviate significantly from the prediction, whereas the Belle Collaboration data \cite{Uehara:2012ag} match well.  In fact, focusing on data at $Q^2>10\,$GeV$^2$, one finds $\chi^2/{\rm datum} = 4.14$ \cite{Aubert:2009mc} and $\chi^2/{\rm datum} = 0.64$ \cite{Uehara:2012ag}.  Thus, it is premature to suggest that Eq.\,\eqref{EqHardScattering} is challenged by existing data; rather, the bulk of such data both tend toward its confirmation and support a picture of the pion DA as a broad, concave function at accessible probe momenta.

Measurements of such transition form factors are difficult.  They typically involve the study of $e^+ e^-$ collisions, with one of the outgoing fermions detected, after a large-angle scattering, whereas the other is scattered through a small angle and, so, undetected.  The detected fermion is supposed to have emitted a highly-virtual photon and the undetected fermion, a soft-photon.  These photons are assumed to fuse and produce the final-state pseudoscalar meson.  Numerous background processes and loss mechanisms are possible in this passage of events, thus providing ample room for systematic error, especially as $Q^2$ increases \cite{Bevan:2014iga}.  It is likely that a full accounting for such errors could reconcile the data from BaBar  \cite{Aubert:2009mc} and Belle \cite{Uehara:2012ag}; and in any event, new data is anticipated from the Belle II experiment \cite{Kou:2018nap}.

Matching Eq.\,\eqref{EqHardScattering} in rigour, QCD also delivers a prediction for the behaviour of the elastic electromagnetic form factor of a $\Pi=f\bar g$ charged pseudoscalar meson \cite{Farrar:1979aw, Lepage:1979zb, Efremov:1979qk, Lepage:1980fj}: $\exists \, Q_0^\prime >\Lambda_{\rm QCD}$ such that
\begin{equation}
\label{EqHardScatteringElastic}
Q^2 F_{\Pi}(Q^2) \stackrel{Q^2 > Q_0^{\prime 2}}{\approx} 16 \pi \hat\alpha(Q^2)  f_{\Pi}^2 \tilde{\mathpzc{w}}_{\Pi}^2(Q^2),
\end{equation}
where
\begin{equation}
\label{weightings}
\tilde{\mathpzc{w}}_\Pi^2 = e_{f}[ \tilde{\mathpzc w}_\Pi^{f}(Q^2)]^2 +
e_{\bar g} [\tilde{\mathpzc w}_\Pi^{\bar g}(Q^2)]^2 \,, \quad
\tilde{\mathpzc w}_{\Pi}^{\mathpzc f} = \tfrac{1}{3}\int_0^1 dx\, {\mathpzc g}_{\mathpzc f}(x) \,\varphi_\Pi^{\mathpzc f}(x;Q^2) \,,
\end{equation}
${\mathpzc g}_f(x) = 1/x$, ${\mathpzc g}_{\bar g}(x) = 1/(1-x)$,  and $e_{f,\bar g}$ are the electric charges of the valence quarks.  Plainly, as before, this hard elastic process is sensitive to the inverse moment of the meson DA.

Compared with the case of neutral meson transition form factors, Eq.\,\eqref{EqHardScattering}, QCD scaling violations are more pronounced in Eq.\,\eqref{EqHardScatteringElastic}: there is the manifest logarithmic suppression introduced by $\hat\alpha(Q^2)$; and that is magnified by QCD evolution, expressed in $\tilde{\mathpzc{w}}_\Pi^2$.  These features highlight that QCD is \emph{not} found in scaling laws.  Instead, it is revealed in the presence and nature of scaling violations, which are a basic feature of quantum field theory in four spacetime dimensions: while different models and theories may predict the same scaling power-law, scaling violations will decide between them.

\subsection{Elastic Electromagnetic Form Factors}
\label{sec:EEFFs}
The status of $\pi$ and $K$ elastic form factor measurements is summarised elsewhere \cite{Horn:2016rip} and is discussed further in Sec.\,\ref{sec:meson-form-factors-exp}.  It is nevertheless worth remarking, as shown by Table~\ref{tab:mesonff-exp}, that precise pion data are available on $Q^2/{\rm GeV^2} \in [0,2.45]$ \cite{Dally:1981ur, Dally:1982zk, Amendolia:1984nz, Amendolia:1986wj, Volmer:2000ek, Horn:2006tm, Tadevosyan:2007yd, Horn:2007ug, Huber:2008id, Blok:2008jy}.   The impetus for the new generation of measurements, made at the Thomas Jefferson National Accelerator Facility (JLab), is a widely held view that information on the $Q^2$ dependence of $F_\pi(Q^2)$ offers the best hope for charting the transition between the strong QCD domain, whereupon observables are determined by EHM and its corollaries and must be calculated using newly developed and developing nonperturbative methods, and the domain of perturbative QCD (pQCD), in which the familiar methods of perturbation theory can be used to obtain formulae such as Eq.\,\eqref{EqHardScatteringElastic}.

Twenty and more years ago, before the strong QCD phenomena described in Sec.\,\ref{sec:CTP} were widely known and appreciated, the transition to pQCD was expected to take place at $Q^2 \approx m_p^2$.  However, with data now available out to $Q^2 \approx 2.5\,m_p^2$, the JLab $F_\pi$ Collaboration has concluded that extant empirical coverage \cite{Horn:2006tm, HornQuote} ``\ldots is still far from the transition to the $Q^2$ region where the pion looks like a simple quark-antiquark pair \ldots'', \emph{viz}.\ far from the $Q^2$ domain upon which Eq.\,\eqref{EqHardScatteringElastic} can be tested.
%%%https://www.jlab.org/research/pion_form
With the challenge posed by Eq.\,\eqref{EqHardScatteringElastic} thus remaining, experiments aimed at reaching $Q^2=6\,$GeV$^2$ were proposed for the 12\,GeV-upgraded JLab facility (JLab\,12) \cite{Dudek:2012vr}.  Now the upgrade is completed, the experiments \cite{E12-06-101, E12-07-105} are running on schedule and the data are being analysed as soon as they are obtained.

Using the information provided in Sec.\,\ref{sec:CTP}, it is possible to develop an estimate of $Q_0^\prime$ in Eq.\,\eqref{EqHardScatteringElastic} following the ideas in Ref.\,\cite{Maris:1998hc}.  In an elastic scattering process, both valence degrees-of-freedom in the pion will most often share the incoming probe momentum equally, \emph{i.e}.\ each will receive $Q/2$.  
Comparing the chiral limit and $u=d$-quark mass functions in Fig.\,\eqref{FigMp2}, it is clear that the perturbative tail only becomes evident at $k^2 \approx 2\,$GeV$^2$, where the ratio of the two curves begins to deviate significantly from unity.  With each quark carrying $Q/2$, it follows that no results calculated using perturbative quark propagators can be valid unless
\begin{equation}
\label{Q0boundary}
(Q/2)^2 > 2\,{\rm GeV}^2 \Rightarrow Q^2 > 8\,{\rm GeV}^2 \approx: Q_0^{\prime 2}\,.
\end{equation}

%%%\begin{figure}[t]
%%%\centerline{%
%%%\includegraphics[clip, width=0.7\textwidth]{F12CDR.pdf}}
%
%%%\caption{\label{FigFpiEIC}
%
%%%$F_\pi(Q^2)$.
%
%%%Solid black curve within grey band -- continuum theory prediction, which bridges large and short distance scales, with estimated uncertainty, calculated using the methods described in Ref.\,\cite{Chen:2018rwz};
%
%%%dot-dashed blue and dotted purple curves -- result obtained with Eq.\,\eqref{EqHardScatteringElastic}, comparing, respectively, the result obtained using a modern EHM-hardened DA like that in Eq.\,\eqref{varphipilinear} with that produced by the asymptotic profile.
%
%%%Projected JLab\,12 data, to be obtained using a Rosenbluth-separation technique -- orange diamonds and green triangle.
%
%%%Data projections for a US electron ion collider (EIC), as anticipated using extraction from a combination of electron-proton and electron-deuteron scattering, each with an integrated luminosity of $20\,{\rm fb}^{-1}$ -- black stars with error bars \cite{Aguilar:2019teb}.
%
%%%(\emph{N.B}.\ The normalisation of all projected data is arbitrary.)
%
%%%The long-dashed green curve is a monopole form factor whose scale is determined by the pion radius.
%%%}
%%%\end{figure}

\begin{figure}[t]
\hspace*{-1ex}\begin{tabular}{lcl}
{\sf A} &\hspace*{2em} & {\sf B} \\[-2ex]
\includegraphics[clip, width=0.46\textwidth]{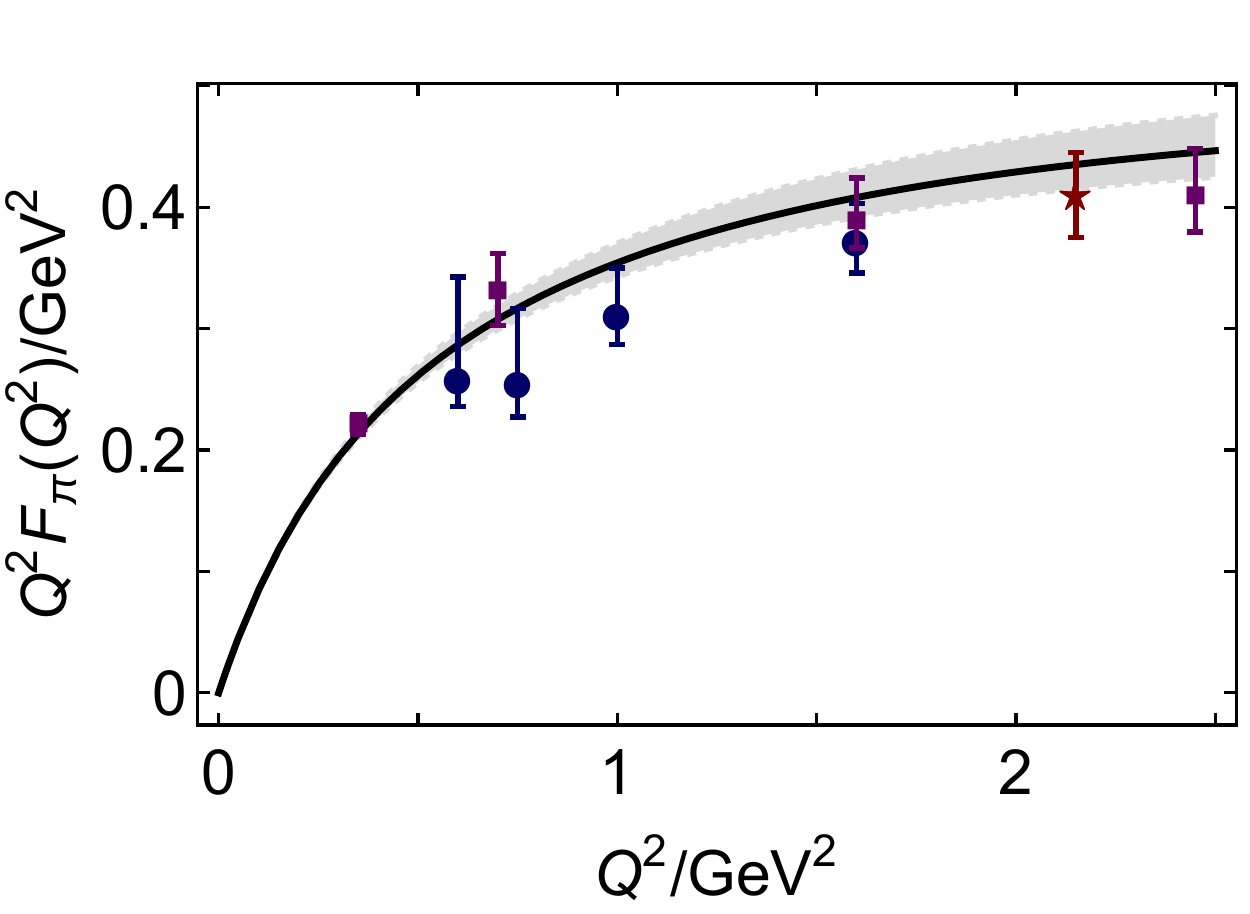} & \hspace*{2em} &
\includegraphics[clip, width=0.46\textwidth]{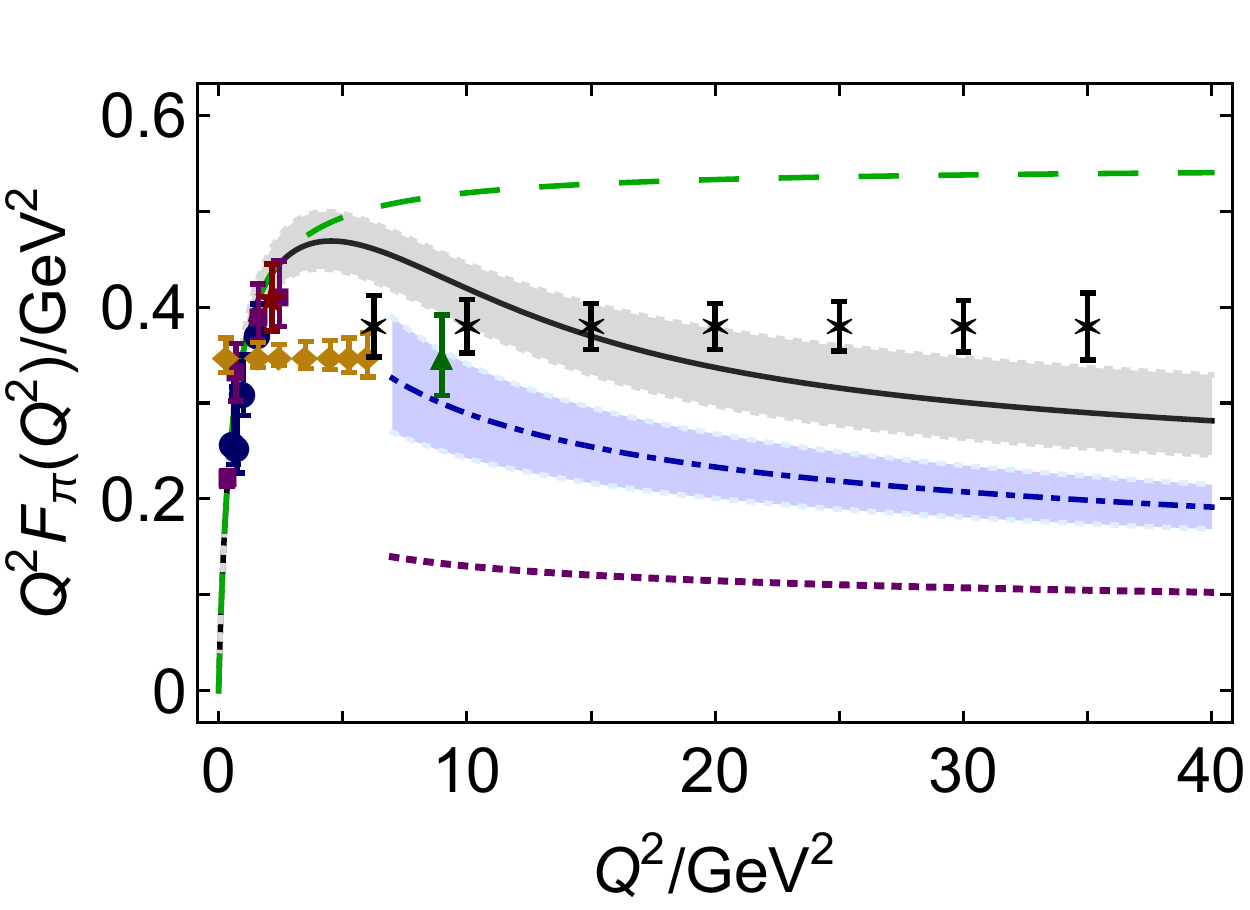}
\end{tabular}
\caption{\label{FigFpiEIC}
$F_\pi(Q^2)$.  Solid black curve within grey band -- continuum theory prediction, which bridges large and short distance scales, with estimated uncertainty, calculated using the methods described in Ref.\,\cite{Chen:2018rwz}.
\emph{Left panel}\,--\,{\sf A}.  Comparison with data analysed by the JLab $F_\pi$ Collaborations \cite{Horn:2007ug, Huber:2008id}.  The $\chi^2/{\rm datum} = 1.0$.
\emph{Right panel}\,--\,{\sf B}.  Dot-dashed blue and dotted purple curves -- results obtained with Eq.\,\eqref{EqHardScatteringElastic}, comparing, respectively, that produced by a modern EHM-hardened DA like that in Eq.\,\eqref{varphipilinear} with that given by the asymptotic profile.
Projected JLab\,12 data, to be obtained using a Rosenbluth-separation technique -- orange diamonds and green triangle.
Black stars with error bars -- data projections for a US electron ion collider (EIC) \cite{Aguilar:2019teb}, as anticipated using extraction from a combination of electron-proton and electron-deuteron scattering, each with an integrated luminosity of $20\,{\rm fb}^{-1}$.  Projections for the electron ion collider in China (EicC), currently under discussion, may be found elsewhere \cite[Sec.\,4]{Chen:2020ijn}.
(\emph{N.B}.\ The normalisation of all projected data is arbitrary.)
The long-dashed green curve is a monopole form factor whose scale is determined by the pion radius.
}
\end{figure}

Within the past decade, the algorithms used for continuum calculations of $F_\pi(Q^2)$ have been comprehensively improved.  This progress capitalised on the new techniques that delivered the DA results in Figs.\,\ref{FigphiDB}, \ref{FigDACriticalMass}, \ref{FigNewKPDAForm}.  It led to a single calculation applicable on the entire domain of spacelike $Q^2$ \cite{Chang:2013nia, Gao:2017mmp}, unifying that covered empirically with the deep ultraviolet.  The result is illustrated in Fig.\,\ref{FigFpiEIC} and the following features are noteworthy.
\begin{description}
\item[JLab pion data] The solid black curve in Fig.\,\ref{FigFpiEIC}A appeared after the JLab data were collected.  Notwithstanding that, it is a parameter-free prediction, which derives from and expresses the features of EHM detailed in Sec.\,\ref{sec:CTP}.  Hence, the result $\chi^2/{\rm datum} = 1.0$ provides meaningful support for the concepts described therein.

\item[Scaling and scaling violations] Fig.\,\ref{FigFpiEIC}\,B shows that the continuum theory prediction tracks a monopole form factor with scale determined by the pion radius until $Q^2 \approx 6\,$GeV$^2$.  Thereafter, the two curves separate, growing further apart with increasing $Q^2$ as QCD scaling violations become increasingly more important in understanding this hard exclusive process.

    If the JLab\,12 measurement at $Q^2 \approx 9\,$GeV$^2$ achieves the anticipated precision, then it will be sufficient to validate this prediction.  If the prediction is correct, then the measurement will be the first to have uncovered QCD scaling violations in a hard exclusive process.

\item[pQCD] Fig.\,\ref{FigFpiEIC}\,B displays qualitative and semiquantitative agreement between the black solid and dot-dashed blue curves on $Q^2 \gtrsim 8\,$GeV$^2$.  This indicates that when used with a pion DA appropriate to the scale of the experiment, Eq.\,\eqref{EqHardScatteringElastic} provides a qualitatively sound understanding of this hard exclusive process at such momentum transfers.  The location of this boundary matches the value predicted in Eq.\,\eqref{Q0boundary}.

\item[large $\mathbf Q^{\mathbf 2}$] Comparison between the black stars and black solid curve in Fig.\,\ref{FigFpiEIC}\,B suggests that EIC (or a similar high-luminosity, high-energy facility) will be capable of delivering quantitative verification of the anomalous dimension predicted by QCD for this hard exclusive process.
\end{description}
It is also worth remarking that the solid black curve in Fig.\,\ref{FigFpiEIC}\,B, drawn from Ref.\,\cite{Chen:2018rwz}, is one of a set that aids in understanding contemporary lQCD calculations of heavy-pion form factors at large $Q^2$ \cite{Chambers:2017tuf, Koponen:2017fvm}.  Such lQCD results are discussed further in connection with Fig.\,\ref{fig:pion_high_Q2}.

\begin{figure}[t]
\hspace*{-1ex}\begin{tabular}{lcl}
{\sf A} &\hspace*{2em} & {\sf B} \\[-2ex]
\includegraphics[clip, width=0.46\textwidth]{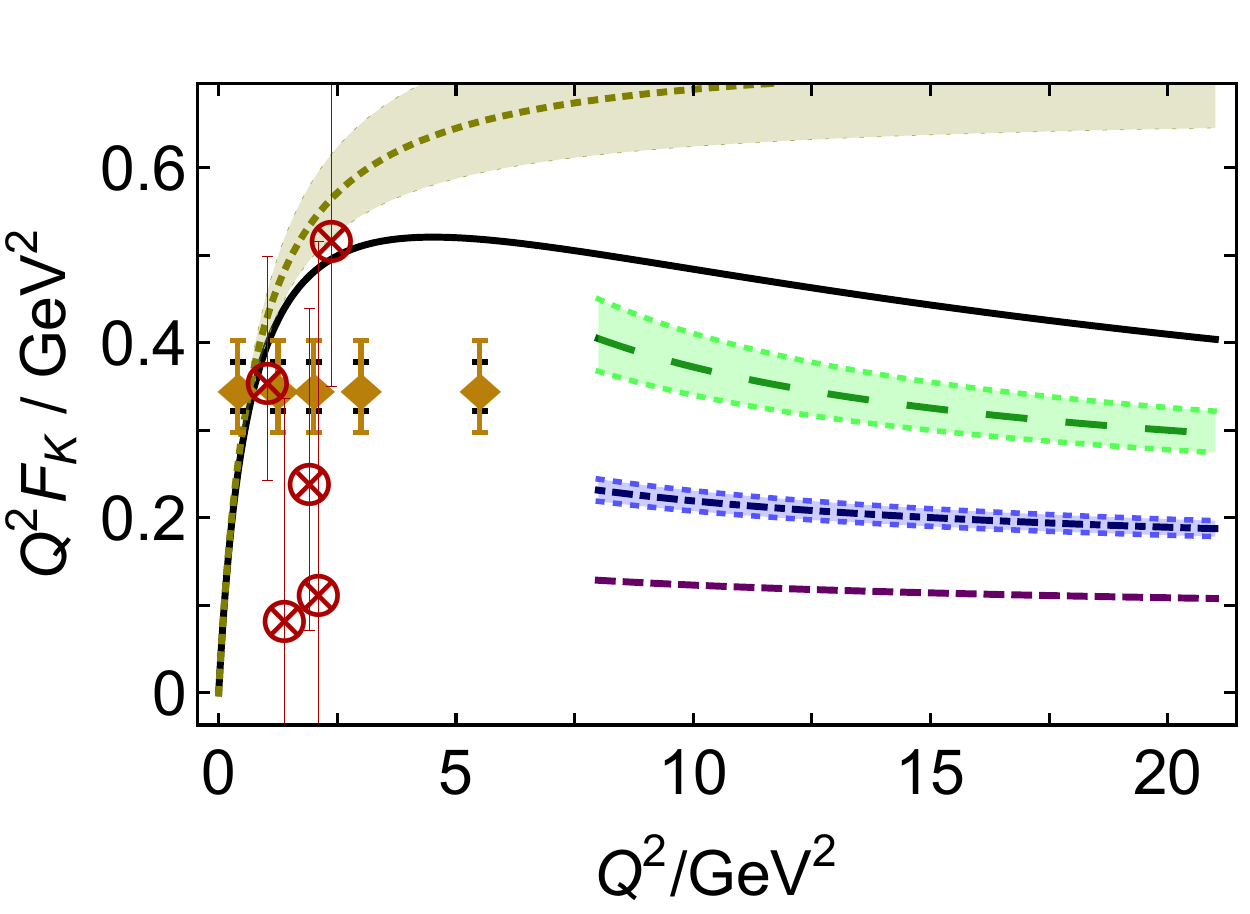} & \hspace*{2em} &
\includegraphics[clip, width=0.46\textwidth]{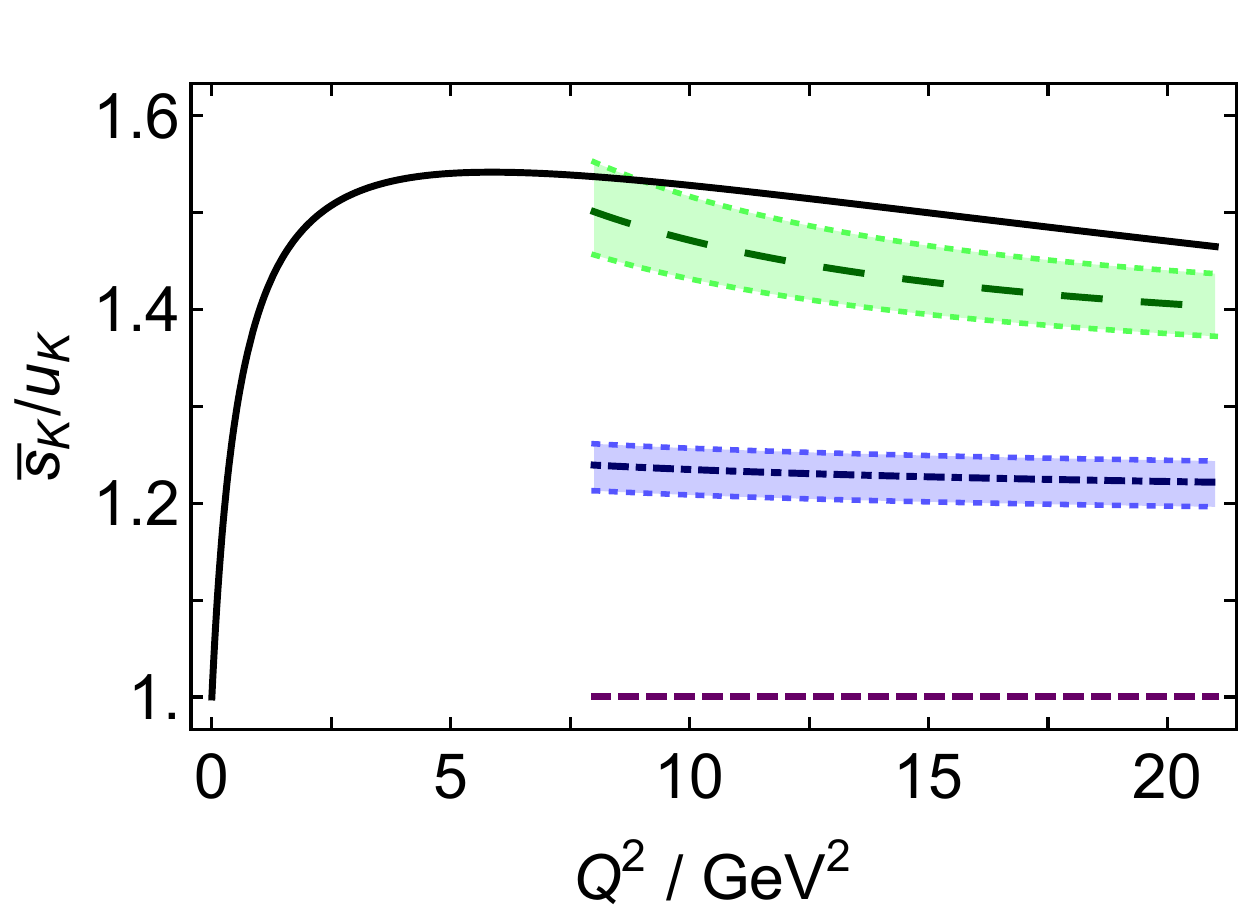} \\
{\sf C} &\hspace*{2em} & {\sf D} \\[-2ex]
\includegraphics[clip, width=0.46\textwidth]{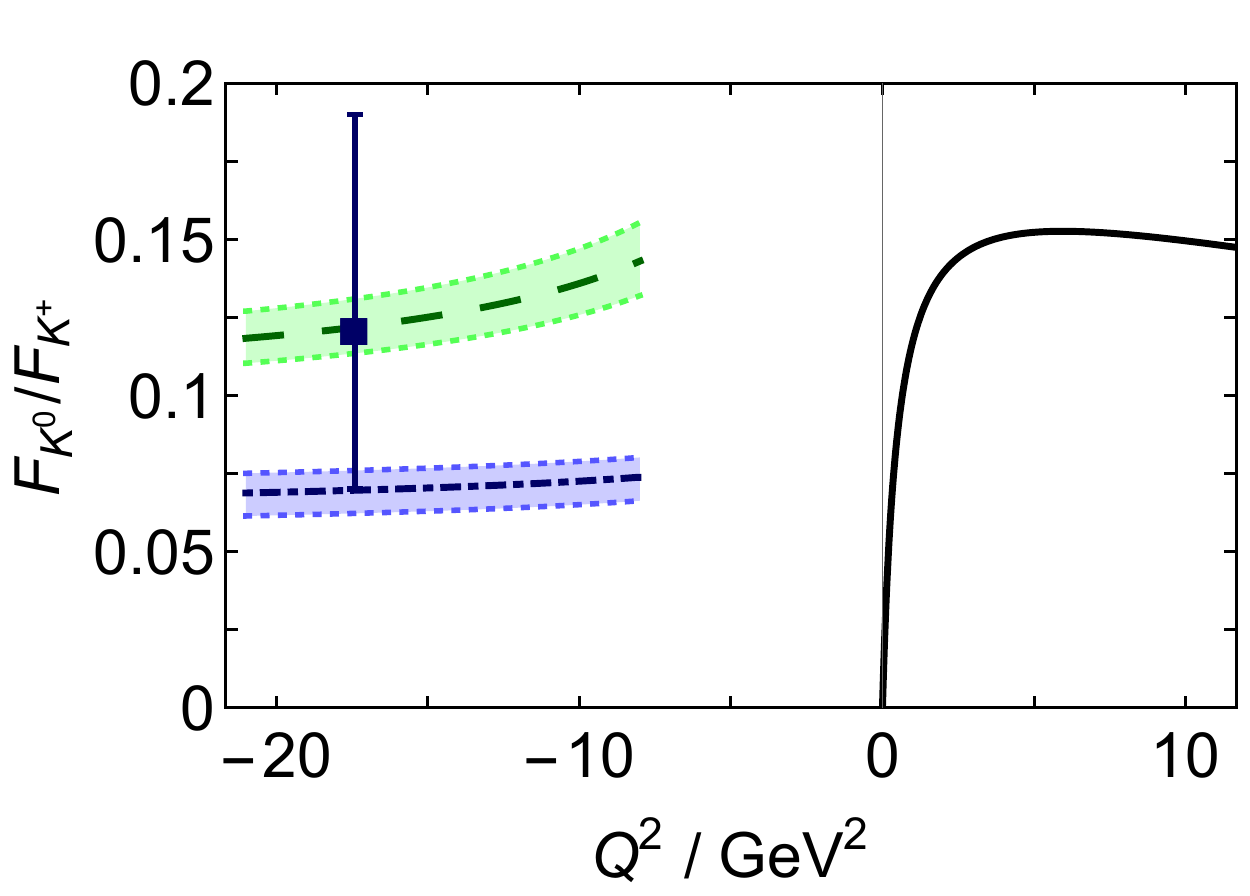} & \hspace*{2em} &
\includegraphics[clip, width=0.46\textwidth]{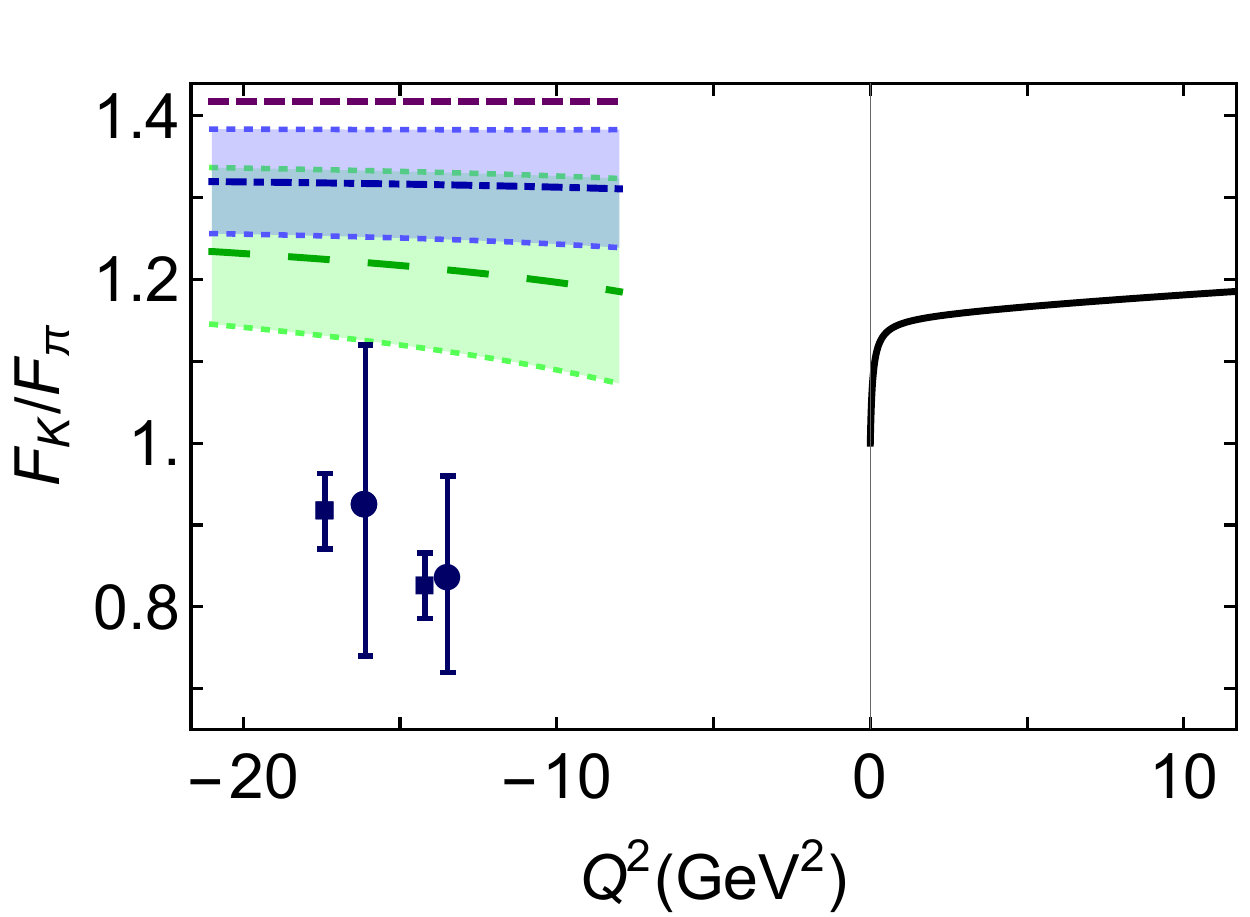}
\end{tabular}
\caption{\label{FigKaonFFs}
Continuum predictions for selected kaon form factors and related ratios drawn from Ref.\,\cite{Gao:2017mmp} -- solid black curve in each panel.
All panels present results obtained from Eq.\,\eqref{EqHardScatteringElastic} using empirical values for $f_{K,\pi}$ and different estimates of $K$ and $\pi$ DAs: dashed green curve --  DAs estimated in Ref.\,\cite{Gao:2017mmp}, for which $\langle \xi^2 \rangle_K^{\zeta_H}=0.27(1)$, $\langle \xi^2 \rangle_\pi^{\zeta_H}=0.28$; dot-dashed blue curve -- DAs from Secs.\,\ref{SecDApion}, \ref{SecDAkaon}, which exhibit 35\% less dilation with respect to $\varphi_{\rm as}(x)$; and purple dashed curve, $\varphi_{\rm as}(x)$.  (The asymptotic profile produces $F_{K^0}\equiv 0$.)
\emph{Legend} notes.
\emph{Upper left}\,--\,{\sf A}.  Dotted olive curve within like-coloured band -- monopole based on kaon charge radius $r_K^+=0.56(3)$\,fm \cite{Zyla:2020zbs};
crossed circles -- data from Ref.\,\cite{Carmignotto:2018uqj}, representing analyses of the $^1\!$H$(e,e^\prime K^+)\Lambda$ reaction; and filled diamonds -- data anticipated from a forthcoming experiment \cite{E12-09-011}, where the two error estimates differ in their assumptions about the $t$- and model-dependence of the form factor extractions.  (Normalisation of projected data is arbitrary.)
\emph{Upper right}\,--\,{\sf B}.   Strange-to-normal-matter charge distribution ratio.
\emph{Lower right}\,--\,{\sf C}.  Prediction from Ref.\,\cite{Gao:2017mmp}: $r_{K^0}^2 = -(0.21\,{\rm fm})^2$ \emph{cf}.\ experiment \cite{Molzon:1978py}: $r_{K^0}^2 = -(0.24 \pm 0.08\,{\rm fm})^2$. Datum from Ref.\,\cite{Seth:2013eaa}, with the error bar marking the 90\% confidence interval.
\emph{Lower left}\,--\,{\sf D}.  Data from Ref.\,\cite{Seth:2012nn}.
}
\end{figure}

Eq.\,\eqref{EqHardScatteringElastic} applies equally to kaon elastic electromagnetic form factors; and in concert with Fig.\,\ref{FigNewKPDAForm}, it suggests that the $u$- and $s$-quark charge distributions in the $K^+$ must differ.  It follows that the neutral kaon has a nonzero charge form factor.  These features are seen in contemporary phenomenology and theory, \emph{e.g}.\ Refs.\,\cite{Chen:2012txa, Gao:2017mmp, Mecholsky:2017mpc}.  Such characteristics, too, are expressions of EHM modulation by the  Higgs-boson.  They are illustrated in Fig.\,\ref{FigKaonFFs}; and the following aspects are worth highlighting.
\begin{description}
\item[Forthcoming kaon data] Fig.\,\ref{FigKaonFFs}\,A illustrates that precise data do not yet exist for the charged-kaon form factor; but it is anticipated that JLab\,12 will remedy this situation, providing valuable information out to $Q^2\approx 6\,$GeV$^2$.  Referred to the prediction in Ref.\,\cite{Gao:2017mmp}, which is the only QCD-connected result which covers the entire domain of spacelike $Q^2$, it appears that the reach of the expected JLab\,12 data should enable scaling violations to be detected in $F_{K^+}(Q^2)$.

\item[DA sensitivity] The sensitivity of the results produced by Eq.\,\eqref{EqHardScatteringElastic} to the endpoint behaviour of $K$ and $\pi$ DAs, through their $\langle 1/x \rangle$ moments, is plain in every panel of Fig.\,\ref{FigKaonFFs}.
    This does not diminish the potential for data to reveal scaling violations, but may affect the ability for such data to be used in determining the anomalous dimensions characterising $F_{K}(Q^2)$.
    %
   % The value of $Q^2$ that marks the beginning of the domain upon which data is sensitive to the value of the anomalous dimension characterising $F_{K}(Q^2)$ the domain

\item[Flavour separation] Fig.\,\ref{FigKaonFFs}\,B depicts the strange-to-normal-matter charge distribution ratio, $\bar s_K/u_K$, in the $K^+$, as predicted in Ref.\,\cite{Gao:2017mmp}.
    With the static electric charges factored out, the ratio is unity at $Q^2=0$, owing to current conservation; and Eq.\,\eqref{EqHardScatteringElastic} predicts that it is also unity on $\Lambda_{\rm QCD}^2/Q^2 \simeq 0$.  Thus, the interesting features are displayed between these limits.
    $\bar s_K/u_K$ rises to a peak value of roughly $1.5$ at $Q^2\approx 6\,$GeV$^2$.  Recalling that $(f_K/f_\pi)^2 \approx 1.4 \approx (M_s(0)/M_{\overline ud}(0))^2$, then this value is evidently typical for Higgs-boson modulation of EHM.
    Owing to the logarithmic nature of DA evolution, the deviation of $\bar s_K/u_K$ from unity must remain significant on a very large part of the domain $Q^2 \gtrsim 6\,$GeV$^2$.

    The anticipated JLab data \cite{E12-09-011} should be capable of testing some of these predictions, including the projected peak height.  Other features, including the persistence of $\bar s_K/u_K >1$, will need the energy and luminosity of EIC or EicC \cite{Aguilar:2019teb, Chen:2020ijn}.

\item[Timelike form factors] On any domain within which Eq.\,\eqref{EqHardScatteringElastic} provides a reasonable approximation, spacelike and timelike form factors are equal at leading order (LO) in $\hat\alpha$.
    Using this fact, one can obtain an estimate for $F_{K^0}/F_{K^{+}}$ at timelike momenta beyond the resonance region, \emph{i.e}.\ on $t=-Q^2 \gtrsim 8\,$GeV$^2$.  Such a projection is drawn in Fig.\,\ref{FigKaonFFs}\,C.  Evidently, the prediction obtained in this way is consistent with the only existing measurement \cite{Seth:2013eaa}, so long as one uses a kaon DA that is qualitatively consistent with that drawn in Fig.\,\ref{FigNewKPDAForm}.

    The picture is less clear for $F_{K^+}/F_{\pi^+}$, depicted in Fig.\,\ref{FigKaonFFs}\,D.  The calculations all indicate that $F_{K^+}/F_{\pi^{+}}>1$, whereas existing data lie below unity \cite{Seth:2012nn}.  This is puzzling because: (i) charge conservation means $F_{K^+}/F_{\pi^{+}}=1$ at $Q^2=0$; (ii) the ordering of charge radii ensures the ratio rises as $|Q^2|$ increases; and (iii) the absence of another set of mass-scales suggests that the asymptotic limit ($f_K^2/f_\pi^2\approx 1.42$) should be approached monotonically from below.
    Notably, considering the separate data for the $\pi$ and $K$ form factors at timelike momenta \cite{Seth:2012nn}, one might review their normalisations because, mapped straightforwardly to spacelike momenta and compared with the calculations in Ref.\,\cite{Gao:2017mmp}, the $\pi$ measurements are roughly twice as large and those for the $K$ are $\sim 1.5$-times greater.
    A mismatch of relative normalisations would cancel in $F_{K^0}/F_{K^{+}}$.
\end{description}

\begin{figure}[t]
\hspace*{-1ex}\begin{tabular}{lcl}
{\sf A} &\hspace*{2em} & {\sf B} \\[-2ex]
\includegraphics[clip, width=0.46\textwidth]{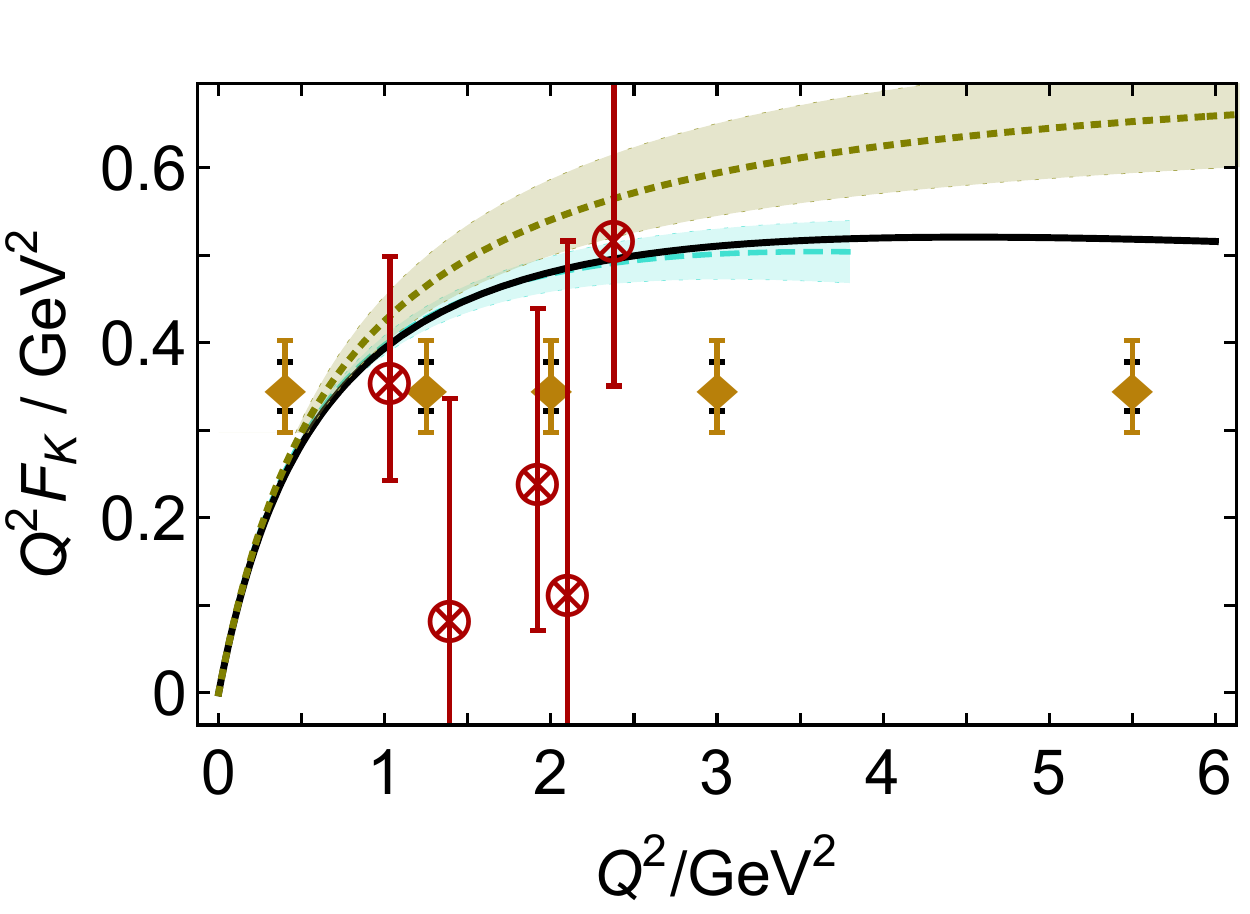} & \hspace*{2em} &
\includegraphics[clip, width=0.46\textwidth]{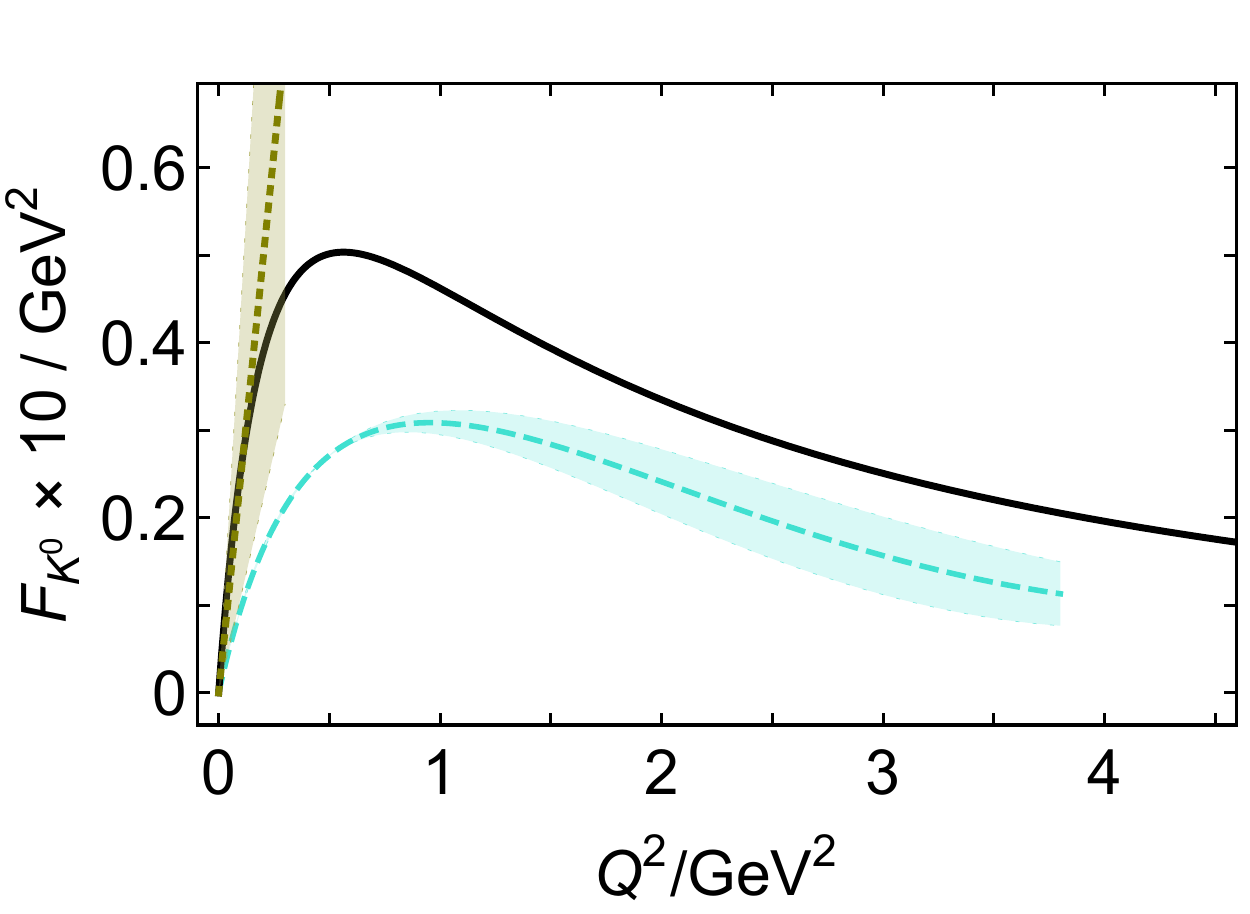}
\end{tabular}
\caption{\label{FigFK0}
\emph{Left panel}\,--\,{\sf A}.  Results for the charged kaon form factor: solid black curve -- prediction in Ref.\,\cite{Gao:2017mmp}; dashed turquoise curve within like coloured band -- lQCD result from Ref.\,\cite{Davies:2019nut}; dotted olive curve within band -- monopole based on kaon charge radius;
crossed circles -- data from Ref.\,\cite{Carmignotto:2018uqj}, representing analyses of the $^1\!$H$(e,e^\prime K^+)\Lambda$ reaction; and filled diamonds -- data anticipated from a forthcoming experiment \cite{E12-09-011}, where the two error estimates differ in their assumptions about the $t$- and model-dependence of the form factor extractions.  (Normalisation of projected data is arbitrary.)
\emph{Right panel}\,--\,{\sf B}.  Neutral kaon form factor (\emph{N.B}.\ multiplied by a factor of 10): solid black curve -- prediction in Ref.\,\cite{Gao:2017mmp}; dashed turquoise curve within like coloured band -- lQCD result from Ref.\,\cite{Davies:2019nut}; dotted olive curve within band -- straight line drawn to indicate the empirical constraint imposed by information on the neutral kaon's charge radius.
Evidently, evaluated at the peak of each function, the continuum analysis predicts $Q^2 F_{K^+} \approx 10 F_{K^0}$.
}
\end{figure}

First lQCD results for the kaon elastic electromagnetic form factors have recently become available \cite{Davies:2019nut}.  As discussed further in Sec.\,\ref{lQCDpionFF}, in completing a lQCD computation of meson form factors, one must meet competing demands, \emph{e.g}.\ large lattice volumes are required to represent light-quark systems, small lattice spacings are needed to reach large $Q^2$, and high statistics are necessary to compensate for a decaying signal-to-noise ratio as form factors drop rapidly with increasing $Q^2$.  For reasons such as these, the results in Ref.\,\cite{Davies:2019nut} only reach to $Q^2\approx 3.8\,$GeV$^2$.  They are reproduced in Fig.\,\ref{FigFK0} in comparison with predictions from Ref.\,\cite{Gao:2017mmp}.

Fig.\,\ref{FigFK0}\,A reveals that the continuum and lattice results for the charged kaon form factor are almost identical on $Q^2\lesssim 4\,$GeV$^2$.  This adds further support to the suggestion made above, \emph{viz}.\ the reach of the anticipated JLab\,12 data should enable scaling violations to be detected in $F_{K^+}(Q^2)$; and they will certainly be seen in precision experiments at EIC or EicC.
Turning to Fig.\,\ref{FigFK0}\,B, the new lQCD result indicates a neutral kaon charge radius $r_{K^0}^{2\,{\rm lQCD}} \approx -(0.16\,{\rm fm})^2$, a magnitude which lies at the lower extreme of the empirical range.
Moreover, there is a clear qualitative likeness and semiquantitative agreement between the lQCD result and the continuum prediction: the momentum dependence is similar and the lQCD result is a roughly uniform two-thirds of the size of the continuum prediction.  Combined with the accord displayed in Fig.\,\ref{FigFK0}\,A, confidence in both results is increased; especially because the $K^0$ charge form factor is determined by a destructive interference between two terms that are identical at $Q^2=0$ and similar in magnitude thereafter, so that any loss of precision is magnified in the difference.
%Furthermore, the lQCD curve differs significantly from the continuum prediction.
% (Similar issues are seen in lQCD computations of the neutron electric form factor \cite[Fig.\,2.2.9]{Barabanov:2020jvn}.)

\subsection{Diffractive Dissociation}
\label{SecDiffractive}
\begin{figure}[t]
\centerline{%
\includegraphics[clip, width=0.46\textwidth]{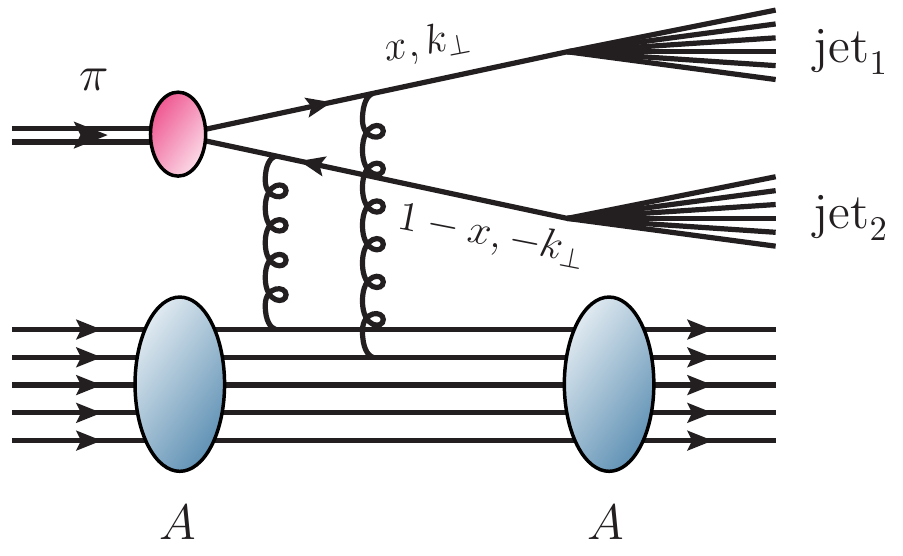}}
\caption{\label{FigDD}
Hard diffractive dissociation of a pion in the gluon field of a heavy nucleus.  The pion's valence-quark--valence-antiquark wave function is $\psi_\pi(x,k_\perp^2;\zeta)$, with $\zeta$ the energy  scale of the interaction.
(Image courtesy of D.\,Binosi.)}
\end{figure}

The exclusive reactions described in Secs.\,\ref{sec:ETFFs}, \ref{sec:EEFFs} provide access to an integrated property of pseudoscalar meson DAs, \emph{viz}.\ the $\langle 1/x \rangle$ moment, which is particularly sensitive to a DA's endpoint behaviour.  It may be possible to obtain information on the $x$-dependence of a meson's DA using the diffractive dissociation process depicted in Fig.\,\ref{FigDD}.  Notionally, a high energy meson dissociates in the colour field of a heavy nucleus, transferring no energy to the target, $A$, so that it remains intact.  Meanwhile in this coherent process, the meson breaks up, with its valence-quark and \mbox{-antiquark} constituents seeding two jets.  So long as the momentum of a given valence constituent is transferred (almost) completely into just one of the two separate jets, then measurement of the jet momenta provides access to the quark and antiquark momenta:
\begin{equation}
x = \frac{p_{{\rm jet}_1}}{p_{{\rm jet}_1}+p_{{\rm jet}_2}}\,.
\end{equation}
Assuming factorisation is valid for the kinematics of a given experiment, then one may argue that the energy scale of the interaction is
\begin{equation}
\zeta^2 = \frac{k_\perp^2}{x(1-x)}\,,
\end{equation}
where $k_\perp$ is the intrinsic light-front transverse momentum of the valence-quark.

Following the analysis in, \emph{e.g}.\ Ref.\,\cite{Frankfurt:1993it}, the differential cross-section for the process depicted in Fig.\,\ref{FigDD} takes the following form:
\begin{equation}
\label{eqDDmeson}
\frac{d\sigma}{dk_\perp^2} \propto
\left| \hat\alpha(k_\perp^2) x_N G(x_N,k_\perp^2)\right|^2
\left| \tilde\varphi_P(x) \frac{\partial}{\partial k_\perp^2} \tilde\psi_P(k_\perp^2)\right|^2\,,
\end{equation}
where $x_N$ is the momentum fraction (Bjorken $x$) of the interacting gluons, $G(x_N,k_\perp^2)$ is the gluon distribution in the target, and a product \emph{Ansatz} has been employed for the meson's light-front wave function, \emph{viz}.\ $\psi_P(x,k_\perp^2) \approx \tilde\varphi_P(x) \tilde\psi_P(k_\perp^2)$, as discussed in Sec.\,\ref{sec:ELFWFs}.

\begin{figure}[t]
\hspace*{-1ex}\begin{tabular}{lcl}
{\sf A} &\hspace*{2em} & {\sf B} \\[-2ex]
\includegraphics[clip, width=0.46\textwidth]{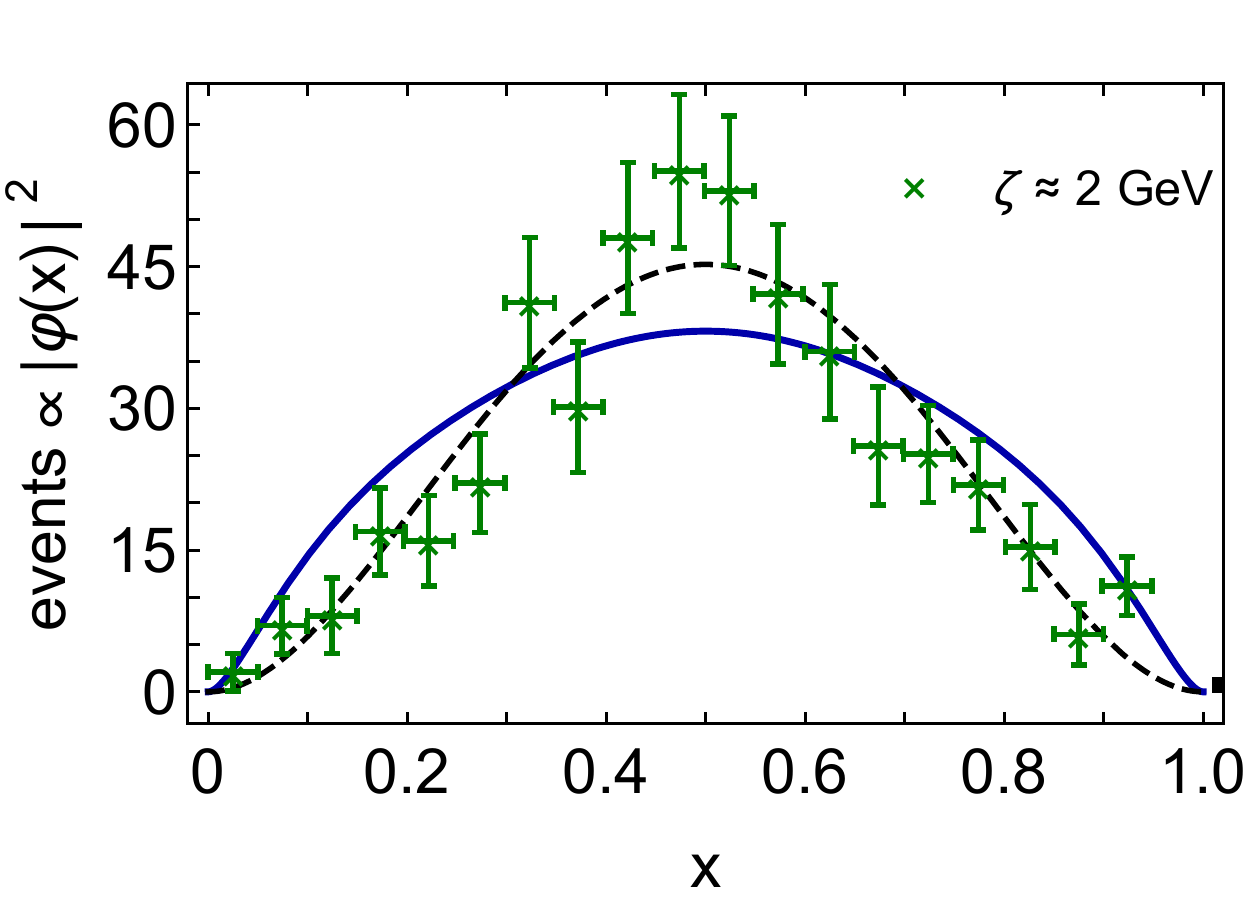} & \hspace*{2em} &
\includegraphics[clip, width=0.46\textwidth]{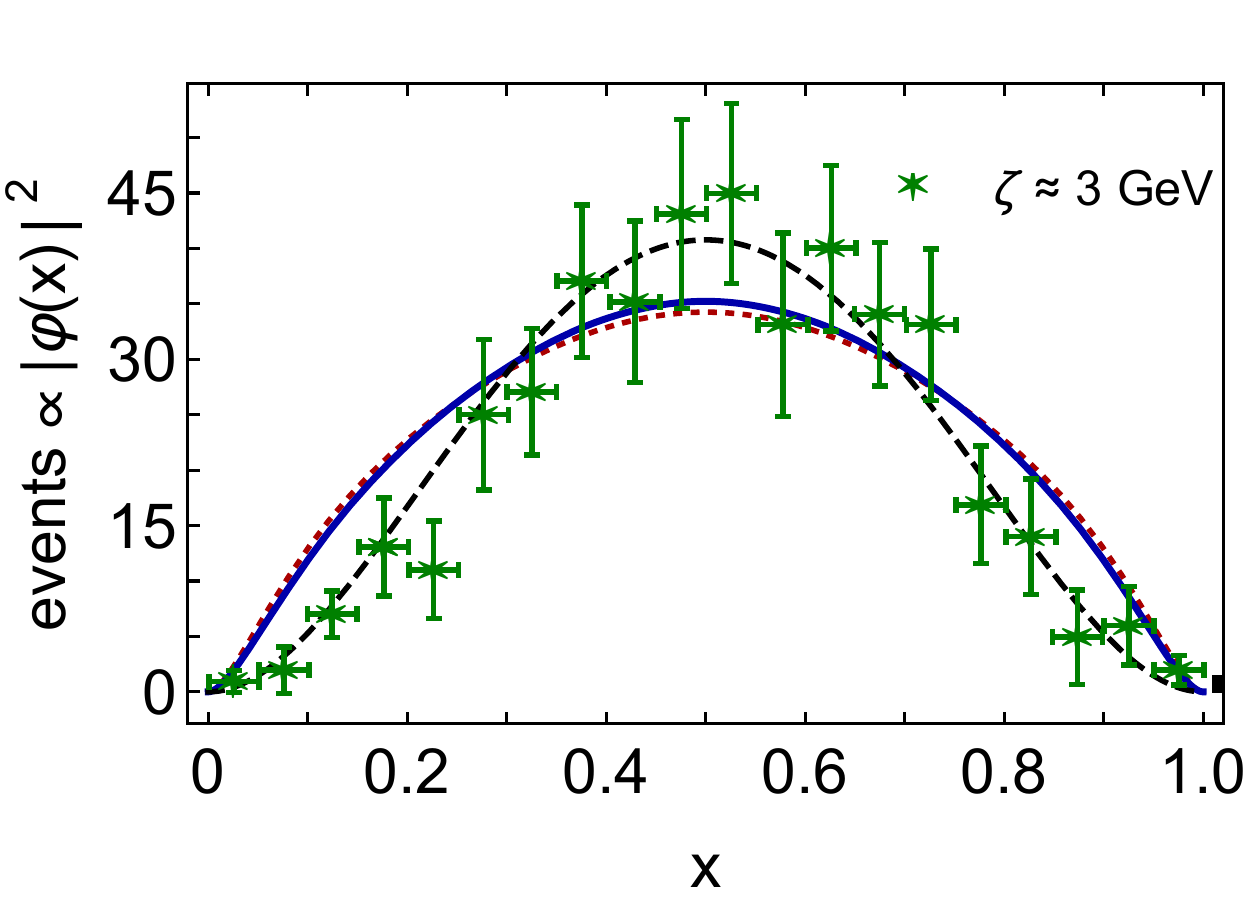}
\end{tabular}
\caption{\label{FigpiDAE791}
\emph{Left panel} -- {\sf A}, $\zeta \approx 2\,$GeV data from Ref.\,\cite{Aitala:2000hb}.  Dashed black curve: fit to data based on $\varphi_{\rm as}(x)$; and solid blue curve, fit using Eq.\,\eqref{varphipilinear}.  $\chi^2/{\rm datum} = 1.1, 2.6$, respectively.
\emph{right panel} -- {\sf B}. $\zeta \approx 3\,$GeV data from Ref.\,\cite{Aitala:2000hb}.  Dashed black curve: fit to data based on $\varphi_{\rm as}(x)$; and solid blue curve, fit using Eq.\,\eqref{varphipilinear}, one-loop evolved to $\zeta = 3\,$GeV.  $\chi^2/{\rm datum} = 0.54, 2.7$, respectively.  The solid blue curve is obtained via one-loop evolution of the dotted red curve, which is based on the solid curve in Panel A.
}
\end{figure}

An experiment (E791) to measure the process sketched in Fig.\,\ref{FigDD} is described in Ref.\,\cite{Aitala:2000hb}.  Two domains of $k_\perp^2$ were identified therein, separated by their perceived $k_\perp^2$ dependence, and connected to $\zeta^2/{\rm GeV}^2 \approx 4, 10$.  The data are presented in Fig.\,\ref{FigpiDAE791}.  If one chooses to interpret them using Eq.\,\eqref{eqDDmeson}, as done in Ref.\,\cite{Aitala:2000hb}, then the number of events is proportional to $|\varphi_\pi(x;\zeta)|^2$.  Two primary curves are drawn in each panel of Fig.\,\ref{FigpiDAE791}: dashed black, derived from $\varphi_{\rm as}(x)$; and solid blue, from Eq.\,\eqref{varphipilinear}.

Taken at face value, $\varphi_{\rm as}(x)$ provides a better description of the data's $x$-dependence than the realistic DA discussed in Sec.\,\ref{SecDApion}.  Moreover, in switching from the $\zeta \approx 2$\,GeV data to the $\zeta \approx 3\,$GeV set, the $\varphi_{\rm as}(x)$ $\chi^2/{\rm datum}$ decreases by a factor of two, with $\varphi_{\rm as}(x)$ unchanging under evolution.  On the other hand, that associated with Eq.\,\eqref{varphipilinear} is practically unchanged, even though the DA evolves.  Notably, as shown by Fig.\,\ref{FigpiDAE791}\,B, evolution is practically nugatory in passing from $2\to 3\,$GeV.

These remarks highlight the conundrum attendant upon any interpretation of the data in Ref.\,\cite{Aitala:2000hb} as a measure of $|\varphi_{\pi}(x;\zeta)|^2$.  There are two marked inconsistencies.
Existing data on the neutral pion electromagnetic transition form factor and new-generation pion elastic form factor data, depicted in Figs.\,\ref{FigFpiEIC}, suggest very strongly that $\varphi_{\rm as}(x)$ does not provide a valid description of pion structure at $\zeta \approx 2\,$GeV; yet it matches the E791 diffractive dissociation data very well.
QCD evolution from $2\to 3\,$GeV does not have any measurable impact on $\varphi_{\pi}(x;\zeta)$; yet the data reported in Ref.\,\cite{Aitala:2000hb} are described therein as changing significantly under this small step.  So, the connection with pion form factor predictions does not become any closer, yet the description of E791 data is claimed to improve.

It is now plain that one cannot be certain whether the E791 division of data into the two sets displayed in the separate panels of Fig.\,\ref{FigpiDAE791} is realistic.  Furthermore, updated analyses of the process illustrated in Fig.\,\ref{FigDD} have demonstrated that the simple dependence on the meson DA expressed in Eq.\,\eqref{eqDDmeson} becomes more complicated \cite{Braun:2001ih, Chernyak:2001ph}.  At best, many representations of the pion DA are consistent with the E791 data \cite{Bakulev:2003cs}.  Hence, one must conclude that whilst hard diffractive dissociation of a meson does provide information on the $x$-dependence of the bound-state's DA, experiments with better precision than E791 are necessary before that connection can profitably be exploited. 

%% file: S5_Continuum.tex
\section{Pion Distribution Functions}
\label{sec:PiKDFs}
\subsection{Forward Compton Scattering Amplitude}
\label{SecFCSA}
A description of the status of experiment and theory for pion and kaon distribution functions as it was ten years ago may be found in Ref.\,\cite{Holt:2010vj}.   An update on experiment is presented in Sec.\,\ref{sec:meson-distribution-functions-exp}.  Here it is worth recapitulating a few facts in order to establish the context for that update and recent developments in theory.  Thus, the hadronic tensor relevant to inclusive deep inelastic lepton+pseudoscalar-meson ($\ell \Pi$) scattering can be expressed via two invariant structure functions \cite{Jaffe:1985je}.  With the incoming photon possessing momentum $q$ and the target meson, momentum $P$, then in the deep-inelastic (Bjorken) limit \cite{Bjorken:1968dy}, \emph{viz}.\
%$q^2\to\infty$, $P\cdot q \to -\infty$ but $x:= - q^2/[2 P \cdot q]$ fixed,
%
\begin{equation}
\label{BJlim}
q^2\to\infty\,,\; P\cdot q \to -\infty, \;\mbox{but}\;
x:= - q^2/[2 P \cdot q] \; %\frac{q^2}{2 P\cdot q}\;\;
\mbox{fixed}\,,
\end{equation}
that tensor is $(t_{\mu\nu} = \delta_{\mu\nu}-q_\mu q_\nu/q^2, P_\mu^{\,t}= t_{\mu\nu}P_\nu)$:
\begin{equation}
W_{\mu\nu}(q;P)  = F_1(x)\, t_{\mu\nu} - \frac{F_2(x)}{P\cdot q}
\,P_\mu^{\,t} P_\nu^{\,t}\,, \quad
F_2(x)  = 2 x F_1(x)\,.
\end{equation}
Bjorken-$x$ measures the struck parton's share of the meson's light-front momentum.
$F_1(x)$ is the meson structure function, which provides access to the meson's quark distribution functions:
\begin{equation}
\label{qPDF}
F_1(x) = \sum_{{\mathpzc q}\in \Pi} \, e_{\mathpzc q}^2 \, {\mathpzc q}^\Pi(x)
\Rightarrow F_2(x) = \sum_{{\mathpzc q}\in \Pi} \, e_{\mathpzc q}^2 \,2 x\, {\mathpzc q}^\Pi(x)\,,
\end{equation}
where $e_{\mathpzc q}$ is the quark's electric charge.  The sum in Eq.\,\eqref{qPDF} runs over all quark flavours; but, \emph{e.g}.\ in the $\pi^+$ it is naturally dominated by ${\mathpzc u}^\pi(x)$, $\bar {\mathpzc d}^\pi(x)$.  Moreover, in the $\mathpzc{G}$-parity symmetric limit, which is a good approximation in Nature, ${\mathpzc u}^\pi(x)=\bar {\mathpzc d}^\pi(x)$.

\begin{figure}[t]
\includegraphics[clip, width=0.99\textwidth]{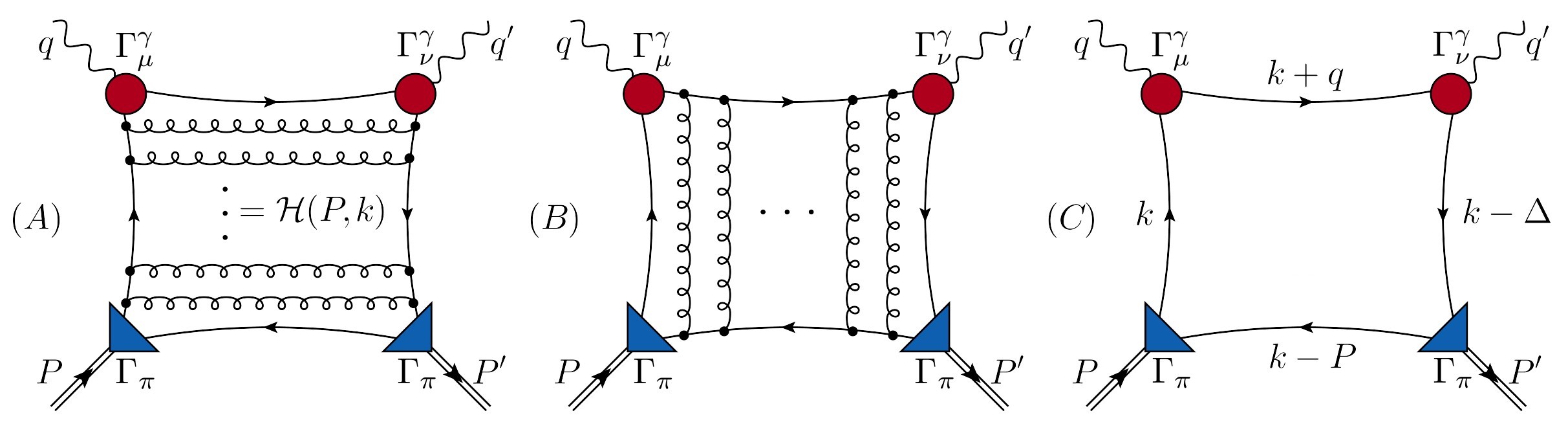}
\caption{\label{figCompton}
Collection of diagrams required to complete a sym\-metry-preserving leading-order (RL rainbow-ladder truncation) calculation of meson Compton scattering.
Amplitude-One (\emph{S1}) = (\emph{A})+(\emph{B})-(\emph{C}).  The ``dots'' in (\emph{A}) and (\emph{B}) indicate summation of infinitely many ladder-like rungs, beginning with zero rungs.
The other two amplitudes are obtained as follows: (\emph{S2}) -- switch vertices to which $q$ and $q^\prime$ are attached; and (\emph{S3}) -- switch vertex insertions associated with $q^\prime$ and $P^\prime$.
In all panels: triangles (blue) -- meson Bethe-Salpeter amplitudes, drawn here for the pion $\Gamma_{\Pi=\pi}$; circles (red) -- amputated dressed-photon-quark vertices, $\Gamma^\gamma$; and interior lines -- dressed-quark propagators.  $\Delta = q^\prime - q$.
Poincar\'e-covariance and electromagnetic current conservation, \emph{inter alia}, are guaranteed so long as each of these elements is computed in RL truncation.
%
%For later use, using (\emph{C}), we define line (\emph{a}) to be that carrying momentum $k$; line (\emph{b}), $k+q$; line (\emph{c}), $k-\Delta$; and line (\emph{d}), $k-P$.
%
}
\end{figure}

Using the optical theorem, the structure function is given by the imaginary part of the virtual-photon--meson forward Compton scattering amplitude: $\gamma^\ast(q) + \Pi(P) \to \gamma^\ast(q) + \Pi(P)$.  This understanding provides one direct way to compute it.  Namely, within the framework of continuum Schwinger function methods and at leading-order in a systematic, symmetry preserving truncation scheme, computation of the collection of diagrams in Fig.\,\ref{figCompton} is both necessary and sufficient to obtain a sound result \cite{Chang:2014lva, Ding:2019qlr, Ding:2019lwe}.  Notably, as first demonstrated in connection with $\pi \pi$ scattering \cite{Bicudo:2003fp, Bicudo:2001jq} and amplified in subsequent discussions of numerous other ``box-diagram'' processes, \emph{e.g}.\ Refs.\,\cite{Cotanch:2002vj, Cotanch:2003xv, Eichmann:2012mp, Eichmann:2015nra}, if any one of the contributions depicted in Fig.\,\ref{figCompton} is neglected in a given calculation of the Compton amplitude, then the result explicitly violates an array of crucial symmetries.
Typical consequences include the following:
overestimation of the sea and gluon content of a given bound-state;
incorrect estimates for the relative size of valence-quark momentum fractions within different but related bound-states;
misidentification of $\zeta_H$, if this scale is used as a parameter to fit some empirically-determined distribution \cite{Jaffe:1980ti};
and because these errors are communicated into the evolved distributions, a lack of credibility in the conclusions and interpretations drawn from the distributions.
Furthermore, the symmetry violations and connected errors are accentuated by including the ${\cal H}(P,k)$ resummation in Fig.\,\ref{figCompton}(\emph{A}) alone because this disrupts the balance of interferences that a fully-consistent truncation is guaranteed to preserve.
Consequently, less harm is done by working solely with Fig.\,\ref{figCompton}(\emph{C}), as done in the earliest studies \cite{Holt:2010vj}.

Working with the virtual-photon--meson forward Compton scattering amplitude specified by the diagrams in Fig.\,\ref{figCompton} and using the Ward-Green-Takahashi identity for the dressed-photon-quark vertex, Ref.\,\cite{Chang:2014lva} derived the following expression for the quark distribution function in a $\Pi = f\bar g$ pseudoscalar meson:
\begin{equation}
{\mathpzc f}^\Pi(x;\zeta_H)  = N_c {\rm tr}\!
\int_{dk}\! \delta_n^{x}(k_\eta)  n\cdot\partial_{k_\eta} \left[ \Gamma_\Pi(k_\eta,-P) S_{f}(k_\eta) \right]
\Gamma_\Pi(k_{\bar\eta},P)\, S_{\bar g}(k_{\bar\eta})\,,
\label{qFULL}
\end{equation}
where the derivative acts only on the bracketed terms.
One may prove algebraically that the result obtained using Eq.\,\eqref{qFULL} is:
independent of $\eta$, \emph{i.e}.\ the definition of the relative momentum within the bound state;
ensures
\begin{equation}
\label{xoneminusx}
\bar {\mathpzc g}^\pi(x;\zeta_H) = {\mathpzc f}^\pi(1-x;\zeta_H)\,;
\end{equation}
and satisfies the baryon number and momentum sum rules, \emph{viz}.\,
\begin{equation}
\label{eqSumRules}
\int_0^1dx\, {\mathpzc f}^\Pi(x;\zeta_H) = 1
= \int_0^1dx\, \bar{\mathpzc g}^\Pi(x;\zeta_H) = 1\,,
\quad
\int_0^1dx\,x \left[ {\mathpzc f}^\Pi(x;\zeta_H) + \bar {\mathpzc g}^\Pi(x;\zeta_H)\right]= 1\,.
\end{equation}

Regarding Eq.\,\eqref{qFULL}, it is natural to explore whether anything can be stated \emph{a priori} about the $x$ dependence of the distribution function it defines.  To this end, consider that the integrand depends on the quark propagator and the bound-state amplitude.
Independent of the theory and the number of spacetime dimensions, $D_{\rm st}$, the large momentum scaling behaviour of a fermion propagator is always $S(k) \approx 1/i\gamma\cdot k$.  Thus, all information about bound-state formation and properties is encoded in the meson's Bethe-Salpeter amplitude.
%The anomalous dimension generated by QCD interactions multiplies into this behaviour with an additional momentum-dependent logarithmic suppression ... could be enhancement, if Z is monotonic.  But lQCD doesn't produce monotonic result.

Beginning with the homogeneous Bethe-Salpeter equation derived from Fig.\,\ref{FigMesonBSE}, it is straightforward to show, using little more than dimensional counting, that a $D_{\rm st}=4$ vector-boson exchange theory, characterised by boson exchange with large-momentum scaling behaviour $1/[k^2]^\nu$, produces a bound-state amplitude with the following feature:
\begin{equation}
\label{eqBSAUV}
 \Gamma_\Pi(k,-P) \stackrel{k^2\gg m_0^2}{\approx} i\gamma_5 \frac{1}{[k^2]^\nu}\,.
\end{equation}
This result is also valid in the symmetry-preserving treatment of a contact interaction, $\nu=0$.  (These remarks extend those made following Eq.\,\eqref{M0OPE}.)  The analysis can be made more rigorous by adapting, \emph{e.g}.\ what has been called the Higashijima-Miransky approximation \cite{Higashijima:1983gx, Miransky:1983vj} for treating the integrals that appear in quantum field theory bound-state equations.  In this way, one can readily reproduce the QCD prediction obtained in Ref.\,\cite{Politzer:1976tv}, which leads to Eq.\,\eqref{eqBSAUV} via Eq.\,\eqref{GTRE} and its corollaries.

The scaling patterns just described are expressed in the following equations:
\begin{subequations}
\label{NakanishiASY}
\begin{align}
\label{eq:sim1}
S(k) & = [-i\gamma\cdot k+M]\Delta_M(k^2)\,,
\quad \Delta_M(s) = 1/[s+M^2]\,, \\
%
%\mathpzc{n}_\pi \Gamma_\pi(k;P) & = & i\gamma_5\int^1_{-1}dz\, \rho_\nu(z) \, \hat\Delta^\nu_M(k^2_{+z})\,,
%
\mathpzc{n}_\Pi \Gamma_\Pi(k_{\bar\eta/\eta};\pm P) & =  i\gamma_5\int^1_{-1}dz\, \rho_\nu(z) \, \hat\Delta^\nu_M(k^2_{z})\,,
\quad
\rho_\nu(z)  =  \frac{1}{\sqrt{\pi}}\frac{\Gamma(\nu+3/2)}{\Gamma(\nu+1)}(1-z^2)^\nu\,,
\label{eq:sim2}
\end{align}
\end{subequations}
where $M$ is a dressed-quark mass-scale, $\hat\Delta_M(s)=M^2\Delta_M(s)$, $k_{z}=k_{\bar\eta/\eta} + (z \pm 1) P/2$, and $\mathpzc{n}_\Pi $ is the canonical normalisation constant.  To simplify the algebra, consider the chiral limit, so that$P^2=0$.  Then inserting Eqs.\,\eqref{NakanishiASY} into Eq.\,\eqref{varphiresult}, one obtains \cite{Chang:2013pq}
\begin{equation}
\label{eqvarphinu}
\varphi_\Pi(x) = \frac{\Gamma(2\nu+2)}{[\Gamma(\nu + 1)]^2} \, [x(1-x)]^\nu\,.
\end{equation}
Evidently, the scaling power of the bound-state's DA is completely determined by that of the exchanged boson which produces the bound state.

Turning now to the DF, consider $\nu=0$ in Eqs.\,\eqref{NakanishiASY}.  In this case $\varphi_\Pi(x) = 1$; and implementing a translationally invariant regularisation of the integral in Eq.\,\eqref{qFULL}, one finds, in accordance with Eq.\,\eqref{PDFeqPDA2}:
\begin{equation}
{\mathpzc q}_{\nu=0}^\Pi(x;\zeta_H) = 1 = [\varphi_\Pi^{\nu=0}(x)]^2\,.
\end{equation}
This well-known result was first obtained in Ref.\,\cite{Davidson:1994uv}.

\begin{figure}[t]
\centerline{%
\includegraphics[clip, width=0.46\textwidth]{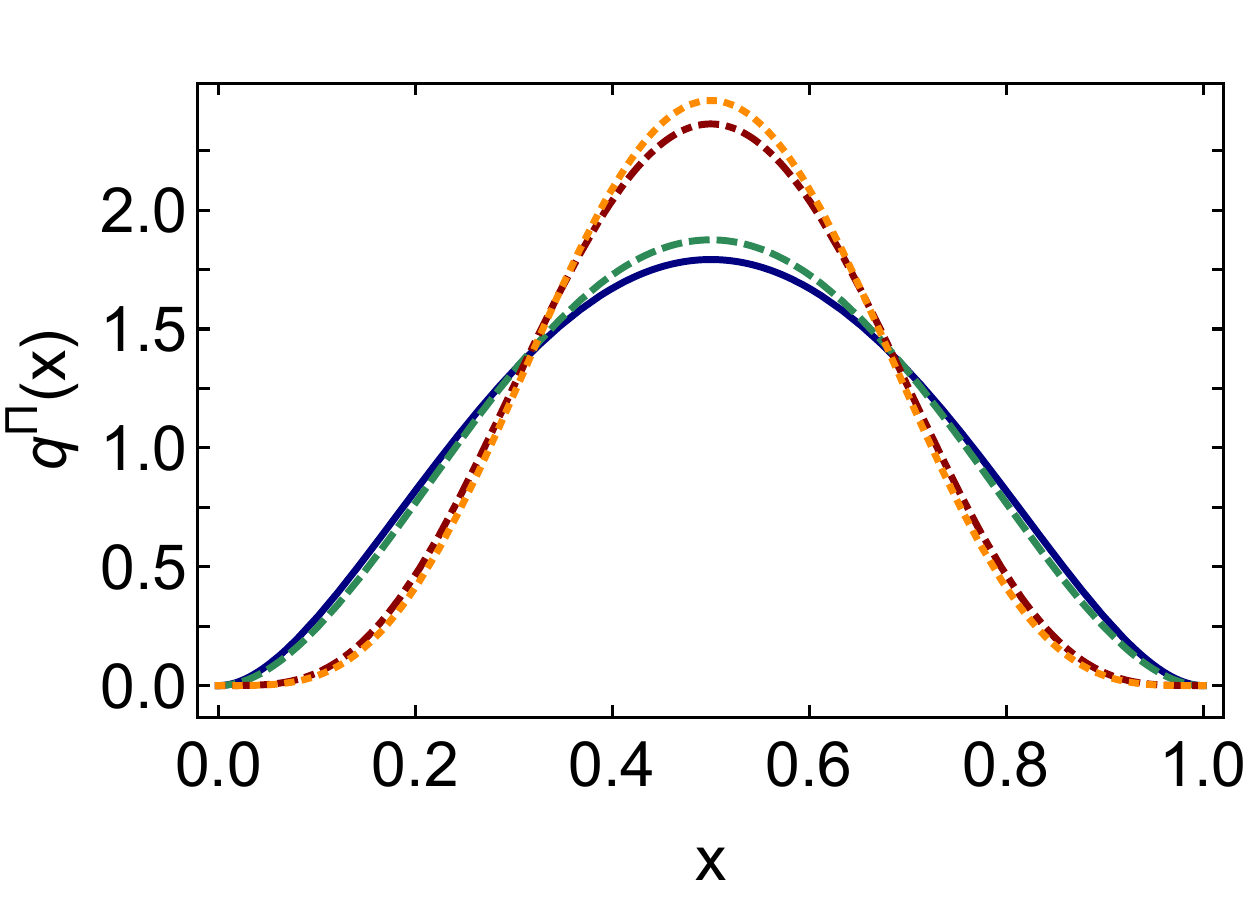}}
\caption{\label{figDFcfDA2}
Comparison: ${\mathpzc q}^\Pi(x;\zeta_H)$ vs.\ $[\tilde\varphi_\Pi(x;\zeta_H)]^2$.  This figure establishes the veracity of Eqs.\,\eqref{eqnu1DFDA1} and \eqref{eqnu2DFDA2}, respectively: solid blue curve \emph{cf}.\ dashed green curve for $1/k^2$ interaction; and dot-dashed red curve \emph{cf}.\ dotted orange for $1/k^4$ interaction.}
\end{figure}

The steps can be repeated using $\nu=1$.  Then \cite{Chang:2013pq, Chang:2014lva}:
\begin{subequations}
\label{nu1All}
\begin{align}
\varphi_\Pi^{\nu=1}(x) &= 6 x (1-x) = \varphi_{\rm as}(x)\,,\\
{\mathpzc q}_{\nu=1}^\Pi(x;\zeta_H) & =
\frac{72}{25} \left[x^3 (x [2 x-5]+15) \ln x
+\left(2 x^2+x+12\right) (1-x)^3 \ln (1-x) \right.\nonumber \\
 & \qquad  \left. + 2 x (6-(1-x) x) (1-x)\right] \\
 & \approx \frac{5}{6} [\varphi_\Pi^{\nu=1}(x)]^2 =: [\tilde \varphi_\Pi^{\nu=1}(x)]^2 \,. \label{eqnu1DFDA1}
\end{align}
\end{subequations}
The veracity of the last statement is established in Fig.\,\ref{figDFcfDA2}: solid blue curve \emph{cf}.\ dashed green curve, for which the ${\mathpzc L}_1$ difference is 4.1\%.  Notably, using Eq.\,\eqref{xoneminusx}, ${\mathpzc q}_{\nu=1}^\Pi(x;\zeta_H)$ is symmetric under $x \leftrightarrow (1-x)$; and
\begin{equation}
{\mathpzc q}_{\nu=1}^\Pi(x;\zeta_H) \stackrel{x\simeq 1}{=} \frac{216}{5} (1-x)^2 + {\rm O}((1-x)^4)\,.
\label{eqqPinu1}
\end{equation}

Using $\nu=2$, one obtains
\begin{subequations}
\begin{align}
\varphi_\Pi^{\nu=2}(x) &= 30 [x (1-x)]^2 \,, \\
{\mathpzc q}_{\nu=2}^\Pi(x;\zeta_H) & =
\frac{15}{7} \left[
  30 x^5 (x [x (x [2 x-9]+20)-28]+35) \ln x \right. \nonumber \\
&   \qquad -30 (x [x (2 x^2+x+5)+7]+20 ) (x-1)^5 \ln (1-x) \nonumber\\
& \qquad \left.
-5 x ( [x-1] x ([x-1] x (12 [x-1] x+25)+378)+120)
   (x-1)\right] \\
&  \approx \frac{7}{10} [\varphi_\Pi^{\nu=2}(x)]^2 =: [\tilde \varphi_\Pi^{\nu=2}(x)]^2\,. \label{eqnu2DFDA2}
\end{align}
\end{subequations}
Eq.\,\eqref{eqnu2DFDA2} is illustrated in Fig.\,\ref{figDFcfDA2}: dot-dashed red curve \emph{cf}.\ dotted orange curve, for which the ${\mathpzc L}_1$ difference is 3.8\%.  Naturally, in this case, too,
${\mathpzc q}_{\nu=2}^\Pi(x;\zeta_H)$ is symmetric under $x \leftrightarrow (1-x)$; and
\begin{equation}
{\mathpzc q}_{\nu=2}^\Pi(x;\zeta_H) \stackrel{x\simeq 1}{=} 1125 (1-x)^4 + {\rm O}((1-x)^6)\,.
\end{equation}

Explicit calculations for ${\mathpzc q}_{\nu}^{\Pi}(x;\zeta_H)$ in Eq.\,\eqref{qFULL} using Eqs.\,\eqref{NakanishiASY} are not available for $\nu\geq 3$.  On the other hand, Eq.\,\eqref{eqvarphinu} is valid for any $\nu$; and using Eq.\,\eqref{PDFeqPDA2} and Fig.\,\ref{figDFcfDA2}, the pattern is clear:
\begin{equation}
{\mathpzc q}_{\nu}^\Pi(x;\zeta_H) \stackrel{x\simeq 1}{\propto} (1-x)^{2\nu} +  {\rm O}((1-x)^{2\nu+2})\,.
\end{equation}
Such a connection between the $k^2$-dependence of the exchange interaction and the $x$-dependence of the meson's DA and DF is found in all analyses of the continuum bound state problem.  Recent examples may be found in Refs.\,\cite{Xu:2018eii, Ding:2019qlr, Ding:2019lwe, Kock:2020frx, Cui:2020dlm, Cui:2020tdf, Broniowski:2020had, Zhang:2020ecj}.

Eq.\,\eqref{eqqPinu1} is an explicit illustration of a long known result in QCD.  Namely, any calculation that capitalises on the known behaviour of pseudoscalar meson wave functions at large valence-quark relative momenta \cite{Lane:1974he, Politzer:1976tv, Lepage:1980fj}, predicts the following large-$x$ behavior of the valence-quark DF (see, e.g.\ Refs.\,\cite{Ezawa:1974wm, Farrar:1975yb, Berger:1979du, Brodsky:1994kg, Yuan:2003fs}):
\begin{equation}
\label{pionPDFlargex}
%{\mathpzc q}^{\Pi}(x;\zeta_H) \stackrel{x\simeq 1}{\sim} (1-x)^{\beta} \,, \text{ with } \beta=2 \,.
{\mathpzc q}^{\Pi}(x;\zeta_H) \stackrel{x\simeq 1}{=} {\mathpzc c}(\zeta_H)\, (1-x)^{\beta_\Pi(\zeta_H)}\,,\; \beta_\Pi(\zeta_H)=2\,,
\end{equation}
${\mathpzc c}(\zeta_H)$ is a constant, \emph{i.e}.\ independent of $x$.  The hadronic scale is not accessible in experiment because specific kinematic conditions must be met before data can be interpreted in terms of ${\mathpzc q}^{\Pi}(x,\zeta)$ \cite{Ellis:1991qj}.  Hence, any prediction of ${\mathpzc q}^{\Pi}(x;\zeta_H)$ must be evolved to $\zeta_E \gtrsim m_p$ for comparison with experiment.  Under such evolution, the exponent becomes $2+\gamma$, where the anomalous dimension $\gamma\gtrsim 0$ and increases logarithmically with $\zeta$.\footnote{This feature can be seen via Eqs.\,\eqref{NakanishiASY}: an anomalous dimension on the meson's wave function can be approximated by inclusion of a small additional power, $\nu =1 \to (1+\epsilon)$, with a controllable level of precision \cite[Eq.\,(22)]{Chang:2013epa}, so that the exponent in Eq.\,\eqref{eqqPinu1} becomes $2(1+\epsilon)$.  For QCD, referring to Eqs.\,\eqref{M0OPE}, \eqref{GTRE}, $\epsilon > 0$.}
Eq.\,\eqref{pionPDFlargex} means that any analysis of data that can reasonably be interpreted in terms of ${\mathpzc q}^{\Pi}(x;\zeta_E)$ must produce $\beta_\Pi > 2$.  If it does not, then either: (\emph{a}) strong interactions in Nature are not described by a $1/k^2$ vector-boson exchange theory; or (\emph{b}) the analysis of the data is flawed in some way, \emph{e.g}.\ by having neglected important contributions to the hard scattering kernel, which is a critical piece in any attempt to connect data with DFs \cite{Ellis:1991qj}.

It is worth noting here that what has come to be known as the Drell-Yan-West relation provides a connection between the large-$x$ behaviour of DFs and the large-$Q^2$ dependence of hadron elastic form factors \cite{Drell:1969km, West:1970av}.  In its original form, the relation was discussed for $J=1/2$ targets, in fact, the proton.  It has long been known that this original form is not valid when the target is a pseudoscalar meson $(J=0)$ and the valence-parton scatterers are $J=1/2$ objects \cite{Ezawa:1973qx, Landshoff:1973pw}.  The generalisation to spin-$J$ targets constituted from $J=1/2$ quarks may be found in Ref.\,\cite{Brodsky:1994kg}, \emph{viz}.\ for a hadron ${\mathpzc H}$ defined by $n+1$ valence $J=1/2$ partons, so that its leading elastic electromagnetic form factor scales as $(1/Q^2)^n$:
\begin{equation}
\label{DYWrelation}
{\mathpzc q}^{{\mathpzc H}}(x;\zeta_H) \stackrel{x\simeq 1}{\sim} (1-x)^{\beta_{\mathpzc H}} \,, \; \beta_{\mathpzc H}= 2 n - 1 + 2 \Delta S_z\,,
\end{equation}
where $\Delta S_z = |S_z^q - S_z^{\mathpzc H}|$.  For a pseudoscalar meson, $n=1$, $S_z^{{\mathpzc H}=\Pi}=0$, and consequently $\beta_{{\mathpzc H}=\Pi}=2$.  Hence, one recovers Eq.\,\eqref{pionPDFlargex}.
For the proton, $n=2$ and $\Delta S_z=0$ for a non-spin-flip transition; so, $\beta_{{\mathpzc H}=p}=3$ for the valence-quark DF.  Within fitting uncertainties, this behavior has been confirmed \cite{Ball:2016spl}.

\subsection{Hadron Scale Pion Distribution Function}
\label{SecHSPDF}
As explicated in Table~\ref{tab:meson-sf-exp}, data that can be interpreted in terms of the pion's valence-quark DF were obtained thirty and more years ago at CERN \cite[WA39, NA3, NA10]{Corden:1980xf, Badier:1983mj, Betev:1985pg, Falciano:1986wk, Guanziroli:1987rp} and Fermilab \cite[E615]{Conway:1989fs}.  The E615 publication included a LO pQCD analysis of their data, which produced the result:
\begin{equation}
\label{qpiE615}
{\mathpzc q}_{\rm E615}^{\pi}(x; \zeta_5 = 5.2\,{\rm GeV}) \stackrel{x\simeq 1}{\sim} (1-x)^{1}\,.
\end{equation}
As highlighted and discussed in Ref.\,\cite{Holt:2010vj}, owing to the striking conflict between this result and Eq.\,\eqref{pionPDFlargex}, the E615 analysis triggered a continuing controversy.  Following that discussion, a new analysis was completed, working at next-to-leading order (NLO) in pQCD and including next-to-leading-logarithm (NLL - soft gluon) resummation \cite{Aicher:2010cb}.  It was known that NLO corrections to the hard-scattering kernel produce some softening of the DF \cite{Wijesooriya:2005ir}; but the inclusion of NLLs in the kernel was new and their effect was striking.  Indeed, the Ref.\,\cite{Aicher:2010cb} analysis delivered
\begin{equation}
\label{pionDFAicher}
{\mathpzc q}_{\mathrm E615\,\mbox{\footnotesize \cite{Aicher:2010cb}}}^{\pi}(x; \zeta_4 = 4\,{\rm GeV}) \stackrel{x\simeq 1}{\propto} (1-x)^{2.4}\,,
\end{equation}
in agreement with Eq.\,\eqref{pionPDFlargex}.

One might have supposed that Ref.\,\cite{Aicher:2010cb} would have resolved the controversy.  However, this has not proved true.  Instead, all subsequent, published analyses of data relevant to the determination of ${\mathpzc q}^\pi(x)$ \cite{Barry:2018ort, Novikov:2020snp, Han:2020vjp} have ignored NLL resummation in constructing the hard-scattering kernel, despite its clearly established importance \cite{Aicher:2010cb, Aicher:2011ai, Bonvini:2015ira, Westmark:2017uig, werner_vogelsang_2020_4019432}.  Consequently, the more recent analyses yield results like that in Eq.\,\eqref{qpiE615}; hence, in conflict with the prediction in Eqs.\,\eqref{pionPDFlargex}, \eqref{DYWrelation}.  In this connection, it is notable that recent steps toward an updated analysis of ${\mathpzc q}^{\pi}$-relevant data are exploring the inclusion of NLL resummation.  They all yield softened large-$x$ behaviour and typically lead to agreement with Eqs.\,\eqref{pionPDFlargex}, \eqref{DYWrelation} \cite{patrick_barry_2020_4019411, NobuoSato}.

Given the persistent concerns over the large-$x$ behaviour of the pion's valence-quark DF and the impact this has as a test of QCD, the past decade has seen numerous continuum calculations of ${\mathpzc q}^\pi(x)$, \emph{e.g}.\ Refs.\,\cite{Nguyen:2011jy, Nam:2012vm, Nam:2012af, Gutsche:2014zua, deTeramond:2018ecg, Lan:2019rba, Chang:2020kjj, Kock:2020frx}.  The problem was tackled using Eq.\,\eqref{qFULL} in Refs.\,\cite{Chang:2014lva, Chen:2016sno}, wherein algebraic models were employed for the required elements, \emph{i.e}.\ propagators for the dressed valence-quarks, Bethe-Salpeter amplitudes for the mesons, and dressed-photon-quark vertices.  These calculations were improved in Refs.\,\cite{Ding:2019qlr, Ding:2019lwe}, with all elements determined by solving the continuum bound-state equations defined by the QCD inputs and features described in Sec.\,\ref{sec:CTP}.  Whilst this makes the analysis a more challenging numerical problem, it delivers results with a tighter link to QCD.

With numerical solutions in hand for the terms in the integrand of Eq.\,\eqref{qFULL}, Refs.\,\cite{Ding:2019qlr, Ding:2019lwe} considered the DF's Mellin moments:
\begin{subequations}
\label{MellinMoments}
\begin{align}
 \langle   x^m  \rangle_{\zeta_H}^\pi  & = \int_0^1dx\, x^m {\mathpzc q}^\pi(x;\zeta_H) \label{MellinA}\\
& = \frac{N_c}{n\cdot P} {\rm tr}\! \int_{dk}\! \left[\frac{n\cdot k_\eta}{n\cdot P}\right]^m \Gamma_\pi(k_{\bar\eta},P)\, S(k_{\bar\eta})\,  n\cdot\partial_{k_\eta} \left[ \Gamma_\pi(k_\eta,-P) S(k_\eta) \right].
\end{align}
\end{subequations}
%%\begin{equation}
%%\langle   x^m  \rangle_{\zeta_H}^\pi  = \int_0^1dx\, x^m {\mathpzc q}^\pi(x;\zeta_H) =
%%\frac{N_c}{n\cdot P} {\rm tr}\! \int_{dk}\! \left[\frac{n\cdot k_\eta}{n\cdot P}\right]^m \Gamma_\pi(k_{\bar\eta},P)\, S(k_{\bar\eta})\,  n\cdot\partial_{k_\eta} \left[ \Gamma_\pi(k_\eta,-P) S(k_\eta) \right].
%%\end{equation}
%
Owing to Eq.\,\eqref{xoneminusx}, the value of any particular odd moment, $\langle x^{m_{\rm o}}\rangle_{\zeta_H}^{\pi}$, $m_{\rm o}=2 \bar m +1$, $\bar m \in \mathbb Z$, is known once all lower even moments are computed.  The identities are readily determined and can be used to validate the numerical methods employed to compute the Mellin moments.  In this way, Refs.\,\cite{Ding:2019qlr, Ding:2019lwe} computed and checked results for the $m=0,\ldots,5$ Mellin moments in Eq.\,\eqref{MellinMoments}.

Although every moment defined by Eq.\,\eqref{MellinMoments} is finite, direct calculation of $m\geq 6$ moments using numerically determined inputs for the propagators and Bethe-Salpeter amplitudes is difficult in practice because the $[n\cdot k_\eta]^m$ factor introduces oscillations that are increasingly more difficult to track using brute force.  In any perfect procedure, the oscillations cancel; but that is hard to achieve numerically.  This problem can be solved by using the Schlessinger point method (SPM) \cite{Schlessinger:1966zz, PhysRev.167.1411, Tripolt:2016cya, Chen:2018nsg, Binosi:2019ecz, Xu:2019ilh, Yao:2020vef, Huber:2020ngt} to construct an analytic function, $M_S(z)$, whose $z=0,1,\ldots,5$ values agree with the moments computed directly and for which $M_S(7)$ is related to $M_S(0)$, $M_S(2)$, $M_S(4)$, $M_S(6)$ as prescribed by Eq.\,\eqref{xoneminusx}.

Based on the Pad\'e approximant, the SPM is a powerful tool from numerical mathematics.  It is able to reliably reconstruct a function $f$ in the complex plane within a radius of convergence which is specified by that one of the branch points of $f$ which lies nearest to the real domain containing the sample points.  Furthermore, owing to the procedure's discrete nature, the reconstruction may also define a sound continuation on a larger domain.  However, this cannot be guaranteed.  Hence, each case must be treated individually.

\begin{figure}[t]
\hspace*{-1ex}\begin{tabular}{lcl}
{\sf A} &\hspace*{2em} & {\sf B} \\[-2ex]
\includegraphics[clip, width=0.46\textwidth]{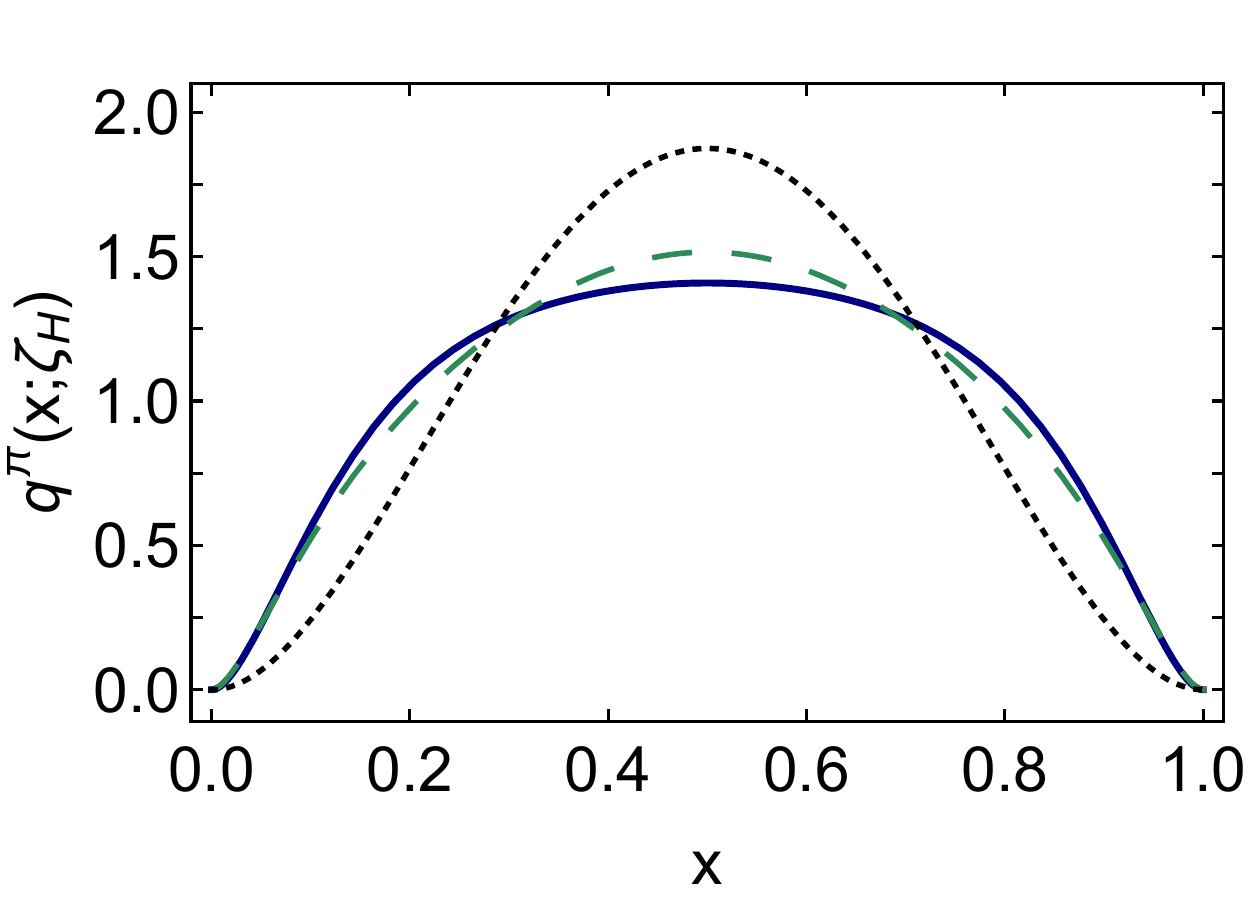} & \hspace*{2em} &
\includegraphics[clip, width=0.46\textwidth]{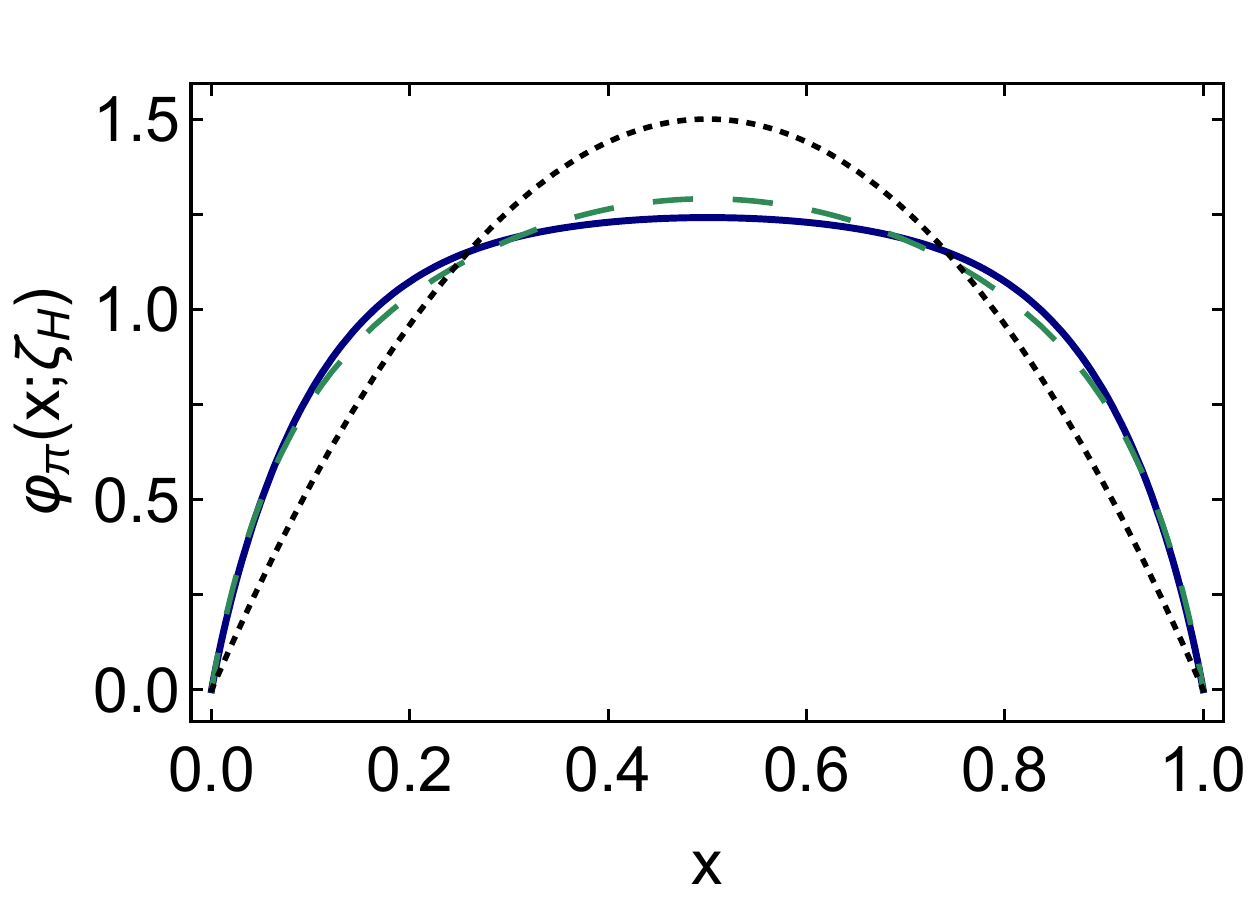}
\end{tabular}
\caption{\label{plotqpizetaH}
\emph{Left panel}\,--\,{\sf A}. ${\mathpzc q}^\pi(x;\zeta_H)$ in Eq.\,\eqref{qpizetaH} (solid navy curve) compared with the result in Eq.\,\eqref{qpizetaHphi2} (long-dashed green) obtained through Eq.\,\eqref{PDFeqPDA2} with the pion DA computed using the same framework, Eq.\,\eqref{varphipilinear}.  Dotted black curve, scale-free result ${\mathpzc q}_{\rm sf}(x)=30 x^2(1-x)^2$, connected with $\varphi_{\rm as}(x)$ in Eq.\,\eqref{phias} via Eq.\,\eqref{nu1All}.
\emph{Right panel}\,--\,{\sf B}. Reverse comparison; namely, $[{\mathpzc q}^\pi(x;\zeta_H)]^{1/2}$ (solid navy) compared with $\tilde\varphi_\pi(x)$ obtained from Eq.\,\eqref{varphipilinear} (long-dashed green).  Dotted black curve, $\varphi_{\rm as}(x)$ in Eq.\,\eqref{phias}.
}
\end{figure}

Working as just described, Refs.\,\cite{Ding:2019qlr, Ding:2019lwe} reconstructed the following valence-quark DF for the pion using eleven SPM-approximant Mellin moments of the pion DF:
\begin{equation}
{\mathpzc q}^\pi(x;\zeta_H)  = 213.32 \, x^2 (1-x)^2  [1 - 2.9342  \sqrt{x(1-x)} + 2.2911 \,x (1-x)]\,, \label{qpizetaH}
\end{equation}
%The mean absolute relative error between its first eleven moments and those of the separate reconstructed distributions is 4(3)\%.
%253.62003173893243*(1 - x)^2*x^2*(1 + 2.6974201897292325*(1 - x)*x - 3.171989803894321*Sqrt[(1 - x)*x])
%213.31662630896608*(1 - x)^2*x^2*(1 + 2.291114178403205*(1 - x)*x - 2.934232803738537*Sqrt[(1 - x)*x])
which is depicted in Fig.\,\ref{plotqpizetaH}\,A.  Also drawn in this panel is the DF obtained via Eq.\,\eqref{PDFeqPDA2} when the pion DA computed using precisely the same framework is inserted as the base, \emph{viz}.
\begin{equation}
\label{qpizetaHphi2}
{\mathpzc q}_{\tilde\varphi^2}^\pi(x;\zeta_H)  =
301.66  x^2 (1-x)^2 [1 - 2.3273 \sqrt{x(1-x)} + 1.7889 \, x (1-x) ]^2\,.
%301.657 (1-x)^2 x^2 \left(1.7889 (1-x) x-2.32734 \sqrt{1-x} \sqrt{x}+1\right)^2
\end{equation}
The ${\mathpzc L}_1$ difference between the two curves is 5.2\%, \emph{i.e}.\ practically identical to that obtained in every example presented above.
Fig.\,\ref{plotqpizetaH}\,B displays the companion image; namely, the DA obtained from Eq.\,\eqref{qpizetaH} using Eq.\,\eqref{PDFeqPDA2} in comparison with the directly computed DA in Eq\,\eqref{varphipilinear}.  Here the ${\mathpzc L}_1$ difference between the two curves is 2.7\%.

Following upon all previous examples, this outcome in the realistic case confirms the statement made above; to wit, with a level of pointwise discrepancy that will be imperceptible in any foreseeable experiment, one may write
\begin{equation}
\label{PDFeqPDA2Repeat}
{\mathpzc q}^\Pi(x;\zeta_H) = \tilde\varphi_{\Pi}^2(x;\zeta_H)\,.
\end{equation}
Given the relative ease of calculating pseudoscalar meson DAs in comparison with using Eq.\,\eqref{qFULL} to obtain the related DF, the value of Eq.\,\eqref{PDFeqPDA2Repeat} is manifest.  Accepting this, then it follows immediately that all expressions of EHM and Higgs modulation thereof in pseudoscalar meson leading-twist DAs are embedded equivalently in the meson valence-quark DFs.

\subsection{Pion Distribution Function at \mbox{$\zeta_2=$}2\,GeV}
\label{SecpiDF2}
It is common to present results for DFs and their moments at $\zeta_2=2\,$GeV; and usually, with a given DF specified at some scale $\zeta_0 <\zeta_2$, the perturbatively defined DGLAP equations \cite{Dokshitzer:1977sg, Gribov:1972ri, Lipatov:1974qm, Altarelli:1977zs} are employed to obtain a new result at the higher scale.  A prescription must nevertheless be specified because the standard DGLAP equations involve QCD's running coupling.  Practitioners typically take a purely pQCD perspective, implementing evolution with a DGLAP kernel calculated at a given order in perturbation theory.  If the scale at which evolution begins is large enough, then LO evolution kernels may be adequate, at least in practice.  If they fail, then NLO can be implemented, and so on, in principle.

An alternative scheme supposes that \cite{Grunberg:1982fw, Dokshitzer:1998nz, Dokshitzer:2004ie, Deur:2016tte, Hoyer:2018hdj}:
(\emph{i}) in connection with a given process, a nonperturbative effective charge (running coupling) exists;
(\emph{ii}) being matched to experiment, this coupling is free of a Landau pole;
and (\emph{iii}) using this charge, the associated leading-order DGLAP equations are exact.
Such effective charges are process dependent (PD); and even though those obtained from distinct observables can in principle be connected through an expansion of one coupling in terms of the other, the PD character is somewhat unsettling.  Typically, knowledge of one PD charge does not enable global predictions to be made for another process; and the relation between two such charges at infrared momenta can only be determined after both are independently constructed.

The PD charge alternative was reinterpreted and refashioned in Refs.\,\cite{Ding:2019qlr, Ding:2019lwe, Cui:2019dwv, Cui:2020dlm, Cui:2020tdf}, wherein it is advocated that evolution should be implemented by using the PI effective discussed in Sec.\,\ref{SecPICharge} to integrate the one-loop DGLAP equations.  Such a procedure leads, \emph{e.g}.\ to the following relationship between the  moments of a meson's valence-quark DF:
\begin{equation}
\label{EqMellin}
\langle x^n {\mathpzc q}^M \rangle_\zeta := \int_0^1 dx\, x^n  {\mathpzc q}^M(x;\zeta)\,,
\end{equation}
\begin{align}
\label{EqMellin2}
{\mathpzc x}^n_M(\zeta_H,\zeta) & :=
\frac{\langle x^n {\mathpzc q}^M\rangle_{\zeta}}{\langle x^n {\mathpzc q}^M\rangle_{\zeta_H}}
=\exp\left[ \frac{\gamma_0^n}{4\pi} \int_{\ln\zeta^2}^{\ln\zeta_H^2}dt\,\hat\alpha(t)\right]\,,
\end{align}
where $t=\ln k^2$ and
\begin{equation}
\label{Eqgamma0n}
\gamma_0^n = -\frac{4}{3} \left[ 3 + \frac{2}{(n+1)(n+2)} - 4 \sum_{k=1}^{n+1}\frac{1}{k}\right]\,.
\end{equation}
Since $\gamma_0^0=0$, then baryon-number does not change with $\zeta$; and $\gamma_0^1 = 32/9$.

Numerous reasons were offered in support of this scheme.  They included the qualities of $\hat\alpha(k^2)$ listed on page~\pageref{alphaPIlist}, for instance, as Fig.\,\ref{FigalphaPI} shows, $\hat \alpha(k^2)$ decreases monotonically on $k^2\geq 0$ and can serve to mark the boundary between soft and hard physics.  In addition, the fact that $\hat\alpha(k^2)$ is known to provide a foundation for the unification of many observables, including: hadron static properties \cite{Wang:2018kto, Qin:2019hgk, Xu:2019sns, Souza:2019ylx}; light- and heavy-meson distribution amplitudes \cite{Shi:2014uwa, Shi:2015esa, Ding:2015rkn, Chouika:2017rzs, Binosi:2018rht}; and related elastic and transition form factors \cite{Raya:2016yuj, Gao:2017mmp, Chen:2018rwz, Ding:2018xwy, Xu:2019ilh}.
%\footnote{Even if one chooses to doubt the implied global character of our evolution scheme, then given that $\hat\alpha(\zeta_H)/(2\pi)=0.25$, $[\hat\alpha(\zeta_H)/(2\pi)]^2=0.06$, it can otherwise be viewed as expressing a self-stabilising one-loop approximation to the DGLAP equations.}

Within the $\hat\alpha$ evolution scheme, an analysis of the large-$n$ behaviour of Eqs.\,\eqref{EqMellin}\,--\,\eqref{Eqgamma0n} yields \cite[Appendix\,1]{Cui:2020tdf}:
\begin{equation}
\label{TextBookbetac}
\beta_{M}(\zeta) = \beta_{M}(\zeta_H) +
\frac{3}{2} \ln {\mathpzc x}_M^1(\zeta,\zeta_H) \,, \quad
{\mathpzc c}(\zeta)  = {\mathpzc c}(\zeta_H)\frac{\Gamma(1+\beta_M(\zeta_H))}{\Gamma(1+\beta_M(\zeta))}
\left[{\mathpzc x}_M^1(\zeta,\zeta_H)\right]^{\frac{3}{2}[\frac{3}{4}-\gamma_E]},
\end{equation}
where $\gamma_E=0.5772\ldots$ is Euler's constant.  When expressed at one-loop order in pQCD, the first identity in Eq.\,\eqref{TextBookbetac}, is a textbook result, \emph{e.g}.\ Ref.\,\cite[Eq.\,(4.137)]{Ellis:1991qj}; yet, it is often overlooked.  Eqs.\,\eqref{TextBookbetac} connect the large-$x$ exponent of a meson's valence-quark DF at scale $\zeta$ with the momentum fraction carried by valence-quarks at this scale.  As might have been anticipated, with decreasing valence-quark momentum fraction, the multiplicative coefficient also decreases and the large-$x$ exponent increases logarithmically.

Following Ref.\,\cite{Jaffe:1980ti}, the initial scale for evolution, $\zeta_0$, came to be viewed as a free parameter, with a value chosen \emph{a posteriori} after securing agreement with the results obtained for some DF Mellin moments in phenomenological analyses of data.  The $\hat\alpha$ evolution scheme begins from a markedly different position.  Namely, referring to the first point on page~\pageref{alphaPIlist}, the initial scale for DF evolution is the hadronic scale, whose value is fixed \emph{a priori} at the position of the Landau pole screening mass: $\zeta_H \approx 1.4\,\Lambda_{\rm QCD}$.  At this scale, the bound state is defined solely in terms of the quasiparticle degrees-of-freedom used to solve the continuum bound-state equations.  Consequently, glue and sea distributions vanish identically at $\zeta_H$, being sublimated into the dressed-quark and
dressed-antiquark quasiparticles, thus explaining the second identity in Eq.\,\eqref{eqSumRules}.

\begin{figure}[t]
\hspace*{-1ex}\begin{tabular}{lcl}
{\sf A} &\hspace*{2em} & {\sf B} \\[-2ex]
\includegraphics[clip, width=0.46\textwidth]{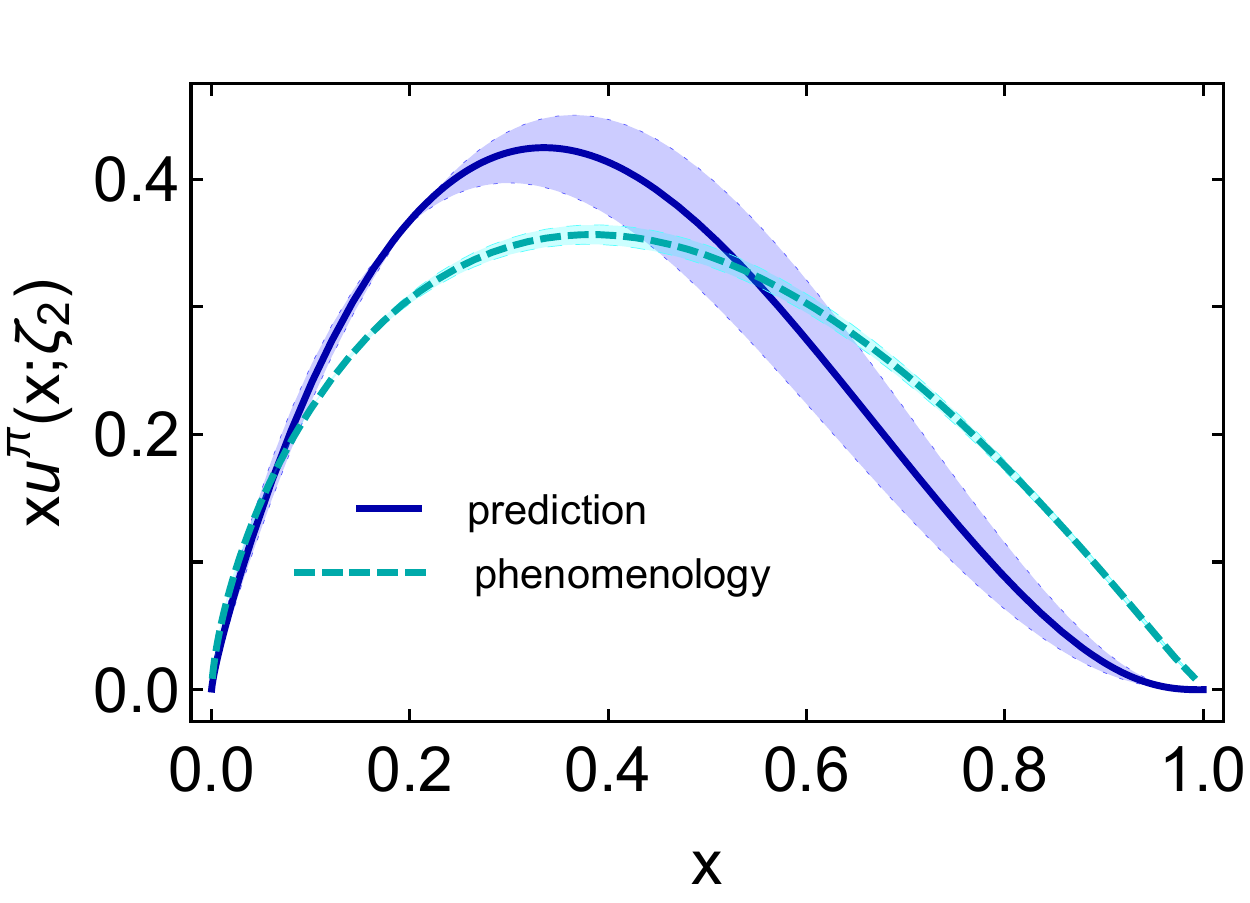} & \hspace*{2em} &
\includegraphics[clip, width=0.46\textwidth]{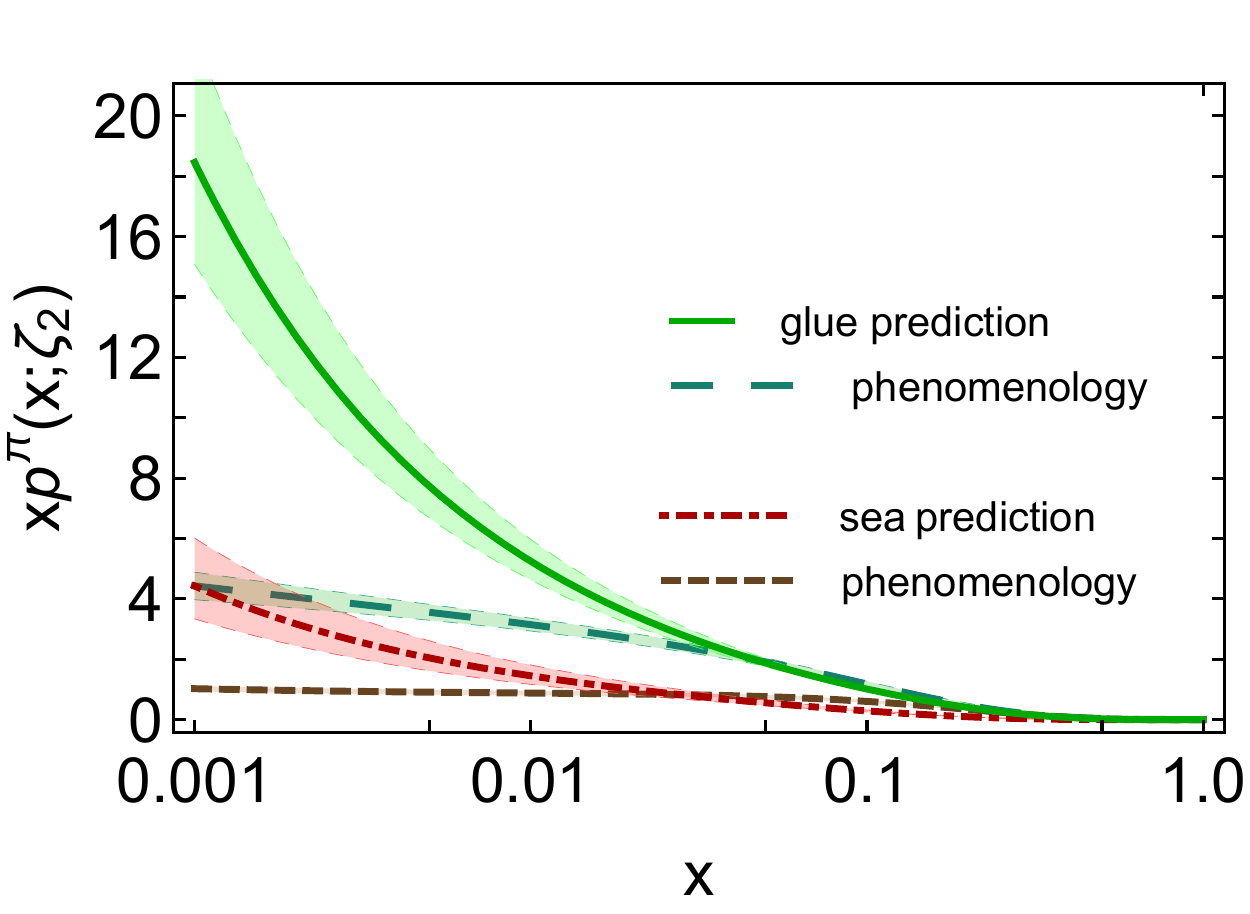}
\end{tabular}
\caption{\label{qpizeta2}
\emph{Left panel}\,--\,{\sf A}.
Solid blue curve -- valence-quark distribution defined by Eqs.\,\eqref{qHDA}, \eqref{varphipilinear}, \eqref{PDFeqPDA2Repeat} evolved to $\zeta=\zeta_2$, using the $\hat\alpha$ evolution scheme explained in connection with Eq.\,\eqref{setzetaH};
and short-dashed cyan curve -- phenomenological result from Ref.\,\cite{Barry:2018ort}.
\emph{Right panel}\,--\,{\sf B}.
Drawn from Ref.\,\cite{Cui:2020tdf}: solid green curve, $p=g$ -- prediction for the pion's glue distribution; and dot-dashed red curve, $p=S$ -- predicted sea-quark distribution.  Phenomenological results from Ref.\,\cite{Barry:2018ort} are plotted for comparison:
$p=\,$glue -- long-dashed dark-green; and
$p=\,$sea -- dashed brown.
Normalisation convention: $\langle x[2 {\mathpzc u}^\pi(x;\zeta_2)+g^\pi(x;\zeta_2)+S^\pi(x;\zeta_2)]\rangle=1$.
Notably, $2 {\mathpzc u}^\pi(x;\zeta_2) > [{g}^\pi(x;\zeta_2)+S^\pi(x;\zeta_2)]$ on $x>0.2$, marking this as the valence domain within the pion.
(The uncertainty bands bracketing the theory predictions are explained following Eq.\,\eqref{setzetaH}.)
}
\end{figure}

Beginning with ${\mathpzc q}^\pi(x;\zeta_H)$ defined by Eqs.\,\eqref{qHDA}, \eqref{varphipilinear}, \eqref{PDFeqPDA2Repeat} and a $n_f=4$ $\overline{\rm MS}$ value of $\Lambda_{\rm QCD}=0.234\,$GeV, used in computing the result in Eq.\,\eqref{varphipilinear}, so that
\begin{equation}
\label{setzetaH}
\zeta_H = 0.331(2)\,{\rm GeV},
\end{equation}
Ref.\,\cite{Cui:2020tdf} delivered the prediction for ${\mathpzc u}^\pi(x;\zeta_2)$ in Fig.\,\ref{qpizeta2}\,A -- solid blue curve.  In order to express a conservative estimate of uncertainty arising from that in the value of $\hat\alpha(0)$, which implicitly determines $\zeta_H$, all predictions in Refs.\,\cite{Cui:2020dlm, Cui:2020tdf} are embedded in bands determined by changing $\zeta_H \to \zeta_H(1\pm 0.01)$.
The prediction in Fig.\,\ref{qpizeta2}\,A exhibits the large-$x$ behaviour prescribed by Eqs.\,\eqref{TextBookbetac}.  For practical purposes, a measurable large-$x$ exponent, $\beta_\pi^{\rm eff}(\zeta)$, can be determined by plotting $\ln  {\mathpzc u}^\pi(x;\zeta)$ against $\ln(1-x)$ on $x\in[0.9,1.0]$ and extracting the slope, with the result
\begin{equation}
\label{Knownbeta2}
\beta_\pi^{\rm eff}(\zeta_2) = 2.63(8)\,.
\end{equation}

Since it remains common for lQCD computations to focus on low-order moments of meson DFs -- Sec.\,\ref{SeclQCDDF}, it is worth comparing some obtained recently with those calculated from the prediction in Fig.\,\ref{qpizeta2}\,A:
\begin{equation}
\label{momentslQCD}
\begin{array}{l|lll}
  \zeta=\zeta_2  & \langle x \rangle_u^\pi & \langle x^2 \rangle_u^\pi & \langle x^3 \rangle_u^\pi\\\hline
%\mbox{\cite{Best:1997qp}} & 0.28(8) & 0.11(3) & 0.048(20)\\
%\mbox{Ref.\,\cite{Detmold:2003tm}} & 0.24(2) & 0.09(3) & 0.053(15)\\
%\mbox{Ref.\,\cite{Brommel:2006zz}} & 0.27(1) & 0.13(1) & 0.074(10)\\
%\mbox{lQCD\,\cite{Oehm:2018jvm}} & 0.21(1) & 0.16(3) & \\
\mbox{lQCD\,\cite{Sufian:2019bol}} & 0.21(3) & 0.082(13) & 0.041(07) \\
\mbox{lQCD\,\cite{Joo:2019bzr}} & 0.254(03) & 0.094(12) & 0.057(04) \\\hline
%{\rm average} & 0.26(8) & 0.11(4) & 0.058(27)\\\hline
%{\rm average} & 0.24(2) & 0.13(4) & 0.064(18)\\\hline
%\mbox{\cite{Oehm:2018jvm, Joo:2019bzr}\,average} & 0.23(1) & 0.127(16) & 0.057(04)\\\hline
%
%\mbox{Ref.\,\cite{Hecht:2000xa}} & 0.24 & 0.098 & 0.049 \\
%
%\mbox{Refs.\,\cite{Ding:2019qlr, Ding:2019lwe}} & 0.24(2) & 0.098(10) & 0.049(07)\\\hline
%
\mbox{continuum\,\cite{Cui:2020tdf}} & 0.24(2) & 0.094(13) & 0.047(08)
% {\rm herein} & 0.26 & 0.11 & 0.052 ... Chen
\end{array}\,.
\end{equation}
%<x>= 0.2075(53)stat(20)sys(90)Z a
% <x2> = 0.163(23)stat(25)sys
%
%{1, 0.259544}, {2, 0.108411}, {3, 0.0560606},
%{1, 0.224122}, {2, 0.0861982}, {3, 0.0420332},
(Here, $\langle x^n\rangle_u^\pi = \langle x^n {\mathpzc u}^\pi(x)\rangle$, with the scale specified separately.  A larger array of comparisons, including model results, may be found in Ref.\,\cite[Table~III]{Lan:2019rba}.)
An uncertainty-weighted average of results in Eq.\,\eqref{momentslQCD} yields the following value of the light-front momentum fraction carried by valence-quarks in the pion at $\zeta=\zeta_2$:
\begin{equation}
\label{pionvalence}
\langle 2 x {\mathpzc u}^\pi(x;\zeta_2)\rangle = 0.49(2)\,.
\end{equation}
%\emph{i.e}.\ one-half.

Fig.\,\ref{qpizeta2}\,A also shows a fit obtained by analysing data on $\pi$-nucleus DY and leading neutron electroproduction data \cite{Barry:2018ort} -- short-dashed cyan curve.  Although this fit produces a consistent momentum fraction, \emph{viz}.\ $\langle 2 x \rangle_{\mathpzc q}^\pi  = 0.49(1)$, its $x$-profile is markedly different from the continuum prediction: using a ${\mathpzc L}_1$ measure, the two curves differ by 40\%.  Furthermore, the phenomenological DF conflicts with the prediction in Eqs.\,\eqref{pionPDFlargex}, \eqref{DYWrelation}.  As noted above, the analysis in Ref.\,\cite{Barry:2018ort} ignored NLL resummation, which is known to be important at large-$x$ \cite{Aicher:2010cb, Aicher:2011ai, Bonvini:2015ira, Westmark:2017uig, werner_vogelsang_2020_4019432}; and an overestimate of the DF at large-$x$ entails an underestimate at intermediate $x$.

\begin{figure}[t]
\hspace*{-1ex}\begin{tabular}{lcl}
{\sf A} &\hspace*{2em} & {\sf B} \\[-2ex]
\includegraphics[clip, width=0.46\textwidth]{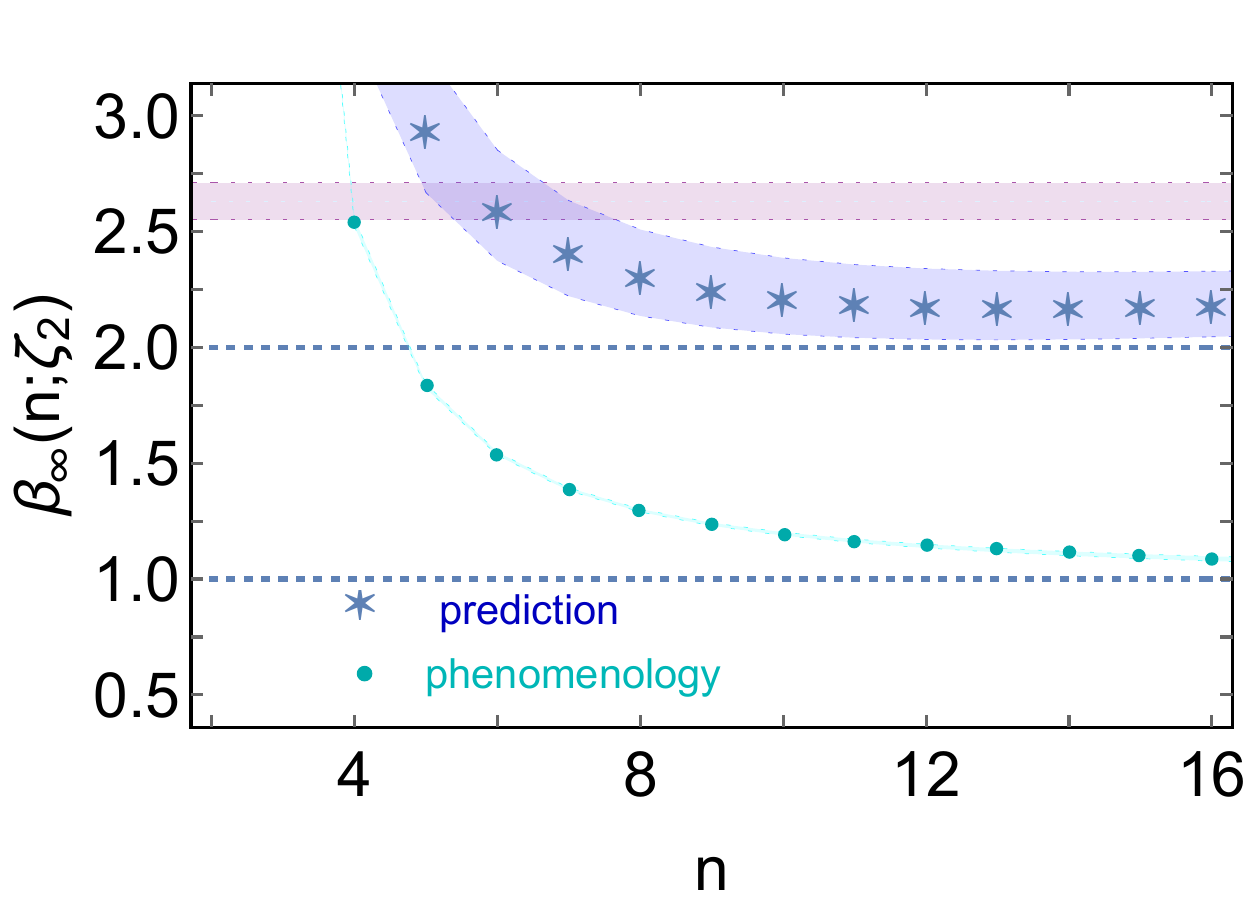} & \hspace*{2em} &
\includegraphics[clip, width=0.46\textwidth]{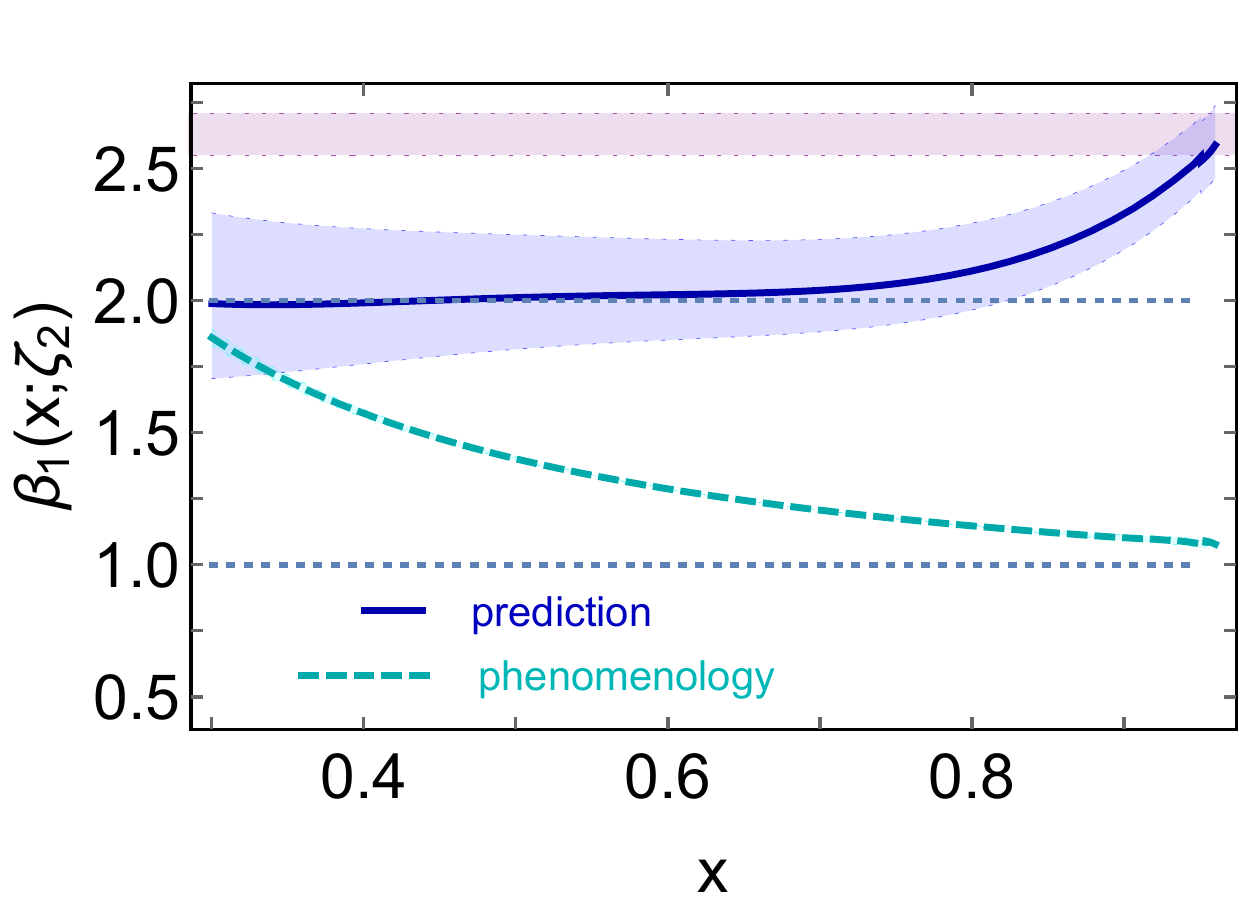}
\end{tabular}
\caption{\label{Figmoments}
\emph{Left panel}\,--\,{\sf A}.
Effective large-$x$ exponent defined via moments of the pion's valence-quark DF, Eq.\,\eqref{betainfinity}:
blue stars -- computed from the solid blue curve in Fig.\,\ref{qpizeta2}\,A; and cyan circles -- obtained from the dashed cyan curve.
\emph{Right panel}\,--\,{\sf B}.
Effective local ($x$-dependent) DF exponent defined in Eq.\,\eqref{beta1}:
solid blue curve -- calculated from the analogous curve in Fig.\,\ref{qpizeta2}\,A; and dashed cyan curve,  computed likewise.
In both panels, the purple band marks the known large-$x$ exponent in Eq.\,\eqref{Knownbeta2}; and to guide comparisons, horizontal dotted lines are drawn at $\beta_{\infty,1} = 1,2$.
(The uncertainty bands bracketing the theory predictions are explained following Eq.\,\eqref{setzetaH}.)
}
\end{figure}

These remarks highlight that low-order moments of a valence-quark DF contain practically no information about its large-$x$ behaviour.  To expand upon this, consider that for a DF with the behaviour in Eq.\,\eqref{pionPDFlargex}, with $\beta_\Pi$ undetermined \cite[Appendix\,1]{Cui:2020tdf}:
\begin{equation}
\langle x^n \rangle \stackrel{n\gg 1}=\frac{ {\rm constant}}{n^{1+\beta_\Pi}} \times [1 + {\rm O}(1/n)]\,,
\end{equation}
which entails \cite{Gao:2020ito}:
\begin{equation}
\label{betainfinity}
\beta_\Pi(n) = \beta_\infty(n) + {\rm O}(1/n)\,, \quad
\beta_\infty(n) =
-1 + \frac{\langle x^{n-2} \rangle - \langle x^{n+2} \rangle}{\langle x^n \rangle}\,.
\end{equation}
Fig.\,\ref{Figmoments}\,A plots $\beta_\infty(n)$ as obtained from both curves in Fig.\,\ref{qpizeta2}\,A. Evidently, one must possess a reliable determination of at least six DF Mellin moments in order to distinguish between these results.  Moreover, moments with very high order are required for extraction of the large-$x$ exponent $\beta_\Pi$ from a realistic DF, such as that calculated in Ref.\,\cite{Cui:2020tdf}, because $\beta_\infty(n)$ does not typically approach its $n\to\infty$ limit from above \cite{Gao:2020ito}.  This remark refers to the fact that the analysis in Ref.\,\cite{Barry:2018ort} employed a very simple DF model, \emph{viz}.\ ${\mathpzc q}(x) \propto x^{\alpha_\Pi}(1-x)^{\beta_\Pi}$, which has long been known to be too inflexible to describe real data \cite{Hecht:2000xa, Wijesooriya:2005ir}.  Additional discussion of this and related issues can be found in Ref.\,\cite{Courtoy:2020fex}.

One may also consider an effective $x$-dependent exponent; namely,
\begin{equation}
\label{beta1}
\beta_1(x) = \frac{(x-1) }{{\mathpzc q}(x)} \frac{d{\mathpzc q}(x)}{dx}\,.
\end{equation}
For the simple DF parametrisation:
\begin{equation}
{\mathpzc q}(x) \propto x^{\alpha_\Pi}(1-x)^{\beta_\Pi} \Rightarrow
\beta_1(x) = \alpha_\Pi - \frac{\alpha_\Pi}{x} + \beta_\Pi \,.
\end{equation}
At scales for which data may properly be interpreted in terms of DFs, one typically has $\alpha_\Pi<0$; hence, $\beta_1(x) \to \beta_1(1)$ from above when such a fitting form is used.  This is not the case for realistic DFs, as shown in Fig.\,\ref{Figmoments}\,B, which depicts $\beta_1(x)$ as obtained from both curves in Fig.\,\ref{qpizeta2}\,A.  Panel\,B indicates that precise data on $0.5 \lesssim x\lesssim 0.7$ should be sufficient to qualitatively confirm the behaviour in Eqs.\,\eqref{pionPDFlargex}, \eqref{DYWrelation}.  It also suggests that a quantitatively reliable extraction of the large-$x$ exponent, which means exposing scaling violations, would require equally good data on $0.7 \lesssim x\lesssim 0.9$.

In the $\hat\alpha$ evolution scheme, the pion's glue and sea distributions are generated by evolution on $\zeta>\zeta_H$; and the predictions from Ref.\,\cite{Cui:2020tdf} are drawn in Fig.\,\ref{qpizeta2}\,B.  It is worth recording that the glue and sea distributions in Refs.\,\cite{Ding:2019qlr, Ding:2019lwe, Cui:2020dlm, Cui:2020tdf} are the first parameter-free predictions to become available.  The associated momentum fractions are ($\zeta=\zeta_2$):
\begin{equation}
\langle x\rangle^\pi_g = 0.41(2)\,, \quad
\langle x\rangle^\pi_{\rm sea} = 0.11(2)\,.
\end{equation}
Evidently, considering Eq.\,\eqref{pionvalence}, the momentum sum rule is preserved.
Adopting a similar procedure, the model in Ref.\,\cite{Lan:2019rba} yields momentum fractions at $\zeta_2$ that are consistent with these predictions, \emph{viz}.\ \cite{Lan:2020hyb}: $\langle x\rangle^\pi_g =0.40$, $\langle x\rangle^\pi_{\rm sea}=0.11$, despite the fact that the valence-quark DF therein conflicts with Eqs.\,\eqref{pionPDFlargex}, \eqref{DYWrelation}.

The glue distribution inferred via data fitting in Ref.\,\cite{Barry:2018ort} is also shown in Fig.\,\ref{qpizeta2}\,B.  It agrees semiquantitatively with the prediction on $x\gtrsim 0.05$; but is markedly different on the complementary domain.  In gross terms, it produces a measurably smaller gluon momentum fraction: $\langle x\rangle^\pi_g=0.35(3)$.  Notably, both glue DFs in Fig.\,\ref{qpizeta2}\,B disagree with those inferred in earlier analyses \cite{Gluck:1999xe, Sutton:1991ay}.  These observations serve to stress the need for modern experiments that are directly sensitive to the pion's gluon content, \emph{e.g}.\ prompt photon and $J/\Psi$ production \cite{Denisov:2018unjF, Chang:2020rdy}.

The sea DF extracted in Ref.\,\cite{Barry:2018ort} is drawn as the short-dashed brown curve in Fig.\,\ref{qpizeta2}\,B.  It produces a large sea momentum fraction, $\langle x\rangle^\pi_{\rm sea}=0.16(1)$, and differs from the Ref.\,\cite{Cui:2020tdf} prediction on the entire $x$-domain.  Plainly, empirical information on the pion's sea distribution is sorely needed.  This can potentially be secured through the collection and analysis of DY data with $\pi^\pm$ beams on isoscalar targets \cite{Londergan:1995wp, Denisov:2018unjF}.

\begin{figure}[t]
\hspace*{-1ex}\begin{tabular}{lcl}
{\sf A} &\hspace*{2em} & {\sf B} \\[-2ex]
\includegraphics[clip, width=0.46\textwidth]{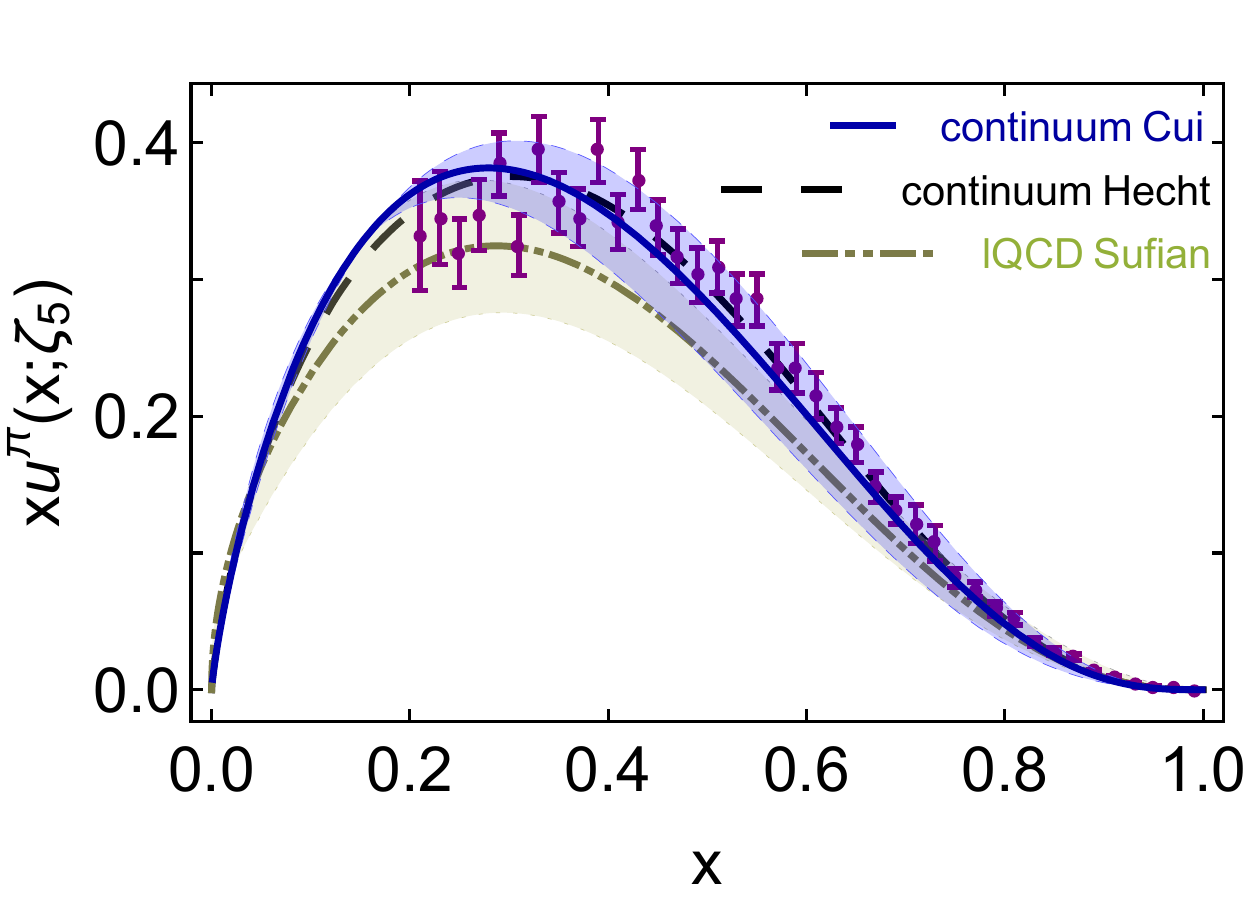} & \hspace*{2em} &
\includegraphics[clip, width=0.46\textwidth]{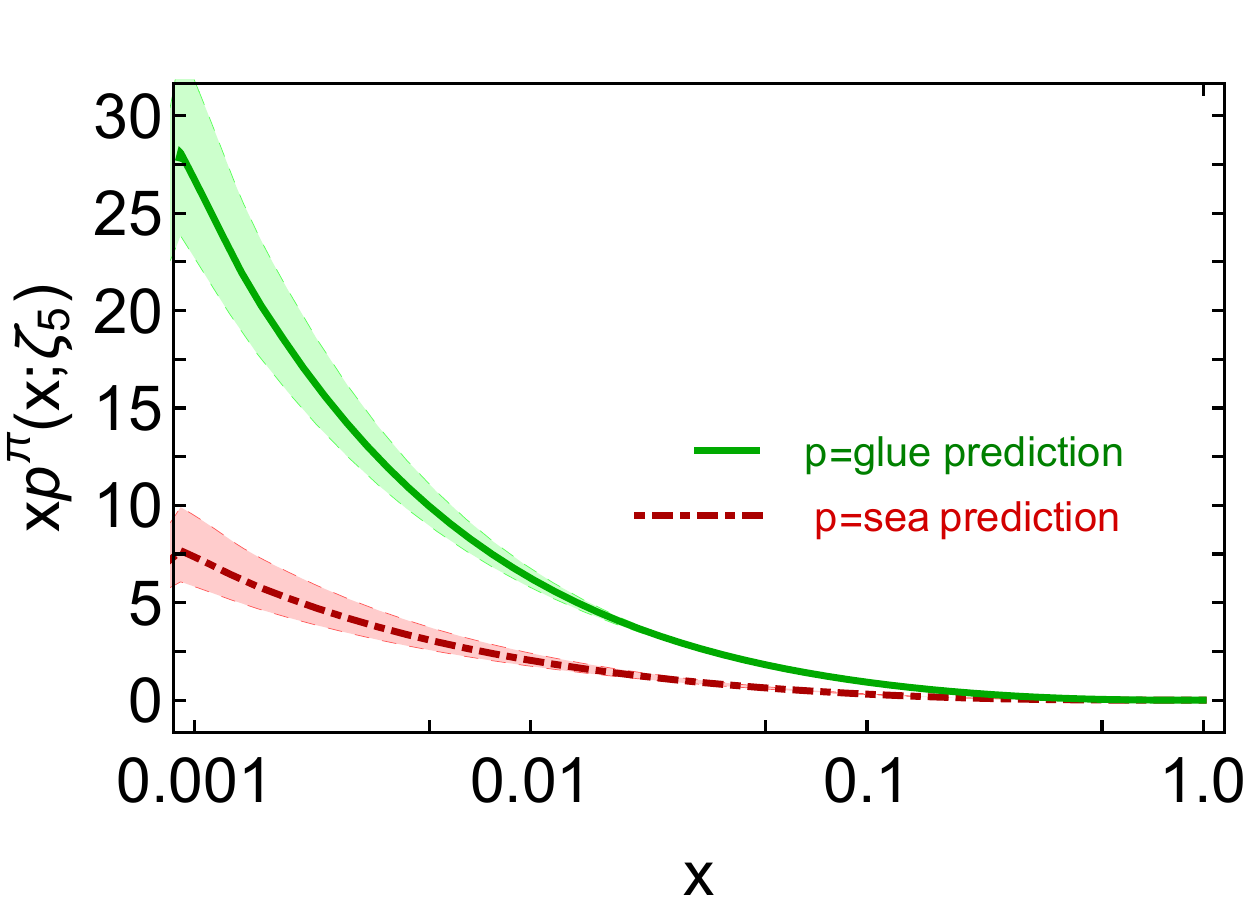}
\end{tabular}
\caption{\label{qpizeta5}
\emph{Left panel--A}.
Solid blue curve -- $\zeta=\zeta_5=5.2\,$GeV prediction for pion's valence-quark DF drawn from Ref.\,\cite{Cui:2020tdf}; and long-dashed black curve, result from Ref.\,\cite{Hecht:2000xa}.
Dot-dot-dashed (olive-green) curve within like-coloured band -- lQCD result \cite{Sufian:2019bol}.
Data (purple) from Ref.\,\cite{Conway:1989fs}, rescaled according to the analysis in Ref.\,\cite{Aicher:2010cb}.
Comparing the central prediction from Ref.\,\cite{Cui:2020tdf} with the plotted data, one obtains $\chi^2/{\rm datum} = 1.66$.
\emph{Right panel--B}. Solid green curve, $p=g$ -- Ref.\,\cite{Cui:2020tdf} prediction for the pion's glue distribution; and dot-dashed red curve, $p=S$ -- kindred predicted sea-quark distribution.
Normalisation convention: $\langle x[2 {\mathpzc u}^\pi(x;\zeta_5)+g^\pi(x;\zeta_5)+S^\pi(x;\zeta_5)]\rangle=1$.
(The uncertainty bands bracketing the theory predictions are explained following Eq.\,\eqref{setzetaH}.)
}
\end{figure}

\subsection{Pion Distribution Function at \mbox{$\zeta_5=$}5.2\,GeV}
\label{SecpiDF5}
The E615 data are linked with the scale $\zeta_5=5.2\,$GeV \cite{Conway:1989fs, Wijesooriya:2005ir}; and working with ${\mathpzc q}^\pi(x;\zeta_H)$ defined by Eqs.\,\eqref{qHDA}, \eqref{varphipilinear}, \eqref{PDFeqPDA2Repeat}, evolution $\zeta_H\to \zeta_5$ in the $\hat\alpha$ scheme delivers the prediction drawn as the solid blue curve in Fig.\,\ref{qpizeta5}\,A.  Here,
\begin{equation}
\label{betaeffupi}
\beta_\pi^{\rm eff}(\zeta_5) = 2.81(8)\,,
%\beta_{\rm eff}(\zeta_5) = 2.66(12)\,.
\end{equation}
a result consistent with Refs.\,\cite{Ding:2019qlr, Ding:2019lwe}: $\beta_\pi(\zeta_5) = 2.66(12)$\,.  Notably, despite significant advances in the implementation and understanding of continuum Schwinger function methods during the past twenty years, the calculated form of ${\mathpzc u}^\pi(x;\zeta)$ is practically unchanged: compare the solid-blue and long-dashed black curves.

The data set plotted in Fig.\,\ref{qpizeta5} is that reported in Ref.\,\cite{Conway:1989fs} after rescaling according to the analysis in Ref.\,\cite{Aicher:2010cb}, which is NLO and includes NLL resummation.  The rescaled data was first presented in Ref.\,\cite{Chang:2014lva} and is now commonly drawn in such figures.  The prediction in Ref.\,\cite{Cui:2020tdf} agrees with the rescaled data ($\chi^2/{\rm datum} = 1.66$); and it is worth highlighting that the EHM-induced broadening of the pion DF, described in Sec.\,\ref{SecDApion}, is crucial to this outcome.

The lQCD result for the pointwise behaviour of ${\mathpzc u}^\pi(x;\zeta_5)$ computed in Ref.\,\cite{Sufian:2019bol} is also drawn in Fig.\,\ref{qpizeta5}\,A.  Within uncertainties, it agrees with the continuum prediction: using a ${\mathpzc L}_1$ measure, the difference between the two central curves is 17\%.   Moreover, one finds $\beta_\pi^{\rm lQCD}(\zeta_5) = 2.45(58)$, consistent with Eq.\,\eqref{betaeffupi}; and also agreement between low-order moments:
\begin{equation}
\label{momentslQCD5}
\begin{array}{l|lll}
\zeta=\zeta_5  & \langle x \rangle_u^\pi & \langle x^2 \rangle_u^\pi & \langle x^3 \rangle_u^\pi\\\hline
%\mbox{\cite{Best:1997qp}} & 0.28(8) & 0.11(3) & 0.048(20)\\
\mbox{lQCD\,\cite{Sufian:2019bol}} & 0.18(3) & 0.064(10) & 0.030(5)\\
\mbox{continuum\,\cite{Cui:2020tdf}}  & 0.20(2) & 0.074(10) & 0.035(6)
% {\rm herein} & 0.26 & 0.11 & 0.052 ... Chen
\end{array} \,.
\end{equation}
%  <x> = 0.177007 +/- 0.0272311
% <x^2> = 0.0643769 +/- 0.010328
% <x^3> = 0.0303173 +/- 0.00516203
%%%
% <x> = 0.193934 +/- 0.0162595
% <x^2> = 0.0682108 +/- 0.00884223
% <x^3> = 0.0310891 +/- 0.00503393
%%% Hecht {0.206723, 0.0762906, 0.0359629}
An uncertainty weighted average of these results yields
\begin{equation}
\label{pionvalence}
\langle 2 x {\mathpzc u}^\pi(x;\zeta_5)\rangle = 0.40(2)\,,
\end{equation}
\emph{i.e}.\ only 40\% of the pion's momentum is carried by valence quarks at the E615 scale.  (Taken alone, Ref.\,\cite{Cui:2020tdf} predicts $\langle 2 x {\mathpzc u}^\pi(x;\zeta_5)\rangle = 0.41(4)$.)

Predictions for the pion's glue and sea DFs at $\zeta_5$ are displayed in Fig.\,\ref{qpizeta5}B, from which one obtains the following momentum fractions ($\zeta=\zeta_5$):
\begin{equation}
\label{pionGS}
\langle x\rangle^\pi_g = 0.45(2)\,, \quad
\langle x\rangle^\pi_{\rm sea} = 0.14(2)\,.
%
%\langle x\rangle^\pi_g = 0.45(1)\,, \quad
%\langle x\rangle^\pi_{\rm sea} = 0.14(2)\,.
\end{equation}
Plainly, the momentum sum rule continues to be preserved.  In this context, another textbook result is worth recalling.  Namely, on $\Lambda_{\rm QCD}^2/\zeta^2 \simeq 0$, for any hadron \cite{Altarelli:1981ax}: $\langle x\rangle_q =0$, $\langle x\rangle_g = 4/7\approx 0.57$, $\langle x\rangle_S = 3/7\approx 0.43$.  Consequently, there is a scale, $\zeta_I$, beyond which DFs cannot provide information that distinguishes between different hadrons: for each one, the valence distribution is a $\delta$-function located at $x=0$ \cite{Georgi:1951sr, Gross:1974cs, Politzer:1974fr}.  Of course, since evolution is logarithmic, $\zeta_I$ is very large, \emph{e.g}.\ even at $\zeta=10^6\,$GeV$=1\,$PeV, valence-quarks still carry roughly 20\% of the pion's light-front momentum.

In closing this section it is worth reviewing Fig.\,\ref{qpizeta5}\,B.  Plainly, the sea and especially the glue parton densities within the pion become very large on $x<0.001$.  The results depicted were obtained using evolution equations based on Refs.\,\cite{Dokshitzer:1977sg, Gribov:1972ri, Lipatov:1974qm, Altarelli:1977zs}.  Missing, therefore, are considerations expressing the fact that when such parton densities become large, new physical effects become crucial, \emph{e.g}.\ parton+parton interactions, like rescattering and recombination, can act to limit DF growth.  Interactions like these lead to modified evolution equations \cite{Gribov:1984tu, Mueller:1985wy, Kovchegov:1999yj, Zhu:2016qif}, whose use in the determination of NG mode sea and glue distributions is yet to be explored.

%% file: S6_Continuum.tex
\section{Kaon Distribution Functions}
\label{sec:kaonDFs}
So far as kaon DFs are concerned, the only information available is a forty year old DY measurement of the $K^- / \pi^-$ structure function ratio \cite{Badier:1980jq}, just eight points of data; and the past decade has seen a raft of model and theory calculations compared with the data on ${\mathpzc u}^K(x;\zeta_5)/{\mathpzc u}^\pi(x;\zeta_5)$ inferred therefrom, \emph{e.g}.\ Refs.\,\cite{Nguyen:2011jy, Alberg:2011yr, Nam:2012vm, Nam:2012af, Chen:2016sno, Peng:2017ddf, Xu:2018eii, Lan:2019rba, Kock:2020frx, Cui:2020dlm, Cui:2020tdf, Lin:2020ssv}.
Mathematically, Eqs.\,\eqref{EqMellin}\,--\,\eqref{TextBookbetac} ensure that the large-$x$ power-law exponents of $u^{\pi}(x,\zeta)$ and $u^{\rm K}(x,\zeta)$ evolve at the same rate.  Hence, the ratio $u^{\rm K}(x,\zeta_5)/u^{\pi}(x,\zeta_5)$ must be nonzero and finite on $x\simeq 1$.
(A similar statement holds for the $d(x)/u(x)$ ratio on $x\simeq 1$ in the proton \cite[Sec.\,3.6]{Chen:2020ijn}.)

A picture of the current status of kaon DF studies can be sketched from Refs.\,\cite{Cui:2020dlm, Cui:2020tdf, Roberts:2021xnz}, which alone provide simultaneous predictions for the kaon's valence, glue and sea distributions.  Working from the kaon DA depicted in Fig.\,\ref{FigNewKPDAForm} and using Eq.\,\eqref{PDFeqPDA2Repeat}, Refs.\,\cite{Cui:2020dlm, Cui:2020tdf} determined
\begin{equation}
{\mathpzc u}^K(x;\zeta_H) =
299.18\,  x^2 (1-x)^2\, [1 + 5.00   x^{0.032} (1-x)^{0.024}
-5.97 \, x^{0.064} (1-x)^{0.048} ]^2, \label{KaonPDF}
\end{equation}
with $\bar{\mathpzc s}^K(x;\zeta_H) = {\mathpzc u}^K(1-x;\zeta_H)$ and the glue and sea distributions vanishing at this scale.

For comparison with the analogous $\pi$ DF at $\zeta_2$ in Fig.\,\ref{qpizeta2}\,A, the $K$ DF in Eq.\,\eqref{KaonPDF} must be evolved.  In this connection, Refs.\,\cite{Cui:2020dlm, Cui:2020tdf} observed that any symmetry-preserving study which begins at $\zeta_H$ with a bound-state built wholly from dressed quasiparticles and enforces physical constraints on meson wave functions will generate kaon glue and sea distributions that are practically the same as those in the pion.  Physically, however, the $\bar s$ quark is heavier than the $u$ quark.  Consequently, \cite{Landau:1953um, Migdal:1956tc}: valence $\bar s$ quarks should generate less gluons than valence $u$ quarks; and gluon splitting should generate less $\bar s s$ pairs than light-quark pairs.  Such Higgs-generated flavour-symmetry breaking effects can be realised in the splitting functions that define the evolution kernels.

The effect of mass-dependent evolution can be estimated by modifying $s\to s$ and $g\to s$ splitting functions \cite{Cui:2020dlm, Cui:2020tdf}:
\begin{subequations}
\label{splittingfunctions}
\begin{align}
P_{s\leftarrow s}(z) & \to P_{q\leftarrow q}(z) - \Delta_{s\leftarrow s}(z,\zeta)\,,\quad
\Delta_{s\leftarrow s}(z,\zeta)  = \sqrt{3}  (1 - 2 z) \sigma(\zeta)\,, \label{Pss}\\
P_{s\leftarrow g}(z)  & \to P_{s\leftarrow g}(z) + \Delta_{s\leftarrow g}(z,\zeta) \label{Pgs}\,,\quad
\Delta_{s\leftarrow g}(z,\zeta) = \sqrt{5} (1-6 z+6 z^2)\sigma(\zeta)\,,
\end{align}
\end{subequations}
with $\sigma(\zeta) = \delta^2/[\delta^2 +(\zeta-\zeta_H)^2]$, $\delta = 0.1\,{\rm GeV} \approx M_s(0)-M_u(0)$, where $M_f(k^2)$ is the running mass of a $f$-quark.  (See Fig.\,\ref{FigMp2}.)  All physical constraints are preserved by Eqs.\,\eqref{splittingfunctions}; and their effects are clear: Eq.\,\eqref{Pss} acts to limit the number of gluons emitted by $\bar s$-quarks and Eq.\,\eqref{Pgs} reduces the density of $s\bar s$ pairs produced by gluons.  Both effects grow with quark quasiparticle mass difference, $\delta$, and diminish as $\delta^2/\zeta^2$ with increasing resolving scale.

\begin{figure}[t]
\hspace*{-1ex}\begin{tabular}{lcl}
{\sf A} &\hspace*{2em} & {\sf B} \\[-2ex]
\includegraphics[clip, width=0.46\textwidth]{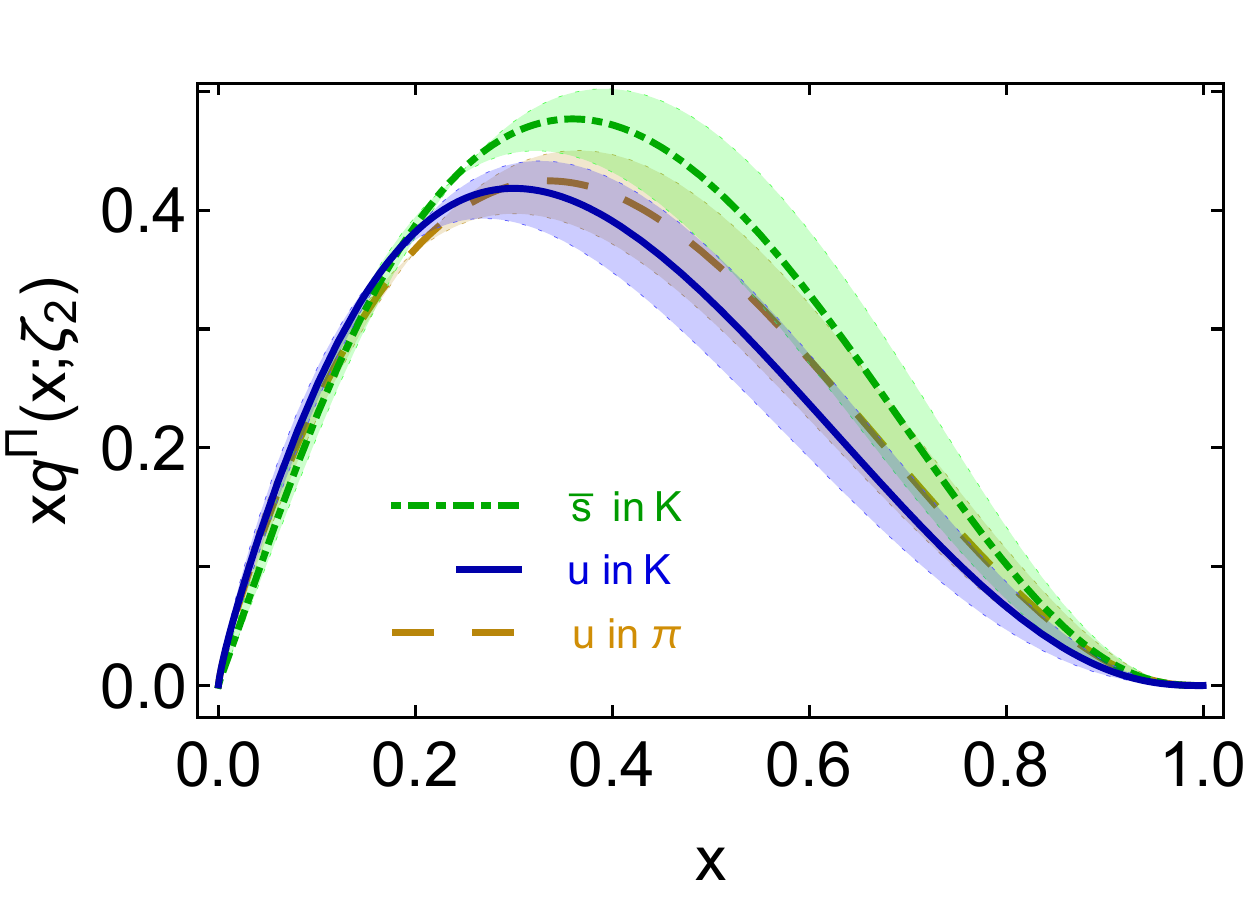} & \hspace*{2em} &
\includegraphics[clip, width=0.46\textwidth]{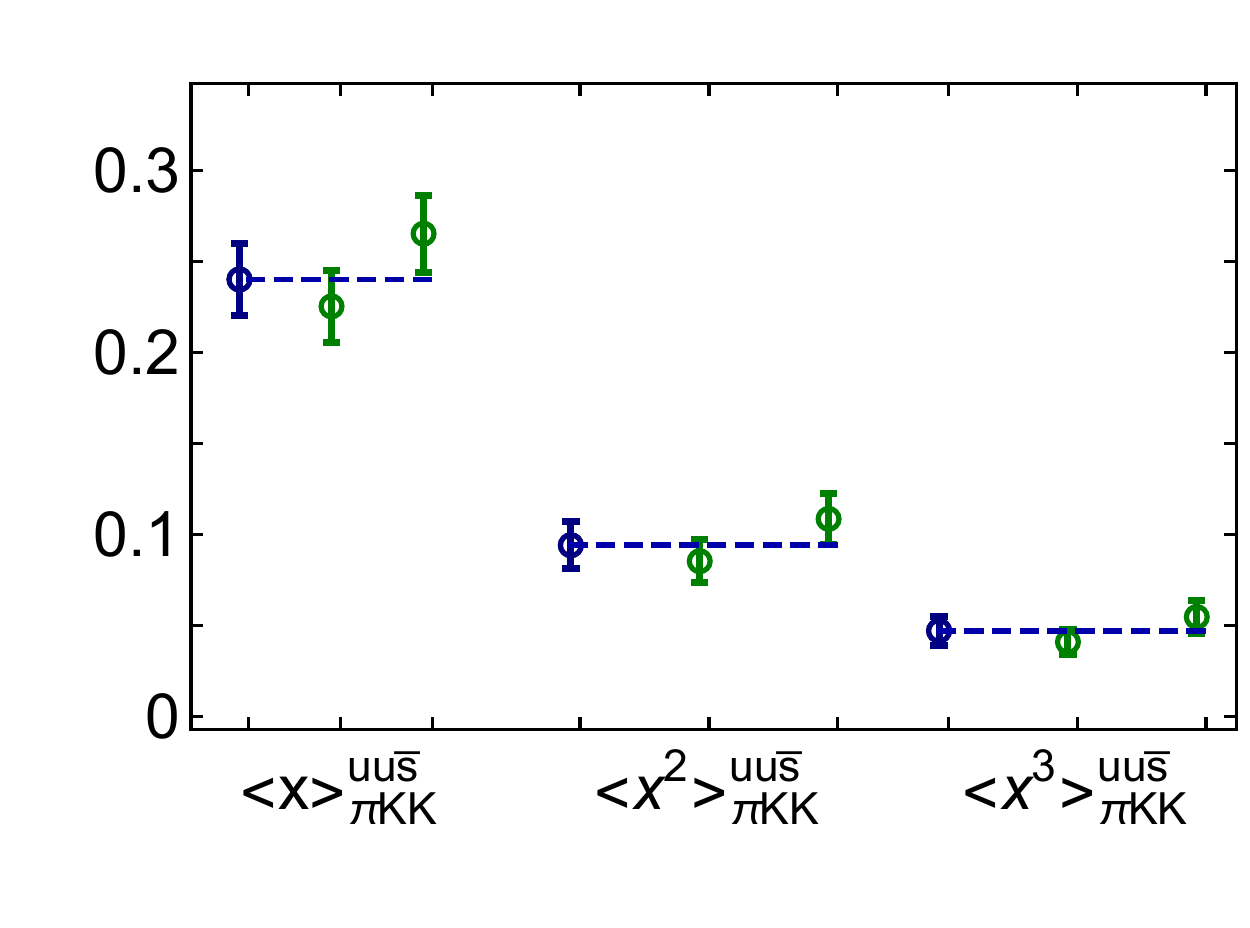}
\end{tabular}
\caption{\label{qKzeta2}
\emph{Left panel}\,--\,{\sf A}.
Solid blue curve -- ${\mathpzc u}^K(x;\zeta_2)$;
dot-dashed green curve -- $\bar{\mathpzc s}^K(x;\zeta_2)$; and
long-dashed gold curve -- ${\mathpzc u}^\pi(x;\zeta_2)$ from Fig.\,\ref{qpizeta2}\,A.
\emph{Right panel}\,--\,{\sf B}.
Low-order Mellin moments of kaon valence-quark DFs compared with those in the pion.  The dashed horizontal lines mark the central in-$\pi$ values.
%and are drawn to highlight the shift induced by Higgs modulation of kaon properties.
%
(The uncertainty bands bracketing the theory predictions are explained following Eq.\,\eqref{setzetaH}.)
}
\end{figure}

With splitting functions modified according to Eqs.\,\eqref{splittingfunctions}, the $\hat\alpha$ evolution scheme produces ${\mathpzc u}^K(x;\zeta_2)$, $\bar{\mathpzc s}^K(x;\zeta_2)$ depicted in Fig.\,\ref{qKzeta2}\,A.  These DFs yield the following low-order moments in comparison with those in the $\pi$:
\begin{equation}
\label{momentsK}
\begin{array}{l|lll}
  \zeta=\zeta_2  & \langle x \rangle_q^\Pi & \langle x^2 \rangle_q^\Pi & \langle x^3 \rangle_q^\Pi\\\hline
\bar{\mathpzc s}^K & 0.27(2) & 0.108(14) & 0.055(09) \\
{\mathpzc u}^K & 0.23(2) & 0.085(12) & 0.041(07) \\\hline
{\mathpzc u}^\pi & 0.24(2) & 0.094(13) & 0.047(08)
% {\rm herein} & 0.26 & 0.11 & 0.052 ... Chen
\end{array}\,.
\end{equation}
This comparison is drawn in Fig.\,\ref{qKzeta2}\,B, which highlights the shifts induced by Higgs modulation of kaon properties relative to those of the pion, an effect introduced in Fig.\,\ref{F1CDR}.
%Notably, $\langle x \bar s \rangle^K/\langle x u \rangle^K=1.18(1)$ For instance, $f_K/f_\pi = 1.19$, and the continuum result $\langle x \bar s \rangle^K/\langle x u \rangle^K=1.18(1)$.

\begin{figure}[t]
\hspace*{-1ex}\begin{tabular}{lcl}
{\sf A} &\hspace*{2em} & {\sf B} \\[-2ex]
\includegraphics[clip, width=0.46\textwidth]{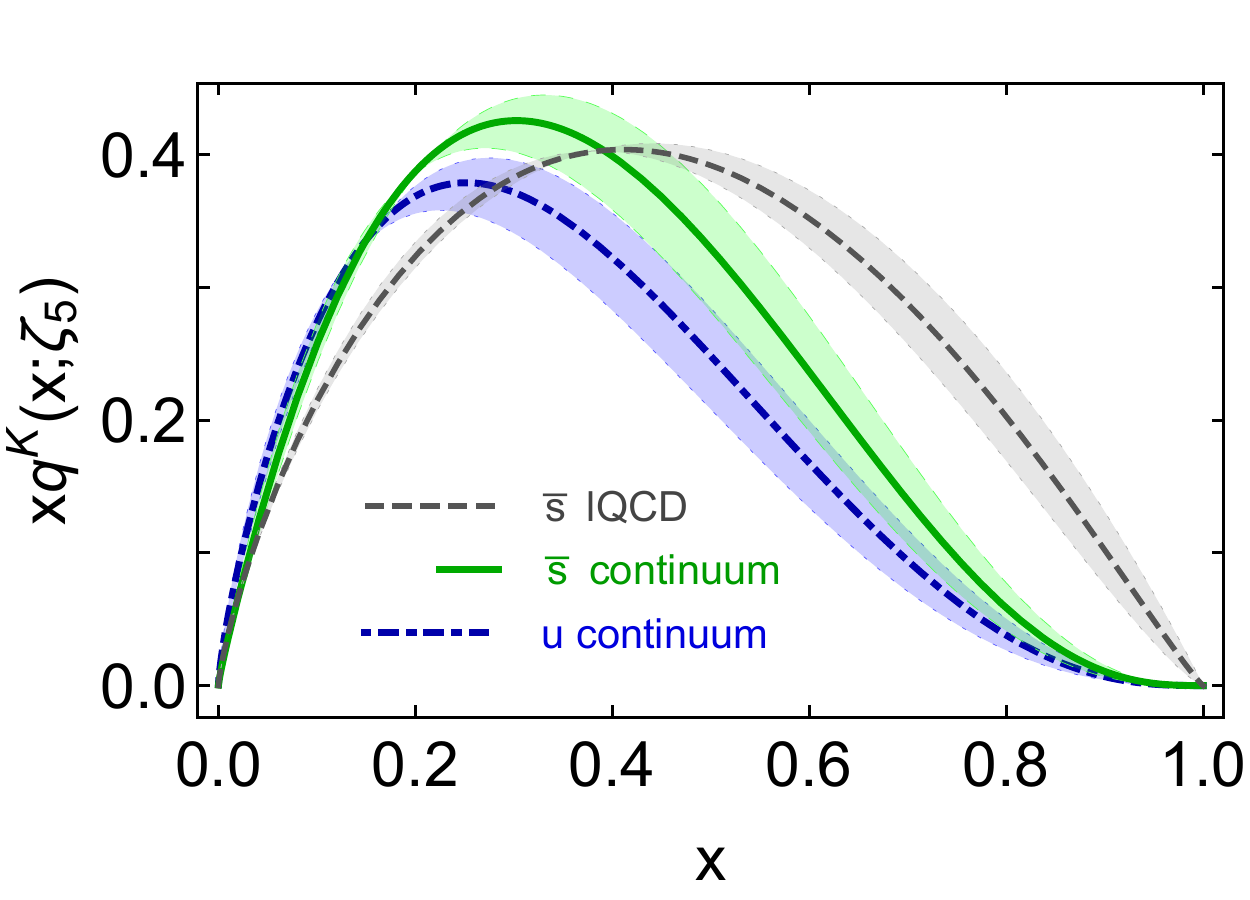} & \hspace*{2em} &
\includegraphics[clip, width=0.46\textwidth]{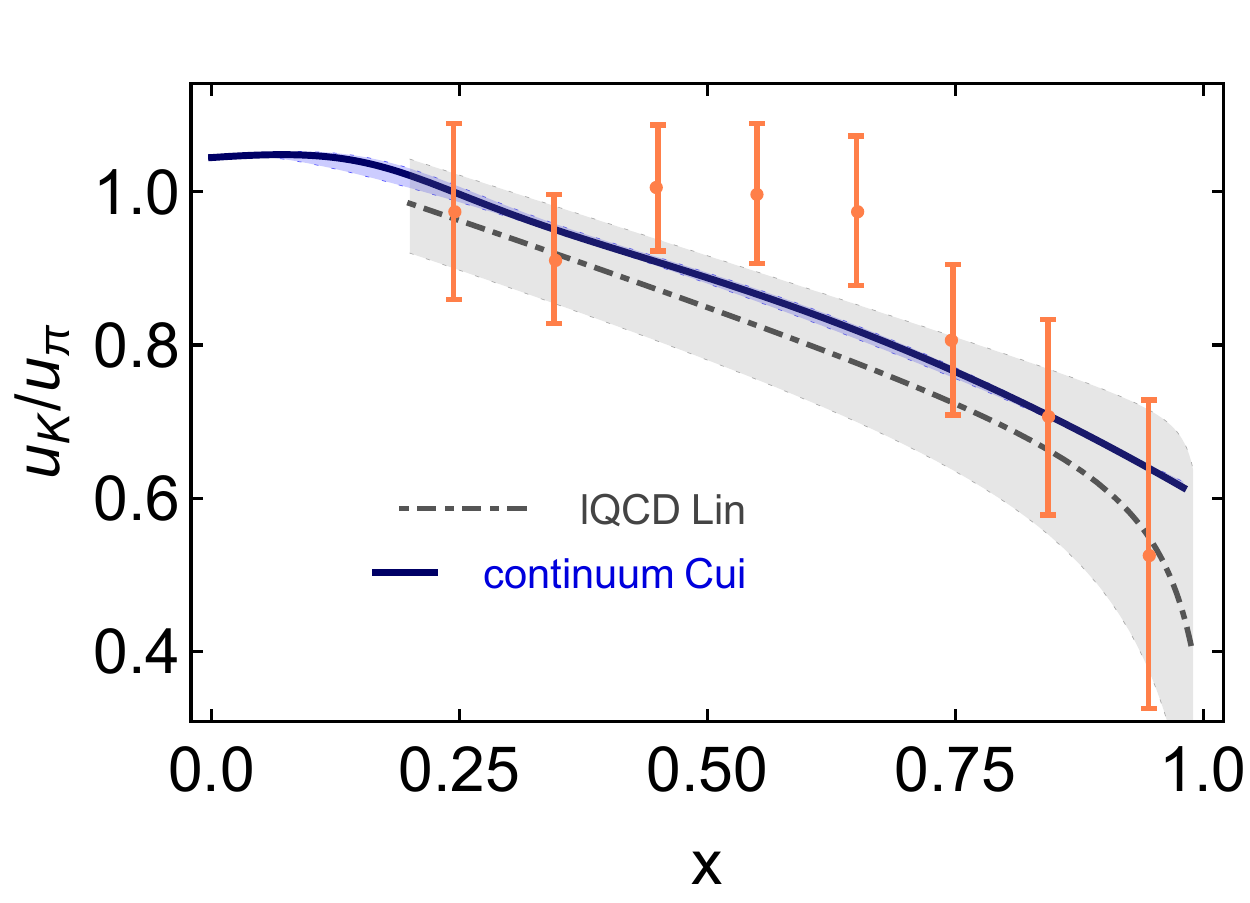}
\end{tabular}
\caption{\label{qKzeta5}
\emph{Left panel}\,--\,{\sf A}.
Continuum results \cite{Cui:2020tdf}: solid green curve -- $\bar{\mathpzc s}^K(x;\zeta_5)$; and dot-dashed blue curve -- ${\mathpzc u}^K(x;\zeta_2)$.
LQCD result \cite{Lin:2020ssv}: grey curve -- $\bar{\mathpzc s}^K(x;\zeta_5)$.
\emph{Right panel}\,--\,{\sf B}.
Ratio ${\mathpzc u}^K(x;\zeta_5)/{\mathpzc u}^\pi(x;\zeta_5)$: solid blue curve -- continuum prediction \cite{Cui:2020tdf}; and dot-dashed grey curve -- lQCD result \cite{Lin:2020ssv}.
Data (orange) as represented in Ref.\,\cite{Badier:1980jq}.
(The uncertainty bands bracketing the continuum theory predictions are explained following Eq.\,\eqref{setzetaH}.  The uncertainty is negligible in the ratio.)
}
\end{figure}

Evolution of Eq.\,\eqref{KaonPDF} to $\zeta=\zeta_5$ is necessary for comparison with results inferred from the data in Ref.\,\cite{Badier:1980jq}, a process which yields the kaon distributions drawn in Fig.\,\ref{qKzeta5}\,A.  This panel also depicts the first lQCD computation of $\bar{\mathpzc s}^K(x;\zeta_5)$ \cite{Lin:2020ssv}.   Evidently, it is significantly harder (more pointlike) than the continuum result.  In fact, the lQCD DF behaves as $(1-x)^{\beta_{\bar{\mathpzc s}^K}}$, $\beta_{\bar{\mathpzc s}^K} = 1.13(16)$, incompatible with Eqs.\,\eqref{pionPDFlargex}, \eqref{DYWrelation}.

The kaon DFs just described produce the following low-order moments:
\begin{equation}
\begin{array}{ll|c|c|c}
&
{\mathpzc q}(\zeta_5) & \langle x{\mathpzc q}^K \rangle & \langle x^2{\mathpzc q}^K \rangle & \langle x^3{\mathpzc q}^K \rangle\\\hline
\mbox{continuum\,\cite{Cui:2020tdf}} &
u & 0.19(2) & 0.067(09) & 0.030(5)\\
&
{\bar s} & 0.23(2) & 0.085(11) & 0.040(7) \\\hline
\mbox{lattice\,\cite{Lin:2020ssv}} &
u & 0.19(1) & 0.080(07) & 0.042(6)\\
&
{\bar s} & 0.27(1) & 0.123(07) & 0.070(6) \\\hline
\end{array}\,.
\label{uKsKmoments}
\end{equation}
Working with the continuum results, accounting for $\zeta_H\to\zeta_H (1.0\pm 0.1)$,
\begin{equation}
\langle x[{\mathpzc u}^K(x;\zeta_5) + \bar{\mathpzc s}^K(x;\zeta_5)] \rangle = 0.42(3)\,;
\end{equation}
so in comparison with the pion, valence quarks carry 5\% more of the kaon's light-front momentum.

%% $u$ -- $0.193(8)$, $0.080(7)$, $0.042(6)$; and
%% $\bar s$ -- $0.267(8)$, $0.123(7)$, $0.070(6)$.

Notably, the lQCD results in Eq.\,\eqref{uKsKmoments} are systematically larger than the continuum predictions, especially for the $\bar s$, \emph{viz}.\ the excesses are: $u$ -- $0.6(4.8)$\%, $21(6)$\%, $40(4)$\%; and $\bar s$ -- $24(7)$\%, $53(13)$\%, $84(16)$\%.  Again, this is because the lQCD DFs are much harder than the continuum DFs.
Furthermore, using the lQCD results, one finds $\langle x \bar s \rangle^K/\langle x u \rangle^K=1.38(7)$, which may be compared with the natural scale for Higgs-modulation of EHM, \emph{i.e}.\ $f_K/f_\pi = 1.19$, and the continuum result $\langle x \bar s \rangle^K/\langle x u \rangle^K=1.18(1)$.
Given these observations, it may reasonably be anticipated that future refinements of lQCD setups, algorithms and analyses will move the lattice and continuum DFs closer together.

The ratio ${\mathpzc u}^K(x;\zeta_5)/{\mathpzc u}^\pi(x;\zeta_5)$ is depicted in Fig.\,\ref{qKzeta5}, displaying both the continuum prediction from Ref.\,\cite{Cui:2020tdf} and the lQCD result from Ref.\,\cite{Lin:2020ssv}.
Evidently, the relative difference between the central lQCD result and the continuum prediction is $\approx 5$\%, in spite of the fact that the individual lQCD DFs are pointwise markedly different from the continuum DFs, as illustrated by Eq.\,\eqref{uKsKmoments} and Fig.\,\ref{qKzeta5}\,A.
This outcome highlights a long known feature, \emph{i.e}.\ ${\mathpzc u}^K(x)/{\mathpzc u}^\pi(x)$ is forgiving of even large differences between the individual DFs used to produce the ratio.  Higher precision data is crucial if ${\mathpzc u}^K(x)/{\mathpzc u}^\pi(x)$ is to be used effectively to inform and test the modern understanding of SM NG modes; and greater discriminating power is provided by separate results for ${\mathpzc u}^\pi(x)$, ${\mathpzc u}^K(x)$ \cite{JlabTDIS1, JlabTDIS2, Denisov:2018unjF}.

\begin{table}[t]
\caption{\label{fittingparametersallV}
When used in Eq.\,\eqref{PDFform}, the coefficients and powers listed here provide interpolations for the $\pi$ and $K$ valence-quark DFs evolved from Eqs.\,\eqref{qpizetaH}, \eqref{KaonPDF} to the scale $\zeta_3=3.1\,$GeV \cite{Roberts:2021xnz}, appropriate for analyses of $J/\Psi$ production.  In each case, the central curve is identified by the $\zeta_3$ row label and is bracketed by the parameters that specify the limits of the uncertainty bands.
}
\begin{center}
\begin{tabular*}%{|c|c|c|c|c|c|c|}\hline
{\hsize}
{
l@{\extracolsep{0ptplus1fil}}|
c@{\extracolsep{0ptplus1fil}}
c@{\extracolsep{0ptplus1fil}}
c@{\extracolsep{0ptplus1fil}}
c@{\extracolsep{0ptplus1fil}}
c@{\extracolsep{0ptplus1fil}}
c@{\extracolsep{0ptplus1fil}}
c@{\extracolsep{0ptplus1fil}}}\hline\hline
%
% M {\[Alpha] -> 0.0442687, \[Alpha]1 -> 0.129375, \[Beta]1 ->  0.905553, \[Rho] -> -1.92696, \[Gamma] -> 0.949586}
 ${\mathpzc u}^\pi$
 & ${\mathpzc n}_{{\mathpzc u}^\pi}\ $ & $\alpha_\pi\ $ & $\beta_\pi\ $ & $\alpha_1\ $ & $\beta_1\ $ & $\rho\ $ & $\gamma\ $ \\\hline
             & $137\phantom{.9}\ $ & $\phantom{-}0.119\phantom{3}\ $ & $3.09\ $ & $0.145\phantom{9}\ $& $0.903\ $& $-1.95\ $ & $0.971\phantom{3}\ $ \\
$\zeta_3\ $ & $118\phantom{.9}\ $ & $\phantom{-}0.0443\ $ & $3.21\ $ & $0.129\phantom{9}\ $& $0.906\ $& $-1.93\ $ & $0.950\phantom{9}\ $ \\
             & $\phantom{1}96.9\ $ & $-0.0450\ $ & $3.35\ $ &$0.109\phantom{9}\ $& $0.911\ $& $-1.90\ $ & $0.925\phantom{9}\ $  \\\hline
%----------
 ${\mathpzc u}^K$
 & ${\mathpzc n}_{{\mathpzc u}^K}\ $ & $\alpha_{{\mathpzc u}^K}\ $ & $\beta_{{\mathpzc u}^K}\ $ & $\alpha_1\ $ & $\beta_1\ $ & $\rho\ $ & $\gamma\ $ \\\hline
             & $65.8\ $ & $\phantom{-}0.179\phantom{7}\ $ & $3.09\ $ & $0.358\phantom{9}\ $& $1.39\phantom{3}\ $& $-2.08\ $ & $1.16\phantom{3}\ $ \\
$\zeta_3\ $ & $57.1\ $ & $\phantom{-}0.119\phantom{7}\ $ & $3.21\ $ & $0.375\phantom{9}\ $& $1.46\phantom{3}\ $& $-2.11\ $ & $1.20\phantom{9}\ $ \\
             & $47.4\ $ & $\phantom{-}0.0421\ $ & $3.35\ $ &$0.374\phantom{9}\ $& $1.52\phantom{9}\ $& $-2.11\ $ & $1.21\phantom{9}\ $  \\\hline
$\bar{\mathpzc s}^K$
& ${\mathpzc n}_{\bar {\mathpzc s}^K}\ $ & $\alpha_{\bar{\mathpzc s}^K}\ $ & $\beta_{\bar{\mathpzc s}^K}\ $ & $\alpha_1\ $ & $\beta_1\ $ & $\rho\ $ & $\gamma\ $ \\\hline
             & $79.7\ $ & $\phantom{-}0.259\phantom{7}\ $ & $3.03\ $ & $0.235\phantom{5}\ $& $1.39\phantom{9}\ $& $-1.92\phantom{3}\ $ & $0.975\ $ \\
$\zeta_3\ $ & $69.0\ $ & $\phantom{-}0.199\phantom{7}\ $ & $3.14\ $ & $0.228\phantom{5}\ $& $1.39\phantom{9}\ $& $-1.90\phantom{3}\ $ & $0.960\ $ \\
             & $58.8\ $ & $\phantom{-}0.132\phantom{7}\ $ & $3.27\ $ &$0.222\phantom{5}\ $& $1.39\phantom{9}\ $& $-1.89\phantom{3}\ $ & $0.956\ $  \\
\hline\hline
\end{tabular*}
\end{center}
\end{table}

Refs.\,\cite{Cui:2020dlm, Cui:2020tdf} also delivered predictions for the kaon's glue and sea distributions at $\zeta=\zeta_5$.  However, given that $J/\Psi$ production may best provide access to gluon DFs in NG modes \cite{Denisov:2018unjF, Chang:2020rdy}, it would be better to have all DFs at $\zeta\approx m_{J/\Psi}\approx\zeta_3=3.1\,$GeV.  This information is reported elsewhere \cite{Roberts:2021xnz} and is readily summarised here.  Namely, using the following interpolating function,
\begin{equation}
{\mathpzc q}^M(x)  = {\mathpzc n}_{{\mathpzc q}^M} \,x^{\alpha_M} (1-x)^{\beta_M}  [1 + \rho\, x^{\alpha_1/4} (1-x)^{\beta_1/4} + \gamma \,x^{\alpha_1/2} (1-x)^{\beta_1/2} ]\,,
\label{PDFform}
\end{equation}
the pion and kaon DFs are obtained by inserting the coefficients and powers listed in Table~\ref{fittingparametersallV}.  These predictions express the $x\simeq 1$ behaviour prescribed by Eqs.\,\eqref{pionPDFlargex}, \eqref{DYWrelation}.  At $\zeta_3$, the pion's valence degrees-of-freedom carry 45(3)\% of its light-front momentum, whereas the analogous result is 46(3)\% in the $K$.

\begin{figure}[t]
\hspace*{-1ex}\begin{tabular}{lcl}
{\sf A} &\hspace*{2em} & {\sf B} \\[-2ex]
\includegraphics[clip, width=0.46\textwidth]{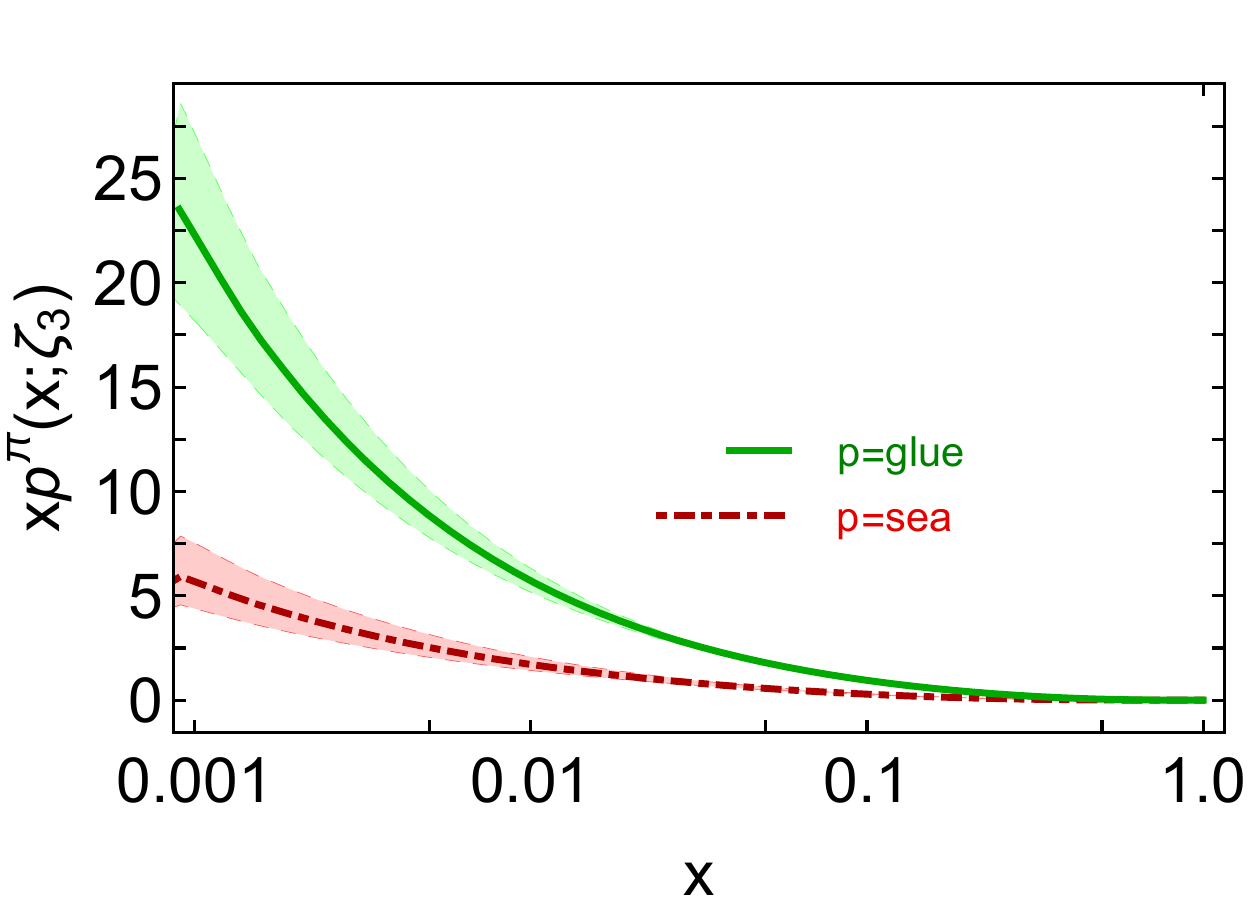} & \hspace*{2em} &
\includegraphics[clip, width=0.46\textwidth]{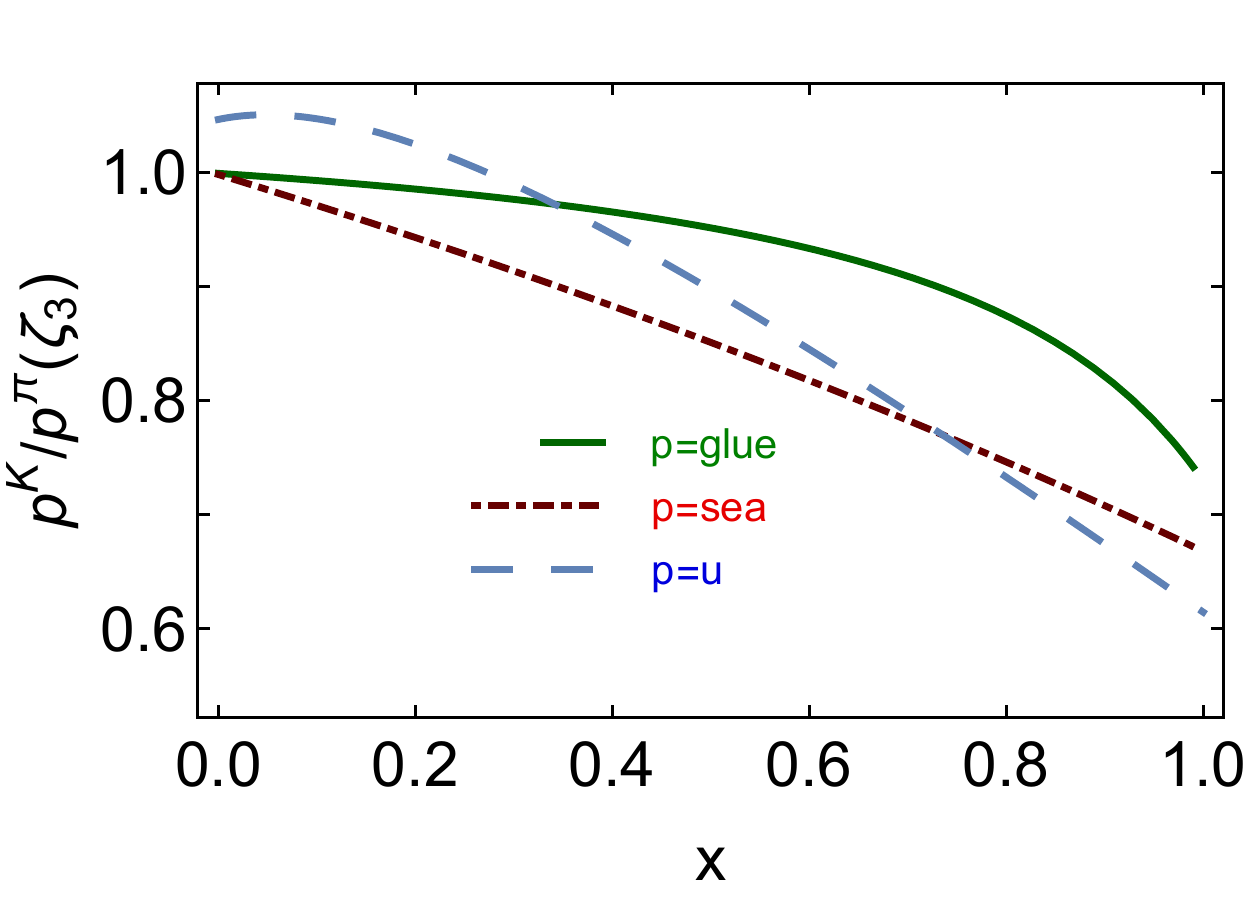}
\end{tabular}
\caption{\label{FigGSpi}
\emph{Left panel}\,--\,{\sf A}.  Solid green curve, $p=g$ -- prediction for the pion's glue distribution; and dot-dashed red curve, $p=S$ -- predicted sea-quark distribution.
\emph{Right panel}\,--\,{\sf B}.  $K/\pi$ DF ratios: $p=g$ -- solid green curve; $p=S$ -- dot-dashed red curve; and $p={\mathpzc u}$ -- long-dashed blue curve.
Normalisation convention: $\langle x[2 {\mathpzc u}^\pi(x;\zeta_3)+g^\pi(x;\zeta_3)+S^\pi(x;\zeta_3)]\rangle=1$.
(The uncertainty bands bracketing the results in Panel\,A are explained following Eq.\,\eqref{setzetaH}.  This uncertainty effectively cancels in the ratios depicted in Panel\,B.  Results at $\zeta=\zeta_3=3.1\,$GeV.  All curves drawn from Ref.\,\cite{Roberts:2021xnz}.)
}
\end{figure}

Using the valence DFs already presented and the $\hat\alpha$ evolution scheme, the $\zeta=3.1\,$GeV $\pi$ glue and sea-quark DFs can be calculated.  Depicted in Fig.\,\ref{FigGSpi}\,A, the predictions are effectively interpolated using \cite{Gluck:1999xe}:
\begin{equation}
x {\mathpzc p}(x)  = {\mathpzc A} \,x^{\alpha_{\mathpzc p}} (1-x)^{\beta_{\mathpzc p}}  [1 + \rho\, x^{1/2} + \gamma \,x ]\,,
\label{PDFformgS}
\end{equation}
${\mathpzc p} = g, S$, with the coefficients in Table~\ref{GlueSeaCoefficients}.  The related momentum fractions are ($\zeta=\zeta_3$):
\begin{equation}
\langle x\rangle^\pi_g = 0.43(2)\,, \quad
\langle x\rangle^\pi_{\rm sea} = 0.12(2)\,.
\end{equation}

Kaon glue and sea-quark distributions at $\zeta_3$ glue can likewise be obtained.  The predictions may usefully be described via comparison with the analogous $\pi$-meson results in Fig.\,\ref{FigGSpi}\,A.  Therefore, Fig.\,\ref{FigGSpi}\,B plots the ratios ${\mathpzc p}^K(x)/{\mathpzc p}^\pi(x)$, ${\mathpzc p}={\mathpzc g},{\mathpzc S}$, which are described by the following functions:
\begin{equation}
{\mathsf R}_{\mathpzc g}^{K\pi} = \frac{1.00\, -0.842 x}{1-0.786 x} \,, \;
{\mathsf R}_{\mathpzc S}^{K\pi}  = \frac{1.00\, -0.462 x}{1-0.197 x}\,.
\end{equation}
The $K$ and $\pi$ glue and sea-quark DFs are quite similar on $x\lesssim 0.2$; but differences are notable on $x\gtrsim 0.2$, \emph{i.e}.\ the domain of valence-quark/antiquark dominance.  These features were to be expected given that mass-dependent splitting functions act primarily to modify the valence DF of the heavier quark; such DFs are small at low-$x$, where glue and sea-quark DFs are large, and vice versa; so, the predominant impact of a change in the valence DFs must lie at large-$x$.  Generated by Higgs-boson couplings into QCD, the observed differences are on the order of $\approx 33$\% at $x=1$ \emph{cf}.: $1-f_\pi^2/f_K^2 \approx 0.3$; and $1-[M_u(0)/M_s(0)]^2\approx 0.3$, where $M_q(k)$ is the dressed-quark mass function in Fig.\,\ref{FigMp2}.  Evidently, they express the natural scale for Higgs-boson modulation of EHM.

\begin{table}[t]
\caption{\label{GlueSeaCoefficients}
Coefficients and powers that reproduce the computed pion glue and sea-quark distribution functions depicted in Fig.~\ref{FigGSpi}A when used in Eq.\,\eqref{PDFformgS}.
}
\begin{center}
\begin{tabular*}%{|c|c|c|c|c|c|c|}\hline
{\hsize}
{
l@{\extracolsep{0ptplus1fil}}|
c@{\extracolsep{0ptplus1fil}}
c@{\extracolsep{0ptplus1fil}}
c@{\extracolsep{0ptplus1fil}}
c@{\extracolsep{0ptplus1fil}}
c@{\extracolsep{0ptplus1fil}}}\hline\hline
 $\zeta_3\ $ & ${\mathpzc A}\ $ & $\alpha_{\mathpzc p}\ $ & $\beta_{\mathpzc p}\ $ & $\rho\ $ & $\gamma\ $ \\\hline
             % & $10.98\ $ & $-0.052\ $ & $2.29\ $ & $-1.40\ $ & $0.637\ $  \\
             & $0.462\ $ & $-0.539\ $ & $4.09\ $ & $-0.296\ $& $\phantom{-}0.229\ $\\
%$\zeta_2\ $ & $\phantom{1}9.26\ $ & $-0.096\ $ & $2.37\ $ & $-1.32\ $ & $0.594\ $ \\
  $g\ $ & $0.735\ $ & $-0.494\ $ & $4.21\ $ & $-1.54\phantom{1}\ $ & $\phantom{-}1.36\phantom{9}\ $\\
            % & $\phantom{1}7.38\ $ & $-0.140\ $ & $2.44\ $ & $-1.21\ $ & $0.538\ $ \\\hline
             & $0.295\ $ & $-0.638\ $ & $4.35\ $ &$\phantom{-}2.23\phantom{1}\ $& $-5.08\phantom{9}\ $\\\hline
             & $0.144\ $ & $-0.488\ $ & $5.09\ $ & $\phantom{-}0.956\ $& $-2.36\phantom{9}\ $\\
%$\zeta_2\ $ & $\phantom{1}9.26\ $ & $-0.096\ $ & $2.37\ $ & $-1.32\ $ & $0.594\ $ \\
$S\ $ & $0.127\ $ & $-0.538\ $ & $5.21\ $ & $\phantom{-}2.20\phantom{8}\ $ & $-4.82\phantom{9}\ $\\
            % & $\phantom{1}7.38\ $ & $-0.140\ $ & $2.44\ $ & $-1.21\ $ & $0.538\ $ \\\hline
             & $0.108\ $ & $-0.595\ $ & $5.35\ $ &$\phantom{-}3.54\phantom{8}\ $& $-7.50\phantom{9}\ $\\
\hline\hline
\end{tabular*}
\end{center}
\end{table}

%% file: S7_Continuum.tex
\section{Three Dimensional Structure of Nambu-Goldstone Modes}
\label{3DNG}
\subsection{Generalised Transverse-Momentum Dependent Parton Distribution Functions}
\label{SecGTMD}
As highlighted already, experiment, phenomenology and theory have long focused on drawing one dimensional (1D) imaging of hadrons.  This effort continues because many puzzles and controversies are unresolved, \emph{e.g}.\ the large-$x$ behaviour of meson structure functions and the glue and sea content of NG modes.  Yet, notwithstanding the need for new, precise data on 1D distributions and related predictions with a traceable connection to QCD, the attraction of generalised parton distributions (GPDs) and transverse momentum dependent parton distributions (TMDs) is great.  Such functions offer the possibility of drawing three-dimensional (3D) images of hadrons and may therefore enable entirely new facets of hadron structure to be revealed.  They probe the multidimensional structure of hadron LFWFs and thereby provide access to, \emph{inter alia}: mass, pressure and spin distributions within a hadron, both in longitudinal and transverse directions; the sharing of these properties amongst the various bound-state constituents, which distributions are frame and scale dependent; and to the volume of spacetime occupied by these constituents, measuring their potentially different  ``confinement'' radii.

In order to fully profit from 3D imaging data collected at modern and anticipated facilities, using them to understand the nature and corollaries of SM mass generation, sound QCD-linked tools must be developed for the calculation of GPDs and TMDs.  Within the past decade or so, numerous models and theory tools have been deployed in this effort, \emph{e.g}.\ Refs.\,\cite{Theussl:2002xp, Brommel:2007xd, Brommel:2007zz, Broniowski:2007si, Nam:2010pt, Nam:2011yw, Dorokhov:2011ew, Carmignotto:2014rqa, Mezrag:2014jka, Son:2015bwa, Engelhardt:2015xja, Mezrag:2016hnp, Fanelli:2016aqc, Lorce:2016ugb, Chouika:2017rzs, Ahmady:2019yvo, Kaur:2020vkq}.  One unifying approach is provided by generalised parton correlation functions (GPCFs) \cite{Meissner:2008ay}, which link GPDs and TMDs to a single source amplitude; and the potential of this approach was recently explored \cite{Zhang:2020ecj}.  Whilst the entire field of 3D structure is too wide to be canvassed herein, it is worth drawing some important connections with EHM.  This is readily achieved by highlighting features of the pion's mass, pressure and transverse spin distributions, whose calculation capitalises upon the analyses in Refs.\,\cite{Brommel:2007xd, Brommel:2007zz, Broniowski:2007si, Fanelli:2016aqc, Chouika:2017rzs, Zhang:2020ecj}.

\begin{figure}[t]
\centerline{%
\includegraphics[clip, width=0.45\textwidth]{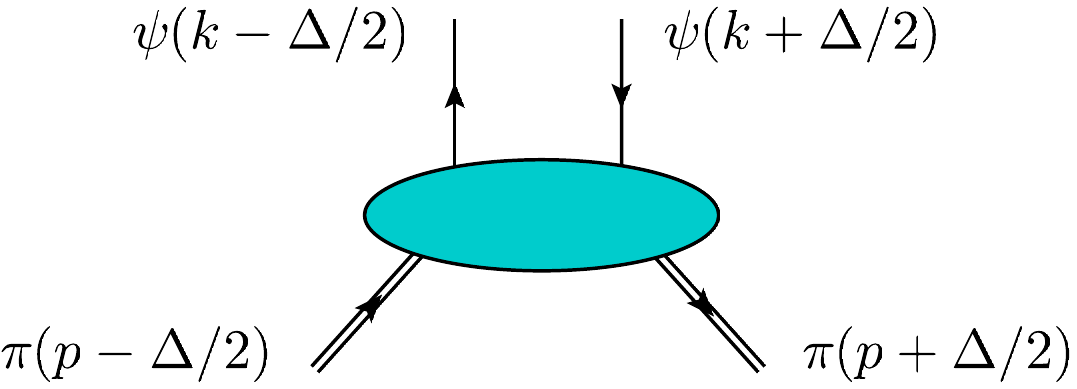}}
\caption{\label{GPCFkinematics}
Momentum-space conventions used when defining the in-pion quark-quark correlator in Eq.\,\eqref{piGPCF}.
}
\end{figure}

One may begin with the following in-pion quark-quark correlator \cite{Meissner:2008ay}:
\begin{equation}
W_{ij}(P,k,\Delta,\bar N;\eta)  = \int\frac{d^4 z}{(2\pi)^4} \, {\rm e}^{i k\cdot z}  \langle\pi(p^\prime)|\,\bar q_j(-\tfrac{1}{2} z){\mathpzc W}(-\tfrac{1}{2} z,\tfrac{1}{2} z;\bar n)\,
q_i(\tfrac{1}{2} z) \,
|\pi(p)\rangle\,,
\label{piGPCF}
\end{equation}
where:
${\mathpzc W}(-\tfrac{1}{2} z,\tfrac{1}{2} z;\bar n)$ is a Wilson line, 
with $\bar n$ a light-like four-vector, conjugate to $n$ in Eq.\,\eqref{varphiresult}, $n\cdot\bar n = -1$, $\bar n^2=0$, $\bar n\cdot P =: P^-$;
$\bar N = \bar n / \bar n \cdot \hat P$, $\hat P^2=1$; 
\begin{equation}
P=(p^\prime+p)/2\,,\;
\Delta = p^\prime - p\,,\;
P\cdot \Delta = 0\,;
\end{equation}
$k$ is the relative quark-antiquark momentum; and the ``skewness'' $\xi = [-n\cdot \Delta]/[2 n\cdot P]$, $|\xi|\leq 1$.  These kinematic conventions are illustrated in Fig.\,\ref{GPCFkinematics}.

A generalised transverse-momentum dependent parton distribution function (GTMD) is obtained from Eq.\,\eqref{piGPCF} by first considering the following partially integrated quantity:
\begin{equation}
W_{ij}(P,x,\vec{k}_\perp,\Delta,\bar N;\eta)=  \int\frac{d^4 z}{(2\pi)^4} \, {\rm e}^{i k\cdot z}\,\delta(n\cdot z) \langle\pi(p^\prime)|\,\bar\psi_j(-\tfrac{1}{2} z){\mathpzc W}(-\tfrac{1}{2} z,\tfrac{1}{2} z;\bar n)\,
\psi_i(\tfrac{1}{2} z) \,
|\pi(p)\rangle\,.
\label{piGPCF1}
\end{equation}
This is a Dirac-matrix valued function, from which contributions at various orders in a twist expansion are obtained by appropriate projection operations.  That is, with $\mathpzc H$ being some aptly chosen combination of Dirac matrices, the scalar functions of interest -- GTMDs -- are obtained via
\begin{align}
W^{[{\mathpzc H}]}(P&,x,\vec{k}_\perp,\Delta,\bar N;\eta)  = \frac{1}{2} W_{ij}(P,x,\vec{k}_\perp,\Delta,\bar N;\eta) {\mathpzc H}_{ji}\nonumber \\
& =  \int\frac{d^4 z}{2(2\pi)^4} \, {\rm e}^{i k\cdot x}\,\delta(n\cdot z)   \langle\pi(p^\prime)|\,\bar\psi_j(-\tfrac{1}{2} z){\mathpzc H}_{ji}{\mathpzc W}(-\tfrac{1}{2} z,\tfrac{1}{2} z;\bar n)\,
\psi_i(\tfrac{1}{2} z) \,
|\pi(p)\rangle\,. \label{GTMD}
\end{align}
This operation corresponds to the insertion of ${\mathpzc H}$ as a link between the open quark and antiquark lines in Fig.\,\ref{GPCFkinematics}: $\psi(k\mp \Delta/2)$, respectively.  The GTMDs defined by Eq.\,\eqref{GTMD} are complex-valued functions.  Their real part is even under the time-reversal operation ($T$-even) and the imaginary part is $T$-odd.  Beginning with Eq.\,\eqref{GTMD}, integration over $\vec{k}_\perp$ yields the GPDs; and TMDs are obtained by setting $\Delta =0$, which entails $\xi=0$.

There are three twist-two pion GTMDs, obtained from Eq.\,\eqref{GTMD} through the following insertion choices:
\begin{align}
{\mathpzc H} \to \{ {\mathpzc H}_1 = i n\cdot\gamma \,,\,
{\mathpzc H}_2 = i n\cdot \gamma \gamma_5\,,\,
%{\mathpzc H}_3 = i \gamma_5 \sigma_{j\mu} n_\mu \}.
{\mathpzc H}_3 = i \sigma_{j\mu} n_\mu \}.
\label{T2Insert}
\end{align}
The simplest is that derived from ${\mathpzc H}_1$, which delivers, \emph{e.g}.\ the pion's valence-quark distribution function and electromagnetic and gravitational form factors.  The GTMD obtained with ${\mathpzc H}_3$ is also of keen interest because it provides access to the pion's quark transversity distributions, \emph{i.e}.\ the dependence of the in-pion quark distributions on their polarisation perpendicular to the pion's direction of motion \cite{Polyakov:2018zvc}.

It is worth writing an explicit form for the GTMD obtained from ${\mathpzc H}_1$ in Ref.\,\cite{Zhang:2020ecj}:
\begin{equation}
F_1(x,k_\perp^2,\xi,t)  = 2 N_c {\rm tr}_{\rm D}\int \frac{dk_3 dk_4}{(2\pi)^4} \delta_n^x(k) \,\Gamma_\pi(-p^\prime)\,  S(k_{+\Delta})\, n\cdot \Gamma^\gamma(\Delta)\, S(k_{-\Delta})\,\Gamma_\pi(p)\, S(k-P)\,,
\label{F1start}
\end{equation}
where $k_{\pm \Delta} = k\pm \Delta/2$,
$t=-\Delta^2$,
$p\cdot\Delta = -\Delta^2/2= - p^\prime\cdot\Delta$.
%%
%${\rm tr}_{\rm D}$ indicates a trace over spinor indices,
%%\begin{equation}
%%k_{\pm \Delta} = k\pm \Delta/2\,\;
%%t=-\Delta^2,\;
%%p\cdot\Delta = -\Delta^2/2= - p^\prime\cdot\Delta\,.
%%\label{EqKinematics}
%%\end{equation}
This formula is particularly simple because Ref.\,\cite{Zhang:2020ecj} employed a symmetry-preserving treatment of a vector$\,\times\,$vector contact interaction; so that, \emph{e.g}.\ the pion Bethe-Salpeter amplitude, $\Gamma_\pi$, and photon-quark vertex, $\Gamma^\gamma$, do not here depend on the quark+antiquark relative momentum. Nevertheless, it illustrates a qualitatively robust point, \emph{viz}.\ both the size and shape of every one of the pion's GTMDs are largely determined by the character of EHM.  This is plain because the integrand in Eq.\,\eqref{F1start} involves a product of two pion Bethe-Salpeter amplitudes and Eq.\,\eqref{GTRE} entails that these amplitudes are a direct measure of EHM.  It follows, in addition, that studies of kaon GTMDs can reveal novel expressions of Higgs-boson induced modulations of EHM.

\subsection{Twist-Two Generalised Parton Distribution Functions}
\label{SecTTGPD}
GPDs are obtained by integrating GTMDs over $\vec{k}_\perp$; and at leading twist, the pion has two GPDs:
\begin{equation}
\label{EqGPDFH}
{\mathsf H}_\pi(x,\xi,t)  = \int d^2\vec{k}_\perp F_1(x, k_\perp^2,\xi,t) \,, \quad
{\mathsf E}_{\pi}^{\rm T}(x,\xi,t)  = \int d^2\vec{k}_\perp H_1^\Delta(x, k_\perp^2,\xi,t) \,,
\end{equation}
where $H_1^\Delta(x, k_\perp^2,\xi,t) $ is an ${\mathpzc H}_3$-generated analogue of $F_1^\Delta(x, k_\perp^2,\xi,t)$ in Eq.\,\eqref{F1start}, and ${\mathsf H}_\pi$, ${\mathsf E}_\pi^{\rm T}$ are typically called the vector (no spin-flip) and tensor (spin-flip) GPDs.
%% theta_1
%% theta_2

Working with ${\mathsf H}_\pi(x,\xi,t)$, one can compute an array of physically important pion elastic form factors.  For instance,
\begin{equation}
\int_{-1}^{1}dx\, {\mathsf H}_\pi(x,\xi,-\Delta^2) = F_\pi(\Delta^2) \,, \quad
\int_{-1}^1 dx\, 2x \, {\mathsf H}_\pi(x,\xi,-\Delta^2) = \theta_2^\pi(\Delta^2) - \xi^2 \, \theta_1^\pi(\Delta^2)\,,
\end{equation}
where $F_\pi$ is the elastic electromagnetic form factor and $\theta_{1,2}^\pi$ are gravitational form factors: $\theta_1^\pi$ relates to pressure distributions within the pion and $\theta_2$ is linked to the distribution of mass.  In the neighbourhood of the chiral limit, $\theta_1^\pi(0) - \theta_2^\pi(0) = {\rm O}(m_\pi^2)$ \cite{Polyakov:1999gs, Mezrag:2014jka}.  It was found in Ref.\,\cite{Zhang:2020ecj} that the pion's mass distribution form factor is harder (more pointlike) than its electromagnetic form factor, which is harder than the gravitational pressure distribution form factor.  Such ordering is consistent with available lQCD results \cite{Brommel:2007xd, Brommel:2007zz}.

With the gravitational form factor $\theta_1^\pi(\Delta^2)$  in hand, one can compute the following Breit-frame distributions \cite{Polyakov:2002yz, Polyakov:2018zvc}:
\begin{subequations}
\label{EqPressure}
\begin{align}
p_\pi(r)  & = \frac{1}{3} \int \frac{d^3\vec{\Delta}}{(2\pi)^3}\frac{1}{2E(\Delta)}\, {\rm e}^{i\vec{\Delta}\cdot \vec{r}}\, [\Delta^2\theta_1^\pi(\Delta^2)]
 = \frac{1}{6\pi^2 r} \int_0^\infty d\Delta \,\frac{\Delta}{2 E(\Delta)} \, \sin(\Delta r) [\Delta^2\theta_1^\pi(\Delta^2)] \,, \label{EqPressureA}\\
 s_\pi (r)  & = -\frac{3}{4} \int \frac{d^3\vec{\Delta}}{(2\pi)^3}\frac{ {\rm e}^{i\vec{\Delta}\cdot \vec{r}}\,}{2E(\Delta)}\, P_2(\hat{\Delta}\cdot \hat{r}) [\Delta^2\theta_1^\pi(\Delta^2)]
  = \frac{3}{8 \pi^2} \int_0^\infty d\Delta \,\frac{\Delta^2}{2 E(\Delta)} \, {\mathpzc j}_2(\Delta r) \, [\Delta^2\theta_1^\pi(\Delta^2)] \,, \label{EqPressureB}
\end{align}
\end{subequations}
%% BesselJ[1, k*r]*(FlQCD\[Theta]1[-k^2, M]) k^2  (k/Sqrt[4 \[Pi]m^2 + k^2])/12/Pi/r
%% (-2*k*r*BesselJ[1, k*r] + 9*BesselJ[2, k*r]))
where
$2E(\Delta)=\sqrt{4 m_\pi^2+\Delta^2}$,
$P_2$ is the degree-two Legendre polynomial, and ${\mathpzc j}_2(z)$ is a spherical Bessel function.
Here, $p_\pi(r)$ is the pressure and $s_\pi (r)$ is the shear force.  (One could alternatively compute distributions in the light-front transverse plane, but such two-dimensional Fourier-transform analogues deliver results of similar magnitude.)

\begin{figure}[t]
\hspace*{-1ex}\begin{tabular}{lcl}
{\sf A} &\hspace*{2em} & {\sf B} \\[-2ex]
\includegraphics[clip, width=0.46\textwidth]{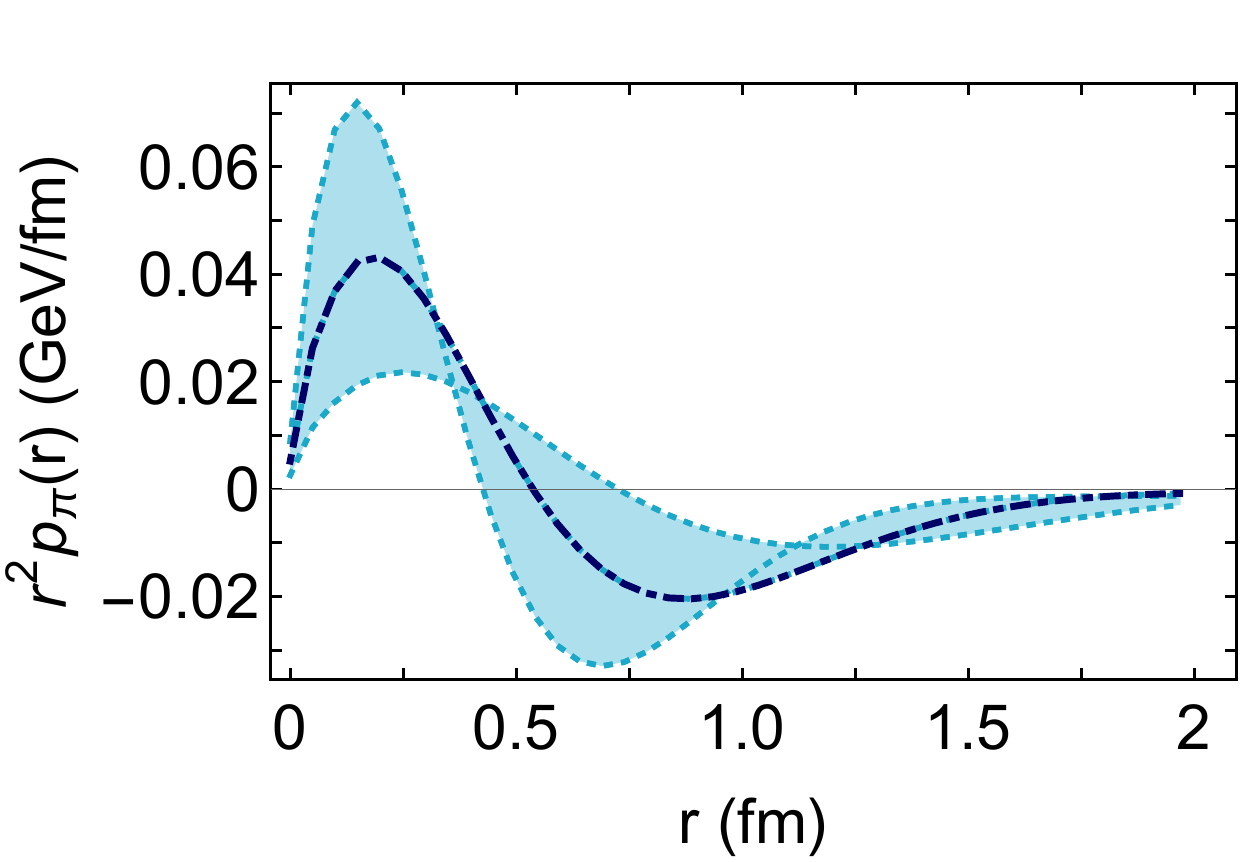} & \hspace*{2em} &
\includegraphics[clip, width=0.46\textwidth]{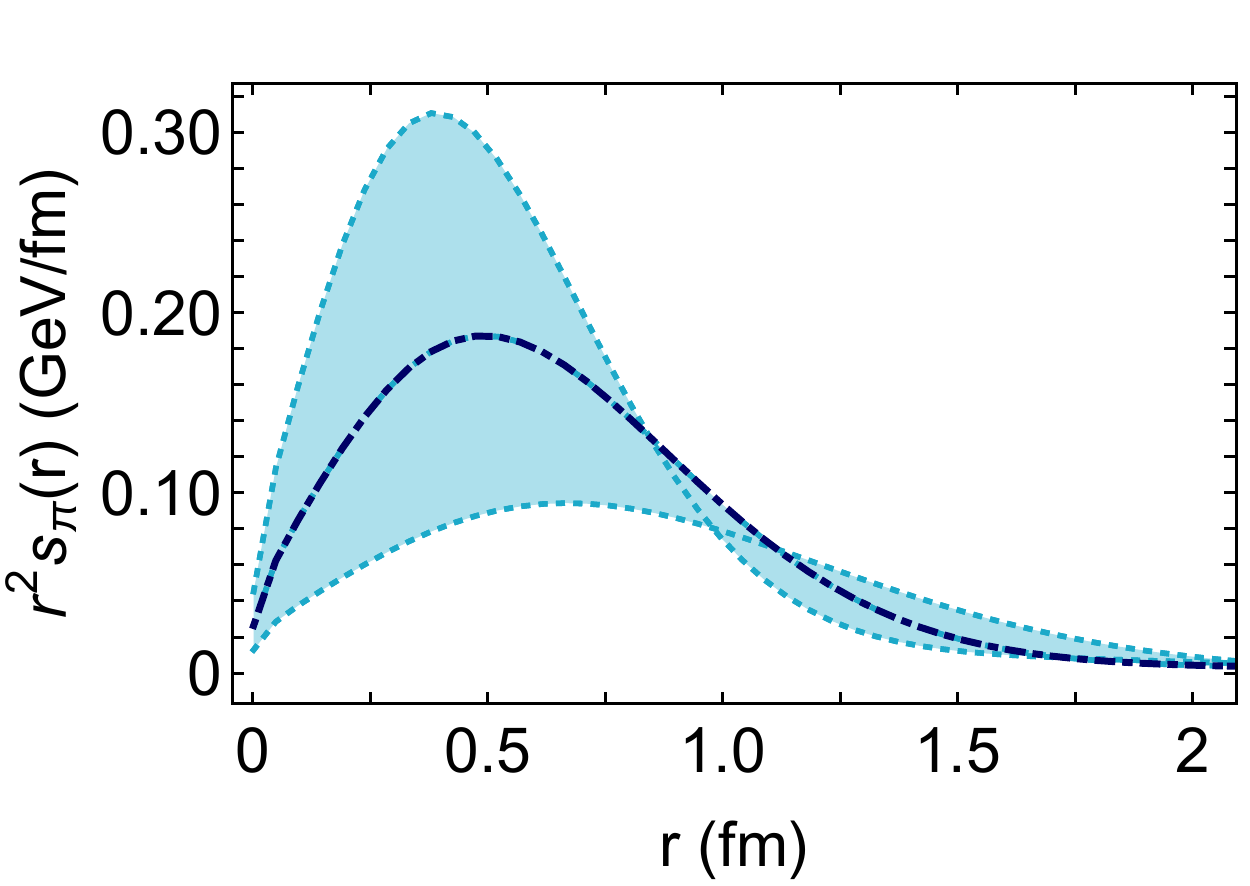}
\end{tabular}
\caption{\label{FigPressure}
\emph{Left panel--A}.  Pressure distribution in the pion, Eq.\,\eqref{EqPressureA}; and \emph{right panel--B} shear force distribution, Eq.\,\eqref{EqPressureB}.
The embedding bands express the uncertainty owing to limitations on the knowledge of $\theta_1^\pi(\Delta^2)$, as detailed in Ref.\,\cite{Zhang:2020ecj}.
}
\end{figure}

The distributions in Eqs.\,\eqref{EqPressure} were estimated in Ref.\,\cite{Zhang:2020ecj} using contact-interaction form factors that were corrected through comparison with lQCD analyses and improved by imposition of physical constraints.
The results are drawn in Fig.\,\ref{FigPressure} and the profiles accord with an  intuitive physical interpretation.  For instance, considering Fig.\,\ref{FigPressure}\,A, the pressure is positive and large on $r\simeq 0$, indicating that the dressed-quark and dressed-antiquark strongly repel each other at small separation; but $p_\pi(r)$ changes sign when the quark-antiquark separation exceeds roughly $1/[2 \Lambda_{\rm QCD}]$, marking a transition into the domain of quark-antiquark attraction produced by confinement forces.  Naturally, since the system is bound:
\begin{equation}
\int_0^\infty dr\,r^2 p_\pi(r) = 0\,;
\end{equation}
and at $\zeta_H$, this is a statement about the quark quasiparticle pressure distributions.

An analogue of Eq.\,\eqref{EqPressure} was used in Ref.\,\cite{Burkert:2018bqq} to infer the proton's quark pressure distribution from extant data on deeply virtual Compton scattering.  Comparing that result with those in Fig.\,\ref{FigPressure}\,A, one notes that: (\emph{a}) the pressure within the pion on $r\simeq 0$ is around five-times larger than that in the proton; and (\emph{b}) the two pressure profiles have similar radial extents.
(The results in Fig.\,\ref{FigPressure} are not affected by issues concerning the analysis in Ref.\,\cite{Burkert:2018bqq} which are canvassed in Refs.\,\cite{Kumericki:2019ddg, Moutarde:2019tqa}.)
Perhaps more striking, profiles like Fig.\,\ref{FigPressure}\,A for neutron stars indicate $r\simeq  0$ pressures therein of roughly $0.1\,$GeV/fm \cite{Ozel:2016oaf}; hence, the core pressures in pions and neutron stars are commensurate.
%% 1.6 10^34 Pa -- neutron star

Turning to Fig.\,\ref{FigPressure}\,B, the results can be interpreted after recognising that $r^2 s_\pi(r)$ measures the strength of QCD forces within the pion that work to deform it.  These forces peak at the zero of $r^2 p_\pi(r)$, whereat attractive confinement pressure begins to win over the forces driving the quark and antiquark away from the core.  Using the results in Fig.\,\ref{FigPressure}, this point is located at
\begin{equation}
r_{\rm c}^p = 0.52^{+0.19}_{-0.11}\, {\rm fm.}
\end{equation}
Qualitatively and semi-quantitatively equivalent results are obtained when working with a more sophisticated expression for the pion's LFWF than that provided by a contact interaction \cite{Zhang:2021mtn}.
%%% \cite{PepeRQ}.

Turning to the tensor GPD in Eq.\,\eqref{EqGPDFH}, it is worth highlighting the physical significance of the two leading Mellin moments:
\begin{equation}
\label{ETmomentsB}
{B}^\pi_{10} (-\Delta^2) = \int_{-1}^1 dx \, {\mathsf E}_\pi^{\rm T}(x,\xi,-\Delta^2) \,,
\quad
{B}^\pi_{20} (-\Delta^2) = \int_{-1}^1 dx \, x\,{\mathsf E}_\pi^{\rm T}(x,0,-\Delta^2) \,.
\end{equation}
These quantities are subject to QCD evolution; and using the $\hat\alpha$ scheme, Ref.\,\cite{Zhang:2020ecj} reports the following values at $\zeta=\zeta_2$:
\begin{equation}
\label{EqB1B2zeta2}
m_\pi\,B^\pi_{10}(0)  = 0.053\,, \quad  m_\pi\,B^\pi_{20}(0) = 0.012\,, \quad
B^\pi_{10}(0) /B^\pi_{20}(0) = 4.57\,.
\end{equation}
This is the scale used in the lQCD study described in Ref.\,\cite{Brommel:2007xd}, discussed in connection with Fig.\,\ref{fig:ns_gpd}\,B, which reports the following values for these $\Delta^2=0$ quantities after an extrapolation to the physical pion mass: $0.22(3)$, $0.039(10)$, $5.66(60)$, in qualitative agreement with the contact interaction results.  Similar conclusions are drawn elsewhere, \emph{e.g}.\ Refs.\,\cite{Nam:2010pt, Dorokhov:2011ew, Fanelli:2016aqc}.

The renormalised form factors ${B}^\pi_{i0} (-\Delta^2)/{B}^\pi_{i0} (0)$, $i=1,2$, are independent of the renormalisation scale \cite{Fanelli:2016aqc}.  Here, therefore, comparison with the lQCD results in Ref.\,\cite{Brommel:2007xd} is meaningful; albeit, quantitative agreement cannot be expected because the $\Delta^2$-dependent lQCD form factors are only available at $m_\pi^2 \approx 20 \, m_\pi^{2\,{\rm empirical}}$ (see Fig.\,\ref{fig:ns_gpd}\,B).  Notwithstanding this, there is semiquantitative agreement between the lQCD study and Ref.\,\cite{Zhang:2020ecj}, \emph{e.g}.: the radii have the same ordering, $r_{B_{10}^\pi}/r_{B_{20}^\pi} = 1.48(17)$ (lQCD) vs.\ 1.14; and $B_{10}(t)$ is softer than $B_{20}(t)$.

\begin{figure}[t]
\hspace*{-1ex}\begin{tabular}{lcl}
{\sf A} &\hspace*{2em} & {\sf B} \\[-2ex]
\includegraphics[clip, width=0.46\textwidth]{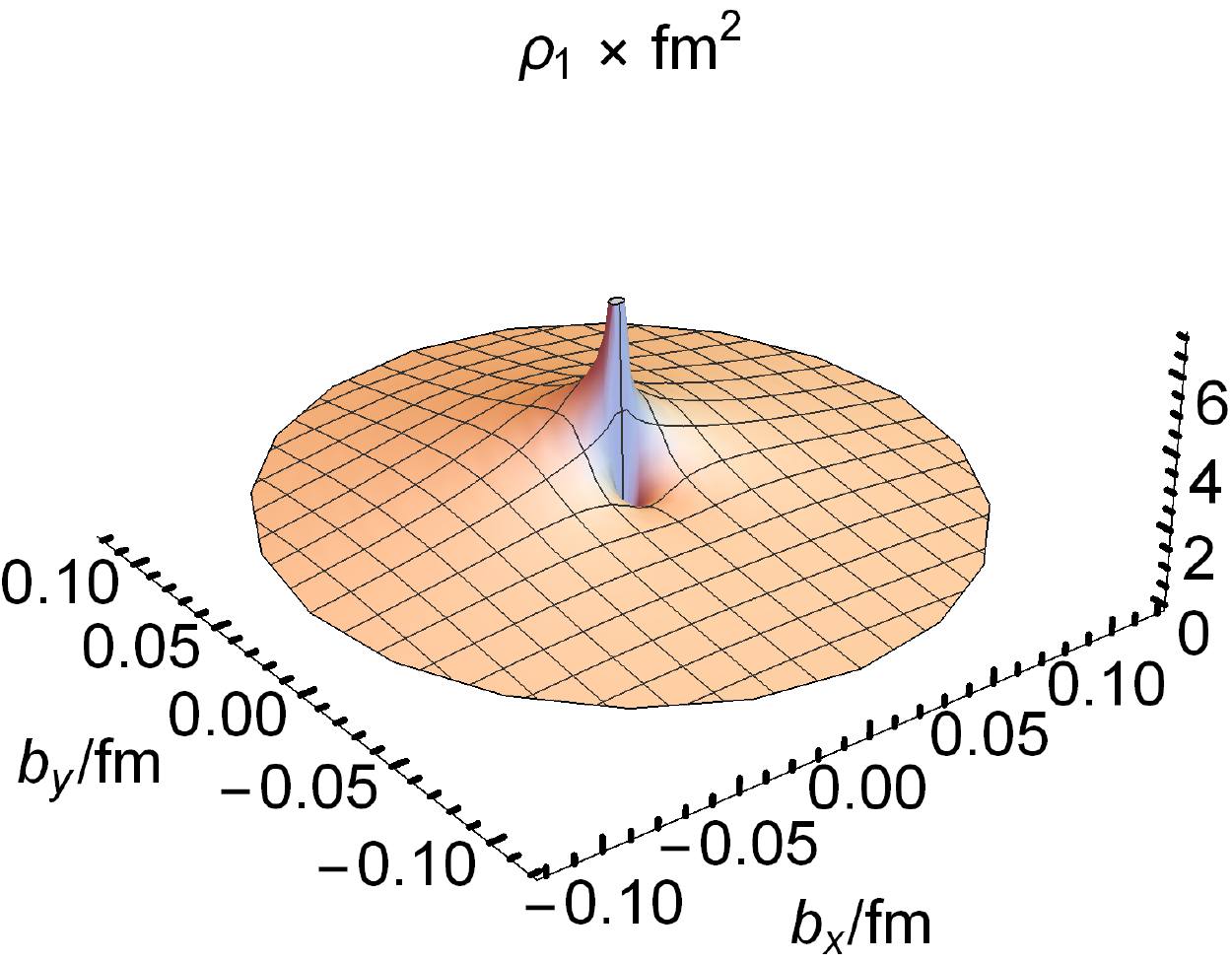} & \hspace*{2em} &
\includegraphics[clip, width=0.46\textwidth]{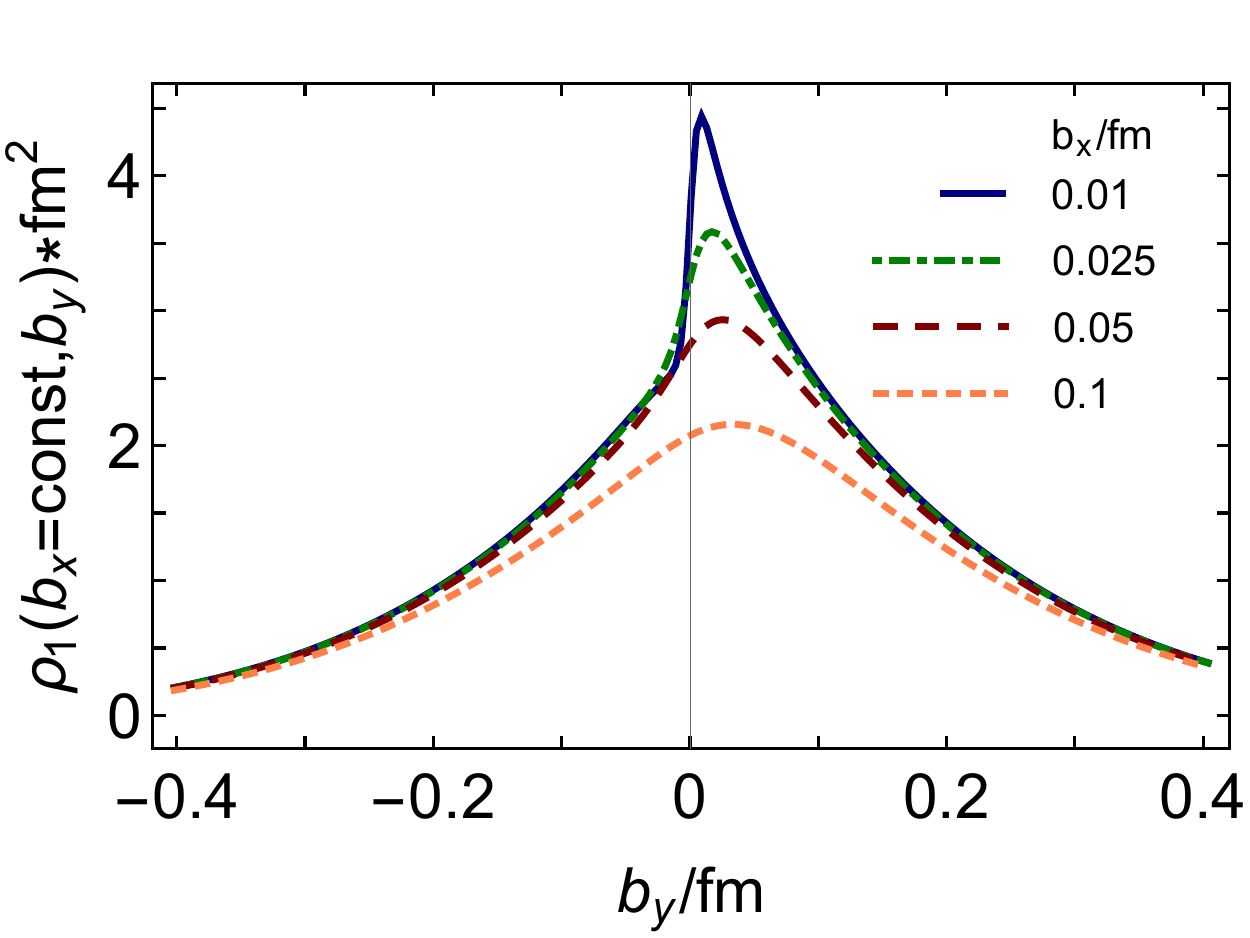}
\end{tabular}
\caption{\label{FigTSpin}
Light-front transverse-spin distribution of dressed valence quarks in the pion
$\rho_1(b_\perp,s_\perp \propto \hat x)$, Eq.\,\eqref{EqSpinDensity}.
\emph{Left panel}\,--\,{\sf A}. Full three-dimensional image.
\emph{Right panel}\,--\,{\sf B}. Slices at constant $b_x/$fm: solid blue -- $b_x=0.01$; dot-dashed green -- $b_x=0.025$; dashed dark-red -- $b_x=0.05$; and short-dashed orange -- $b_x=0.1$.
In both panels, the scale is $\zeta_H$, Eq.\,\eqref{setzetaH}.
}
\end{figure}

Combining the leading Mellin-moments of the pion's vector and tensor GPDs and transforming to impact parameter space, one arrives at the light-front transverse-spin ($s_\perp$) distribution of dressed-quarks within the pion \cite{Brommel:2007xd}:
\begin{equation}
\label{EqSpinDensity}
\rho_1(b_\perp,s_\perp) = \tfrac{1}{2}\tilde q_\pi(|b_\perp|) - \tfrac{1}{2}\varepsilon^{ij} s_\perp^i b_\perp^j B^{\prime\pi}_{10}(|b_\perp|)\,,
\end{equation}
with
\begin{subequations}
\begin{align}
\tilde q_\pi(|b_\perp|) & =
\int_0^\infty\!\frac{ d|\Delta|}{2\pi} \,\Delta\, J_0(|b_\perp ||\Delta|)\,F_\pi(-\Delta^2)\,,\\
B^{\prime\pi}_{10}(|b_\perp|) & = -\frac{1}{4\pi |b_\perp|}
\int_0^\infty \! d |\Delta|\, \Delta^2\, J_1(|b_\perp||\Delta|) B_{10}^\pi(-\Delta^2)\,,
\end{align}
\end{subequations}
where $J_{0,1}$ are cylindrical Bessel functions.  For a dressed-quark polarised in the $+x$ direction and $\hat s_\perp\cdot \hat b_\perp= \cos\phi_\perp$, $\varepsilon^{ij} s_\perp^i b_\perp^j = |b_\perp| \sin\phi_\perp$.  Using similar contact-interaction form factors, improved through comparison with lQCD analyses and implementation of physical constraints, as employed to draw Figs.\,\ref{FigPressure}, Ref.\,\cite{Zhang:2020ecj} delivered the estimate for the pion's light-front dressed-quark transverse-spin distribution illustrated in Fig.\,\ref{FigTSpin}.

It is evident from Fig.\,\ref{FigTSpin} that the transverse-spin density associated with a dressed valence-quark polarised in the light-front-transverse $+x$ direction is not symmetric around $\vec{b}_\perp=(b_x=0,b_y=0)$.  Rather, strength is transferred from $b_y<0$ to $b_y>0$ and the peak shifted to $(b_x=0,b_y>0)$.  The $b_y$ profile remains symmetric around the line $b_x=0$; and using the formula in Ref.\,\cite{Brommel:2007xd}, the result in Fig.\,\ref{FigTSpin} describes an average transverse shift
\begin{equation}
\langle b_y \rangle = \frac{1}{2} B_{10}(0)/m_\pi = 0.049\,{\rm fm} \approx 0.1\,r_c^p \,.
\end{equation}
This distortion vanishes logarithmically with $B_{10}^\pi(0) \to 0$ under QCD evolution; but as noted in connection with Eq.\,\eqref{pionGS}, that process is very slow.

%Regarding transverse momentum dependent distribution functions (TMDs), these studies indicate that their magnitude and domain of support decrease with increasing twist~\cite{Zhang:2020ecj}. Consistent with intuition~\cite{Zhang:2020ecj}, at $\zeta_H$, the simplest Wigner distribution associated with the pion's twist-two dressed-quark GTMD is sharply peaked on the kinematic domain associated with valence-quark dominance; has a domain of negative support; and broadens as the transverse position variable increases in size.  With improved studies in the offing, now is the right time to plan on exploiting the capacities of the EIC to probe these higher-dimensional aspects of pion and kaon structure.

Given the information about the SM's two mass generating mechanisms and the interference between them that is accessible via pion and kaon 1D distributions, it is certain that a wide range of additional, novel insights could be drawn from comparisons between the 3D distributions for which those 1D distributions are, effectively or literally, boundary values.  Thus, extensions of kaon studies such as those in Refs.\,\cite{Nam:2011yw, Son:2015bwa, Kaur:2020vkq} are to likely attract much attention.

\subsection{Meson Fragmentation Functions}
\label{SecMFFs}
%The measurement of cross-sections from which one can extract generalised parton distributions (GPDs) and transverse-momentum-dependent parton distributions (TMDs), and, therefrom, 3-D images of hadrons, presents a fascinating new frontier within the Standard Model, promising to deliver tomographic pictures of hadron structure.  EicC could be a leader in this field.
In closing this section it is worth stressing that many new challenges are faced in drawing 3D images from new-generation experiments.  Sound calculations and models of a wide variety of parton distribution functions will be crucial.  They can inform estimates of the size of the cross-sections involved and indicate the best means by which to analyse them \cite{Berthou:2015oaw}.  Notwithstanding that, experiences with meson structure functions have shown that full capitalisation on such experiments requires the use of calculational tools which can reliably connect experiments with qualities of QCD.  In this, too, continuum calculations can provide valuable insights \cite{Mezrag:2014jka, Mezrag:2016hnp, Chouika:2017rzs, Xu:2018eii}.

An additional complication arises in connection with TMD extractions: every cross-section that can yield hadronic TMDs involves related parton fragmentation functions (FFs) \cite{Field:1977fa}, the structure of which must be known in detail.  The future of momentum imaging therefore depends critically on making real progress with the measurement and calculation of FFs.  It is a serious issue, therefore, that no realistic computations of FFs currently exist.  Even a formulation of the problem remains uncertain.

These issues are greatly amplified by the connection between FFs and confinement because an elucidation of confinement is central to solving QCD.  With light quarks in the mix, confinement is a dynamical process.  A scattering event produces a gluon or quark that begins to propagate in spacetime.  After a short interval, an interaction occurs.  In this way, the parton loses its identity, sharing it with others.  Finally, after a cascade of such events, a cloud of partons is produced, which fuses into color-singlet final states.  (Additional discussion may be found, \emph{e.g}.\ in Refs.\,\cite[Sec.\,3]{Horn:2016rip} and \cite[Sec.\,2.2]{Cloet:2013jya}.)  These are the processes captured in FFs, which chart how QCD partons, nearly massless when produced in a high-energy event, convert into a shower of massive hadrons.  Namely, they describe how hadrons with mass emerge from massless partons.

Such qualities indicate that FFs may be the cleanest expression of dynamical confinement in QCD.  In addition, owing to Gribov-Lipatov reciprocity \cite{Gribov:1971zn}, DFs and FFs are linked by crossing symmetry, one being the analytic continuation of the other at their common boundary of support.  Hence, like DFs, FFs provide basic insights into the origin of mass: they serve as timelike analogues, providing a basic counterpoint to the DFs.  Thus, modern facilities that can supply precise data on quark fragmentation into a pion or kaon may provide the means to directly test those facets of QCD calculations which embody and express the SM's most fundamental emergent phenomena: confinement, DCSB, and bound-state formation.

%% file: S8_Lattice.tex
\section{Developments in Lattice QCD}
\label{SeclQCD}
\subsection{Formulation}
\label{SecFormalism}
Discretising QCD on a Euclidean lattice, first proposed more than forty years ago \cite{Wilson:1974sk}, provides a first-principles means of solving QCD in the strong-coupling regime.  The starting point for a lattice computation is the discretised, Euclidean path integral, wherein physical ``observables" ${\cal O}$ are computed as
\begin{equation}
    \langle {\cal O} \rangle= \frac{1}{\cal Z}\prod_{n,\mu} d U_\mu(n) \prod_n d\psi(n) \prod_n d \bar{\psi} (n) {\cal O}(U,\psi, \bar{\psi}) e^{-(S_G[U] + S_F[U,\psi,\bar{\psi}])},
\end{equation}
where $U_\mu(n)$ are $3 \times 3$ unitary matrices representing the gauge fields, 
$\psi, \bar{\psi}$ are Grassmann variables representing the fermion and antifermion fields respectively, 
$S_G$ and $S_F$ are the gauge and fermion discretisations of the action, 
$n = (n_1, n_2, n_3, n_4)$ represent the sites on a discrete Euclidean lattice,
and ${\cal Z}$ ensure the result is unity when ${\cal O}$ is an identity operator.  The Grassmann variables can then be integrated to yield
\begin{equation}
   \langle {\cal O} \rangle = \frac{1}{\cal Z} \prod_{n,\mu} dU_\mu(n) {\cal O}(U, G[U]) \det M[U] e^{-S_G[U]}\,,
\end{equation}
where $G$ are the quark propagators
\[
G(U,m,n)^{ij}_{\alpha\beta} \equiv \langle \psi^i_\alpha(m) \bar{\psi}^j_\beta(n) \rangle
\]
and $M$ is the fermion determinant.  A lattice calculation proceeds as follows:
\begin{enumerate}
    \item Generate an ensemble of $N_{\rm cfg}$ equilibrated, statistically independent, gauge configurations with probability
    \[
    P[U] \propto \det M[U] e^{-S_G[U]}\,.
    \]
    \item Calculate the expectation of a physical observable on those configurations as
    \[
        \langle {\cal O} \rangle = \frac{1}{N_{\rm cfg}} \sum_{s = 1}^{N_{\rm cfg}} {\cal O}( U^s, G[U^s])\,.
        \]
\end{enumerate}

By formulating the theory in Euclidean space, the integrand is rendered real, thereby enabling $P[U]$ to be interpreted as a probability distribution amenable to importance sampling.  The first step above proceeds through a Markov process, whereby each gauge configuration $U^s$ is generated from its previous gauge configuration; this is a \textit{capability} computing task, requiring the largest-available leadership-class computers.
The second step in calculating observables can be performed independently on each gauge configuration, and therefore is a \textit{capacity} computing task, albeit one for which the total computational demand might exceed that of the capability stage. As will be seen below, the formulation in Euclidean space and in finite volume imposes challenges.

It is beyond the scope of this article to describe the theoretical formalism and computational implementation of lattice gauge theory in general, and the reader is referred to several pedagogical textbooks \cite{Rothe:1992nt, Montvay:1994cy, DeGrand:2006zz, Gattringer:2010zz}.  In addition to the statistical uncertainties outlined above, lattice computations are subject to a variety of systematic uncertainties: the lattice cutoff expressed through the lattice spacing $a$; the finite physical volume $V$; the quark masses, and in particular those of the light $u,d,s$ quarks employed in the calculation; and delineating the properties of the lowest-lying stable pion, kaon and nucleon, from those of their excitations.  A review of the calculation of key quantities of low-energy hadronic physics in contained in Ref.\,\cite{Aoki:2016frl}.  As described below, for the most widely studied quantities discuss herein, such as the moments of the quark distribution amplitudes, an important effort has aimed at gaining control of these systematic uncertainties.  For some more recently studied quantities, computations are often performed at unphysically large light-quark masses, and at a single physical spatial volume or lattice spacing, in anticipation of full control over systematic uncertainties in the years ahead.

\begin{figure}[t]
    \includegraphics[width=0.44\textwidth]{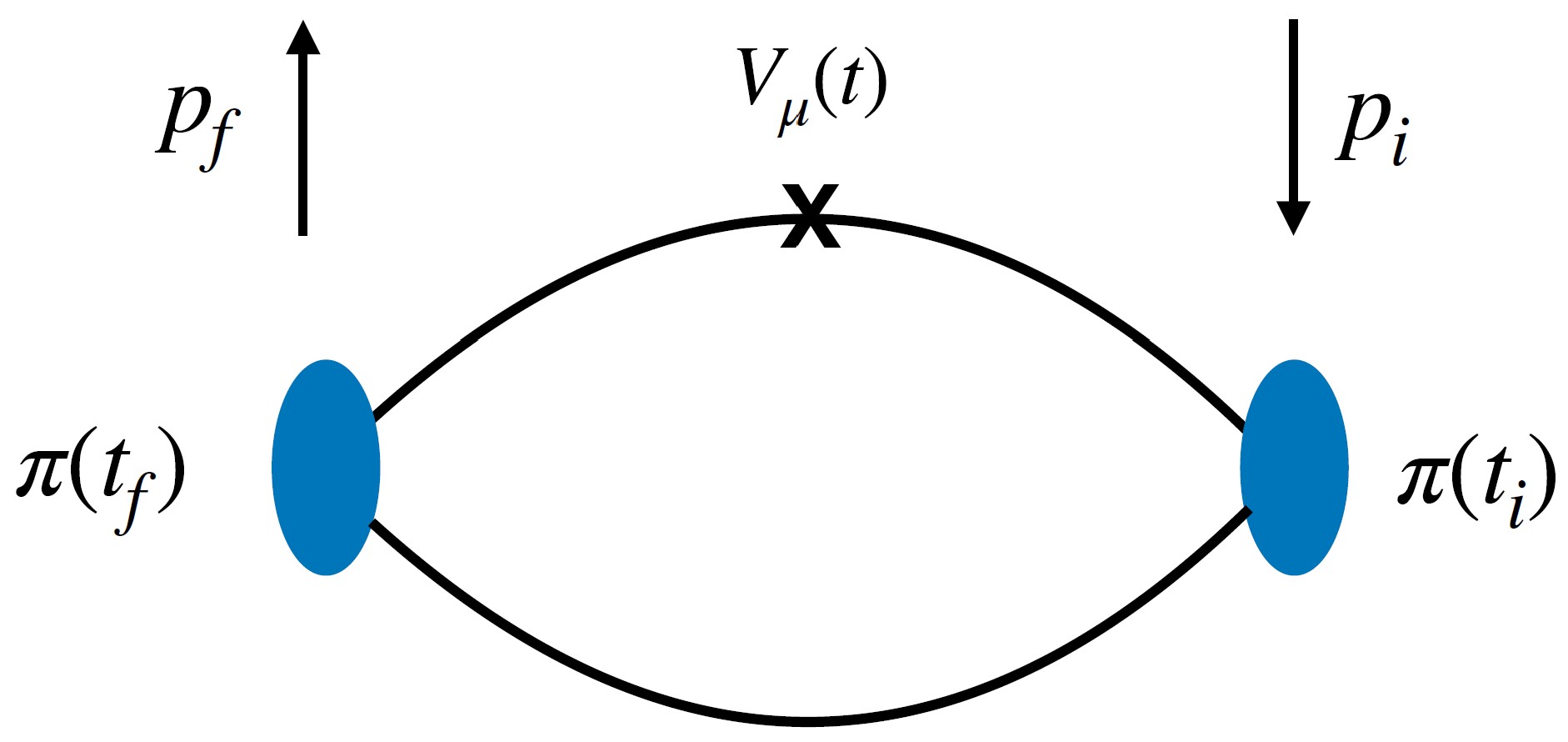} %\vspace{-1.3in}
\caption{\label{fig:threept} Three-point function defining the pion form factor in lQCD analyses: $t_i, t_f$ and $t$ denote the time slices of the initial pion, final pion and current insertion, computed to obtain the vector-current matrix element of Eq.\,\eqref{eq:vec}. }
\end{figure}

\subsection{Pion Form Factor}
\label{lQCDpionFF}
The study of the structure of hadrons, and in particular the nucleon and pion, has been a essential component of lQCD calculations since its inception.  The nucleon and pion are the only QCD states composed of the light $u$ and $d$ quarks that are stable under strong interactions; and properties such as the electromagnetic form factors, which can be expressed as matrix elements of a local external current, are the most straightforwardly accessible to lattice computations.  Indeed the pion and kaon can be directly studied as single, isolated states in contrast to the experimental situation.  The form factor, for the case of the charged pion, is expressed as
\begin{equation}
    \langle \pi (\vec{p}_f) \mid V_\mu(0) \mid \pi(\vec{p}_i) \rangle = F_\pi(Q^2) (p_i + p_f)_\mu . \label{eq:vec}
\end{equation}
%where $Q^2$ is the four-momentum transfer.  
For the case of spacelike $Q^2$, the matrix element is straightforward to evaluate in lQCD through the three-point function depicted in Fig.\,\ref{fig:threept}.

The low-$Q^2$ slope of the form factor is related to the pion's charge radius, while its behaviour at large $Q^2$ is a measure of the evolution from hadronic to partonic degrees of freedom.  The pion form factor has been extensively studied in a progression of increasingly precise calculations, see \emph{e.g}.\ Refs.\,\cite{Martinelli:1987bh, Draper:1988bp, Bonnet:2004fr, Brommel:2006ww, Frezzotti:2008dr, Aoki:2009qn, Nguyen:2011ek, Chambers:2017tuf, Koponen:2017fvm, Alexandrou:2017blh, Wang:2020nbf}.  Computations of the form factor in the two regimes of small and large $Q^2$ demonstrate the conflicting demands on high-precision lattice calculations.

\begin{figure}[t]
    \includegraphics[width=4in]{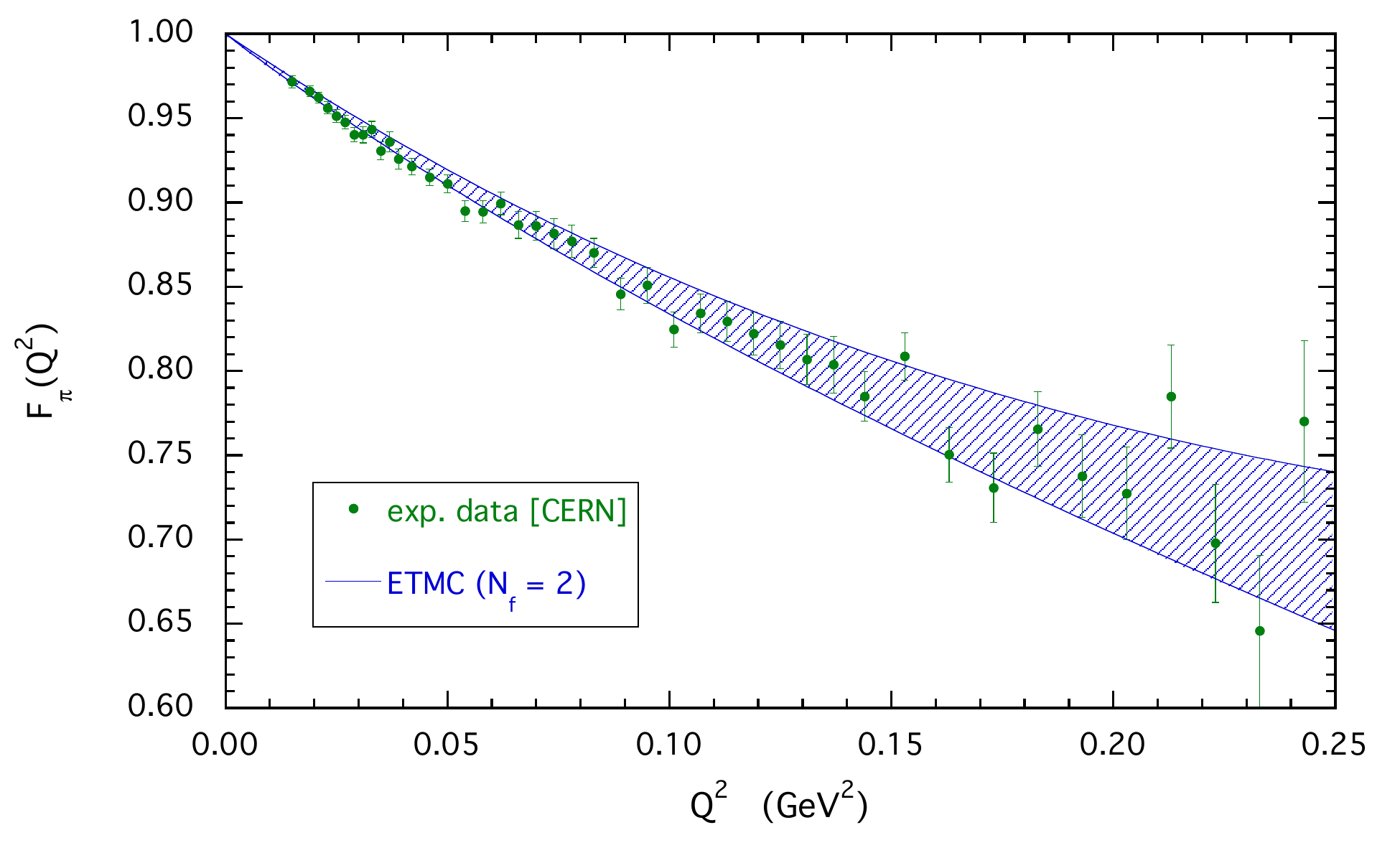}
    \caption{Pion form factor $F_\pi(Q^2)$ at the physical pion mass in a theory with 2 degenerate flavours of light quarks, together with experimental data \cite{Amendolia:1986wj}.  (Figure adapted from Ref.\,\cite{Alexandrou:2017blh}.)\label{fig:pion_ff}}
\end{figure}

The $\pi^+$ charge radius is defined via
\begin{equation}
    \langle r_\pi^2 \rangle = - 6 \left. \frac{d F_\pi(Q^2)}{dQ^2} \right|.\label{eq:charge_rad}
\end{equation}
On a discretised lattice, momentum is also discrete; hence, $\langle r_\pi^2\rangle$ is obtained through a fit to data away from $Q^2 = 0$.  An important development has been the adoption of model-independent fits to the data through the use of a $z$-expansion \cite{Hill:2010yb, Epstein:2014zua}, rather than performing a simple dipole fit to the data.  Nevertheless, the ability to reach the low momentum range needed to reliably determine the charge radius is constrained by the spatial volume of the lattice, in an analogous manner to which the charge radius in electron scattering is limited by the smallest energy transfer in experiment.  In lQCD, this can be ameliorated through the use of non-periodic, or twisted, spatial boundary conditions \cite{Alexandrou:2017blh}.  The resulting form factor, with two degenerate $u,d$ quark flavours and extrapolated to the physical pion mass, is shown in Fig.\,\ref{fig:pion_ff}.  However, there are alternative, coordinate-space approaches that enable the charge radius to be computed directly \cite{Lellouch:1994zu, Aglietti:1994nx, Bouchard:2016gmc, Feng:2019geu}; applied to the pion charge radius, a recent analysis finds \cite{Feng:2019geu} $\langle r_\pi^2 \rangle = (0.69(8)(6)\,{\rm fm})^2$, consistent with experiment $\langle r_\pi^2 \rangle=  (0.659(4)\,{\rm fm})^2$ \cite{Zyla:2020zbs}.

\begin{figure}[t]
    \includegraphics[width=0.52\linewidth]{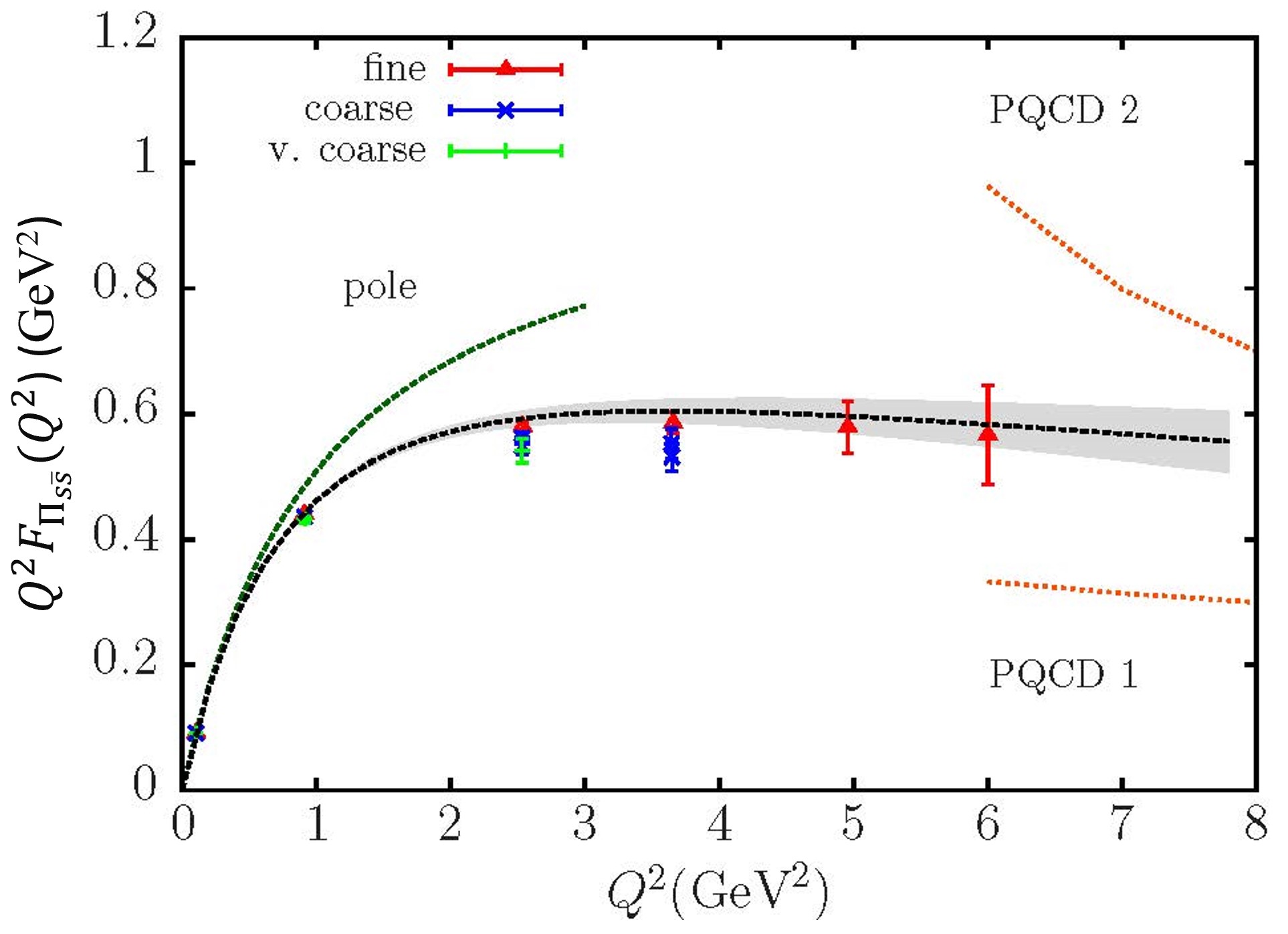}
    \caption{\label{fig:pion_high_Q2}
    Elastic electromagnetic form factor of the fictitious $\Pi_{s\bar s}$ meson.  Lattice QCD results compared with expectations based on Eq.\,\eqref{EqHardScatteringElastic}: PQCD\,1 uses $\varphi_{\rm as}(x)$, whereas PQCD\,2 is based on a realistic EHM-broadened pion DA -- Refs.\,\cite{Arthur:2010xf, Braun:2015axa} and Sec.\,\ref{SecDApion}.  (Figure adapted from Ref.\,\cite{Koponen:2017fvm}.)}
\end{figure}

At high momenta, in contrast, the constraints are two-fold, one statistical and the other systematic.  While at low momentum, the degradation with temporal separations of the signal-to-noise ratio for pion and kaon correlation functions is less severe than for the case of the nucleon; indeed, at zero momentum the signal-to-noise ratio remains constant with the temporal separation, the situation is reversed at high momenta.  Thus the most precise computation of the high-$Q^2$ pseudoscalar form factor has been performed not at the physical pion mass, but rather for the $\Pi_{s\bar s}$, containing a strange quark and antiquark, attaining $Q^2 \simeq 6\,{\rm GeV}^2$ \cite{Koponen:2017fvm}, covering some of the domain that will be explored by forthcoming JLab\,12 experiments, as highlighted in Fig.\,\ref{FigFpiEIC}.  (Additional discussion of this result, including comparisons with other lattice and continuum studies may be found in Ref.\,\cite{Chen:2018rwz}.)
The resulting form factor is shown in Fig.\,\ref{fig:pion_high_Q2}.  Evidently, $Q^2 F_{\Pi_{s\bar s}}(Q^2)$ exhibits the approach to a plateau, breaking away from expectations based on single-pole vector meson dominance and thereby indicating a transition to a description based on partonic degrees of freedom in the manner described in Sec.\,\ref{sec:EEFFs}.  Likewise, with connections to the hard-scattering formula in Eq.\,\eqref{EqHardScatteringElastic}, where expectations based on a broad, concave DA are most realistic: PQCD\,1 is obtained with $\varphi_{as}(x)$, whereas PQCD\,2 is based on a modern understanding of the NG mode DAs -- Refs.\,\cite{Arthur:2010xf, Braun:2015axa} and Sec.\,\ref{SecDApion}.

The discussion above relates only to the spacelike pion form factor.  The formulation in Euclidean space might seem to preclude calculations in the timelike region; but a method was proposed for the domain $\sqrt{s} < 4 m_\pi$ \cite{Meyer:2011um}, and there have since been several calculations \cite{Feng:2014gba, Andersen:2018mau, Erben:2019nmx}.  These calculations are restricted to low values of $Q^2$, but are important for an \textit{ab initio} understanding of the hadronic contributions to the muon $g-2$ \cite{Aoyama:2020ynm}.

\subsection{Parton Distribution Functions}
\label{SeclQCDDF}
Some of the key measures of hadron structure, such as DAs, DFs, GPDs and TMDs, are characterised by matrix elements of operators that are separated along the light cone.  Thus, for the case of the pion valence-quark DF, ${\mathpzc q}^\pi$, one writes
\begin{equation}
    {\mathpzc q}^\pi(x) = \int \frac{d \eta^-}{4\pi} e^{i x p^+ \eta^-} \langle \pi(\vec{p}) \mid \bar{\psi} (\eta^-/2) \gamma^+ W(\xi^-/2,-\eta^-/2) \psi(-\eta^-/2) \mid \pi(\vec{p}) \rangle,\label{eq:pdf_defn}
\end{equation}
where $W$ is a line along the light cone that ensures gauge invariance.  The computation of such light-cone-separated, and thereby time-dependent, operator matrix elements is precluded in Euclidean-space  lQCD.  A solution is to appeal to the operator-product expansion, whereby the Mellin moments can be expressed as matrix elements of local operators that are computable within Euclidean-space lQCD.  Specifically, for the case of the pion DF, one introduces
\begin{equation}
    \langle x^n\rangle = \int_0^1 dx \, x^n ({\mathpzc q}^\pi(x) - (-1)^n \bar{{\mathpzc q}}^\pi(x)),
\end{equation}
and writes
\begin{equation}
    \langle \pi(\vec{p}) \mid {\cal O}_q^{\mu_1,\dots,\mu_{n+1}}  \mid \pi(\vec{p})\rangle = p^{\mu_1}\dots p^{\mu_{n+1}} \langle x^n \rangle,
\end{equation}
where, for the flavour-non-singlet case, the ${\cal O}_q^{\mu_1,\dots,\mu_n}$ are twist-two operators given by
\begin{equation}
{\cal O}_q^{\mu_1,\dots,\mu_n} = i^{n-1} \bar{\psi} \gamma^{\{ \mu_1} D^{\mu_2}\dots D^{\mu_n\}} \psi.\label{eq:local_op}
\end{equation}
Such matrix elements of quasi-local operators can be computed on a Euclidean lattice, in the manner of the vector current in Eq.\,\eqref{eq:vec}.

As described elsewhere \cite{Holt:2010vj}, there have been several studies of the lowest moments $\langle x\rangle$ and $\langle x^2\rangle$ of the pion valence DF.  However, a limited number of moments is insufficient to constrain the $x$-dependent DF, even under the strong constraints imposed by  a selected parametrisation \cite{Detmold:2003rq}, \emph{e.g}.\ as observed in Ref.\,\cite{Hecht:2001fr} and in connection with Fig.\,\ref{Figmoments}, precise results for at least the first six Mellin moments are necessary before one begins to gain sensitivity to the large-$x$ exponent, $\beta_\Pi$, in Eq.\,\eqref{pionPDFlargex}.
The ability to compute higher moments using lQCD is limited both by statistical precision, and more fundamentally by the breaking of $O(4)$ symmetry introduced by the lattice discretisation.  This introduces power-divergences, in the lattice spacing $a$, mixing with lower-dimensional operators, restricting the accessible moments to $\langle x^n \rangle$, $n\leq 3$.

A major advance with the ability to study the internal structure of hadrons using lQCD was a realisation that parton DFs, and the other quantities represented as matrix elements of operators separated along the light cone, could be related to quantities computable in lQCD within the framework of large-momentum effective theory (LaMET) \cite{Ji:2013dva, Ji:2014gla}.  Since then, two additional frameworks have been proposed; namely, parton pseudo-distributions (pseudo-PDFs) \cite{Radyushkin:2017cyf} and good lattice cross sections (GLCS) \cite{Ma:2017pxb}.

Both the LaMET and pseudo-PDF frameworks start with the same lattice building blocks $h(z,p_z)$, namely the matrix elements of operators separated along a spatial axis, taken here to be the $z$ direction:
\begin{equation}
    h(z,p_z) = \frac{1}{2 p_\alpha}\langle \pi(p_z)\mid \bar{\psi}(z) \gamma_\alpha W(z,0) \psi(0) \mid \pi(p_z)\rangle.\label{eq:wilsonop}
\end{equation}
Both the LaMET and pseudo-PDF frameworks require the renormalisation of the Wilson-line operator $W(z,0)$, where a variety of different approaches have been proposed and adopted \cite{Ji:2015jwa, Stewart:2017tvs, Orginos:2017kos, Constantinou:2017sej, Green:2017xeu, Monahan:2017hpu, Izubuchi:2018srq}. The difference between the approaches in essence lies in the relationship between the matrix element computed on the lattice and the parton DF \cite{Lin:2017snn}.  Specifically, within LaMET, one introduces a parton quasi-distribution function (quasi-PDF) \cite{Ji:2013dva}
\begin{equation}
\tilde{{\mathpzc q}}(x,\Lambda,p_z) = \int \frac{dz}{2\pi} p_z h(z,p_z),
\end{equation}
where $\Lambda$ is an ultraviolet scale.  The DF is then related through
\begin{equation}
    \tilde{{\mathpzc q}}(x,\Lambda,p_z) = \int_{-1}^1 \frac{dy}{\mid y \mid} Z\left(\frac{x}{y},\frac{\zeta}{p_z},\frac{\Lambda}{p_z}\right) {\mathpzc q}(y,Q^2) + {\cal O}\left(\frac{\Lambda^2_{\rm QCD}}{p_z^2}, \frac{M^2}{p_z^2}\right). \label{eq:lamet}
\end{equation}
The coefficient $Z$ is perturbatively calculable \cite{Ji:2014gla}.  The remaining terms are analogous to higher-twist or mass corrections; and for mesons, attempts to understand their magnitude using QCD-inspired models may be found in Refs.\,\cite{Hobbs:2017xtq, Xu:2018eii}.

The pseudo-PDF approach expresses the matrix element of Eq.\,\eqref{eq:wilsonop} in terms of Lorentz invariant (pseudo-)Ioffe time $\nu = p \cdot z$ \cite{Ioffe:1969kf, Braun:1994jq}, and a short distance scale $z^2$.  A reduced matrix element ${\mathcal M}(\nu,z^2)$, constructed to control the ultraviolet divergences, can then be constructed such that
\begin{equation}
    {\mathcal M}(\nu,z^2) = \int_0^1 du \ K(u,z^2 \zeta^2, \alpha) {\cal Q}(u \nu, z^2) \,, \label{eq:reduced}
\end{equation}
where, once again, $K$ is a perturbatively calculable kernel.  The Ioffe-time distribution ${\cal Q}$ is then the Fourier transform of the DF
\begin{equation}
    {\cal Q}(\nu) = \int_{-1}^1 dx \, {\mathpzc q}(x) e^{i \nu x}. \label{eq:pseudo}
\end{equation}

The final approach is that of so-called good lattice cross-sections \cite{Ma:2017pxb}, where the matrix element of a short-distance operator is computed
\begin{equation}
    \sigma_{j_1 j_2}(\nu,z^2,p^2) \equiv \langle \pi(p) \mid T {\cal O}_{j_1 j_2}(z) \mid \pi(p) \rangle\,,
\end{equation}
where $T$ is a time ordering operation, and $z$ the largest distance scale, which can then be written as a factorisable expression
\begin{equation}
\sigma_{j_1 j_2}(\nu,z^2, p^2) = \int_{-1}^1 \frac{dx}{x} {\mathpzc q}(x, \zeta^2) K_{j_1 j_2}(x \nu, z^2, x^2p^2, \zeta^2) + {\cal O} (z^2 \Lambda_{\rm QCD}^2)\,, \label{eq:glcs}
\end{equation}
where $K_{j_1 j_2}$ is a perturbatively calculable coefficient.  The operator ${\cal O}_{j_1 j_2}$ encompasses the Wilson-line operator employed in the LaMET and pseudo-PDF frameworks; but it is more general and, in particular, can include two gauge-invariant currents separated by the short-distance scale $z$, or gluon field operators.

\begin{figure}[t]
\includegraphics[clip, width=0.46\textwidth]{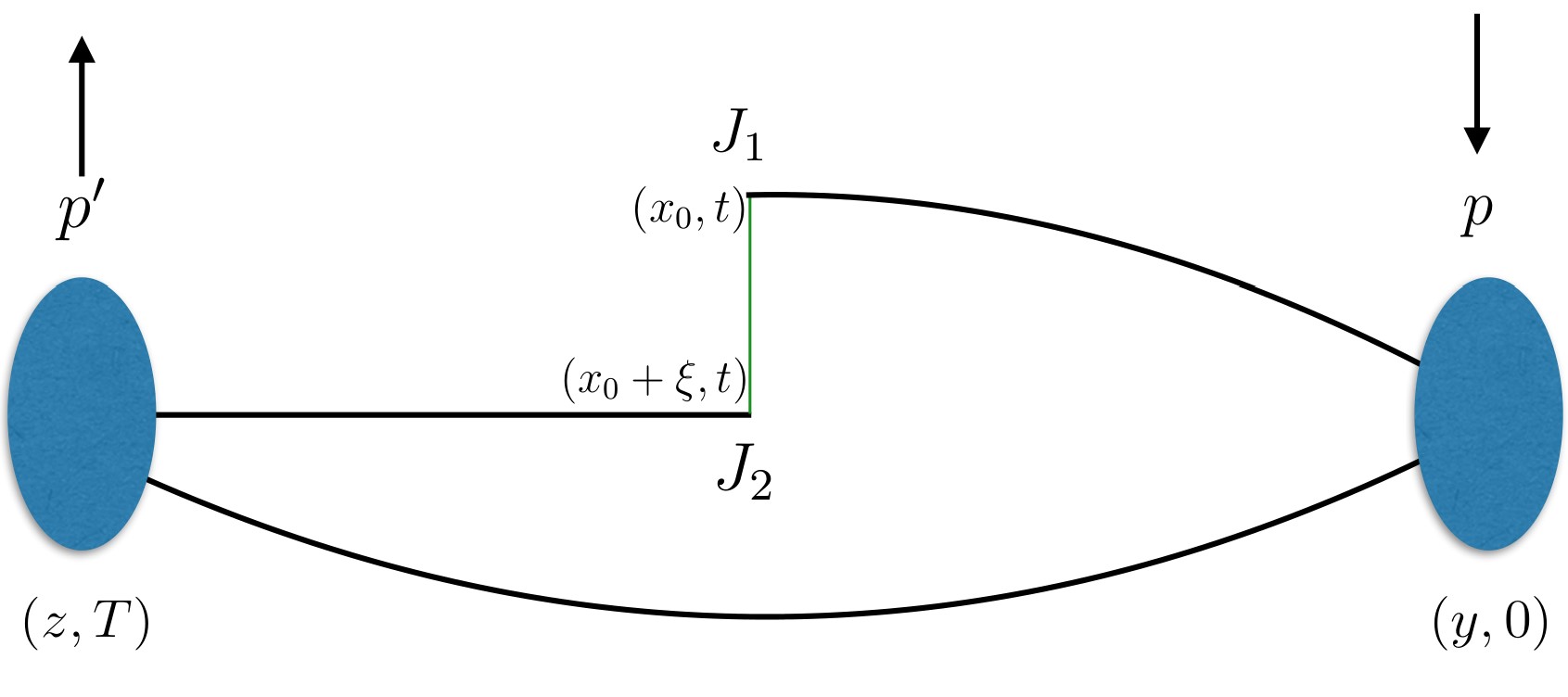}
   \caption{\label{fig:3pt_pdf}
    Matrix element whose lQCD computation yields the pion's valence-quark DF.  For the LaMET and pseudo-PDF approaches, the currents $J_1$ and $J_2$ correspond to the quark and anti-quark fields, respectively, connected by a Wilson line; whereas in the GLCS approach, $J_1$ and $J_2$ are gauge-invariant currents, with the thin, vertical green line corresponding to a quark propagator.}
\end{figure}

There have been studies of the pion DF within each of these approaches, beginning with the earliest studies using the LaMET approach \cite{Chen:2018fwa, Chen:2018xof, Izubuchi:2019lyk}; and more recently extended to the DF of the kaon \cite{Lin:2020ssv}, as discussed in Sec.\,\ref{sec:kaonDFs}.  In each approach, computation of the three-point function in Fig.\,\ref{fig:3pt_pdf} is required.

%%%<<<--- problem figure

\begin{figure}[t]
\hspace*{-1ex}\begin{tabular}{lcl}
{\sf A} &\hspace*{2em} & {\sf B} \\[-2ex]
\includegraphics[clip, width=0.48\textwidth]{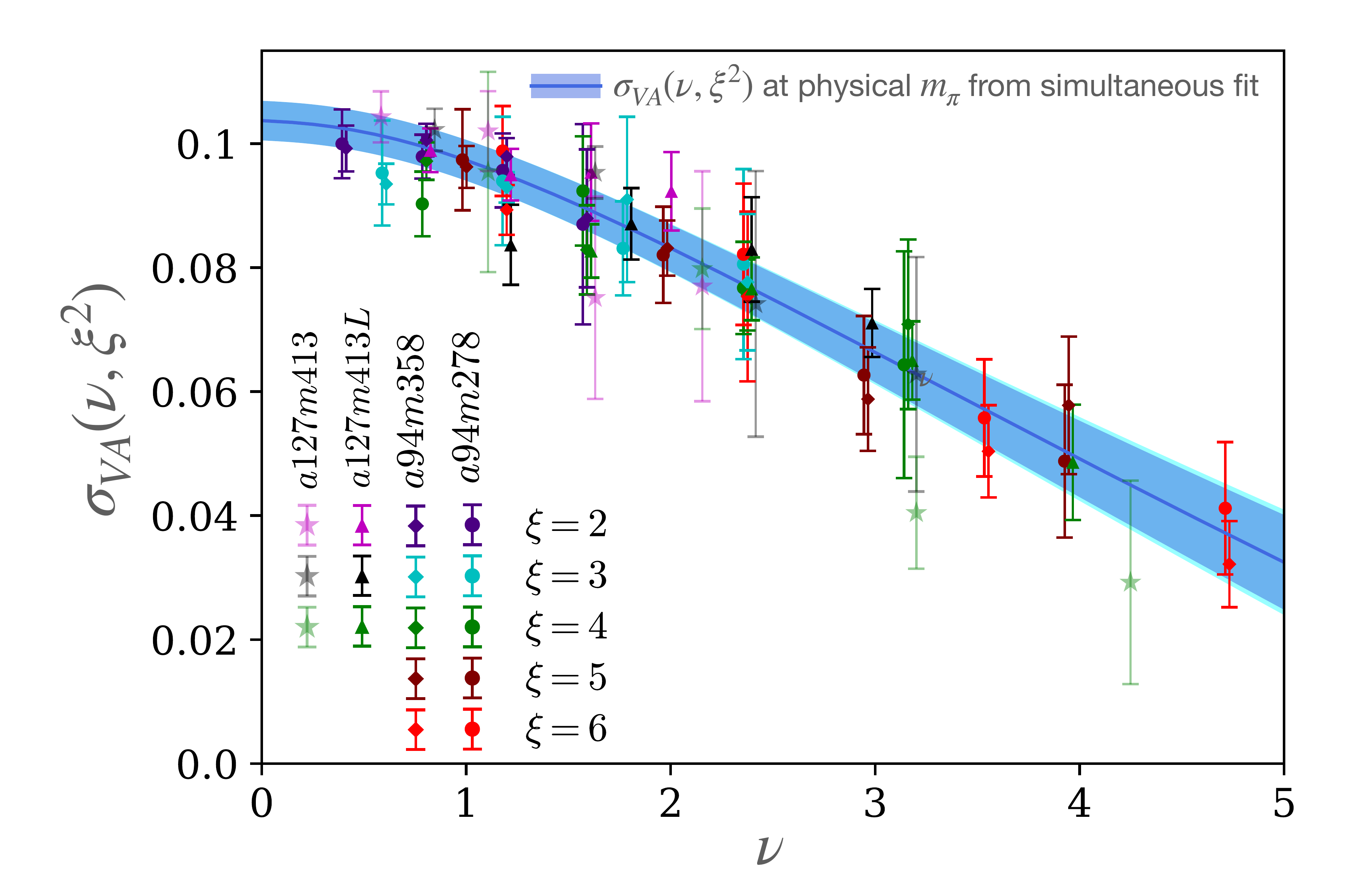} & \hspace*{2em} &
\includegraphics[clip, width=0.46\textwidth]{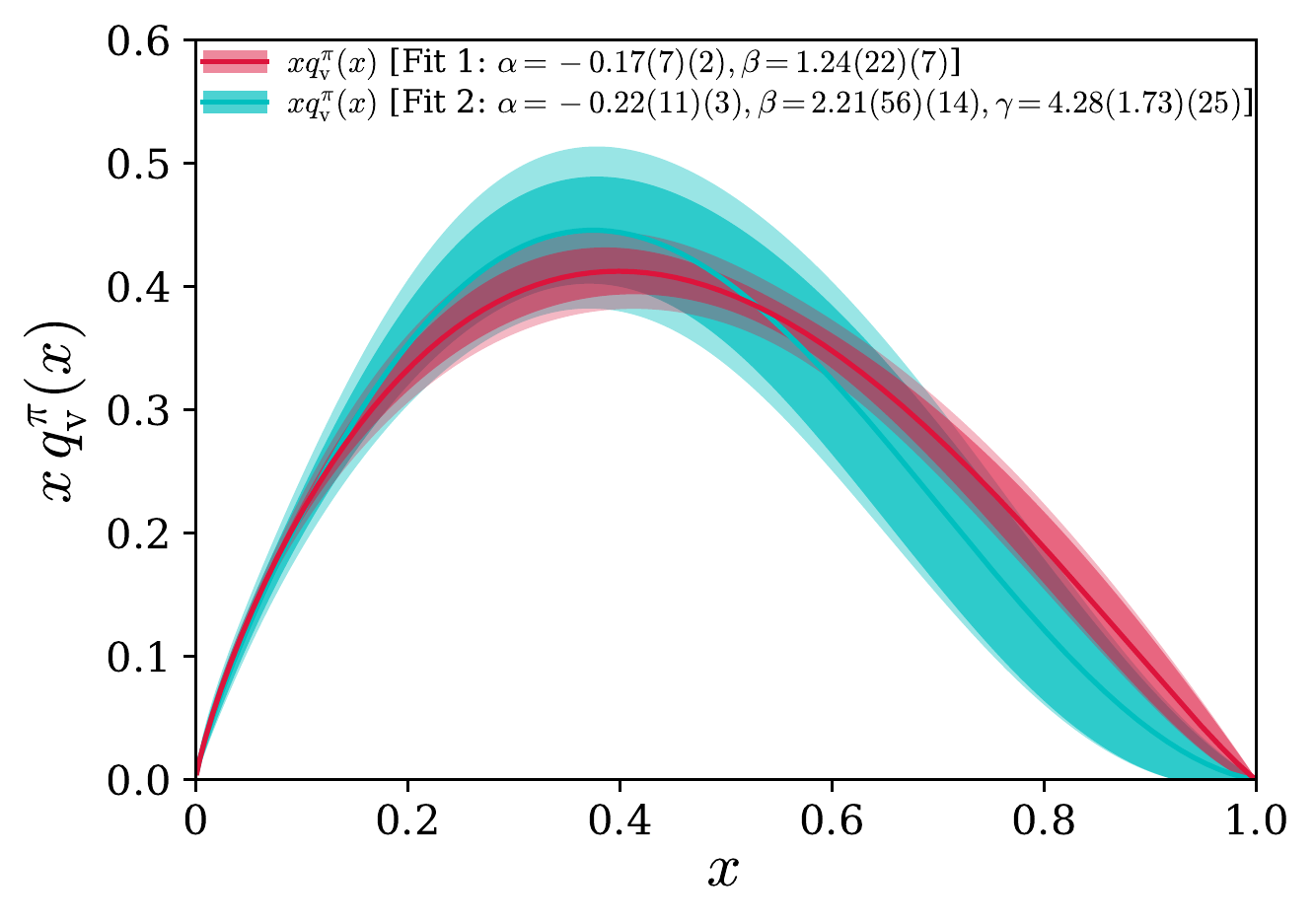}
\end{tabular}
    \caption{\label{fig:GLCS}
    \emph{Left panel}\,--\,{\sf A}.
     Matrix element results obtained from an antisymmetric combination of V and A currents on four different lattice ensembles as a function of the Ioffe time $\nu = p_z \xi$, where $\xi$ is the separation between the currents along a spatial axis \cite{Sufian:2020vzb}.  The filled area denotes a fit to the results based on Eq.\,\eqref{eq:glcs} and using the three-parameter fitting form in Eq.\,\eqref{eq:fit_jam}.
   \emph{Right panel}\,--\,{\sf B}.
   Two- and three-parameter fits to the results in Panel\,{\sf A} using the fitting form in Eq.\,\eqref{eq:fit_jam}.   Figures adapted from Ref.~\cite{Sufian:2020vzb}
    %\label{fig:GLCS} ... {fig:fit_glcs
}
\end{figure}

The valence-quark DF of the pion was discussed within the pseudo-PDF framework in Ref.\,\cite{Joo:2019bzr}, producing the moments listed in Eq.\,\eqref{momentslQCD}; and recently in a high-precision calculation using both the RI-mom renormalisation scheme characteristic of the LaMET approach and the ratio renormalisation scheme adopted for the pseudo-PDF approach \cite{Gao:2020ito}.
The GLCS framework is computationally more demanding.  In fact, the only computations performed so far have been for the pion, rather than for the nucleon; but the use of short-distance-separated gauge-invariant operators requires that the only lattice renormalisation is that of the lattice-discretised currents.  The first computation, using a vector/axial-vector combination of currents, contained only a tree-level determination of the kernel in Eq.\,\eqref{eq:glcs}, precluding the association of a scale with the resulting DF \cite{Sufian:2019bol}.  Low-order moments and the pointwise behaviour of ${\mathpzc u}^\pi(x)$ as determined in this study were discussed in Secs.\,\ref{SecpiDF2}, \ref{SecpiDF2}.  The analysis has since been extended to include the short-distance kernel computed at one loop, as well as control over the systematic uncertainties arising from unphysical quark masses, finite volume, and discretisation \cite{Sufian:2020vzb}.  The GLCS matrix element for the antisymmetric VA current computation is shown in Fig.\,\ref{fig:GLCS}\,A.

The journey from the Ioffe-time distributions shown in Fig.\,\ref{fig:GLCS} to the parton DFs requires more than the computation of the perturbative kernel.  Each of these frameworks involves a quantity computed on the lattice, the left-hand sides of Eqs.\,\eqref{eq:lamet}, \eqref{eq:pseudo} and \eqref{eq:glcs}, that is expressed as a convolution over the desired DF and that kernel, together with corrections.  However, the lattice points that give rise to the left-hand sides are incomplete and, moreover, quite sparse.  The situation mirrors that encountered in the extraction of DFs from experimental cross sections, as are the methods adopted; and there have been recent studies providing a comparison of the different methods aimed at the most ``model-independent" extraction \cite{Karpie:2019eiq, DelDebbio:2020rgv}.  The most straightforward, and widely adopted, means of providing that additional information is through the use of an assumed parametrisation, such as
\begin{equation}
    {\mathpzc q}(x) = \frac{x^{\alpha_\Pi} (1-x)^{\beta_\Pi} ( 1 + \gamma x)}{B(\alpha_\Pi + 1, \beta_\Pi + 1) + \gamma B(\alpha_\Pi + 2, \beta_\Pi + 1)},\label{eq:fit_jam}
\end{equation}
where the denominator ensures the correct normalization of the DF.  Two- and three-parameter fits for the current-current GLCS results in Fig.\,\ref{fig:GLCS}\,A are shown in Fig.\,\ref{fig:GLCS}\,B, matched to a scale $\zeta_2 = 2~{\rm GeV}$. The inflexibility of the simple two-parameter fit, highlighted in Sec.\,\ref{SecpiDF2}, is evident here: the additional freedom provided by $\gamma$ readily enables a softer DF to be extracted.  However, an important feature of the GLCS calculation is that the position-space NLO kernel entering into Eq.\,\eqref{eq:glcs} is well-controlled, with no large threshold logarithms, thereby offering the prospect for a better determination of the exponent $\beta_\Pi$ with the availability of data over a larger range of Ioffe time.

%%%\begin{figure}[t]
%%    \centering
%%    \includegraphics[width=3in]{ITD_sys.pdf}
%%    \caption{Matrix element corresponding to an antisymmetric combination of V and A currents on four different lattice ensembles as a function of the Ioffe time $\omega = p_z \xi$, where $\xi$ is the separation between the currents along a spatial axis \cite{Sufian:2020vzb}.  The filled area denotes a fit to the results based on Eq.\,\eqref{eq:glcs} and using the three-parameter fitting form in Eq.\,\eqref{eq:fit_jam}.
%%    \label{fig:GLCS}}
%%\end{figure}

%%\begin{figure}[t]
%%    \centering
%%    \includegraphics[width=3in]{XlatxPDF.pdf}
%%    \caption{Two- and three-parameter fits to the functional form of eqn.~\ref{eq:fit_jam}}
%%    \label{fig:fit_glcs}
%%\end{figure}

A method that obviates the need to address the inverse problem is provided by computation of the higher moments of the DF directly from the reduced matrix element in Eq.\,\eqref{eq:reduced} \cite{Karpie:2018zaz}.  Specifically, one can write
\begin{equation}
    {\cal M}(\nu,z^2) = \sum_{n=0}^{\infty} i^n \frac{\nu^2}{n\!} b_n(z^2)\,,
    \quad b_n(z^2) = \int_0^1 dx \, x^n {\cal P}(x,z^2)\,,
\end{equation}
%%where
%%\begin{equation}
%%   b_n(z^2) = \int_0^1 dx \, x^n {\cal P}(x,z^2)
%%\end{equation}
where $b_n(z^2)$ is the $n^{\rm th}$ moment of the pseudo-PDF ${\cal P}(x,z^2)$.  The Mellin moments $a_n(\zeta^2)$ of the DF are then given by
\begin{equation}
 b_n(z^2) = K_n(\zeta^2 z^2) a_n(\zeta^2),
\end{equation}
where the $K_n(\zeta^2 z^2)$ are the calculable Wilson coefficients.  With the quite limited range of Ioffe time available in current calculations, the ability to precisely constrain the large-$x$ exponent is limited, reflected both in the difference between the exponents for the two- and three-parameter fits in Fig.\,\ref{fig:GLCS}\,B, and in the relatively large errors on those exponents.

\begin{figure}[t]
    \includegraphics[clip, width=0.46\textwidth]{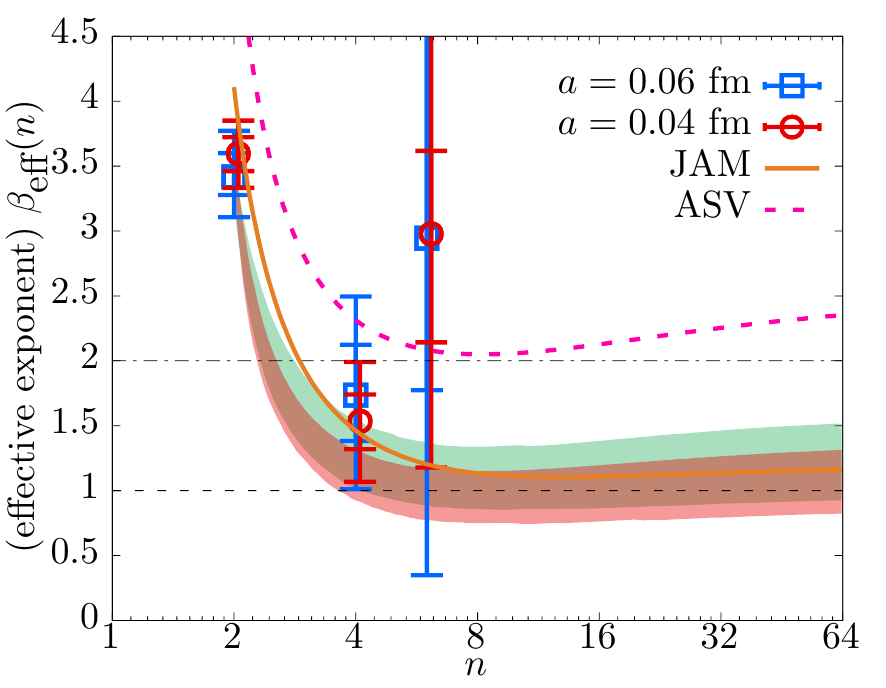}
    \caption{Effective exponent characterising the large-$x$ behavior of the pion valence-quark DF computed at two lattice spacings in Ref.\,\cite{Gao:2020ito} using Eq.\,\eqref{betainfinity}.  Comparisons were provided therein with the phenomenological fits in Ref.\,\cite{Barry:2018ort} (NLLs omitted) and Ref.\,\cite{Aicher:2010cb} (NLLs included).  Figure adapted from Ref.~\cite{Gao:2020ito}.
    \label{fig:moment}}
\end{figure}

As discussed in Sec.\,\ref{SecpiDF2}, particularly in connection with Fig.\,\ref{Figmoments}, calculation of the moments affords a means of obtaining a model-independent value for an effective $\beta_\Pi$ from the behavior of large-$n$ moments through Eq.\,\eqref{betainfinity}, denoted $\beta_{\rm eff}$  in \cite{Gao:2020ito}.  The original analysis is depicted in Fig.\,\ref{fig:moment}, which also includes results obtained using the pion valence-quark DF extracted in Ref.\,\cite{Aicher:2010cb}.

In concluding this subsection, it is worth noting an important feature of the lattice gauge formalism, \emph{i.e}.\ the ability to vary the underlying theory and its parameters as a means of revealing the emergence of the important physics.  A recent example is a study of the pion's valence quark DF in $2+1$-dimensional two-colour QCD, where varying the number of quark flavours, $N$, allows an investigation of how meson structure evolves as the theory changes from being one with strongly broken scale invariance at $N=0$ to a conformal theory as $N$ is increased \cite{Karthik:2021qwz}.
%%\begin{equation}
%%    \beta_{\rm eff}(n) \equiv -1 + \frac{\langle x^{n-2}\rangle - \langle x^{n+2} \rangle}{\langle x^2\rangle} \frac{n}{4},\label{eq:large_n}
%%\end{equation}
%%shown in Figure~\ref{fig:moment}.  (Additional details and analysis are presented in connection with Fig.\,\ref{Figmoments}.)

\subsection{Quark Distribution Amplitude}
\label{SeclQCDDA}
The quark DA is likewise described by the matrix elements of operators separated along the light cone and therefore subject to computational restrictions analogous to those which arise in the case of the DFs.  They are, however, computationally less demanding, requiring only the computation of a two-point correlation function, in contrast to the three-point correlation functions needed to extract a DF.  For the case of a positively charged pion, the leading-twist DA is defined via
\begin{equation}
    \langle 0 \mid \bar{d}(z_2 n)\! \not \! n W(z_2 n, z_1 n)  \gamma_5 u(z_1 n) \mid \pi^+(p)\rangle = i f_\pi (p \cdot n) \int_0^1 dx \, e^{-i(z_1x + z_2(1-x))p \cdot n} \varphi_\pi(x;\zeta),\label{eq:qda}
\end{equation}
where $p$ is the four-momentum of the on-shell pion and W is a Wilson line introduced to ensure gauge invariance.  The overall renormalisation of the wave function is simply given by the matrix element of a local operator
\begin{equation}
\langle 0 \mid \bar{d} \gamma_0 \gamma_5 u \mid \pi^+ (p)\rangle = i f_\pi p_0\,;
\end{equation}
and the determination of the pion decay constant is a benchmark computation in lQCD \cite{Aoki:2019cca}.

In analogy with the Mellin moments of the DFs, one can define the moments of the DA through
\begin{equation}
    \langle (\xi= 2x -1)^n \rangle = \int_0^1 (2x-1)^n \varphi(x;\zeta)\,, \label{eq:xi_mom}
\end{equation}
where $\xi $ can be interpreted as the difference in the momentum fractions carried by the quark $x$ and antiquark $1-x$.  For the case of the pion, only the even moments are non-zero, through charge-conjugation symmetry.  It is these moments that can be related to the pion-to-vacuum matrix elements of the local, twist-two operators
\begin{equation}
    {\cal O}_5^{\mu_1,\dots,\mu_n} = i^{n-1} \bar{d} \gamma_5 \gamma^{\{ \mu_1}\dots \gamma^{\mu_n\}} u.
\end{equation}

The computation of these matrix elements presents the same challenges as that of DF moments; notably, the breaking of $O(4)$ rotational symmetry on the lattice restricts computations to only the lowest few moments.  In the pion case, for which all odd $\xi$-moments vanish, this means that practical computations are limited to only a single moment, \emph{viz}.\ $n=2$.
The computation of this second moment constituted one of the first lattice calculations of hadronic matrix elements \cite{Kronfeld:1984zv, Gottlieb:1986ie}, and there has since been a progression of increasingly precise computations \cite{Martinelli:1987si, Daniel:1990ah, Bali:2019dqc}.  A recent calculation of the second moment using an auxiliary heavy quark field offers a prospect for the calculation of higher moments \cite{Detmold:2020lev}.

The distribution amplitude can also be expressed as
\begin{equation}
    \varphi(x;\zeta) = \varphi_{\rm as}(x) \bigg[ 1 + \sum_{n=1}^\infty a_n(\zeta) C_n^{3/2}(2x-1) \bigg] \,, \label{eq:da_defn}
\end{equation}
where $\varphi_{\rm as}$ is the asymptotic profile, introduced in Eq.\,\eqref{phias}, $C_n^{3/2}$ are Gegenbauer polynomials of degree $3/2$, and the coefficients $\{a_n\}$ are the DA's Gegenbauer moments:
\begin{equation}
    a_n(\zeta) = \frac{2(2n+3)}{3(n+1)(n+2)} \int_0^1 dx \, C^{3/2}_n (2x-1)\,\varphi(x; \zeta)\,,
\label{eq:da_moments}
\end{equation}
which are straightforwardly related to the moments in Eq.\,\eqref{eq:xi_mom}.

%%% <<< --- problem figures
\begin{figure}[t]
\includegraphics[clip, width=0.98\textwidth]{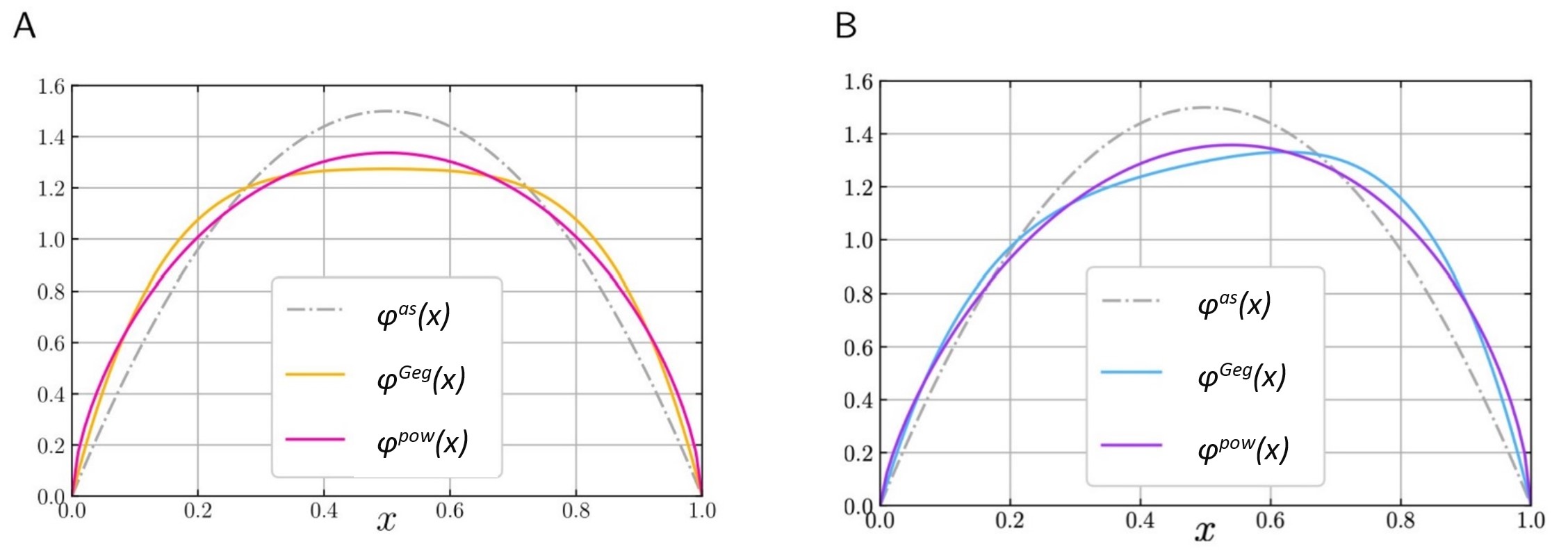}
%%
%%\hspace*{-1ex}\begin{tabular}{lcl}
%%{\sf A} &\hspace*{2em} & {\sf B} \\[0.5ex]
%%\includegraphics[clip, width=0.43\textwidth]{fig_DAs_Ptmp.pdf} & \hspace*{2em} &
%\includegraphics[clip, width=0.43\textwidth]{fig_DAs_Ktmp.pdf}
%\end{tabular}
%
%%      \includegraphics[width=3in]{fig_DAs_P.pdf}\hspace{1cm}\includegraphics[width=3in]{fig_DAs_K.pdf}
%
    \caption{\label{fig:da_P_K}
    Lattice QCD results for DAs of pion, \emph{left panel}\,--\,{\sf A}, and kaon, \emph{right panel}\,--\,{\sf B} inferred using the functional forms in Eqs.\,\eqref{eq:gegen_paramA}.  For comparison, $\varphi_{\rm as}$ is drawn as the dot-dashed grey curve in each panel.  (Figures adapted from Ref.\,\cite{Bali:2019dqc}.)}
\end{figure}

The values of $\langle \xi^2\rangle$, and of the lowest-two Gegenbauer moments, $a_1$ and $a_2$, are a measure of the degree of dissimilarity between the DA, and its asymptotic profile $\varphi_{\rm as}(x)$.
However, as seen with DFs, the finite-scale DA, $\varphi(x, \zeta)$, contains additional physical information, which can be expressed through either a simple truncation of the expansion in Eq.\,\eqref{eq:da_defn} or a parametrisation \cite{Chang:2013pq, Segovia:2013eca}.  Such approaches have been adopted in Ref.\,\cite{Bali:2019dqc}, which used
\begin{subequations}
\label{eq:gegen_paramA}
\begin{eqnarray}
\varphi^{\rm Geg}(x) & = & 6x(1-x) (1 + a_1 C_1^{3/2} (\xi) + a_2 C_2^{3/2} (\xi) )\label{eq:gegen_param}\\
\varphi^{\rm pow}(x) & = & \frac{\Gamma(2 + \alpha^+ + \alpha^-)}{\Gamma(1+\alpha^+)\Gamma(1 + \alpha^-)} x^{\alpha^+}(1-x)^{\alpha^-} \,. \label{eq:power_param}
\end{eqnarray}
\end{subequations}
For the pion, $a_1$ vanishes and $\alpha^+ = \alpha^-$.
Fig.\,\ref{fig:da_P_K} shows the pion and kaon DAs computed using this approach in Ref.\,\cite{Bali:2019dqc}.
%% --- not correct ... eq:power_param is EXACTLY the form used in Segovia:2013eca ... see also figures above.
%%Note that an alternative approach, in which the DA is expanded not in terms of the Gegenbauer polynomials of degree $3/2$, but rather of order $\delta$, where the $\delta$ itself is determined through the lattice computation \cite{Segovia:2013eca} yields a broader distribution.
The results agree quantitatively with those presented in Figs.\,\ref{FigphiDB}, \ref{FigNewKPDAForm}.

\begin{figure}[t]
\includegraphics[clip, width=0.98\textwidth]{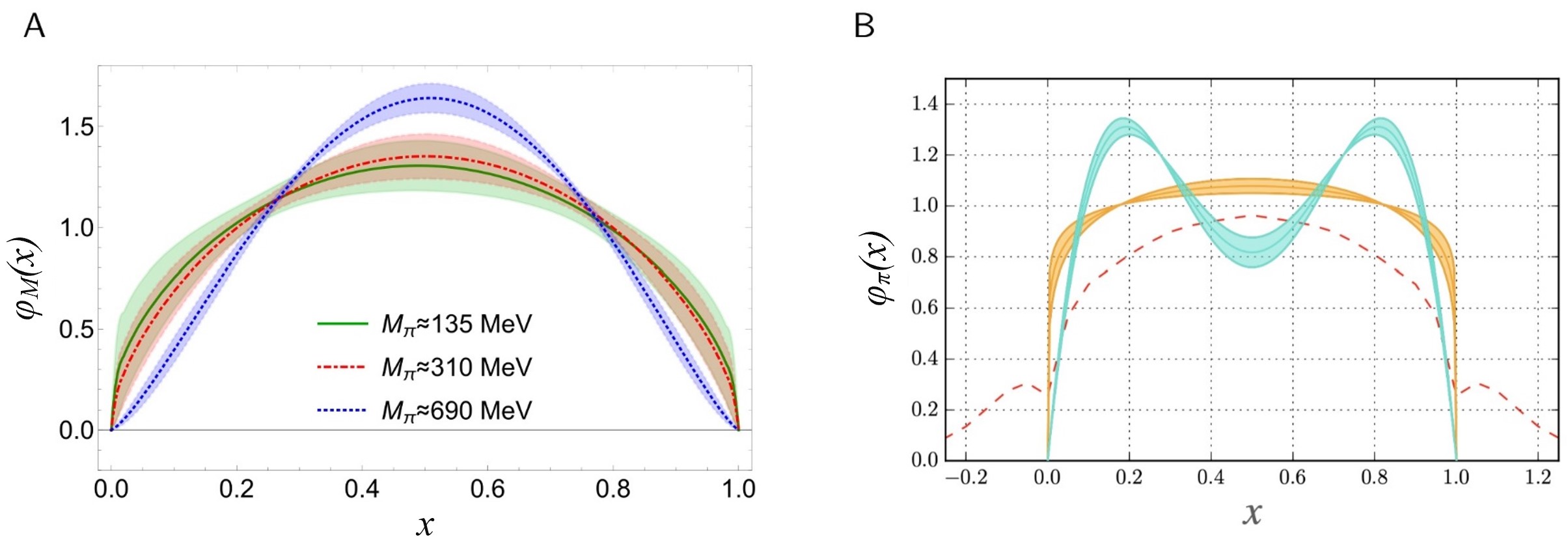}
%%\hspace*{-1ex}\begin{tabular}{lcl}
%%{\sf A} &\hspace*{2em} & {\sf B} \\[0.5ex]
%%\includegraphics[clip, width=0.46\textwidth]{fit_all_one_term_unnorm.pdf} & \hspace*{2em} &
%%\includegraphics[width=0.46\textwidth]{DA_comparisonMOD.pdf}
%\includegraphics[width=0.46\textwidth]{DA_comparison.pdf}
%%\end{tabular}
%
%%    \centering
%%    \includegraphics[width=3in]{fit_all_one_term_unnorm.pdf}
%%    \includegraphics[width=3in]{DA_comparison.jpg}
%
   \caption{
    \emph{Left panel}\,--\,{\sf A}.  Pion DA obtained in the LaMET approach for pion momentum $P_z = 1.73\,{\rm GeV}$ as the pion mass is decreased, using a fit to the form in Eq.\,\eqref{eq:power_param} \cite{Zhang:2020gaj} (Figure adapted from Ref.~\cite{Zhang:2020gaj}).  As described in connection with Fig.\,\ref{FigDACriticalMass}\,B, these results agree quantitatively with continuum calculations.
    \emph{Right panel}\,--\,{\sf B}.
    $x$-dependent pion DA computed using the two-current-correlator approach, with the turquoise (bimodal) and yellow (concave) bands denoting fits to the forms of Eq.\,\eqref{eq:gegen_param} and \eqref{eq:power_param}, respectively \cite{Bali:2018spj}.
     The dashed red curve is the LaMET result from Ref.\,\cite{Chen:2017gck}.  
    (Figure adapted from Ref.\,\cite{Bali:2018spj}).
    \label{fig:x_dep_DA}}
\end{figure}

The direct calculation of the $x$ dependence of the pion and kaon DAs follows the methods described for the DFs.  One of the principal challenges in these computations, namely the renormalisation of the operators, is precisely that described for the DFs.
There have been several calculations within the LaMET framework \cite{Zhang:2017bzy, Zhang:2020gaj}, as well as a computation wherein the vacuum-to-pion matrix element of two-current correlators is computed \cite{Bali:2017gfr, Bali:2018spj}.  As before, progress from the matrix elements computed on the lattice, in the case of the two-current approach expressed as functions of Ioffe-time, to the DA requires additional input, such as an imposed parametrisation of the functional form; and, in the case of Ref.\,\cite{Zhang:2020gaj}, through the adoption of machine-learning methods.  The resulting form for the pion DA using the LaMET and two-current approaches are shown in Fig.\,\ref{fig:x_dep_DA}.  In general, in contemporary lattice computations, the limited range of accessible Ioffe times means that strong constraints can only be placed on the second DA moment in the case of the pion, or the lowest two moments of the kaon DA.

\subsection{Three-Dimensional Imaging of Mesons}
\label{Sec3DlQCD}
%
%Three-dimensional imaging of hadrons can be expressed through projections of the five-dimensional Wigner function in momentum space, the so-called transverse-momentum-dependent distributions (TMDs)\cite{Collins:1992kk}, or in impact-parameter space through the Generalized Parton Distributions (GPDs)\cite{Mueller:1998fv,Ji:1996nm,Radyushkin:1997ki}.
Returning to the themes of Sec.\,\ref{3DNG}, the single leading-twist unpolarised GPD of a pseudoscalar meson can be expressed in the following form, amenable to lQCD analysis:
\begin{equation}
H(x,\xi,t,\zeta) = \int \frac{ d \eta^- \, P^+}{2\pi} e^{-i x \eta^- P^+}
\langle \pi(P+\Delta/2) \mid \bar{q}(\frac{\eta^-}{2}) \gamma^+ W(\frac{\eta^-}{2},-\frac{\eta^-}{2}) q(\eta^-/2) \mid \pi (P-\Delta/2)\label{eq:gpd_defn}
\end{equation}
%where $t = \Delta^2$ is the momentum transfer, $\xi = -\frac{\Delta^+}{2 P^+}$ is the skewness,
where, as usual, W is a Wilson line introduced to ensure gauge invariance.
%, and $\mu$ is the scale; here $x \in [-1,1]$.
As can be seen by inspection, this is a generalisation of Eq.\,\eqref{eq:pdf_defn} to the off-forward ($\Delta\neq 0$) case, and its lQCD computation encounters the same challenges noted earlier.  Consequently, the first lattice studies focused on the Bjorken-$x$ moments of the GPDs, yielding generalised form factors (GFFs), which can be computed as the off-forward matrix elements of the leading-twist operators in Eq.\,\eqref{eq:local_op}.

\begin{figure}[t]
\hspace*{-1ex}\begin{tabular}{lcl}
{\sf A} &\hspace*{2em} & {\sf B} \\[0.5ex]
\includegraphics[clip, width=0.46\textwidth]{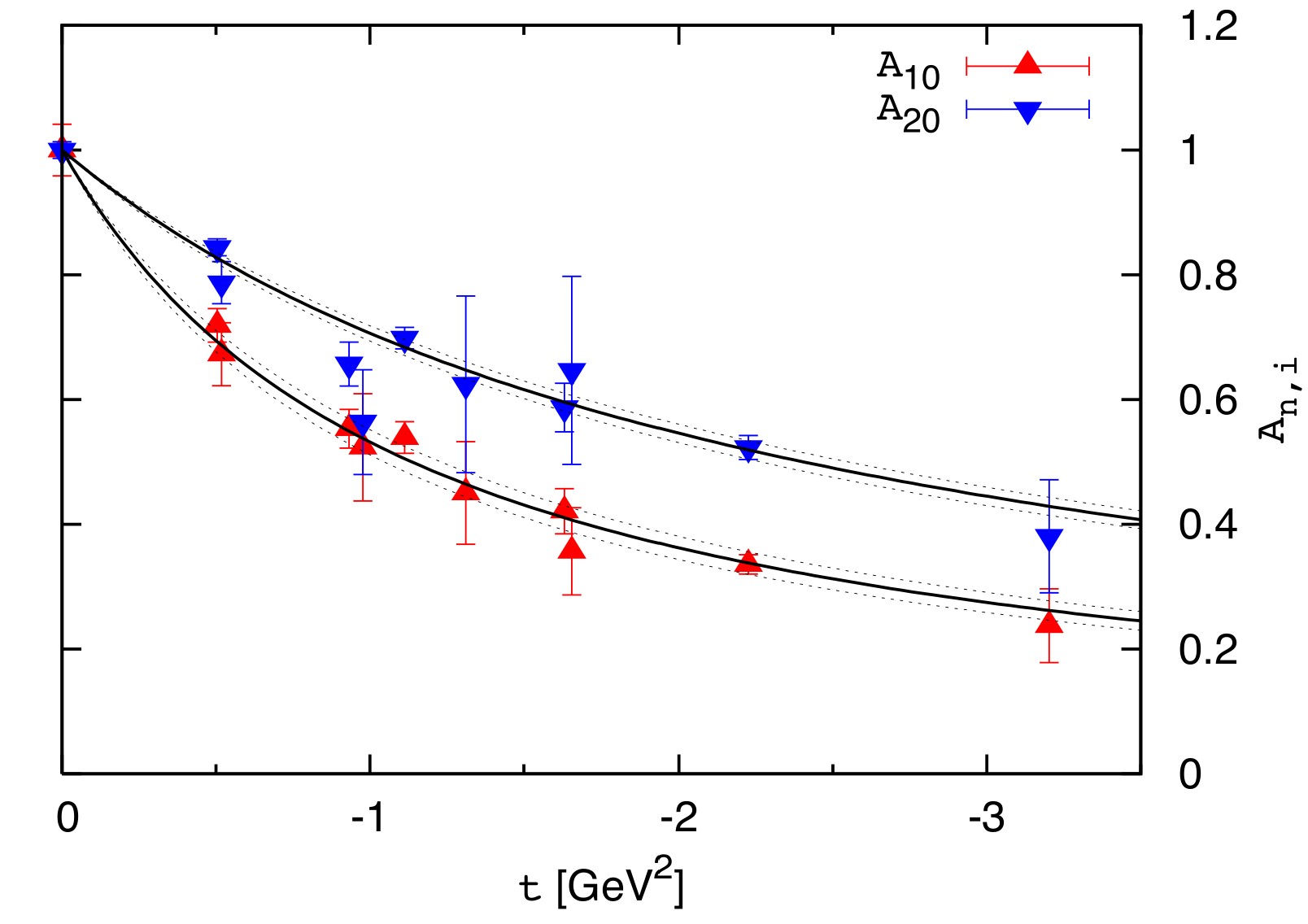} & \hspace*{2em} &
\includegraphics[clip, width=0.46\textwidth,height=12em]{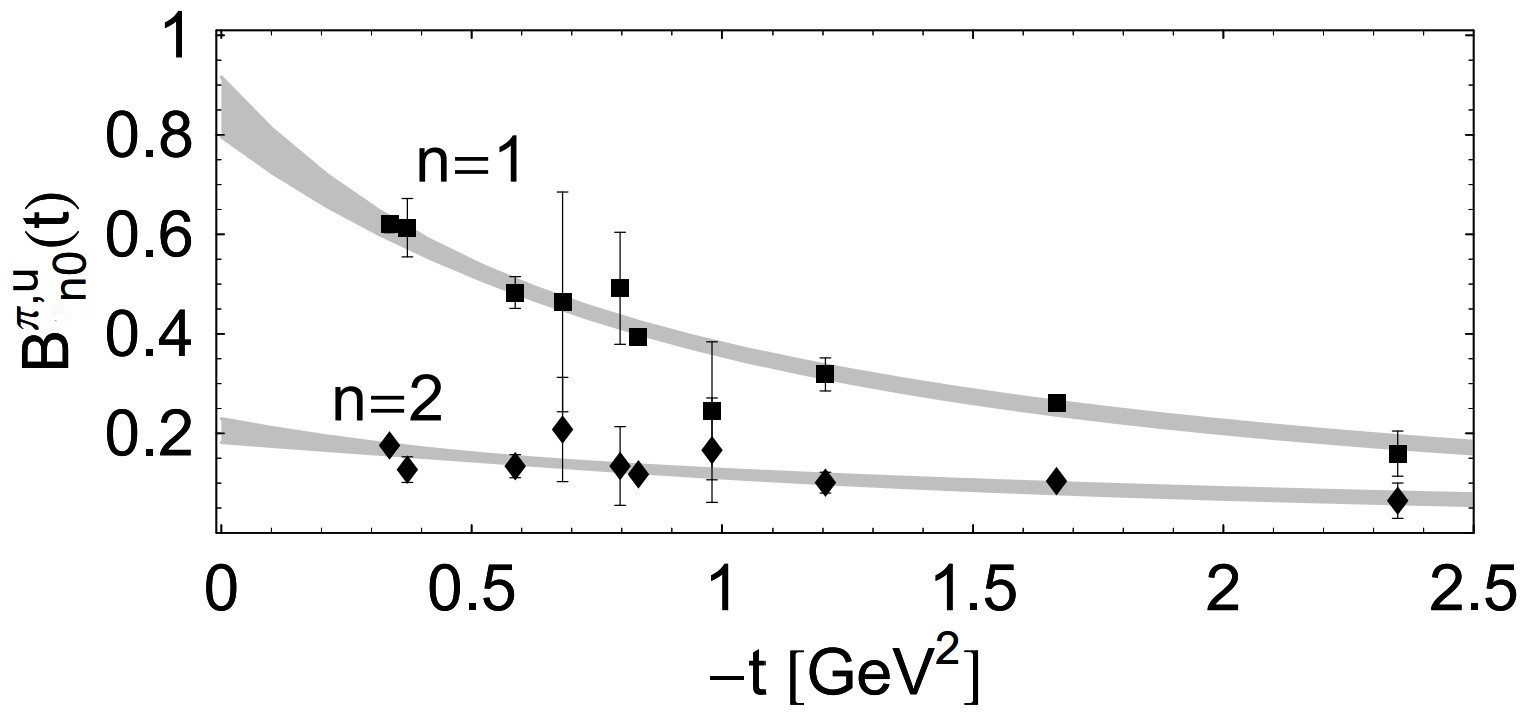}
\end{tabular}
%
%    \centering \vspace{-1.4in}\includegraphics[width=6.5in]{unpol_slope.pdf}\vspace{-1.3in}
    \caption{\label{fig:ns_gpd}
    \emph{Left panel}\,--\,{\sf A}. Pion's lowest two unpolarised GFFs $A_{10}$, $A_{20}$, normalised to unity at $t=0$, obtained at a pion mass $m_\pi \simeq 1.09\, {\rm GeV}$ in a theory with two dynamical quark flavours.  (Adapted from Ref.\,\cite{Brommel:2005ee}.)
    \emph{Right panel}\,--\,{\sf B}.  Lowest two moments of the pion's tensor GPD, Eq.\,\eqref{ETmomentsB}, obtained at a pion mass $m_\pi \simeq 0.6 {\rm GeV}$ in a theory with two dynamical quark flavours.  (Adapted from Ref.\,\cite{Brommel:2007xd}.)
    }
\end{figure}

Specialising to the insertion of the $u$ quark, one obtains \cite{Brommel:2005ee}
\begin{align}
 \langle & \pi^+(P+\Delta/2) \mid \bar{u}(0) \gamma^{\{ \mu} iD^{\mu_1} iD^{\mu_2}\dots iD^{\mu_n\}} u(0) \mid \pi+(P-\Delta/2) \rangle \nonumber\\
& = 2P^{\{ \mu} P^{\mu_1}\dots P^{\mu_n \}} A_{n+1,0}(\Delta^2) + 2 \sum_{i=1,{\rm odd}}^n \Delta^{\{ \mu}
\Delta^{\mu_1}\dots \Delta^{\mu_i} P^{\mu_{i+1}} \dots P^{\mu_n} A_{n+1,i+1}(\Delta^2)\,,\label{eq:pion_gff}
\end{align}
which for $n=0$ simply corresponds to the form factor $A_{1,0} (t) = F_\pi(t)$.  In contrast to the case of the nucleon, there has only been a single lQCD computation of pion flavour-non-singlet unpolarised GFFs  \cite{Brommel:2005ee, Brommel:2007zz}, with the results depicted in Fig.\,\ref{fig:ns_gpd}\,A.  This study was used to inform the analysis described in Sec.\,\ref{SecTTGPD}.

As discussed in Sec.\,\ref{SecTTGPD}, whilst there is no GPD associated with quarks polarised parallel to the pion's direction of motion, the tensor GPD in Eq.\,\eqref{EqGPDFH} provides access to information on transversely polarised quarks.  A lQCD study of this GPD is described in Ref.\,\cite{Brommel:2007xd}, which focused on the following matrix elements:
\begin{align}
 \langle  \pi^+(P+\Delta/2) \mid & \bar{u}(0) i \sigma^{\mu\nu} iD^{\mu_1} iD^{\mu_2}\dots iD^{\mu_{n-1}} u(0) \mid \pi^+(P-\Delta/2) \rangle  \nonumber\\
& = \frac{P^\mu \Delta^\nu - P^\nu \Delta^\mu}{m_\pi} \sum_{i=1,{\rm even}}^{n-1} \Delta^{\mu_1}
\dots \Delta^{\mu_i} P^{\mu_{i+1}} \dots P^{\mu_{n-1}} B^\pi_{ni}(\Delta^2).
\end{align}
(Symmetrisation is suppressed.)  This analysis produced the form factors shown in Fig.\,\ref{fig:ns_gpd}\,B, which were used to inform the analysis described in connection with Fig.\,\ref{FigTSpin}.

%%\begin{figure}[t]
%%    \includegraphics[width=3.5in]{BTn0_pion_b5p29k13590_v2.pdf}
%%    \caption{\label{fig:gpd_trans}
%%    The lowest two transverse polarized Generalized Form Factors of the pion GPD, obtained at a pion mass $m_\pi \simeq 0.6 {\rm GeV}$ in a theory with two dynamical quark flavor.  (Figure adapted from Ref.\,\cite{Brommel:2007xd}.)
%%    {\color{red} DGR: Would be good to put this next to unpolarized figure.  I cannot find any publication with both plots, which is odd}}
%%\end{figure}
%%Brommel:2007xd

\begin{figure}[t]
    \centering
    \includegraphics[width=0.55\linewidth]{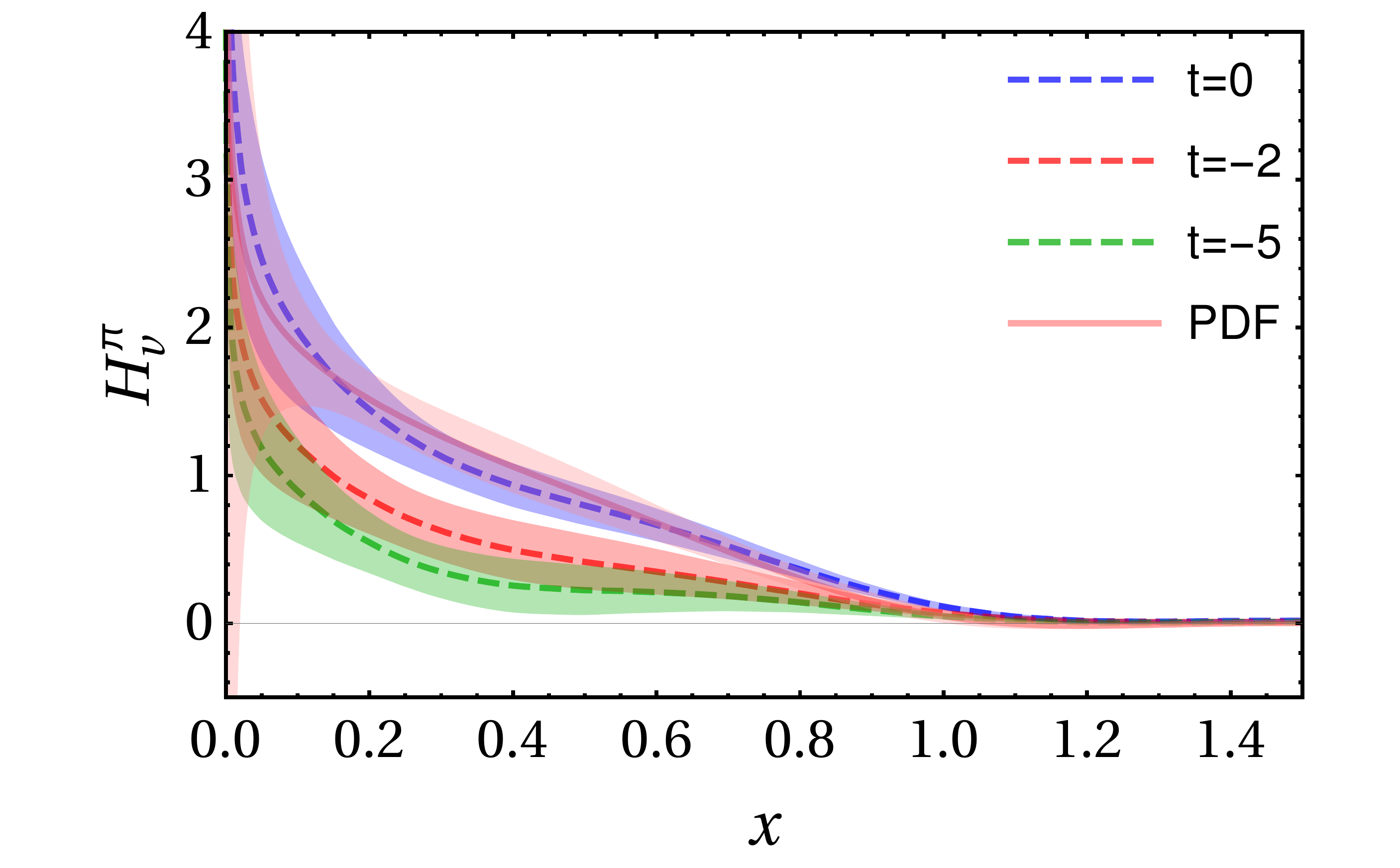}
    \caption{Zero-skewness pion GPD $H(x,0,t \times (2\pi/L), \zeta = 4\,{\rm GeV})$, obtained on a lattice with spatial extent $L=24$ and  spacing $a \simeq 0.12\,{\rm fm}$ \cite{Chen:2019lcm}.  The curve marked ``PDF'' denotes the result obtained in Ref.\,\cite{Chen:2018fwa}.  (Adapted from Ref.\,\cite{Chen:2019lcm}.)    \label{fig:gpd_lattice}}
\end{figure}

The matching of the Generalized quasi-Parton Distributions to the GPDs has been derived in detail \cite{Xiong:2015nua, Liu:2019urm}, but the only published computation of the pion GPD within the LaMET framework \cite{Chen:2019lcm}, shown in Fig.\,\ref{fig:gpd_lattice}, is restricted to zero skewness at a single lattice spacing and a pion mass of $310\,{\rm MeV}$. The matching coefficients have recently been computed in the pseudo-PDF framework \cite{Radyushkin:2019owq}, and the GLCS approach likewise encompasses the calculation of the GPDs.   Importantly, as the calculations mature, each of these approaches will enable the GPD to be charted both as a function of $x$, and for a set of discrete skewness values, $\xi$, and momentum transfers, $t$.  Finally, in concluding this subsection, it is worth noting that three-dimensional momentum-space imaging encoded within the TMDs has also been studied for the pion \cite{Engelhardt:2015xja}, in particular for the $T$-odd Boer-Mulders effect that, experimentally, is a manifestation of long-range initial- or final-state interactions in DY and semi-inclusive deep-inelastic scattering processes, respectively.

\subsection{Gluon and Flavour-Singlet Computations}
\label{SecGluelQCD}
The computations sketched above can be extended to the flavour-singlet sector; in particular, the gluonic contributions to hadron structure.  Beginning with the DFs, one begins by generalising Eq.\,\eqref{eq:pdf_defn} to the gluonic case:
\begin{equation}
    g_\pi(x) = \int \frac{d \eta^-}{\pi x} e^{-i x \eta^- P^+} \langle \pi(\vec{p}) \mid
    G_{\mu +}^a(\eta^-) W(\eta^-/2,-\eta^-/2)_{ab}G_{\mu +}^b(-\eta^-/2) \mid \pi(\vec{p}) \rangle\,,
\end{equation}
where $G^a_{\mu\nu}$ is the field-strength tensor in Eq.\,\eqref{gluonSI}.
The theoretical challenges of, and solution to, evaluating this on a Euclidean lattice follow the discussion of the valence distributions discussed above; specifically, the moments of the pion's unpolarised glue distribution can be expressed in terms of matrix elements of local, traceless interpolating operators:
\begin{equation}
    {\cal O}_g^{\mu_1 \dots \mu_n} = G_{a\alpha}^{\{ \mu_1} D^{\mu_2}\dots D^{\mu_{n-1}}  G_a^{\mu_n \} \alpha}\,.\label{eq:local_op_gluon}
\end{equation}
The lowest-order operator corresponds to the gluonic elements of the energy-momentum tensor
\begin{equation}
    T^{\mu\nu} = \frac{1}{4} \bar{\psi} \gamma^{\{\mu} D^{\nu\}} + G^{\mu \alpha}G^{\nu \alpha} - \frac{1}{4} G^2,\label{eq:lat_energy_mom}
\end{equation}
and the symmetric, traceless component gives rise to the momentum sum rule
\[
\sum_q \langle x \rangle_q + \langle x \rangle_g = 1.
\]

The computation of gluonic matrix elements, and of the disconnected quark contributions with which they mix, are far more computationally demanding than those corresponding to flavour-non-singlet quantities: the former are subject to considerable statistical noise, whilst the latter require the calculation of all-to-all propagators.  To overcome the statistical noise in such calculations, one must:
exploit the freedom afforded by different discretisations of the field-strength tensor;
smooth short-distance fluctuations by, \emph{e.g}.\ employing HYP- or Stout-smeared links \cite{Hasenfratz:2001hp, Morningstar:2003gk};
and use complete sampling of the lattice.

\begin{figure}[t]
\includegraphics[clip, width=1.0\textwidth]{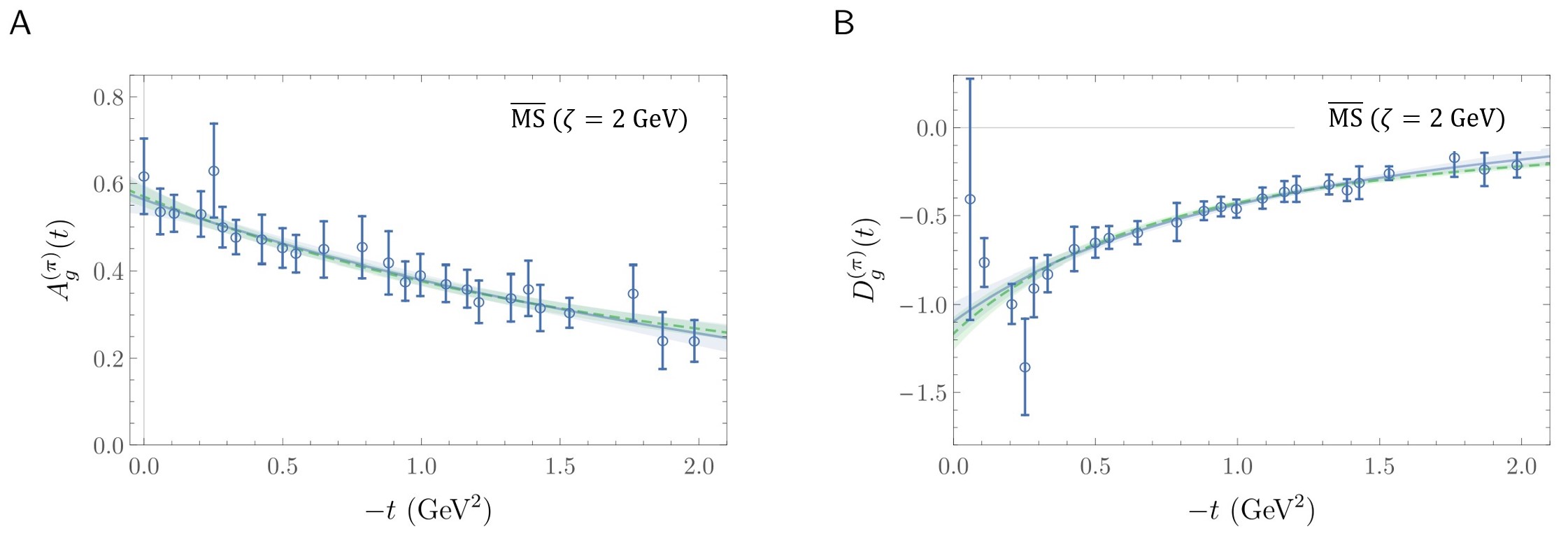}
    %%\centering
    %%\includegraphics[width=3.2in]{PIONAgRenorm.pdf}
    %%\includegraphics[width=3.2in]{PIONDgRenorm.pdf}
    %%
    \caption{Lattice QCD results for pion's glue GFFs defined in Eq.\,\eqref{EqgGFFs}, calculated at a pion mass of roughly $450\,{\rm MeV}$ and matched to a scale $\zeta = 2\,{\rm GeV}$.  Blue and green bands show z-expansion and dipole fits to the data.  Figure adapted from Ref.\,\cite{Shanahan:2018pib}, which contains additional details.
    \label{fig:gluon_gpd}}
\end{figure}

The first effort at exploring gluonic contributions to pion structure was a computation of the momentum fraction carried by glue \cite{Meyer:2007tm}, performed in quenched QCD at a pion mass of between 600 and 1100\,{\rm MeV}.  Whilst there have been numerous computations of the flavour-singlet contribution to the nucleon momentum fractions, there is only one evaluation for the pion in full QCD \cite{Shanahan:2018pib} at a pion mass of $450~{\rm MeV}$.  It reports a value commensurate with the quenched results at the lightest $600~{\rm MeV}$ pion mass -- see $A_g^{(\pi)}(0)$ in Fig.\,\ref{fig:gluon_gpd}\,A, which is discussed further below.  A study of the mass decomposition of mesons, using the overlap formalism for the fermion action, computed the quark contributions to the pion mass, and then used the mass sum rule to infer the gluonic contributions, including that arising from the trace anomaly \cite{Yang:2014xsa}.

The off-forward components of the lowest-order gluon operator
$\langle \pi(P_{+\Delta}) | G_{a\alpha}^{\{ \mu} G_a^{\nu  \alpha\}} |\pi(P_{-\Delta}) \rangle$,
$P_{\pm \Delta}=P \pm \Delta/2$, give rise to the in-pion gluon gravitational form factors through an expansion analogous to that in Eq.\,\eqref{eq:pion_gff}:
\begin{equation}
 \langle \pi(P + \Delta/2) | G_{a\alpha}^{\{ \mu} G_a^{\nu  \alpha\} } \mid \pi(P - \Delta/2) \rangle = 2 P^\mu P^\nu A_g(\Delta^2) + \frac{1}{2} \Delta^\mu \Delta^\nu D_g(\Delta^2)\,.
 \label{EqgGFFs}
\end{equation}
The first form factor, $A_g(\Delta^2)$, drawn in Fig.\,\ref{fig:gluon_gpd}\,A, relates to the momentum fraction carried by the gluons, discussed above; and the second, $D_g(\Delta^2)$, is drawn in Fig.\,\ref{fig:gluon_gpd}\,B \cite{Shanahan:2018pib}.  There is currently no lQCD calculation of the in-pion quark GFFs on the same ensembles.  Hence, the only comparison can be with the quark calculation described in Ref.\,\cite{Brommel:2007xd}, which was performed at a different pion mass and neglected disconnected contributions.  One observation is that the momentum fraction carried by the gluons is similar in the pion and in the nucleon at these heavy pion masses \cite{Shanahan:2018pib}.

\subsection{Era of Exascale Computing}
\label{SecExa}
A raft of key measures of meson structure, most notably the pion form factor and charge radius, and low-order moments of the DAs, are now available with controlled systematic errors and high statistical precision, including results with light-quark masses at their physical values.
Novel algorithmic developments, such as the use of so-called momentum smearing \cite{Bali:2016lva}, which enables states to be studied at high spatial momentum, and distillation \cite{Peardon:2009gh}, enabling the efficient application of the variational method that is essential at high momenta, allow increasingly refined computations of the form factors, DFs and GPDs.
The advent of leadership-class exascale computing \cite{Joo:2019byq} promises to enable these developments to be fully exploited through a programme of computations at a variety of lattice spacings and volumes, and with sufficient statistical precision to resolve not only flavour-non-singlet structure but also the contribution of gluons and sea quarks \cite{Detmold:2019ghl}.

The inclusion of lattice results in global DF fits \cite{Lin:2020rut}, either on the same footing as experimental data, as has been done for the case of the nucleon \cite{Bringewatt:2020ixn}, or as a Bayesian prior in the fit to experimental data, as was accomplished for the nucleon tensor charge \cite{Lin:2017stx}, is an opportunity that will develop as the precision of lattice calculations increases.  Furthermore, the calculation of the perturbative kernel at next-to-next-to-leading-order (NNLO) \cite{Li:2020xml} within the position-space formulation of Eq.\,\eqref{eq:glcs} affords the prospects of a lattice determination of the key measures of hadron structure at NNLO comparable to that attainable from global fitting.
The aim is to use both experiment and lattice results to provide more information about such measures of hadron structure than either can alone.  In connection with pion and kaon measurements via the Sullivan process, discussed in connection with Fig.\,\ref{FigSP}, lQCD computations may provide some additional support for the experimental analysis, potentially providing novel benchmarks to quantify the effects of off-shellness and/or kinematical extrapolations in $t$.

The discussion in this section has focused on the properties of the pion and kaon, particles stable under the strong interactions.  An exciting prospect for the future is found in recent developments that promise to provide a framework for the rigorous study of resonances and multi-hadron states; specifically, providing access to their internal structure \cite{Briceno:2015tza, Baroni:2018iau}.
%%% ---
%%% {\color{red}  Probably need to add more here, but it is largely "stream of consciousness" stuff}. 

%% file: S910_Experiment.tex
\section{Experiments Completed or In Train}
\label{sec:Experiments}

\subsection{Sullivan Process}
\label{sec:sullivan}
\begin{figure}[t]
\hspace*{-1ex}\begin{tabular}{lcl}
%{\sf A} &\hspace*{2em} & {\sf B} \\[-2ex]
\includegraphics[clip, width=0.4\textwidth]{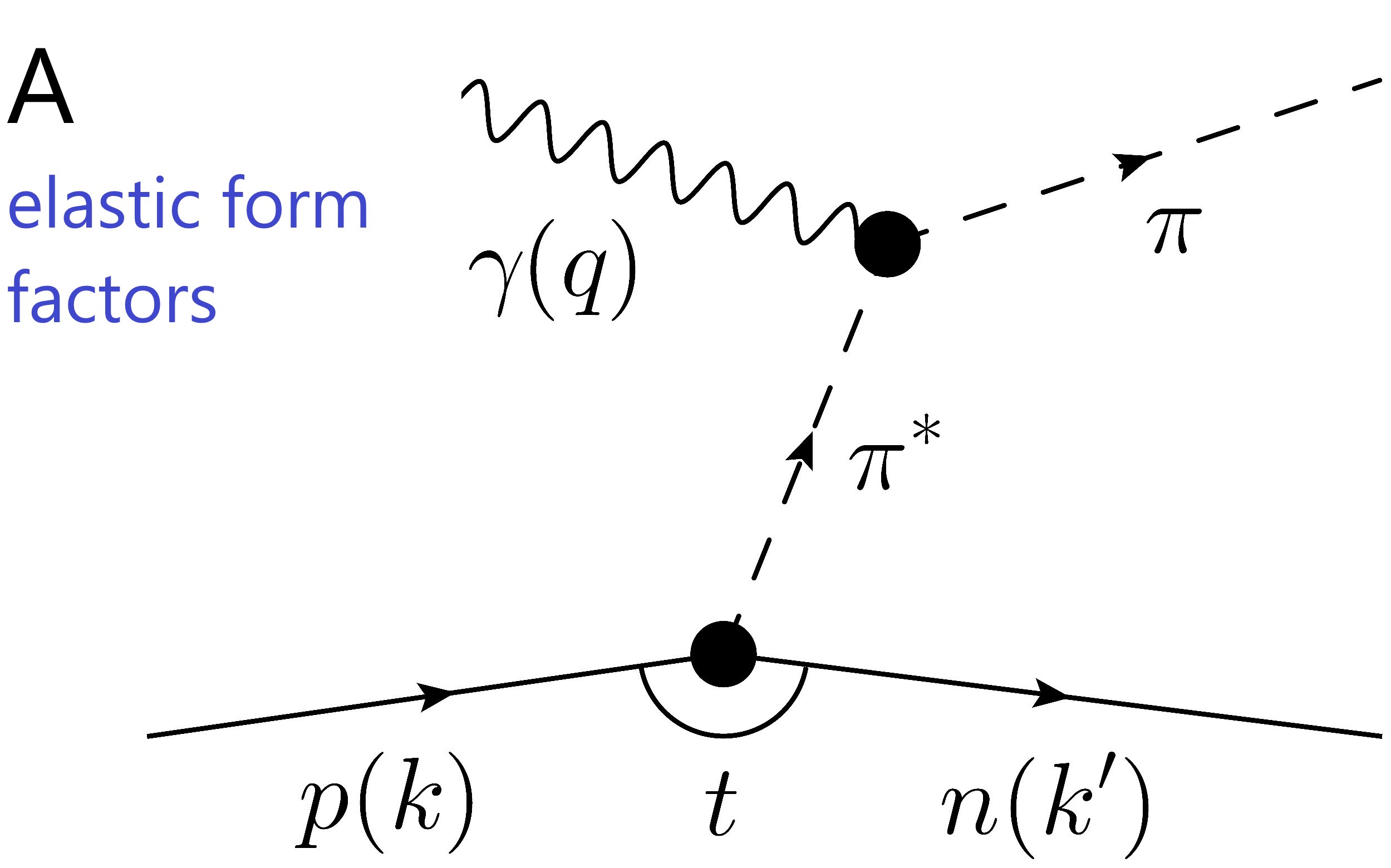} & \hspace*{2em} &
\includegraphics[clip, width=0.4\textwidth]{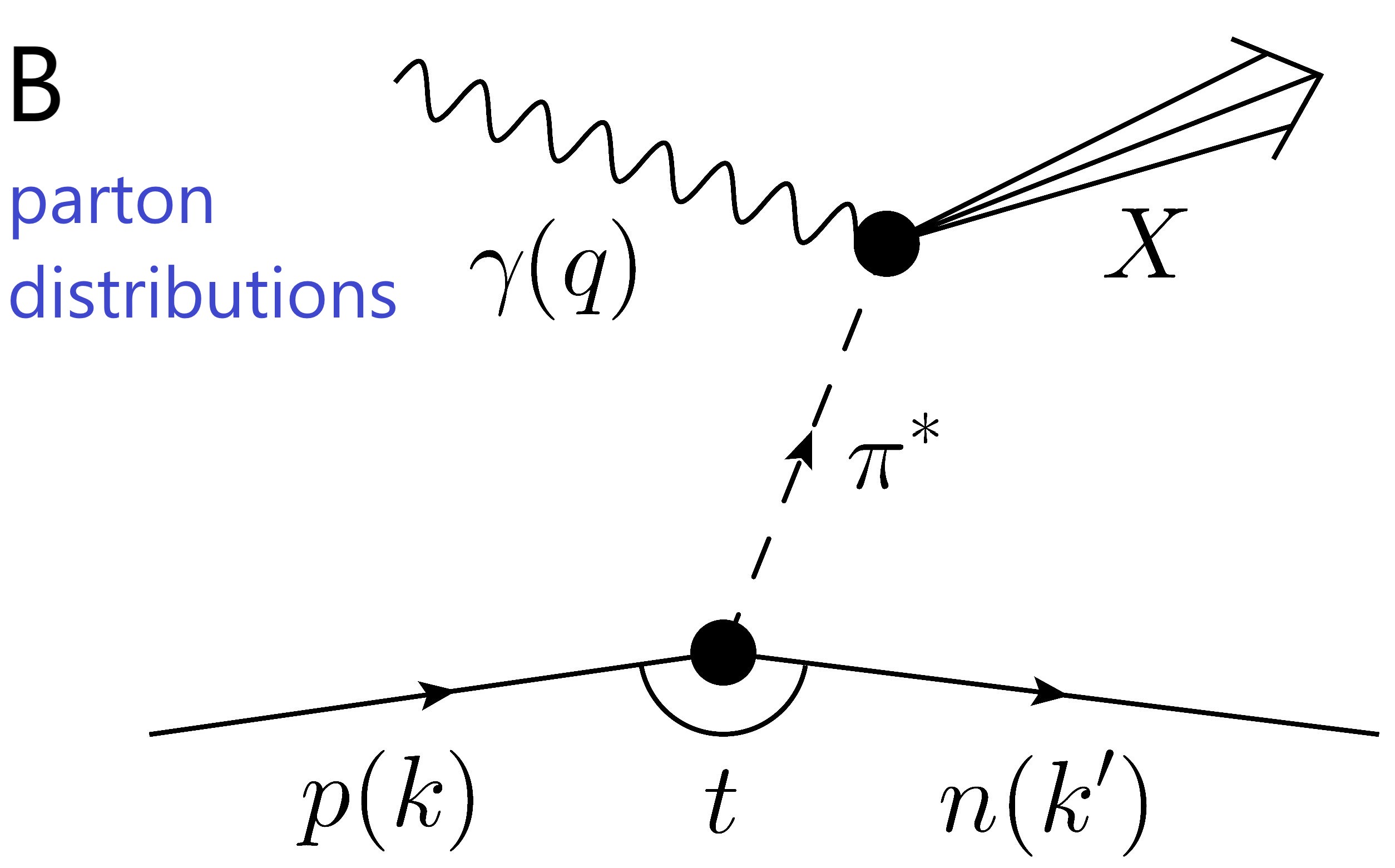}
\end{tabular}
\caption{\label{FigSP}
Sullivan processes.  These examples indicate how a nucleon's pion cloud may be used to provide access to the pion's (A) elastic form factor and (B) distribution functions.  The intermediate pion, $\pi^\ast(P)$, is off-shell, with $P^2= –t$.  Estimates suggest \cite{Qin:2017lcd} that such processes provide reliable access to a pion target on $-t\lesssim 0.6\,$GeV$^2$; and for the kaon, on $-t\lesssim 0.9\,$GeV$^2$.
}
\end{figure}

In specific kinematic regions, the observation of recoil nucleons (N) or hyperons (Y) in the semi-inclusive measurement $e p \to e^\prime (N\,{\rm or}\,Y) X$ can reveal features associated with correlated quark-antiquark pairs in the nucleon, referred to as the ``meson cloud'' or ``five-quark component'' of the nucleon.  At low values of $t$, the four-momentum transfer from the initial proton to the final nucleon or hyperon, the cross-section displays behaviour characteristic of meson pole dominance. Illustrated in Fig.\,\ref{FigSP}, the process in which the electron scatters off the meson cloud of a nucleon target is called the Sullivan process \cite{Sullivan:1971kd}.  For elastic scattering, this process carries information on the meson (pion or kaon) form factor, as discussed in Sec.\,\ref{sec:meson-form-factors-exp}.  For deep inelastic scattering, the typical interpretation is that the nucleon parton distributions contain a mesonic parton content. To access the pion or kaon partonic content via such a structure function measurement requires scattering from a meson target.

The Sullivan process can provide reliable access to a meson target in the space-like $t$ region, if the pole associated with the ground-state meson remains the dominant feature of the process and the structure of the related correlation evolves slowly and smoothly with virtuality. To check whether these conditions are satisfied empirically, one can take data covering a range in $t$, particularly low $|t|$, and compare with phenomenological and theoretical expectations.  A recent calculation \cite{Qin:2017lcd} explored the circumstances under which these conditions should be satisfied.  For the pion (kaon) Sullivan process, low $-t$ equates to $-t <$ 0.6 (0.9) GeV$^2$ to be able to cleanly extract pion (kaon) structure, and data over a range of $-t$ down to the lowest practically accessible are needed to verify pion (kaon) structure extraction.

\subsection{Pion and Kaon Form Factors}
\label{sec:meson-form-factors-exp}
At low values of $Q^2$, $F_{\pi}$ has been determined directly up to photon energies of $Q^2$=0.253 GeV$^2$ at Fermilab \cite{Dally:1982zk, Dally:1981ur} and at the CERN Super Proton Synchrotron (SPS) \cite{Amendolia:1984nz, Amendolia:1986wj} from the scattering of high-energy, charged pions by atomic electrons. These data were used to constrain the charge radius of the pion, with the result $r_\pi=0.657 \pm 0.012\,$fm \cite{Amendolia:1984nz}.  Owing to kinematic limitation in the energy of the pion beam and unfavourable momentum transfer, one must use other experimental methods to reach the higher $Q^2$ regime.  At higher values of $Q^2$, $F_{\pi}$ can be determined through the Sullivan process; specifically, the electroproduction reaction.
%% Dally:1982zk 0.663 \pm  0.023 fm

Electroproduction reactions are of general interest because they allow for measuring photoproduction amplitudes as functions of the photon mass.  The weakness of the electromagnetic interaction allows one to treat these reactions in the one-photon exchange approximation as virtual photoproduction by spacelike photons, $Q^2>$0, whose mass, energy, direction, and polarisation density are tagged by the scattered electron \cite{Hand:1963bb}.  The electroproduction reaction can be described in terms of form factors, which are generalisations of the form factors observed in elastic electron-hadron scattering or in terms of cross-sections that are extensions of the photoproduction cross-sections.
%In general, the virtual photon is polarized. There are two transverse polarization states and a third component, which can be taken as a scalar or as longitudinal with its only component along the direction of the virtual photon.

\begin{figure}[t]
\begin{center}
\includegraphics[width=0.55\textwidth]{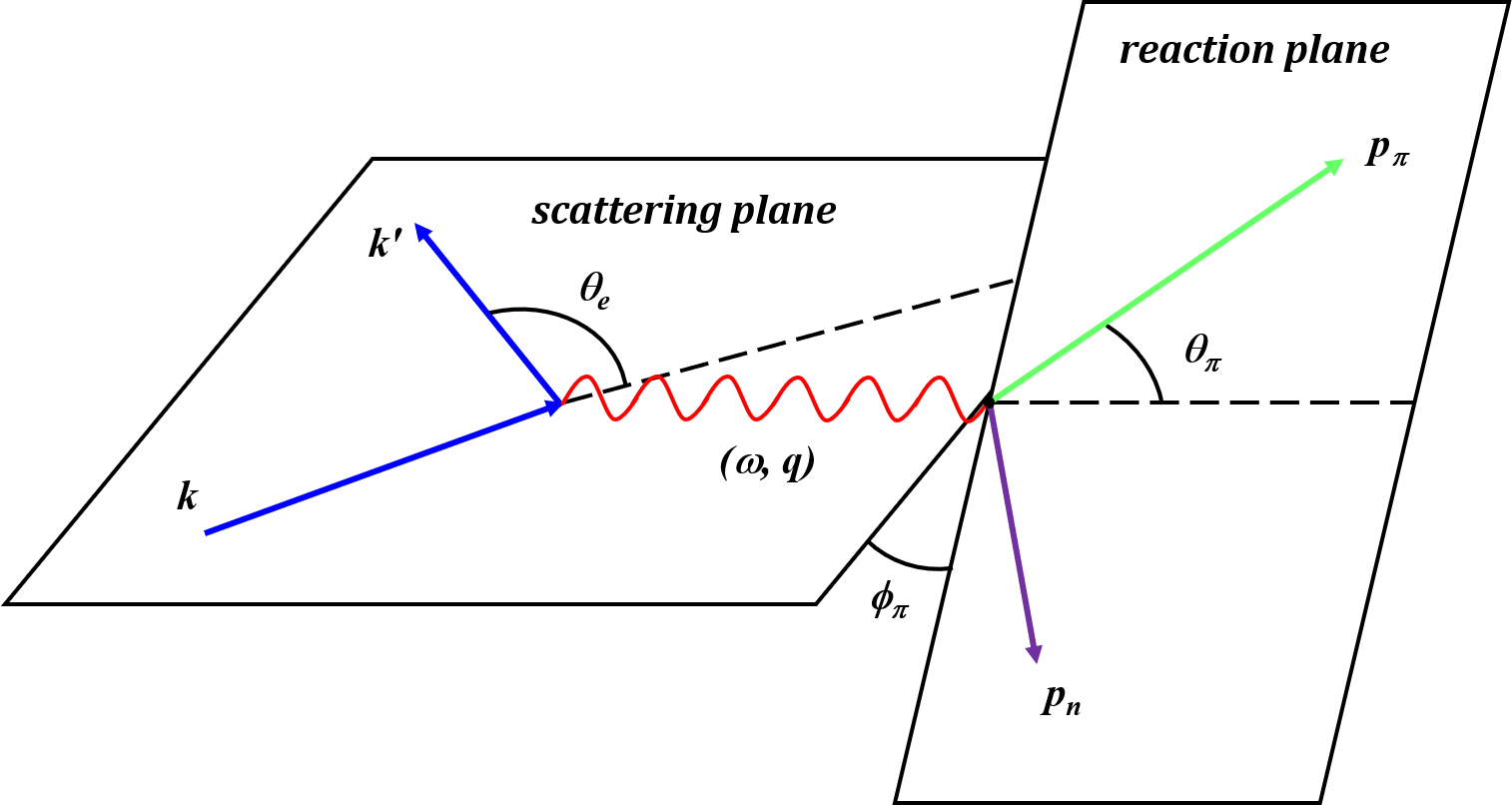}
\vspace{-0.2cm}
\caption{\label{fig:eepi}
Kinematic considerations for electroproduction, here for the pion.}
\end{center}
\end{figure}

For a coincidence experiment in which the scattered electron and the electroproduced charged pion are detected the differential cross-section can be expressed in terms of a known electrodynamic factor and a virtual photoproduction cross-section. The latter can be expressed in terms of the linear combinations of the products of virtual-photoproduction helicity amplitudes, which are the unpolarised transverse production, the purely scalar (longitudinal) production, and the interference terms between the transverse and transverse-scalar states.
This reduced cross-section can be written as a sum of four separate cross-sections or structure functions, which depend on $W$, $Q^2$, $t$,
\begin{equation}
\label{eq:sepsig}
  2\pi \frac{d^2 \sigma}{dt d\phi_\pi}  =  \frac{d \sigma_T}{dt} + \epsilon \frac{d \sigma_L}{dt}
                                    +  \sqrt{2 \epsilon (1 + \epsilon)}  \frac{d \sigma_{LT}}{dt} cos \phi_\pi
                                    +  \epsilon  \frac{d \sigma_{TT}}{dt} cos 2 \phi_\pi.
\end{equation}
Here, $\epsilon=\left(1+\frac{2 |\mathbf{q}|^2}{Q^2} \tan^2\frac{\theta_{e}}{2}\right)^{-1}$ is the polarisation of the virtual photon, where (see Fig.\,\ref{fig:eepi}):
$\mathbf{q}$ denotes the three-momentum of the transferred virtual photon;
$\theta_{e}$ is the electron scattering angle;
and $\phi_\pi$ is the angle between the scattering plane defined by the incoming and scattered electrons and the reaction plane defined by the transferred virtual photon and the scattered meson.  To separate the different structure functions one has to determine the cross-section for at least two sufficiently different values of $\epsilon$ as a function of the angle $\phi_\pi$ for fixed values of the invariant mass of the virtual photon-nucleon system, $W$, $Q^2$, and $t$.

The determination of the pion form factor from pion electroproduction requires that one-meson exchange (the pion pole) dominates the longitudinal cross-section at small values of $t$. The pion pole includes a factor $[-t$/$(t-m^2_\pi)^2]$, which is zero at $t$=0 and reaches a maximum at $t$=$-m^2_\pi$.
The first value is unphysical since forward scattering occurs at $t_{min}$=$-4m_p^2 \xi^2/(1-\xi^2)$, where $\xi=x/(2 - x)$ is the skewness, with $x$ being Bjorken-$x$ throughout this section, while the second can be reached in experiments for $\xi \sim m_{\pi}/2m_p$. The dominance of the pion pole in the longitudinal cross-section and its characteristic $t$ dependence allows for extractions of the electromagnetic pion form factor from these data.

Cross-section data suggest a dominant pion pole in the longitudinal $\pi^+$ cross-section at values of $-t<$0.3 (GeV/c)$^2$ \cite{Favart:2015umi}.  The strength of the pion pole falls off rapidly with increasing values of $t$.  However, the observation of a dominant pion pole alone is not sufficient to make a precise extraction of $F_{\pi}$ from the data.  To minimise background contributions, the longitudinal cross-section, $\sigma_L$, should be isolated via a Rosenbluth L/T(/LT/TT) separation.  Without an explicit L/T separation it is not clear what fraction of the cross-section arises from longitudinal photons and what the contribution of the pole to the cross-section is in these kinematics.  Data from survey experiments, like those from Refs.\,\cite{Airapetian:2007aa, Park:2012rn}, though interesting in their own right, are thus not used in precision form factor extractions.  Past, current, and planned pion and kaon form factor precision measurements are listed in Table~\ref{tab:mesonff-exp}.

% Please add the following required packages to your document preamble:
% \usepackage{graphicx}
% \usepackage[table,xcdraw]{xcolor}
% If you use beamer only pass "xcolor=table" option, i.e. \documentclass[xcolor=table]{beamer}
\begin{table}[t]
\resizebox{\textwidth}{!}{%
\begin{tabular}{|l|l|l|l|c|l|}
\hline
\textbf{Measurement} &
\textbf{Nominal $Q^2$ (GeV$^2$)} &
\textbf{Facility} &
\textbf{Experiment} &
  \multicolumn{1}{l|}{\textbf{First/most recent publication}} &
  \textbf{References} \\ \hline
Pion  FF & 0.0368-0.094                       & Fermilab        &    E456                       & 1981/1982      & \cite{Dally:1982zk,Dally:1981ur}\\
charge radius & & & & &
\\ \hline
Pion  FF & 0.014-0.26                       & CERN        &      NA7                     & 1984/1986      & \cite{Amendolia:1986wj,Amendolia:1984nz}\\
charge radius & & & & &
\\ \hline
Pion   xsec, FF, &
  0.18-1.19    &
  CEA or &
   & 1973
   &
  \cite{Brown:1973wr} \\
charge radius & 0.62-2.01, & Cornell & & 1974 & \cite{Bebek:1974iz} \\
 & 1.20-3.99, & &  & 1976 & \cite{Bebek:1974ww}\\
  & 1.18-9.77 & &  & 1977 & \cite{Bebek:1977pe}
  \\ \hline
Pion   L/T  xsec, & 0.35                       & DESY        &                           & 1977/2008      & \cite{Ackermann:1977rp,Huber:2008id} \\
Pion FF & & & & &
\\ \hline
Pion   L/T  xsec, & 0.70                       & DESY        &                           & 1979/2008      & \cite{Brauel:1979zk,Huber:2008id} \\
Pion FF & & & & &
\\ \hline
Pion   L/T  xsec, & 0.60, 0.75, 1.0, 1.60      & JLab 6 GeV  & E93-021                  & 2000/2008      & \cite{Volmer:2000ek,Tadevosyan:2007yd,Huber:2008id} \\
Pion FF & & & (Fpi1) & &
\\ \hline
Pion   L/T  xsec, & 1.60, 2.45                 & JLab 6 GeV  & E01-004                  & 2006/2008      & \cite{Horn:2006tm,Blok:2008jy,Huber:2008id} \\
Pion FF & & & (Fpi2) & &
\\ \hline
Pion   L/T  xsec, & 2.15, 3.91                & JLab 6 GeV  & E01-107                   & 2008           & \cite{Horn:2007ug}  \\
Pion FF & & & (pionCT) & &
\\ \hline
Pion   L/T  xsec, &
  0.375, 0.425, 0.3, 0.5,    &
  JLab 12 GeV &
  E12-19-006, &
  data taking/under analysis &
  \cite{E12-19-006} \\
Pion FF & 1.6, 2.115, 2.45, 3, & & (PionLT) & & \\
 & 3.85, 5, 6, 8.5 & & E12-09-011 & &
  \\ \hline
Kaon  FF & 0.037-0.119                       & Fermilab        &    E456                       & 1980      & \cite{Dally:1980dj}\\
charge radius & & & & &
\\ \hline
Kaon  FF & 0.015-0.10                       & CERN        &      NA7                     & 1986      & \cite{Amendolia:1986ui}\\
charge radius & & & & &
\\ \hline
Kaon FF                   & 1.0, 1.36, 1.9, 2.07, 2.35 & JLab 6 GeV  & E93-018, & 2018           & \cite{Carmignotto:2018uqj} \\
& &  & E98-008, &  & \\
& & & E01-004 & &
\\ \hline
Kaon   L/T  xsec, & 0.5, 2.115, 3.0, 4.4, 5.5  & JLab 12 GeV & E12-09-001                & under analysis & \cite{E12-09-011} \\
Kaon FF & & & (KaonLT) & &
\\ \hline
\end{tabular}%
}
\caption{Past and current pion and kaon form factor measurements using elastic scattering or meson electroproduction reactions.  (``xsec'' denotes cross-section.)
\label{tab:mesonff-exp}}
\end{table}

Data on $F_{\pi}$ obtained using pion electroproduction have been obtained for values of $Q^2$ up to 10 GeV$^2$ at the Cambridge electron accelerator (CEA) and Cornell University \cite{Brown:1973wr, Bebek:1974iz, Bebek:1974ww, Bebek:1977pe}.  However, those data suffer from relatively large statistical and systematic uncertainties.  More precise data were obtained at the Deutsches Elektronen-Synchrotron (DESY) \cite{Ackermann:1977rp, Brauel:1979zk}.  With the availability of high-intensity electron beams, combined with accurate magnetic spectrometers at JLab, it has been possible to determine longitudinal-transverse (L/T) separated cross-sections with high precision.  The measurement of these cross-sections in the regime of $Q^2$=0.60-1.60 GeV$^2$ (Experiment Fpi1) and $Q^2$=1.60-2.45 GeV$^2$ (Experiments Fpi2 and pionCT) are described in detail elsewhere  \cite{Horn:2006tm, Blok:2008jy, Huber:2008id, Horn:2007ug, Volmer:2000ek, Tadevosyan:2007yd}.
Referring to Eq.\,\eqref{eq:sepsig}, a minimum of two cross-section  measurements, $\sigma_{1,2}$, at fixed ($Q^2$, $W$) and different values of the virtual photon polarisation, $\epsilon_{1,2}$, $|\mathbf{q}_{\epsilon_1}|>|\mathbf{q}_{\epsilon_2}|$, are needed to determine the longitudinal cross-section, $\sigma_L$, which allows for the extraction of the pion form factor.  Using the approximation that $\sigma_L \sim F^2_{\pi}$, the experimental error in $F_{\pi}$ is
\begin{equation}
    \frac{\Delta F_{\pi}}{F_{\pi}} = \frac{1}{2} \frac{1}{\epsilon_1-\epsilon_2} \sqrt{(\frac{\Delta \sigma_1}{\sigma_1})^2 (r+\epsilon_1)^2 + (\frac{\Delta \sigma_2}{\sigma_2})^2 (r+\epsilon_2)^2}\,,
\end{equation}
where $\Delta \sigma_{1,2}$ are the uncorrelated errors on the two cross-sections $\sigma_{1,2}$, respectively.  The correlated error is a normalisation uncertainty and is added in quadrature to the uncorrelated uncertainty.  
The relevant quantities for the extraction of the L/T-separated cross-sections and the form factor are $\Delta \epsilon=\epsilon_1-\epsilon_2$ between the two kinematic settings and the fixed-($Q^2$, $W$) value of $r = \sigma_T / \sigma_L$.
To estimate the uncertainty in $F_{\pi}$, one has to take into account both the variation of counts across the spectrometer acceptance at both low and high values of $\epsilon$ and the variation in the theoretical model across the acceptance.  Magnetic spectrometers with well understood acceptance can provide the facilities for high precision measurements of $F_{\pi}$.

Pion electroproduction experiments are performed at the smallest possible value of $(-t)$, which is still a distance away from the pion pole. The extraction of $F_{\pi}$ from the data therefore requires that the $t$ dependence of $\sigma_L$ be compared to a theoretical model.  In this method, consistency between data and model is essential.  A detailed discussion of the systematic uncertainty in the extraction of $F_{\pi}$ can be found in Ref.\,\cite[Sec.\,2]{Horn:2016rip}.

The next simplest meson available for experimental studies is the kaon.  The kaon contains strangeness; hence, as highlighted above, studies of kaon structure are essential for understanding the origin and character of mass.  The kaon form factor has been determined directly up to $Q^2\approx 0.10\,$GeV$^2$ at Fermilab and at the CERN SPS from the scattering of high-energy charged kaons by atomic electrons \cite{Amendolia:1986ui, Dally:1980dj}. These data were used to constrain the mean square charge radius of the charged kaon, with the result $\langle r_K^2\rangle = (0.58\pm 0.04\,{\rm fm})^2$.

At higher energies, the kaon form factor can, in principle, be extracted from kaon electroproduction data; but there are experimental challenges that need to be addressed.  The extraction of $F_K$ from $\sigma_L$ relies on dominance of the kaon exchange term.  The kaon pole is farther from the physical region than the pion, which may raise doubts about the ability to extract $F_K$ from electroproduction data.  To lend confidence to the method, we note two aspects.  First, the pion form factor was extracted from pion electroproduction data at small $(-t)$ by carefully studying the model dependence of the analysis, not by direct extrapolation.  Second, comparative extractions of the pion form factor from low-$(-t)$ to large-$(-t)$ data suggest only a modest model dependence.  Furthermore, recent calculations suggest that the kaon pole is dominant for $-t <$ 0.9 (GeV/c)$^2$ \cite{Qin:2017lcd}.

A recent electroproduction determination of the kaon form factor at $Q^2$=1.00, 1.36, 1.90, 2.07, 2.35\,GeV$^2$ is discussed in Ref.\,\cite{Carmignotto:2018uqj}, with the results drawn in Figs.\,\ref{FigKaonFFs}\,A and \ref{FigFK0}\,A.  The L/T-separated kaon electroproduction cross-sections were extracted at different values of $(-t)$ using data from JLab \cite{Mohring:2002tr, Coman:2009jk, Horn:2006tm} and the successful method from Refs.\,\cite{Horn:2006tm, Blok:2008jy} was applied to infer the kaon form factor.
Note that the largest $(-t)$ pion data discussed above lie at similar distances from the pole as the kaon data discussed in Ref.\,\cite{Carmignotto:2018uqj}. These data range from $-t$=0.4\,-\,0.7 (GeV/c)$^2$, and so fall into the regime where the kaon pole is expected to contribute. The kaon form factor was extracted at $(-t_{min})$ for each $Q^2$ point. The model dependence was estimated using a comparative study at different values of $(-t)$ and was found to be on the order of about 0.1 on the form factor value. The data sets are internally consistent, lending confidence to the method used for extracting the kaon form factor from longitudinal cross-section data.

The high quality, continuous electron beam at the 12\,GeV JLab, combined with the high momentum spectrometer (HMS) and superconducting super high momentum spectrometer (SHMS), makes JLab\,12 the only facility currently able to pursue a programme of precision pion and kaon form factor measurements.  At the time of writing, the running E12-19-006 (PionLT) experiment \cite{E12-19-006} enables measurements of the pion form factor at low $(-t_{min})$ up to $Q^2$ = 6 GeV$^2$, allows for measurements of the separated $\pi^+$ cross-sections as a function of $Q^2$ at three fixed $x$ values (0.2, 0.3, 0.4), and enables the measurement of the pion form factor to the largest $Q^2$ accessible at JLab\,12, \emph{viz}.\ 8.5 GeV$^2$.  Three data points at $Q^2$=0.38 GeV$^2$ and center of mass energy $W$=2.20 GeV were acquired during the first phase of the experiment in 2019.  The data are under analysis.

The E12-09-011 (KaonLT) experiment \cite{E12-09-011} is an exclusive measurement of the L/T-separated kaon electroproduction cross-section important for understanding the role of strangeness in hadron imaging studies and the kaon form factor. The KaonLT experiment collected data in 2018/19 and the data are presently under analysis. Data were acquired for five $Q^2$ points, 0.5, 2.115, 3.0, 4.4, and 5.5 GeV$^2$, with center of mass energies $W=2.40, 2.95, 3.14, 2.74, 3.02\,$GeV. The data were taken at five beam energies (3.8, 4.9, 6.2, 8.2, 10.2 GeV), which will enable the first precision measurements of the L/T-separated kaon electroproduction cross-sections as a function of $Q^2$  above the resonance region.
%% These data could also provide information about the onset of factorisation in kaon electroproducation.
%% We have proved a factorization theorem for exclusive meson production in high Q electroproduction.
A direct comparison of the scaling properties of the $\pi^+$ and $K^+$ separated cross-sections would provide an important tool for the study of the onset of factorisation in the transition from the hadronic to the partonic regime \cite{Collins:1996fb} and provide a possibility to study effects related to SU(3)-flavour symmetry breaking, \emph{viz}.\ Higgs-induced modulation of EHM.  The L/T-separated cross-sections of pion data simultaneously collected at $Q^2$=0.5, 2.115, and 3.0 GeV$^2$ can provide further constraints on the pion form factor.

\subsection{Empirical Information on Parton Distribution Functions}
\label{sec:meson-distribution-functions-exp}
Experimental knowledge of the partonic structure of the pion is very limited owing to the lack of a stable pion target; and the situation is far worse for the kaon, with data limited to less than 10 points of data worldwide.
Most of the current knowledge about the pion structure function in the valence region was obtained primarily from pionic DY scattering (0.2 $\leq$ $x \leq$ 0.99), and in the pion sea region at low Bjorken-$x$, from hard diffractive processes measured in $e-p$ collisions at the \emph{Hadron-Elektron-Ringanlage} (HERA) H1 and ZEUS experiments (3 $\times$ 10$^{-4} \leq x \leq$ 0.01).  These processes are complementary methods to probe the partonic structure of pions (and kaons). However, at present there is no overlap between the data sets obtained with the two different techniques. Past and anticipated measurements are listed in Table~\ref{tab:meson-sf-exp}.

Pionic DY scattering data were collected by the NA3 \cite{Badier:1983mj}, NA10 \cite{Betev:1985pg}, and WA39 \cite{Corden:1980xf} collaborations at the CERN Super Proton Synchrotron (SPS) and by the E615 \cite{Conway:1989fs} collaboration at Fermilab.
In these experiments one measures a lepton pair produced from hadron-hadron inelastic collisions in the region $s \rightarrow \infty$, $Q^2/s$ finite, where $Q^2$ ($s$) is the invariant mass-squared of the lepton pair (the initial hadrons).
In the CERN SPS experiments, muon pairs were produced by charged meson beams of energies 200 GeV/c ($\pi^+$) and 150-280 GeV/c ($\pi^-$) incident on a heavy target (platinum, tungsten). DY scattering data using $\pi^-$ beams were acquired by NA10, while data with both $\pi^+$ and $\pi^-$ beams were acquired by NA3 and WA39. The muon events were analyzed in a magnetic spectrometer. Charged hadrons from the colliding beams were identified with differential (negative charge) or threshold (positive charge) Cherenkov counters. Events were selected by the muon pair mass and angle to distinguish from the resonance region, secondary interactions in the target, and misidentified $J/\Psi$ events produced by asymmetrical di-muons.
% Add Fermilab experiment overview

% Please add the following required packages to your document preamble:
% \usepackage{graphicx}
\begin{table}[t]
\resizebox{\textwidth}{!}{%
\begin{tabular}{|l|l|l|l|l|l|}
\hline
\textbf{Measurement} & ${\mathbf{x_\Pi}}$ %\textbf{x}$_{\mathbf{Bjorken}}$
& \textbf{Facility} & \textbf{Experiments} & \textbf{First/most recent publication} & \textbf{References}                    \\ \hline
Pion SF              & $0.21-0.99$ % 0.2-0.98
& Fermilab          & E615                 & \multicolumn{1}{l|}{1989/2005}         & \cite{Conway:1989fs, Wijesooriya:2005ir} \\ \hline
Pion SF & $0.25 - 0.75$ % 0.2-0.99
& CERN & WA39  & 1980 & \cite{Corden:1980xf}  \\ \hline
Pion SF & $0.17 - 0.97$ % 0.2-0.99
& CERN & NA3  & \multicolumn{1}{l|}{1983/2005} & \cite{Badier:1983mj, Wijesooriya:2005ir}  \\ \hline
Pion SF & $0.16 - 0.99$ % 0.2-0.99
& CERN & NA10 & \multicolumn{1}{l|}{1985/2005} & \cite{Betev:1985pg, Wijesooriya:2005ir}   \\ \hline
Pion SF & $7\times 10^{-4} - 0.5$ % $3\times 10^{-4}$ - 0.01
& DESY & ZEUS & 1996                           & \cite{Derrick:1996ax} \\ \hline
Pion SF & $1\times 10^{-3} - 0.22$ % $3\times 10^{-4}$ - 0.02
& DESY & H1   & 1999                           & \cite{Adloff:1998yg}  \\ \hline
Pion SF & $0.45 - 0.9$ %0.06-0.28
& JLab\,12 & C12-15-006   & conditionally approved                           &  \cite{JlabTDIS1} \\ \hline
Kaon SF & $0.45 - 0.9$ % 0.06-0.28
& JLab\,12 & C12-15-006A   & conditionally approved                           &  \cite{JlabTDIS2} \\ \hline
Pion SF & $0.16-0.99$ % $x_\pi\in[0.16,0.99]$
& CERN & COMPASS II & data collected 2015/2018 & \cite{Gautheron:2010wva} \\\hline
\end{tabular}%
}
\caption{%Completed or soon to be completed pion and kaon structure function measurements.  
%Completed or soon to be completed 
Pion and kaon structure function measurements that are either completed or whose completion is anticipated.
Notably, the NA3 publication did not report the measured cross-sections; consequently, this work is not included in global fits aimed at determining the pion valence-quark DF.
\label{tab:meson-sf-exp}}
\end{table}

The LO DY cross-section for a pion interacting with a nucleon can be written,
\begin{equation}
    \frac{d^2 \sigma}{dx_{\pi}dx_N} = \frac{4 \pi \alpha_{em}^2}{9 M^2_{\gamma}}  \sum_q e_q^2 [q_{\pi}(x_{\pi})\bar{q}_N(x_N) + \bar{q}_{\pi}(x_{\pi})q_N(x_N)]
\end{equation}
where $\alpha_{em}$ is the QED fine-structure constant, the sum is over quark flavour, $q_{\pi}$ ($q_{N}$) is the PDF for quark flavour $q$ in the pion (nucleon), $e_q$ is the charge of the quark (in units of the positron charge), $M_{\gamma}$ is the mass of the virtual photon, and $x_{\pi}$ ($x_N$) is the momentum fraction (Bjorken $x$) of the interacting quark in the pion (nucleon).  Using symmetry arguments, the cross-section can be expressed in terms of the pion and proton DFs.
The pion valence distribution was extracted from global analyses at LO and NLO using available DY data.  In the global analyses, using the better known proton DFs as input, the pion's sea content was derived from momentum conservation, while the gluon contribution was constrained by prompt photon measurements by the CERN Collaborations WA70 \cite{Bonesini:1987mq} and NA24 \cite{DeMarzo:1986vi}.  The authors of Ref.\,\cite{Sutton:1991ay} provide their own parametrisation of the sea, assuming that the sea contribution amounts to 10\%, 15\%, and 20\%, which leads to three different results for the gluon contribution.

Pion structure was measured at DESY using HERA in deep-inelastic scattering tagged via a leading neutron, $e p \to e^\prime n$.  Four kinematic variables are needed to describe this interaction. They are defined in terms of the four-momenta of the incoming and outgoing particles. Two variables are chosen from the Lorentz invariants used in inclusive measurements: $Q^2$; $x$, the Bjorken scaling variable; $y = Q^2/(sx)$, the inelasticity; and $W^2$, the squared mass of the produced hadronic system.  Here, $\sqrt{s}$ is the energy of the $ep$ center-of-momentum-system.  Two additional variables are required to describe the leading neutron.  They were chosen as the laboratory production angle of the neutron, $\theta_n$, and the energy fraction carried by the produced neutron, $x_L \sim E_p/E_n$, where $E_p$ and $E_n$ are the proton and neutron energy, respectively.  The transverse momentum of the neutron is given by $p_T \sim x_L E_p \theta_n$. The squared momentum transfer from the target proton is $t \sim \frac{p_T^2}{x_L} - t_0$, where $t_0=\frac{1-x_L}{x_L} (m_n^2-x_L m_p^2)$ is the minimum kinematically allowed value of $|t|$ for a given $x_L$.

%%%https://en.wikipedia.org/wiki/ZEUS_(particle_detector)
Events containing a leading baryon were measured with the H1 and ZEUS detectors, featuring a forward neutron calorimeter (FNC) at forward rapidity.  The ZEUS central detector featured a 1.43\,T solenoidal field with a typical vertex resolution of 0.4\,cm (0.1\,cm transverse).  The hadronic calorimeter was a uranium-scintillator device with energy resolution $\sigma(E)/E = 0.35/\sqrt{E}$ for hadrons.  The FNC was a lead-scintillator (compensating) calorimeter located 106\,m from the ZEUS central detector, had seven interaction lengths in the front section and three in the rear section, and achieved a resolution of 0.65/$\sqrt E$.  For 500 GeV neutrons, this results in a 3\% energy resolution.  The FNC acceptance for neutrons was about 20\% of 2$\pi$ up to about 0.8\,mr, limited by beam line elements.  This resulted in a mean transverse momentum acceptance of $p_T^2 < 0.05$ GeV$^2$.  A forward neutron ``tracker'' with 1.5\,cm wide scintillating fingers was installed to measure $t$ $(P_t)$.
The H1 detector featured a 1.16\,T solenoidal field and a tracking system with a resolution of  $\sigma(p_T)/p_T = 0.005p_T + 0.015$ for charged particles.  (Here and following, all mass-dimensioned quantities in GeV.)  The hadronic calorimeter was a liquid argon sampling calorimeter with steel absorber.  The total depth of the calorimeter ranged from 4.5 to 8 interaction lengths and achieved a resolution of 0.50/$\sqrt E + 0.02$ for charged pions. The FNC was located 106\,m from the H1 central detector, consisted of a 1.6 interaction lengths lead-scintillator sandwich preshower and an 8.9 interaction lengths sandwich type main calorimeter, and achieved a resolution of $\sigma(E)/E = 0.63/\sqrt{E} + 0.03$ and $\sigma(x,y) = 10\,{\rm cm}/\sqrt{E} + 0.6\,$cm for hadrons. The electromagnetic energy resolution of the preshower was $\sigma(E)/E = 0.20/\sqrt{E}$.

The four-fold differential cross-section for leading neutron production, denoted $LN(4)$, can be written as
\begin{equation}
\frac{d^4\sigma^{ep \rightarrow e^\prime Xn}}{dx dQ^2 dx_L dp_T} =
%K \,(1+\delta_{LN})\,
%{\mathpzc K}_{\,LN}\,
{\mathsf K}_{LN} \,
F_2^{LN(4)}(x,Q^2,x_L,p_T)\,,
\end{equation}
where ${\mathsf K}_{LN}$ is a combination of kinematic factors -- see, \emph{e.g}.\ Ref.\,\cite[Eq.\,(3)]{Adloff:1998yg}.  Integrating this equation up to the maximum experimentally accessible neutron angle, corresponding to a maximum transverse neutron momentum that varies with $x_L$, results in
\begin{equation}
\frac{d^3\sigma^{ep \rightarrow e^\prime Xn}}{dx dQ^2 dx_L } = {\mathsf K}_{LN}\, F_2^{LN(3)}(x,Q^2,x_L)\,,
\end{equation}
where
\begin{equation}
F_2^{LN(3)}(x,Q^2,x_L) = \int_0^{p_T^{max}} \! dp_T\, F_2^{LN(4)}(x,Q^2,x_L,p_T)
\end{equation}
is the neutron-tagged proton structure function integrated over the measured range in $\theta_n$.
The pion structure function $F_{2}^{\pi}$ can then be extracted from $F_{2}^{LN(3)}$ {\it using models}, such as the Regge model of baryon production \cite[Sec.\,6]{Adloff:1998yg}.  In the Regge model the contribution of a specific exchange $i$ is defined by the product of its flux $f_{i}(y,t)$ and its structure function $F_{2}^{i}$ evaluated at $(x_{i},Q^2)$.

%% - this next text appears to repeat that above ... therefore omitted.
%In order to extract the structure function from the measured tagged cross-sections a model framework is needed, e.g. a Regge framework.  If the data are well described by the predictions for pion exchange, the pion flux from Regge model fits to hadron-hadron data may be used to extract the pion structure function.

For the HERA data, the pion flux was determined from the description of the available data \cite{Erwin:1961ny, Pickup:1961zz, Robinson:1975dt, Engler:1974nz, Flauger:1976ju, Hanlon:1976ct, Hanlon:1979qr, Eisenberg:1977cp, Blobel:1978yj, Abramowicz:1979ca} on charge-exchange processes ($p \rightarrow n$) in hadron-hadron interactions obtained using the exchange of virtual particles.  In such processes, the pion dominates the $p \rightarrow n$ transition amplitude, with its relative contribution increasing as $|t|$ decreases.  In the one-pion approximation, the cross-section for charged hadron production can be written
\begin{equation}
\frac{d \sigma_{hp \rightarrow Xn}}{dx_L dt} = f_{\pi/p}(x_L,t) \sigma_{tot}^{h\pi}(s^\prime),
\end{equation}
where $s^\prime = s (1-x_L)$, the square of the $h p$ centre-of-mass energy, and $f_{\pi/p}(x_L,t)$ is the flux factor that describes the splitting of a proton into a $\pi n$ system.

Charge exchange processes at the CERN intersecting storage rings (ISR) and Fermilab were found to be well described using
\begin{equation}
f_{\pi/p}(x_L, t) = \frac{1}{4 \pi} \frac{2g^2_{\pi pp}}{4 \pi} \frac{-t}{(t-m^2_{\pi})^2} (1-x_L)^{1-2 \alpha^R_{\pi}(t)} (F(x_L,t))^2\,,
\end{equation}
where $\frac{g^2_{\pi pp}}{4 \pi}$=14.5, $\alpha^R_{\pi}(t)=\alpha^\prime t$, and $\alpha^\prime$=1 GeV$^{-2}$. The form factor $F(x_L,t)$ parametrises the distribution of the pion cloud in the proton and accounts for final-state rescattering of the neutron.  For the HERA data, a good description of hadronic charge-exchange experiments was found using the Bishari flux \cite{Bishari:1973sd}, which sets $F(x_L,t)=1$.  The resulting flux factor was interpreted as an effective pion flux, which takes into account processes that occur in hadronic charge-exchange processes, absorption, non-pion exchange, and off-mass-shell effects.

At HERA, the leading neutron data were dominated by pion exchange in the range 0.64 $< x_L <$ 0.82, and the one-pion exchange model was used to determine the structure function of the pion, $F_2^{\pi}$, whose $t$ dependence is absorbed into the flux factor that describes the hadronic charge exchange data \cite{Erwin:1961ny, Pickup:1961zz, Robinson:1975dt, Engler:1974nz, Flauger:1976ju, Hanlon:1976ct, Hanlon:1979qr, Eisenberg:1977cp, Blobel:1978yj, Abramowicz:1979ca}.  The structure function of the real pion is then given by
\begin{equation}
F_2^{\pi}(x_{\pi}, Q^2) = \Gamma (Q^2,x_L) F_2^{LN(3)}(x, Q^2, x_L)\,,
\end{equation}
where $\Gamma=1/[(1-\delta_{abs}(x_L,Q^2)) \int_{t_{\rm min}}^{t_{\rm max}}dt\, f_{\pi/p}(x_L,t)]$ is the inverse of the pion flux factor integrated over the measured $t$ region and corrected for $t$-averaged absorptive effects: theory estimates $\delta_{abs}(x_L,Q^2)< 10$\% for $Q^2>$10 GeV$^2$ \cite{Kopeliovich:1996iw}.

The HERA data show that, in the region of 0.64 $< x_L <$ 0.82, the extracted $F_2^{\pi}$ has the same shape as the proton structure function, $F_2^p$, scaled by 0.361, and approximately the same $x$ and $Q^2$ dependences.  The data do not match expectations based on DY studies \cite{Sutton:1991ay}.  This discrepancy may owe to the fact that the major portion of the HERA data are at $x_\pi<10^{-2}$, whereas the DY data are at $x_\pi>0.2$.  From this perspective, the HERA data provide constraints on the shape of the pion structure function in the low $x_{\pi}$ region.

%\subsubsection{Background Considerations}

The extraction of the pion structure function has to be corrected for non-pion pole contributions, $\Delta$ and $N^*$ resonances, absorptive effects, and uncertainties in the pion flux.  These corrections can be minimised by measuring at the lowest $(-t)$ or tagged nucleon momentum possible. At these lower momenta, the absorptive correction is reduced because the pion cloud is further from the bare nucleon.  In addition, higher meson mass exchanges are suppressed, the charged pion exchange process has reduced background from Pomeron and Reggeon processes \cite{Nikolaev:1998se}, and the charged pion cloud is expected to be roughly double the neutral pion cloud in the proton.  Having data from {\it both} protons and deuterons would provide essential cross checks for the models used in the extraction of the pion structure function.

The largest uncertainty in extracting the pion structure function, however, likely owes to uncertainty in the pion flux in the framework of the pion cloud model.  One of the main issues is whether to use the $\pi NN$ form factor or the Reggeised form factor.  The difference between these two methods can be as much as $20$\% \cite{DAlesio:1998uav}.  From the $NN$ data, the $\pi NN$ coupling constant is known to $5$\% \cite{Stoks:1998xv, Stoks:1999bz}.  The uncertainties on the corrections are somewhat difficult to constrain; but if one assumes them to be $\lesssim 50$\% and assumes a 20\% uncertainty on the pion flux factor, then the overall theoretical systematic uncertainty could reach $25$\%.  An alternative approach is direct measurement of the pion flux factor by comparing to pionic DY data.
%For example the pion structure function at $x=0.5$ has been measured from the pionic Drell-Yan data to an accuracy of $5\%$ (see, for example~\cite{dy1,dy2}). New data from COMPASS could allow to further leverage this possibility and likely reduce projected uncertainty even further.

%\subsection{GPDs: Experimental Meson Data}
\subsection{L/T-Separated Meson Cross-Sections: Toward Flavour Separation with GPDs}
\label{SecGPDExp}
In recent years, hard exclusive processes, like deeply virtual photon (DVCS) and meson (DVMP) leptoproduction, have attracted much interest.  It has been shown that in the generalised Bjorken regime of large $Q^2$ and large $W$ in the virtual-photon--proton centre-of-mass frame, but fixed Bjorken-$x$, these processes factorise into products of hard partonic subprocesses and soft hadronic matrix elements, parametrised through GPDs.  As in DIS, the asymptotically dominant contributions come from longitudinally polarised photons while those arising from transversely polarised photons are suppressed by $1/Q^2$ \cite{Collins:1996fb}.  It is not clear that there is such factorisation for transversely polarised photons \cite{Favart:2015umi}.
The relative contribution of longitudinal and transverse terms to the meson cross-section and their $t$ and $Q^2$ dependences are thus of interest in evaluating the potential of meson production for probing the nucleon's GPDs.  In general, only if experimental evidence for leading-twist behaviour is shown can one begin to be confident about a handbag formalism.  One of the most stringent experimental tests is the $Q^2$ dependence of the longitudinal meson cross-section.

The cross-sections for $\rho^0$, $\omega$, $\phi$, $J/\Psi$, $\Psi(2S)$, and $\Upsilon$ production have been measured over a wide range of energies, from threshold up to $W \sim  200\,$GeV for light vector mesons and, taking recent LHC data into account, up to about 1\,TeV for heavy vector mesons.  For light vector mesons, $Q^2$ ranges from $0 -100\,$GeV$^2$ in the high $W$ domain, where there are data from HERA -- H1 \cite{Aid:1996bs, Adloff:1998st, Aaron:2009xp}, ZEUS \cite{Breitweg:1997ed, Breitweg:1998nh, Chekanov:2007zr, Derrick:1996af, Chekanov:2005cqa}, HERMES \cite{Airapetian:2000ni, Airapetian:2009ad}; and Fermilab -- E665 \cite{Adams:1997bh}.  At $W\lesssim 5\,$GeV, experiments at Cornell University \cite{Cassel:1981sx} and with the CLAS detector cite{Lukashin:2001sh, Hadjidakis:2004zm, Morand:2005ex, Morrow:2008ek, Santoro:2008ai} have measured the $\rho^0$, $\omega$, $\phi$ electroproduction channels up to $Q^2=5 \,{\rm GeV}^2$.  For the heavy mesons, there are electroproduction data only for the $J/\Psi$ from HERA \cite{Chekanov:2004mw, Aktas:2005xu}.  In addition, there are heavy-meson photoproduction data from the H1 \cite{Alexa:2013xxa}, ZEUS \cite{Chekanov:2002xi}, LHCb \cite{Aaij:2013jxj, Aaij:2014rms, Aaij:2015kea, McNulty:2015cpd}, and ALICE \cite{TheALICE:2014dwa} experiments, and some fixed target experiments, \emph{e.g}.\ at JLab, with $W \lesssim 5\,$GeV.

For the $\rho$ and $\omega$ channels, two regimes are clearly apparent, in both lepto- and photoproduction.  Starting from threshold, after a rapid rise owing to the opening of the phase space, the cross-sections decrease from $W \sim 2\,$GeV down to $W \sim 7\,$GeV.  The cross-sections then rise slowly with energy.  For the other vector meson channels, above the threshold effect, there is only one behaviour of the cross-section: a steady rise with $W$ from threshold up to the highest energies measured.  One can clearly see that the slope of the $W$-dependence at large $W$ increases with $Q^2$ for the $\rho$ electroproduction channel and with the mass of the vector mesons for the photoproduction data.  This indicates that the mass of the heavy mesons acts, like $Q^2$, as a hard scale.

In order to determine the contribution of longitudinal and transverse terms to the vector meson cross-section, one can analyse the decay angular distribution giving access to the polarisation states.  Assuming that the helicity between the final-state vector meson and the initial virtual photon is conserved, one can deduce the polarisation of the virtual photon and, therefrom, the longitudinal and transverse cross-sections. This property is referred to as $s$-channel helicity conservation (SCHC).  It can be checked experimentally by looking at specific vector-meson decay matrix elements that are sensitive to $\gamma^\ast_L \rightarrow V_T$ , $\gamma_T \rightarrow V_L$ or $\gamma^\ast_T \rightarrow V_{-T}$ transitions.  In $\rho^0$ and $\phi$ production, SCHC has been found to hold well enough to enable the L/T separation of the cross-sections.

The H1 data for the $\rho^0$ cross-section at fixed $x \sim 0.002$ span more than an order of magnitude in $Q^2$.  The cross-section falls approximately as $1/Q^4$, \emph{i.e}.\ the $Q^2$ slope is milder than that predicted by the na\"{\i}ve leading-twist handbag formalism.  Given the behaviour of the L/T cross-section ratio and with the transverse contribution being more important as $Q^2 \rightarrow 0$, the experimental $Q^2$-dependence of the longitudinal cross-section is even flatter than $1/Q^4$.

Recent results on helicity amplitudes and spin density matrix elements are available from COMPASS \cite{Alexeev:2020rcg, Adolph:2012ht, Adolph:2016ehf, Adolph:2016ehf} and HERMES \cite{Airapetian:2017vit, Airapetian:2010dh, Airapetian:2009af, Airapetian:2014gfp}.  The $\rho^0$ helicity amplitude ratios with nucleon-helicity flip were found to be small and consistent with zero, while the non-flip amplitudes were found to be non-zero \cite{Airapetian:2017vit}.  New spin-dependent matrix elements were found to be in partial agreement with the Goloskokov-Kroll model. Existing DVMP data are consistent with the QCD factorisation prediction over a limited $Q^2$ range.

Confirming the possibility of accessing GPDs through pseudoscalar meson production requires a full separation of the cross-section.  One-meson exchange plays an important role in leptoproduction of pseudoscalar mesons.  Its dominance in the longitudinal cross-section is required for extraction of the electromagnetic meson form factor from electroproduction data and is also part of the GPD $\tilde E$.  Separated cross-sections for $\pi^+$, $K^+$, and $\pi^0$ production have been measured at JLab  \cite{Carmignotto:2018uqj, Mazouz:2017skh, Defurne:2016eiy, E12-06-114, E12-09-011}.
Unseparated data have been measured at several facilities, including HERMES, COMPASS, and JLab \cite{Airapetian:2007aa, Park:2012rn, Bedlinskiy:2012be, Alexeev:2019qvd}

The leading-twist, lowest-order calculation of the $\pi^+$ longitudinal cross-section underpredicts the data obtained with $x_B=0.31, 0.45$, $-t=0.15, 0.41\,$GeV$^2$, $Q^2/{\rm GeV}^2 = 1.45-2.45$, $2.45-3.9$ by an order of magnitude \cite{Horn:2007ug, Favart:2015umi}. This implies that the data are not in the region where the leading-twist result applies.  That current experimental data are not in the region where the leading-twist result applies can also be seen in the $Q^2$ and $t$ dependence of the separated longitudinal and transverse $\pi^+$ cross-sections.  The $1/Q^{6}$ scaling-law prediction is reasonably consistent with the longitudinal data.  However, the transverse cross-section does not follow the scaling expectation; and when compared with the longitudinal cross-section, it is too large for consistency with QCD factorisation.
Regarding the $(-t)$ dependence, the longitudinal contribution is greater than the transverse in the $\pi^+$ cross-section for values of $-t < 0.3\,$GeV$^2$.  This is consistent with a dominant meson pole in this region; but the transverse contribution is greater than the longitudinal for values of $-t >$ 0.3 GeV$^2$, providing further evidence that the leading-twist does not apply in the currently available experimental kinematics.

Similar trends were found in the analysis of the separated $K^+$ cross-sections \cite{Carmignotto:2018uqj}.  The first L/T-separated $\pi^0$ cross-sections were measured in Hall A and cover a $Q^2$ range between 1.5 and 2 GeV$^2$ with $x \sim 0.36$ \cite{Mazouz:2017skh, Defurne:2016eiy}.  The results suggested that in this limited kinematic region transversely polarised photons dominate the total cross-section and also hinted at possible dependences on $Q^2$, $-t$, and $x$.  A more recent experiment at JLab\,12 extended these measurements to higher $Q^2$ and $x$ \cite{Dlamini:2020ulg}.
% add information from new pi0 paper

The recent L/T-separated cross-section data suggest that transversely polarized photons play an important role in charged pion and kaon and in neutral pion electroproduction.  Further experimental evidence for strong transverse virtual-photon transitions comes from the unseparated JLab CLAS $\pi^+$ cross-section data \cite{Park:2012rn} and from the $\sin{\phi_s}$ harmonics measured with a transversely polarized target by HERMES \cite{Airapetian:2009ac}.  Moreover, the JLab CLAS \cite{Bedlinskiy:2012be} and Hall A measurements \cite{Collaboration:2010kna} of the unseparated $\pi^0$ cross-sections reveal a transverse-transverse interference cross-section that amounts to a substantial fraction of the unseparated cross-section. Evidence for significant L-T interference in $\pi^0$ production also comes from JLab CLAS beam spin asymmetry measurements \cite{DeMasi:2007id}.

This review of the data on hard exclusive meson lepto- and photoproduction reveals clear evidence for a common dynamical mechanism underlying such processes, for which a handbag-diagram approach is a viable explanation candidate.  However, most of the currently available data do not lie in a region for which the simple leading-twist result applies.  Future measurements of DVMP cross-sections should make it possible to confirm the estimates of transverse photon contributions, establish the potential for access to GPDs in meson production, and understand the remaining puzzles.

\section{Measurements on the Horizon}
\label{Sec:Prospects}
The broad international science programme aimed at understanding pion and kaon structure and the SM mechanisms behind the emergence of hadron masses requires a strong, constructive interplay between experiment, phenomenology and theory.  Experimental prospects must be matched and guided by new theoretical insights, assisted by rapid advances in computing and high-level QCD phenomenology.  The identification and conduct of those experiments which can best lead to novel theory insights and understanding, and the interpretation of the new experimental results, both require a coherent worldwide effort.

At its $5\,$GeV centre-of-mass energy, JLab\,12 will provide tantalising, precision data relating to the pion (kaon) form factor up to $Q^2\sim 10 (5)\,{\rm GeV}^2$ and measurements of the pion (kaon) structure function at $x>0.5$ through the Sullivan process.  It will thereby deliver the first new-generation information that is capable of addressing numerous key issues highlighted herein, \emph{e.g}.\ discovering scaling violations in pseudoscalar meson elastic form factors, revealing Higgs modulation of EHM, and charting the $x$-dependence of pion and kaon valence-quark DFs.

With recent CERN approval of their Phase-1 plans \cite{Adams:2676885}, one can expect the AMBER Collaboration to play a crucial role because they are uniquely capable of delivering pion (kaon) DY measurements in the centre-of-mass energy region $\sim 10-20\,$GeV.
%
%Some old pion and kaon DY data exist; but for the kaon, this is limited to less than 10 points of data worldwide.
%
This new programme is an essential part of the global effort to measure and understand the pion structure function.  Importantly, the effort will deliver DY cross-sections with both $\pi^-$ and $\pi^+$ beams, enabling the separation of valence and sea quarks in the pion.  It will also place a handle on a determination of the so-called ``pion flux'' factor needed for related EIC Sullivan process measurements; and it is \emph{sine qua non} for any map of the kaon structure function.  One may reasonably expect that AMBER data in themselves will already enable fundamental insights to be drawn into EHM.

An electron-ion collider is under discussion in China (EicC).  With a similar centre-of-mass energy range as AMBER, EicC could both develop a powerful synergy with that Collaboration's plans and neatly fill a gap between JLab and the EIC at Brookhaven National Laboratory.  On its own, and even better in concert with AMBER, EicC could provide access to pion and kaon structure functions on $x \gtrsim 0.01$ and expose impacts of EHM and Higgs-induced modulations on NG-mode valence-quark and gluon structure.  Furthermore, EicC could extend the Rosenbluth L/T-separated cross-section technique, and may therefore be able to extend the kinematic reach of pion and kaon elastic form factor measurements beyond that achievable at JLab\,12.

%"...cross section technique, that may allow for extending the kinematic reach of pion and possibly kaon elastic form factor measurements..."
%providing information on pion and kaon elastic form factors to values of $Q^2$ that are roughly $2-4$ times larger than can be reached at JLab.

%The EIC will play a key role in the experimental program to chart in-pion and in-kaon distributions of, $\sl inter alia$, mass, charge, magnetization and angular momentum.
It is anticipated that EIC will be unique in having the capacity to provide information on pion and kaon structure over a large and tunable centre-of-mass energy range within the domain $\sim 20-140\,$GeV.  With such coverage and reach, EIC will vastly extend the $(x,Q^2)$ range of pion and kaon structure charts and may very well write the final words on many of the issues discussed herein, \emph{e.g}.\
%is there a limit on the gluon density in the neighbourhood $x\simeq 0$,
what is the gluon content of NG modes,
how does it compare with that in the proton,
and, indeed, what makes NG modes \emph{different} in the SM. 

%% file: S11_Epilogue.tex
\section{Epilogue}
\label{Sec:Epilogue}
Existence of the pion was predicted eighty-five years ago \cite{Yukawa:1935xg}; yet, even now, very little is known about its structure.  The pion's mass is measured with precision -- to one-part in a million; but its radius is only constrained to within 1\% and even this is debatable, given the uncertainties now attendant upon radius measurements made using electron+hadron scattering experiments \cite{Karr:2019}.  Regarding the kaon, discovered over seventy years ago \cite{Rochester:1947mi} and the first known particle to possess strangeness \cite{Christy:1957lsa, Yamanaka:2019}, the mass is known to three-parts in one-hundred-thousand, but its radius is at least ten-times more uncertain than that of the pion and precise information on even its low-$Q^2$ electromagnetic form factor is practically nonexistent.  This is a very poor state of affairs, given that pions and kaons are Nature’s most fundamental Nambu-Goldstone modes, whose existence and properties are vital to the formation of everything from baryons, to nuclei, and on to neutron stars.  The coming decades promise to eliminate much of this ignorance as technology reaches the point where achievable terrestrial experiments can use energy and luminosity to overcome the absence of stable pion and kaon targets.

Advances in experimental capabilities place great pressure on phenomenology and theory to keep pace.  Perhaps most promising here are predictions for pion and kaon form factors.  The past decade has seen both continuum and lattice methods agree that pion and kaon distribution amplitudes (DAs) are broadened as a consequence of emergent hadronic mass (EHM).  Hence, attention has shifted away from looking for experimental results to produce agreement with predictions of hard-scattering formulae made using the asymptotic DA profile.  Now it has turned toward identifying the breakaway from simple power-law scaling in recognition both that QCD is found in scaling violations and hard-scattering formulae should be evaluated using broadened distributions if comparisons with experiment are to be realistic.  New results from pion and kaon form factor experiments are therefore eagerly awaited.  Notwithstanding these advances, the hard scattering formulae that provide a strong motivation for this programme are only sensitive to a single inverse moment of the meson DAs.  Experiments able to provide information on their pointwise behaviour would be valuable.  In this connection, hard diffractive dissociation of mesons in the field of a heavy nucleus still appears to offer best promise.

Regarding distribution functions (DFs), the patient is less healthy.  There are many continuum predictions for the pion’s valence-quark DF, ${\mathpzc u}^\pi(x)$.  All agree that the large-$x$ dependence of ${\mathpzc u}^\pi(x)$ is sensitive to the behaviour of the pion’s light-front wave function at large relative momenta; and those connected with the QCD-predicted momentum dependence produce ${\mathpzc u}^\pi(x) \sim (1-x)^\beta$, with $\beta>2$ at any scale for which a rigorous connection may be drawn between experimental cross-sections and DFs.  However, phenomenology is still grappling with many challenges, including the task of building a complete hard-scattering kernel -- hence, studies return conflicting results for $\beta$; and lattice-QCD (lQCD) is only now beginning to contribute information on the pointwise behaviour of DFs and results that can be used to determine the large-$x$ exponent.  Crucially, new empirical data is coming; and this provides impetus for improvements in all areas of theory.

Turning to the kaon’s valence-quark distribution functions, there are again many continuum calculations; the status of which matches the pion case.  On the other hand, lQCD is only beginning to tackle the kaon.  Moreover, since less than ten points of empirical data exist, there is no phenomenology.  Here, new-era data are crucial.

Glue and sea distributions in the pion have long been a subject of phenomenological analyses; but given the issues with such studies of its valence-quark DF, the inferred glue and sea distributions are of uncertain accuracy.  Contemporary continuum theory has begun to make predictions for these DFs, which promise to serve as benchmarks for new-generation experiments and analyses.  Lattice QCD results for the pion’s glue and sea distributions lie in the future.  The state of kaon studies is similar.  Here there is the additional opportunity for exploring the modulation of EHM by Higgs-boson couplings into QCD; and first results indicate that the kaon’s glue and sea distributions are noticeably softer than those in the pion.  Such predictions present an opportunity for future facilities.

Notwithstanding the puzzles that remain with predictions and measurements relating to such studies of the one-dimensional structure of Nature’s Nambu-Goldstone modes, the allure of three-dimensional imaging is great and the challenges are greater still.  For pions and kaons, little has yet been accomplished in this area.  So, there are no controversies and many opportunities; and first results indicate some fascinating possibilities, \emph{e.g}.\ that the mechanical pressure at the pion’s core is greater than that in the proton and, in fact, commensurate with that at the core of a neutron star.  This field is wide open for input from new era facilities, high-level phenomenology, and novel theory.

Understanding the origin of more than 98\% of visible mass in the Universe, charting its distribution throughout the objects within which it is lodged, and identifying those measurements which can be used to validate the answers, are key remaining challenges within SM.  As argued herein, although many observables carry signals of these things, pion and kaon observables provide some of the cleanest signatures.  Now, with a confluence of developments in experiment and theory promising heretofore unparalleled synergies, there is room for optimism in anticipation of uncovering the solutions.